\documentclass[11pt,a4paper]{article}
\usepackage{SGpreprint}

\usepackage{psfrag}
\usepackage{latexsym}
\usepackage{natbib}
\usepackage{pstricks,pst-node,pst-coil,pst-plot}
\usepackage{multicol}        
\usepackage{multido}
\usepackage{appendix}
\usepackage{wrapfig}
\usepackage{algorithm}
\usepackage{algorithmic}
\usepackage{multirow}
\usepackage{slashbox}

\newtheorem{theorem}{Theorem}
\newtheorem{prop}[theorem]{Proposition}
\newtheorem{lemma}[theorem]{Lemma}
\newtheorem{corollary}[theorem]{Corollary}
\newtheorem{example}[theorem]{Example}
\newtheorem{defn}[theorem]{Definition}

\newenvironment{declaration}[1]{\trivlist \item[\hskip \labelsep{\bf #1 }]
\ignorespaces}{\endtrivlist}
\newenvironment{definition}{\begin{declaration}{Definition}}{\end{declaration}}
\newenvironment{proofof}[1]{\begin{declaration}{#1}}{\end{declaration}}
\newenvironment{proof}{\begin{proofof}{Proof.}}{\end{proofof}}

\usepackage{hyperref}

\nologohead{
Michael  D. K{\"o}nig, Stefano Battiston, Frank Schweitzer: 
Modeling Evolving Innovation Networks \\
In: \emph{Innovation Networks - New Approaches in Modeling and
  Analyzing} (Eds. A. Pyka, A. Scharnhorst) \\ Heidelberg: Springer,
  2008}

\begin{document}

\vspace*{0.2cm}
\begin{center}
  \textbf{\Large Modeling Evolving Innovation Networks}\\[5mm]
  {\large Michael D. K\"onig$^{\star}$, Stefano Battiston and Frank
    Schweitzer} 

  \emph{Chair of Systems Design, ETH Zurich, Kreuzplatz 5, 8032 Zurich,
    Switzerland}
    
  $^{\star}$ corresponding author, email: \url{mkoenig@ethz.ch}
   
\end{center}

\begin{abstract}
  We develop a new framework for modeling innovation networks which
  evolve over time. The nodes in the network represent firms, whereas the
  directed links represent unilateral interactions between the firms.
  Both nodes and links evolve according to their own dynamics and on
  different time scales.  The model assumes that firms produce knowledge
  based on the knowledge exchange with other firms, which involves both
  costs and benefits for the participating firms. In order to increase
  their knowledge production, firms follow different strategies to create
  and/or to delete links with other firms. Dependent on the information
  firms take into account for their decision, we find the emergence of
  different network structures. We analyze the conditions for the
  existence of these structures within a mathematical approach and
  underpin our findings by extensive computer simulations which show the
  evolution of the networks and their equilibrium state. In the
  discussion of the results, particular attention is given to the
  emergence of direct and indirect reciprocity in knowledge exchange,
  which refers to the emergence of cycles in the network structure.

  In order to motivate our modeling framework, in the first part of the
  chapter we give a broad overview of existing literature from economics
  and physics. This shows that our framework bridges and extends two
  different lines of research, namely the study of equilibrium networks
  with simple topologies and the dynamic approach of hypercycle models.\\
  
\end{abstract}

\section{Introduction}
\label{sec:introduction}

\subsection{The Importance of Innovation Networks}
\label{sec:importance}

Economists widely agree on technological change and innovation being the
main components of economic growth \citep{aghion98:_endog_growt_theor,
  tirole88:_theor_indus_orgniz}.  In the absence of ongoing technological
improvements economic growth can hardly be maintained
\citep{barro04:_econom_growt}. The close link between innovation and
economic performance has become generally accepted. Following this
insight, in recent years of economic growth, OECD countries have fostered
investments in science, technology and innovation
\citep{scientific06:_oecd_scien_techn_indus_outlook}.

Moreover, technologies are becoming increasingly complex. This increasing
complexity of technologies can make an agent's ``in-house'' innovative
effort insufficient to compete in an R\&D intensive economy.  Thus,
agents have to become more specialized on specific domains of a
technology and they tend to rely on knowledge transfers from other
agents, which are specialized in different domains, in order to combine
complementary domains of knowledge for production (``recombinant growth''
\citep{weitzman98:_recom_growt}).

When one agent benefits from knowledge created elsewhere we speak of
knowledge spillovers. Knowledge spillovers define ``any original,
valuable knowledge generated somewhere that becomes accessible to
external agents ... other than the originator.''\footnote{``Involuntary
  spillovers are a feature of market competition.  Competition not only
  creates incentives to produce new knowledge but it also forces the
  other agents to increase their own performance through imitation,
  adoption and absorption of the new knowledge created elsewhere.''
  \citep[p.  91]{foray04:_econom_knowl}
} \citep[p.  95]{foray04:_econom_knowl} 

The knowledge-based economy is developing towards a state in which the
costs for acquiring, reproducing and transmitting knowledge are
constantly decreasing, spatial and geographical limitations on knowledge
exchange are becoming less important and attitudes change towards more
open behavior of sharing knowledge instead of hiding it from others. In
this state, knowledge externalities will play an increasingly important
role.

When agents are using knowledge that is created elsewhere, they must have
access to other agents across a network whose links represent the
exchange or transfer of knowledge between agents.  The importance of
networks in innovative economies has been widely recognized, e.g.  it has
been observed that ``the development of knowledge within industries is
strongly influenced by the network structure of relations among agents''
\citep[p.  1]{antonelli96:_local}. Subsequently, an ample body of
empirical research has documented the steady growth of R\&D partnerships
among firms \citep{hagedoorn06:_and_patter_in_inter_firm_r}.

\subsubsection{Markets for Knowledge Exchange }
\label{sec:knowledge_markets}

The exchange of knowledge is not unproblematic. Markets for knowledge
exchange can exhibit serious market failures
\citep{geroski95:_appropriability, arora04:_market_techn} which make it
difficult for innovators to realize a reasonable return from trading the
results of their R\&D activities (the problem of appropriability
\citep{geroski95:_appropriability}). This is due to the public good
character of knowledge, which makes it different from products or
services.  Knowledge is non-rival, meaning its use by one agent does not
diminish its usability by another agent, and sometimes (when knowledge
spillovers cannot be avoided) non-excludable, meaning that the creator of
new knowledge cannot prevent non-payers from using it. The problems
associated with trading of knowledge can prevent agents from exchanging
knowledge at all.

There are three generic reasons for failures of markets for technology
\citep{arrow62:_rate_direc_inven_activ, geroski95:_appropriability,
  arora04:_market_techn}: (1) economies of scale/scope, (2) uncertainty
and (3) externalities.

(1) R\&D projects often require huge initial investments and they can
exhibit economies of scale since the cost for useful technological
information per unit of output declines as the level of output increases
\citep{wilson75:_infor_econom_scale}. Besides,
\citet{nelson59:_simpl_econom_basic_resear} has shown that economies of
scope can apply to innovative agents. The broader an agents'
``technological base'', the more likely it is that any outcome of its
R\&D activities will be useful for her. The result is that markets for
knowledge exchange are often dominated by monopolies.

(2) Almost all economic investments bear a risk of how the market will
respond to the new product (commercial success). Innovators face
additional risks. First, their investment into R\&D does not necessarily
lead to a new technology. Second, if such a new technology is discovered,
it has to be put into practice in a new and better product than the
already existing ones. This inherent uncertainty of R\&D projects often
causes agents to invest ``too little''.

(3) Externalities are important when the action of one agent influences
the profits of another agent without compensation through the
market. Public goods are a typical example of creating externalities.
Knowledge is a public good  and the returns innovators can realize often
are far below their investments into R\&D. This can seriously diminish an
agent's incentive to do R\&D.

In order to overcome the above mentioned problems associated with the
returns on investment into R\&D appropriate incentive mechanisms have to
be created that encourage agents to invest into R\&D.  In general,
\citet{von03:_open_sourc_softw_privat_collec_innov_model} suggest three
basic models of encouraging agents to invest into R\&D:

(1), the private investment model assumes that innovation is undertaken
by private agents investing their resources to create an innovation.
Society then provides agents with limited rights to exclusively use the
results of their innovation through patents or other intellectual
property rights (by creating a temporarly
monopoly)\footnote{For
  a more detailed treatment of this issue we recommend
  \citet{scotchmer04:_innov_incen}.}.

(2), the so called collective action model \citep{allen83:_collec}
assumes that agents are creating knowledge as a public good.  Knowledge
is made public and unconditionally supplied to a public pool accessible
to everybody.  The problem is that potential beneficiaries could wait
until others provide the public good and thereby could free-ride. One
solution to this problem is to provide contributors (in this case
innovators) with some form of subsidy. Scientific research is such an
example where reputation based rewards are granted to scientists for
their good performance. 

(3), in the private-collective innovation model participants use their
private resources in creating new knowledge and then make it publicly
available.  This is typically observed in open source projects.  There
are several incentives \citep{lerner02:_some_simpl_econom_open_sourc,
  von03:_open_sourc_softw_privat_collec_innov_model} for agents to
participants in open-source projects. These range from elevated
reputations, the desire of building a community to the expectation of
reciprocity from the community members for their efforts.

The collective action approach (2) gives a possible explanation for the
willingness of agents to share knowledge if there are no costs associated
with it. One can think of a pool of technologies that is accessible to
everybody (``broadcasting'' of technologies) \citep{allen83:_collec}.
This can be the case where agents are non-rivals and shared information
may have no competitive cost.  Additionally, knowledge must be easily
understandable and transferable.  This assumes that knowledge is highly
codified\footnote{The opposite case of codified knowledge is tacit
  knowledge. ``Tacit knowledge is difficult to make explicit for transfer
  and reproduction.  The exchange, diffusion, and learning of tacit
  knowledge require those who have it to take deliberate action to share
  it. This is difficult and costly to implement ... Knowledge can,
  however, be codified. It can be expressed in a particular language and
  recorded on a particular medium.  As such, it is detached from the
  individual. When knowledge is codified, it becomes easily
  transferable.''  \citep[p.  73]{foray04:_econom_knowl}} such that the
transfer of knowledge from one agent to the other is costless.  But, if
these assumptions do not hold, the costs for transferring knowledge can
often be considerable, and agents become more selective about whom to
share their knowledge with. We study this situation in the next sections.

\subsection{Economies as Evolving Networks}
\label{sec:econ-as-evolv}

As we already outlined above, modern economies are becoming increasingly
networked, and this also affects the innovation process where information
and knowledge are exchanged by interactions between agents
\citep{kirman97:_economy_evolving_network,
  gallegati99:_beyon_repres_agent}.  In the agent based view, the
aggregate behavior of the economy (macroeconomics) cannot be investigated
in terms of the behavior of isolated individuals.  Not only there are
different ways in which firms interact, learning over time, based on
their previous experience; also interactions between them take place
within a network and not in a all-to-all fashion.

The standard neoclassical model\footnote{A standard neoclassical model
  includes the following assumptions
  \citep{gabszewicz00:_strategic_interaction}: (1) perfect competition,
  (2) perfect information, (3) rational behavior, (4) all prices are
  flexible (all markets are in equilibrium).  The resulting market
  equilibrium (allocation of goods) is then efficient. See
  \citet{hausman03:_inexac_separ_scien_econom} for a discussion of these
  assumptions.} of the economy assumes that anonymous and autonomous
individuals take decisions independently and interact only through the
price system which they cannot influence at all. However, competition
easily becomes imperfect because, if agents have only a minimal market
power, they will anticipate the consequences of their actions and
anticipate the actions of others. 

Game theorists have tried to integrate the idea of strategically
interacting agents into a neoclassical\footnote{The individual decision
  making process is represented as maximizing a utility function. A
  utility function is a way of assigning a number to every possible
  choice such that more-preferred choices have a higher number than
  less-preferred ones \citep{varian96:_inter_microec}. The gradients of
  the utility function are imagined to be like forces driving people to
  trade, and from which economic equilibria emerge as a kind of force
  balance \citep{farmer05:_economics_next_physical_science}.
}
framework. But still they leave two questions unanswered. First, it is
assumed that the behavior is fully optimizing. This leads to agents with
extremely sophisticated information processing capabilities. Such ability
of passing these enormous amounts of information in short times cannot be
found in any realistic setting of human interaction. Advances in
weakening that assumption are referred to as ``boundedly rationality''
\citep{gigerenzer02:_bound_ration}.  Second, the problem of coordination
of activities is not addressed in the standard equilibrium model of the
economy. Instead it is assumed that every agent can interact and trade
with every other agent, which becomes quite unrealistic for large
systems.

One has to specify the framework within the individual agents take price
decisions and thus limit the environment within which they operate and
reason.  An obvious way is to view the economy as a network in which
agents interact only with their neighbors.  In the case of technological
innovation, neighbors might be similar firms within the same industry,
but these firms will then be linked either through customers or suppliers
with firms in other industries. Through these connections innovations
will diffuse through the network. The rate and extent of this diffusion
then depends on the structure and connectivity of the network. The
evolution of the network itself should be made endogenous where the
evolution of the link structure is dependent on the agents' experience
from using the links available to them. In this framework the individuals
learn and adapt their behavior and this in turns leads to an evolution of
the network structure. The economy then becomes a \textit{complex
  evolving network}.

\subsubsection{Complex Networks}
\label{sec:complex-networks}

Although no precise mathematical definition exists for a \textit{complex
  network}, it is worth to elaborate the notion associated with it. In
general, a network is a set of items some of which are linked together by
pairwise relationships. The structure of the relationships can be
represented mathematically as a graph in which nodes are connected by
links (possibly with varying strength). However a network is usually also
associated with some dynamic process on the nodes which in turn affects
the structure of the relationships to other nodes. A wide variety of
systems can be described as a network, ranging from cells (a set of
chemicals connected by chemical reactions), to the Internet (a set of
routers linked by physical information channels). It is clear that the
structure of the relationships co-evolves with the function of the items
involved.

As a first step, a network can be described simply in terms of its
associated graph\footnote{In general a graph represents pairwise
  relations between objects from a certain collection. A graph then
  consists of a collection of nodes and a collection of links that
  connect pairs of nodes.}. There are two extreme cases of relatively
simple graphs: regular lattices on one side and random graphs\footnote{
  The classical Erd\"{o}s-Reny random graph is defined by the
  following rules  \citep{bollobas85:_random_graphs}:\\
  (1) The total number of nodes is fixed.\\
  (2) Randomly chosen pairs of nodes are connected by links with
  probability $p$. \\
  The construction procedure of such a graph may be thought of as the
  subsequent addition of new links between nodes chosen at random, while
  the total number of nodes is fixed.} on the other side.  During the
last century, graph theory and statistical physics have developed a body
of theories and tools to describe the behavior of systems represented by
lattices and random graphs. However, it turns out that, at least for
physical scales larger than biomolecules, most systems are not structured
as lattices or as random graphs. Moreover, such a structure is not the
result of a design, but it emerges from self-organization. In some cases
self-organization results from the attempt to optimize a global function.
In other cases, as it is typical in economics, it results from nodes
locally trying to optimize their goals, e.g. an individual utility
function.

Large networks are collectively designated as complex networks if their
structure (1) is coupled to the functionality, (2) emerges from
self-organization, and (3) deviates from trivial graphs, This definition
includes many large systems of enormous technological, intellectual,
social and economical impact \citep{frenken06:_techn}.

\subsection{The Statistical Physics Approach}
\label{sec:stat-phys-appr}

As we will discuss later, many of the theoretical tools developed in
economics and specifically in game theory to characterize the stability
of small networks of firms cannot be used for large networks. Asking
which is the optimal set of connections that a firm should establish with
other firms has little meaning in a large network if strategic
interaction is taken into account (with more than say, 100 nodes it is
simply not feasible to compute). On the other hand, it makes sense to ask
what are connectivity properties of the nodes a firm should try to target
in order to improve its utility with a certain probability. It is then
necessary to turn towards a statistical description of these systems,
where one is no longer interested in individual quantities but only in
averaged quantities.

There exists an arsenal of such tools developed within
\textit{statistical physics} in the last century, that allow to predict
the macroscopic behavior of a system from the local properties of its
constituents
\citep{durlauf99:_statistical_mechanics_contribute_social_science,
  stanley:_econop}. Such tools work very well for systems of identical
particles embedded in regular or random network structures in which
interactions depend on physical distance. Both a regular and a random
structure have a lot of symmetries, which one can exploit to simplify the
description of the system. However, in complex networks many of those
symmetries are broken: individuals and interactions are heterogeneous.
Moreover the physical distance is often irrelevant (think for instance of
knowledge exchange via the Internet). Therefore, a satisfactory
description of such systems represents a major challenge for statistical
physics \citep{amaral01:_application_statistical_physics}. 

In the last few years we have thus witnessed an increasing interest and
effort within the field of statistical physics in studying complex
networks that traditionally were object of investigation by other
disciplines, ranging from biology to computer science, linguistics,
politics, anthropology and many others. One of the major contributions of
statistical physics to the field of complex networks has been to
demonstrate that several dynamic processes taking place on networks that
deviate from random graphs, exhibit a behavior dramatically different
from the ones observed on random graphs.  

An example for all is the case of virus spreading: it has been shown that
while for random networks a local infection spreads to the whole network
only if the spreading rate is larger than a critical value, for
scale-free networks\footnote{A scale-free network is characterized by a
  degree distribution which follows a power law, $f(d) = \alpha
  d^{\gamma}$. The degree distribution gives the number of agents with a
  certain number, $d$, of in- or outgoing links (in- or outdegree), see
  the next section for a definition of degree of a node.  } any spreading
rate leads to the infection of the whole network. Now, technological as
well as social networks are much better described as scale free graphs
than as random graphs.  Therefore all vaccination strategies for both
computer and human viruses, which have been so far designed based on the
assumption that such networks were random graphs, need to be revised.
This highly unexpected result goes against volumes previously written on
this topic and is due to the presence of a few nodes with very large
connectivity. In this case, the rare events (infection of highly
connected nodes) and not the most frequent ones matter.

Explaining the macroscopic behavior of a system in terms of the
properties of the constituents has been a major success of the
physicist's reductionist approach. But, while in physical systems the
forces acting on single constituents can be measured precisely, this is
not the case in a socio-economic system where, moreover, each agent is
endowed with high internal complexity. Today, the physicist's approach to
socio-economic systems differs from the nineteenth century positivist
approach in so far as it does not aim at predicting, for instance the
behavior of individual agents. Instead, taking into account the major
driving forces in the interactions among agents at the local level we try
to infer, at a system level, some general trends or behavior that can be
confirmed looking at the data. This is also very different from taking
aggregate quantities and infer a macroscopic behavior from a
``representative'' agent\footnote{The concept of the representative agent
  assumes an economy which consists of a sufficiently homogeneous
  population of agents.  Because all the agents are equivalent, the
  aggregate quantities of the system can be calculated by multiplying the
  average agent, or the representative agent, by the number of agents
  (the system size). For example the total production of an economy is
  obtained by summing up the production levels of the individual firms
  that constitute the economy. To determine the behavior of the system it
  is therefore sufficient to know the characteristics of the
  representative agent.}  as it is done in several approaches in
mainstream economics.

\subsubsection{Dynamics versus Evolution in a Network}
\label{sec:dynam-vers-evol}

After discussing the notion of a complex network which has been strongly
influenced by physics, we now try to classify different complex networks.

The nodes in an economic network are associated with a state variable,
representing the agents' wealth, a firm's output or, in the case of
innovation networks, knowledge. There is an important difference between
the evolution of the network and the dynamics taking place on the state
variables. In the first, nodes or links are added to/removed from the
network by a specific mechanism and in the latter, the state variables
are changed as a result of the interactions among connected nodes (see
also \citet{gross07:_adapt_coevol_networ_review} for a review).
Consequently, there are four aspects that can be investigated in complex
economic networks \citep{battiston03:_dynamics_evolution_networks}.
\begin{enumerate}
\item statistic characterization of the static network topology without
  dynamics of state variables,
\item Network evolution without dynamics of the state variables,
\item Dynamics of state variables in a static network and
\item Dynamics of state variables and evolution of the network at the
  same time.
\end{enumerate}
This can be incorporated in the following table.
\begin{table}[htbp]
\begin{center}
\begin{tabular}[c]{c|c|c} 
	$\quad$ case  $\quad$ 	& $\quad$ state variables $\quad$	
& $\quad$	network $\quad$	\\
	\hline
	1. 	& static 		& 	static 		\\
	2. 	& dynamic 		& 	static 		\\
	3. 	& static 		& 	dynamic 	\\
	4. 	& dynamic 		& 	dynamic 	\\
\end{tabular}
\caption{Overview of the different ways in which a network and the state
  variables of the nodes can be related.\label{tab:network}}
\end{center}
\end{table}

In socio-economic systems as well as in biological systems, dynamics and
evolution are often coupled, but do not necessarily have the same time
scale. In section (\ref{sec:evolution-links}) we will show how the
coupling of fast knowledge growth (dynamics) and slow network evolution can
lead to the emergence of self-sustaining cycles in a network of knowledge
sharing (cooperating) agents.

\subsection{Outline of this Chapter}

In this chapter we focus on (i) the emergence and (ii) the performance of
different structures in an evolving network. The different scenarios we
develop shall be applied to firms exchanging knowledge in a competitive,
R\&D intensive economy. In the existing literature reviewed in the
following section, there are two different lines of research addressing
these problems: (i) Models of network formation were developed based on
individual utility functions, e.g. by
\citet{jackson03:_survey_models_network_formation_stability_efficiency},
in which simple architectures emerge in the equilibrium. (ii) In another
group of models, e.g. \citet{Padgett03:_economic_production_chemistry},
firms have specific skills and take actions based on goals or learning
and innovation is associated with the emergence of self-sustaining cycles
of knowledge production.  Although both lines of work address the problem
of network emergence and performance, they differ significantly in terms
of methods and results.  We try to bridge them by introducing a novel
model of evolving innovation networks that combines the topological
evolution of the network with dynamics associate with the network
nodes\footnote{The approach of combining a dynamics of the network with a
  dynamics in the nodes is discussed in
  \citet{gross07:_adapt_coevol_networ_review}.}.

We start our approach by giving a short introduction to graph theory in
section (\ref{sec:graph-theory}).  Here we restrict ourselves only to the
most important terms and definitions that are necessary in the following
sections\footnote{The reader interested in more details in graph theory
  can consult \citet{west01:_introd_graph_theor}}.

We then proceed by giving an overview of the existing literature on
economic networks.  In the first part of our literature review we explore
some basic models of innovation networks. The selection of these models
is by no means unique nor exhaustive, but points to important
contributions to the growing literature on economic and innovation
networks\footnote{For an excellent introduction see
  \citet{jackson07:_social_econom_networ},
  \citet{vega-redondo07:_compl_social_networ} and
  \citet{goyal07:_connec}.}. Similar to our own approach, these models
make considerably simple assumptions and thus allow for analytical
insights. This holds in particular for the connections model in section
(\ref{sec:connections-model}). The model in section
(\ref{sec:small-world}) can be considered as an extension of the basic
connections model where ``small world'' networks emerge.
In the subsequent section, (\ref{sec:innov-sett-evol}), we discuss a
model that takes heterogeneous knowledge into account, as a further
extension.  In the second part of the literature review we briefly sketch
models in which we observe cyclic network topologies. We show that in
certain cases the stability of a network and its performance depends
critically its cyclic structure. The critical role of cycles in a
networked economy has already been identified by
\citet{rosenblatt57:_linear_model_graph_minkow_leont_matric} and many
succeeding authors, e.g.
\citet{maxfield94:_gener_equil_theor_direc_graph,
  goyal00:_Noncooperative_model_network_formation, kima07:_networ}. In
this chapter we review some recent models in which cyclic networks emerge:
in section (\ref{sec:prod-recip-artif}) a model of production networks
with closed loops is presented and in section (\ref{sec:an-autoc-model})
we introduce a model of cycles of differently skilled agents.

Finally, in section (\ref{sec:evolving_networks}) we develop a novel
framework, which we call \textit{Evolving Innovation Networks}, to study
the evolution of innovation networks. We show how different modalities of
interactions between firms and cost functions related to these
interactions can give rise to completely different equilibrium networks.
We have studied the case of linear cost and bilateral interactions in
\citet{koenig07:_algeb_graph_theor_applied_dynam_innov_networ,
  koenig07:_effic_stabil_dynam_innov_networ}. There we find that,
depending on the cost, the range of possible equilibrium networks
contains complete, intermediate graphs with heterogeneous degree
distributions as well as empty graphs. Here, we focus on a type of
nonlinear cost and both, on unilateral and bilateral interactions. In the
unilateral case, we find that, in a broad range of parameter values,
networks can break down completely or the equilibrium network is very
sparse and consists of few pairwise interactions and many isolated
agents.  Equilibrium networks with a higher density can be reached if (i)
the utility function of the agents accounts for a positive externality
resulting from being part of a technological feedback loop or if (ii) all
interactions are bidirectional (direct reciprocal).  Otherwise the
network collapses and only few, if any, agents can beneficially exchange
knowledge.

The results found in our novel approach to evolving innovation networks
are summarized in section (\ref{sec:summary}). The appendix shall be
useful for the reader interested in more numerical results and the
parameters and explanation of the algorithms used.

\section{Basic Models of Innovation Networks} 
\label{sec:basic_models}

\subsection{Graph Theoretic Network Characterization}
\label{sec:graph-theory}

Before we start to describe specific models of economic networks, we give
a brief introduction to the most important graph theoretic terms used
throughout this chapter to characterize networks. For a broader
introduction to graph theory see \citet{west01:_introd_graph_theor}. In
this chapter we will use the terms graph and network interchangeably,
i.e. both refer to the same object.  The same holds for nodes and
nodes as well as links and links.

A \textit{graph} $G$ is a pair, $G=(V,E)$, consisting of a node set
$V(G)$ and an link set $E(G)$. $K_n$ is the \textit{complete graph} on
$n$ nodes. $C_n$ the \textit{cycle} on $n$ nodes. Nodes $i$ and $j$ are
the endpoints of the link $e_{ij} \in E(G)$.

The \textit{degree}, $d_i$, of a node $i$ is the number of links incident
to it. A graph can either be \textit{undirected} or \textit{directed},
where in the latter case one has to distinguish between
\textit{in-degree}, $d^-_i$, and \textit{out-degree}, $d^+_i$, of node
$i$. In the case of an undirected graph, the neighborhood of a node $i$
in $G$ is $N_i= \{ w \in V(G): e_{wi} \in E(G) \}$. The degree of a node
$i$ is then $d_i = |N_i|$.  The first order neighborhood is just the
neighborhood, $N_i$, of node $i$.  The second order neighborhood is, $N_i
\cup \{ N_v: v \in N_i \}$.  Similarly, higher order neighborhoods are
defined. In the case of a directed graph we denote the out-neighborhood
of node $i$ by $N_i^+$ and the in-neighborhood by $N_i^-$. A graph $G$ is
\textit{regular} if all nodes have the same degree.  A graph $G$ is
\textit{$k-regular$} if every node has degree $k$.

A \textit{walk} is an alternating list, $\{
v_0,e_{01},v_1,...,v_{k-1},e_{k-1k},v_k \}$, of nodes and links. A
\textit{trail} is a walk with no repeated link. A \textit{path} is a walk
with no repeated node. The shortest path between two nodes is also
known as the \textit{geodesic distance}. If the endpoints of a trail are
the same (a closed trail) then we refer to it as a \textit{circuit}. A
circuit with no repeated node is called a \textit{cycle}.

A \textit{subgraph}, $G'$, of $G$ is the graph of subsets of the
nodes, $V(G') \subseteq V(G)$, and links, $E(G') \subseteq E(G)$. A
graph $G$ is \textit{connected}, if there is a path connecting every pair
of nodes.  Otherwise $G$ is disconnected. The \textit{components} of a
graph $G$ are the maximal connected subgraphs.

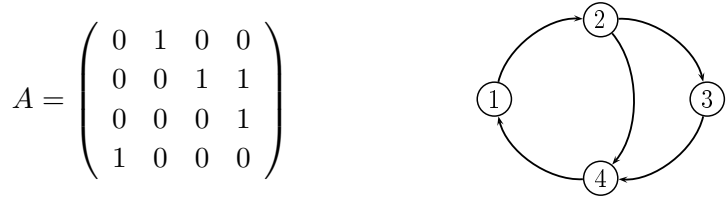
\begin{figure}[htpb]
\begin{minipage}{.45\textwidth}
\begin{displaymath}
A=\left(
\begin{array}{cccc}
0 & 1 & 0 & 0\\
0 & 0 & 1 & 1\\
0 & 0 & 0 & 1\\
1 & 0 & 0 & 0
\end{array}
\right)
\end{displaymath}
\end{minipage}
\begin{minipage}{.45\textwidth}
    \scalebox{0.7}[0.7]{
      \begin{pspicture}(0,1)(6,5) \cnodeput(1,3) {1}{\Large 1}
        \cnodeput(3,4.5){2}{\Large 2} \cnodeput(3,1.5){3}{\Large 4}
        \cnodeput(5,3) {4}{\Large 3}
        \ncarc[linewidth=1pt,arcangle=40]{->}{1}{2}
        \ncarc[linewidth=1pt,arcangle=40]{->}{2}{3}
        \ncarc[linewidth=1pt,arcangle=40]{->}{3}{1}
        \ncarc[linewidth=1pt,arcangle=40]{->}{2}{4}
        \ncarc[linewidth=1pt,arcangle=40]{->}{4}{3}
      \end{pspicture}}
\end{minipage}
\caption{(Right) a directed graph consisting of $4$ nodes and $5$
  links. (Left) the corresponding adjacency matrix $A$
  \label{example_dirnet}}
\end{figure}

The \textit{adjacency matrix}, $\mathbf{A}(G)$, of $G$, is the $n$-by-$n$
matrix in which the entry $a_{ij}$ is $1$ if the link $e_{ij} \in E(G)$,
otherwise $a_{ij}$ is $0$. For an undirected graph $\mathbf{A}$ is
symmetric, i.e. $a_{ij}=a_{ji}$ $\forall i,j \in V(G)$. An example of a
graph and its associated adjacency matrix is given in Fig.
(\ref{example_dirnet}). For example, in the first row with elements,
$a_{11}=0, a_{12}=1, a_{13}=0, a_{14}=0$, the element $a_{12}=1$
indicates that there exist an link from node $1$ to node $2$.

In a \textit{bipartite} graph $G$, $V(G)$ is the union of two disjoint
independent sets $V_1$ and $V_2$.  In a bipartite graph, if $e_{12} \in
E(G)$ then $v_1 \in V_1$ and $v_2 \in V_2$. In other words, the two
endpoints of any link must be in different sets. The \textit{complete
  bipartite graph} with partitions of size $|V_1|=n$ and $|V_2|=m$ is
denoted $K_{n,m}$. A special case is the \textit{star} which is a
complete bipartite graph with one partition having size $n=1$, $K_{1,m}$.

There exists an important class of graphs, \textit{random graphs}, which
are determined by their number of nodes, $n$, and the (independent)
probability $p$ of each link being present in the graph
\citep{bollobas85:_random_graphs}.

We now introduce two topological measures of a graph, the clustering
coefficient and the average path length. For further details see e.g.
\citet{newman03:_struc_funct_compl_networ} and \citet{costa-2007}. The
following definitions assume undirected graphs.

For each node, the \textit{local clustering coefficient}, $C_i$,
is simply defined as the fraction of pairs of neighbors of $i$ that are
themselves neighbors. The number of possible neighbors of node $i$ is
simply $d_i(d_i-1)/2$. Thus we get
\begin{equation}
C_i= \frac{|\{e_{jk} \in E(G): e_{ij} \in E(G) \wedge e_{ik} \in
    E(G)\}|}{d_i(d_i-1)/2}
\label{eq:clustering}
\end{equation}
The \textit{global clustering coefficient $C$} is then given by
\begin{equation}
C=
\frac{1}{n} 
\sum_{i=1}^{n} C_i
\label{eq:clustering-global}
\end{equation}
A high clustering coefficient $C$ means (in the language of social
networks), that the friend of your friend is likely also to be your
friend. It also indicates a high redundancy of the network.

The \textit{average path length} $l$ is the mean geodesic (i.e.
shortest) distance between node pairs in a graph:
\begin{equation}
l=\frac{1}{\frac{1}{2}n(n-1)}
\sum_{i \ge j}^{n} d_{ij}
\label{eq:pathlength}
\end{equation}
where $d_{ij}$ is the geodesic distance from node $i$ to node $j$. 

In section (~\ref{sec:small-world}) we will show a model of innovation
networks that produces ``small-worlds'' which combine the two properties
of a high clustering coefficient and a small average path length.

In the following sections we will describe some basic models of economic
network theory, where we shall use the definitions and notations
introduced above.

\subsection{The Connections Model}
\label{sec:connections-model}

The \textit{connections model} introduced by
\citet{jackson96:_strat_model_social_econom_networ} is of specific
interest since it allows us to compute equilibrium networks analytically.
The succeeding models can then be considered as extension of the
connections model. Since these models are more complicated than the basic
connections model they can, to a large extent, only by studied via
computer simulations\footnote{For the use of computer simulations in
  economics see \citet{axelrod06:_guide}}. Nevertheless they are of
interest because they show a wider range of possible network
configurations and associated performance of the agents in the economy.

In the following we discuss the (symmetric) connections model proposed by
\citet{jackson96:_strat_model_social_econom_networ}\footnote{For a good
  introduction and discussion of related works we recommend the lecture
  notes of \citet{zenou:_cours_networ}. There one can find the proofs
  given here and related material in more detail. For a general
  introduction to economic networks see also
  \citet{jackson06:_advan_econom_econom}.} In this model agents pass
information to those to whom they are connected to.  Through these links
they also receive information from those agents that they are indirectly
connected to, that is, trough the neighbors of their neighbors, their
neighbors, and so on.

The individual incentives to form or severe links determine the addition or
deletion of links. Incentives are defined in terms of the utility of the
agents which depends on the interactions among agents, i.e. the
network. The utility functions assigns a payoff to every agent as a
function of the network the agents are nested in.

The utility, $u_i(G)$, agent $i$ receives from network $G$ with $n$
agents is a function $u_i: \{ G \in \mathcal{G}_n \} \rightarrow
\mathbb{R}$ with
\begin{equation}
u_i(G)=\sum_{j=1}^n \delta^{d_{ij}} - \sum_{j \in N_i} c
\label{eq:utility_connections_model}
\end{equation}
where $d_{ij}$ is the number of links in the shortest path between agent
$i$ and agent $j$. $d_{ij}=\infty$ if there is no path between $i$ and
$j$. $0 \le \delta \le 1$ is a parameter that takes into account the
decrease of the utility as the path between agent $i$ and agent $j$
increases. $N(i)$ is the set of nodes in the neighborhood of agent $i$.
In this model the network is undirected.

A measure of the global performance of the network is introduced by
its efficiency.  The total utility of a network is defined by
\begin{equation}
  U(G)= \sum_{i=1}^n u_i(G)
\end{equation}
A network is considered efficient if it maximizes the total utility of
the network $U(G)$ among all possible networks, $\mathcal{G}(n)$ with $n$
nodes.
\begin{defn}
  A network $G$ is strongly efficient if $U(G)=\sum_{i=1}^n u_i(G) \ge
  U(G')=\sum_{i=1}^n u_i(G')$ for all $G' \in \mathcal{G}(n)$
\end{defn}
Under certain conditions no new links are accepted or old ones deleted.
This leads to the term pairwise stability.
\begin{defn} A network $G$ is \textit{pairwise stable} if and only if
  \begin{enumerate}
  \item for all $e_{ij} \in E(G)$, $u_i(G) \ge u_i(E \backslash e_{ij})$
    and $u_j(G) \ge u_j(E \backslash e_{ij})$
  \item for all $e_{ij} \notin E(G)$, if $u_i(G) < u_i( E \cup e_{ij})$
    then $u_i(G) > u_j(E \cup e_{ij})$
  \end{enumerate}
\end{defn}
In words, a network is pairwise stable if and only if (i) removing any
link does not increase the utility of any agent, and (ii) adding a link
between any two agents, either does not increase the utility of any of the
two agents, or if it does increase one of the two agents' utility then it
decreases the other agent's utility.

The point here is that establishing a new link with an agent requires the
consensus (i.e. a simultaneous increase of utility) of both of them. The
notion of pairwise stability can be distinguished from the one of Nash
equilibrium\footnote{ Considering two agents playing a game (e.g. trading
  of knowledge) and each adopting a certain strategy. A Nash equilibrium
  is characterized by a set of strategies where each strategy is the
  optimal response to all the others.} which is appropriate when each
agent can establish or remove unilaterally a connection with another
agent.

There exists a tension between stability and efficiency in the
connections model. This will become clear, after we derive the following
two propositions.
\begin{prop}
  The unique \textit{strongly efficient} network in the symmetric
  connections model is
  \begin{enumerate}
  \item the complete graph $K_n$ if $c < \delta - \delta^2$,
  \item a star encompassing everyone if $\delta - \delta^2 < c < \delta +
    \frac{n-2}{2} \delta^2$
  \item the empty graph (no links) if $\delta + \frac{n-2}{2} \delta^2
    <c$.
  \end{enumerate}
\end{prop}
\begin{proof}
  \begin{enumerate}
  \item We assume that $\delta^2 < \delta - c$. Any pair of agents that
    is not directly connected can increase its utility (the net benefit
    for creating a link is $\delta - c - \delta^2 >0$) and thus the total
    utility, by forming a link. Since every pair of agents has an
    incentive to form a link, we will end up in the complete graph $K_n$,
    where all possible links have been created and no additional links
    can be created any more.
  
  \item Consider a component of the graph $G$ containing $m$ agents, say
    $G'$. The number of links in the component $G'$ is denoted by $k$,
    where $k \ge m-1$, otherwise the component would not be connected.
    E.g. a path containing all agents would have $m-1$ links. The total
    utility of the direct links in the component is given by $k(s \delta
    - 2 c)$.  There are at most $\frac{m(m-1)}{2}-k$ left over links in
    the component, that are not created yet. The utility of each of these
    left over links is at most $2 \delta^2$ (it has the highest utility
    if it is in the second order neighborhood). Therefor the total
    utility of the component is at most
    \begin{equation}
      k 2 (\delta - c) + \left( \frac{m(m-1)}{2} - k \right) 2 \delta^2
    \end{equation}
    Consider a star $K_{1,m-1}$ with $m$ agents. The star has $m-1$
    agents which are not in the center of the star. An example of a star
    with $4$ agents is given in Fig. (\ref{fig:star}). The utility of any
    direct link is $2 \delta - 2 c$ and of any indirect link $(m - 2)
    \delta^2$, since any agent is $2$ links away from any other agent
    (except the center of the star). Thus the total utility of the star
    is
    \begin{equation}
      \underbrace{(m - 1)(2 \delta - 2 c)}_{\text{direct connections}} +
      \underbrace{(m - 1) (m - 2) \delta^2}_{\text{indirect connections}}
    \end{equation}

    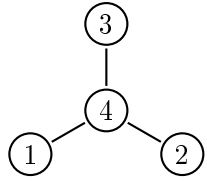
\begin{figure}
        \begin{center}
      \begin{pspicture}(2,2)
        \cnodeput(0,0){A}{1} 
        \cnodeput(2,0){B}{2} 
        \cnodeput(1,1.73){C}{3}
        \cnodeput(1,0.58){D}{4} 
        \psset{nodesep=1pt}
        \ncline{-}{D}{A}
        \ncline{-}{D}{B}
        \ncline{-}{D}{C}
      \end{pspicture}
       \end{center}
       \caption{A star encompassing $4$ agents.}
    \label{fig:star}
    \end{figure}
    The difference in total utility of the (general) component and the
    star is just $2 (k - (m-1)) (\delta - c - \delta^2)$. This is at most
    $0$, since $k \ge m-1$ and $c >\delta - \delta^2$, and less than $0$
    if $k > m-1$. Thus, the value of the component can equal the value of
    the star only if $k=m-1$. Any graph with $k=m-1$ edges, which is not
    a star, must have an indirect connection with a distance longer than
    $2$, and getting a total utility less than $2 \delta^2$.  Therefore
    the total utility from indirect connections of the indirect links
    will be below $(m-1)(m-2) \delta^2$ (which is the total utility from
    indirect connections of the star). If $c< \delta - \delta^2$, then
    any component of a strongly efficient network must be a star.

    Similarly it can be shown
    \citep{jackson96:_strat_model_social_econom_networ} that a single
    star of $m+n$ agents has a higher total utility than two separate
    stars with $m$ and $n$ agents.  Accordingly, if a strongly efficient
    network is non-empty, it must be a star.
  
  \item A star encompassing every agent has a positive value only if
    $\delta + \frac{n-2}{2} \delta^2 > c$. This is an upper bound for the
    total achievable utility of any component of the network. Thus, if
    $\delta + \frac{n-2}{2} \delta^2 < c$ the empty graph is the unique
    strongly efficient network.
  \end{enumerate}
\end{proof}
\begin{prop} In the connections model in which the utility of each agent
  is given by (\ref{eq:utility_connections_model}), we have
  \begin{enumerate}
  \item A pairwise stable network has at most one (non-empty) component.
  \item For $c < \delta -\delta^2 $, the unique pairwise stable network is
    the complete graph $K_n$.
  \item For $\delta - \delta^2 < c < \delta$ a star encompassing every agent is
    pairwise stable, but not necessarily the unique pairwise stable graph. 
  \item For $\delta < c$, any pairwise stable network that is non-empty
    is such that each agent has at least two links (and thus is efficient).
  \end{enumerate}
\end{prop}
\begin{proof}
  \begin{enumerate}
  \item Lets assume, for the sake of contradiction, that $G$ is pairwise
    stable and has more than one non-empty component. Let $u^{ij}$ denote
    the utility of agent $i$ having a link with agent $j$. Then, $u^{ij}
    = u_i(G + e_{ij}) - u_i(G)$ if $e_{ij} \notin E(G)$ and $u^{ij} =
    u_i(G) - u_i(G - e_{ij})$ if $e_{ij} \in E(G)$. We consider now
    $e_{ij} \in E(G)$. Then $u^{ij} \ge 0$. Let $e_{kl}$ belong to a
    different component. Since $i$ is already in a component with $j$,
    but $k$ is not, it follows that $u^{jk}> u^{ij} \ge 0$, because agent
    $k$ will receive an additional utility of $\delta^2$ from being
    indirectly connected to agent $i$. For similar reasons $u^{jk}>
    u^{lk} \ge 0$. This means that both agents in the separate component
    would have an incentive to form a link. This is a contradiction to
    the assumption of pairwise stability.
  
  \item The net change in utility from creating a link is $\delta -
    \delta^2 - c$. Before creating the link, the geodesic distance
    between agent $i$ and agent $j$ is at least $2$. When they create a
    link, they gain $\delta$ but they lose the previous utility from
    being indirectly connected by some path whose length is at least $2$.
    So if $c < \delta - \delta^2$, the net gain from creating a link is
    always positive. Since any link creation is beneficial (increases the
    agents' utility), the only pairwise stable network is the complete
    graph, $K_n$.
  
  \item We assume that $\delta - \delta^2 < c - \delta$ and show that the
    star is pairwise stable. The agent in the center of the star has a
    distance of $1$ to all other agents and all other agents are
    separated by $2$ links from each other. The center agent of the star
    cannot create a link, since she has already maximum degree. She has
    no incentive to delete a link either. If she deletes a link, the net
    gain is $c - \delta$, since there is no path leading to the then
    disconnected agent. By assumption, $\delta - \delta^2 < c < \delta$,
    $c - \delta <0$ and the gain is negative, and the link will not be
    removed. We consider now an agent that is not the center of the
    star. She cannot create a link with the center, since they are both
    already connected. The net gain of creating a link to another agent
    is $\delta - \delta^2 -c$, which is strictly negative by
    assumption. So she will not create a link either. The star is
    pairwise stable.

    The star encompasses all agents. Suppose an agent would not be
    connected to the star. If the center of the star would create a link
    to this agent, the net gain would be $\delta - c > 0$ and the benefit
    of the non-star agent is again $\delta - c > 0$. So both will create
    the link.
  
    The star is not the unique pairwise stable network. We will show that
    for $4$ agents, the cycle, $C_4$ is also a pairwise stable
    network. Consider Fig. (\ref{fig:four_cycle})

    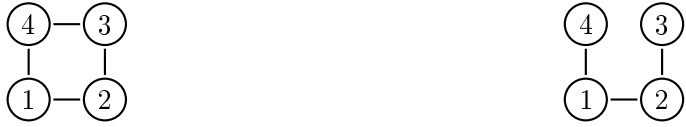
\begin{figure}
      \begin{center}
      \begin{minipage}{.45\textwidth}
        \begin{center}
      \begin{pspicture}(2,1.5)
        \cnodeput(0,0){A}{1} 
        \cnodeput(1,0){B}{2} 
        \cnodeput(1,1){C}{3}
        \cnodeput(0,1){D}{4} 
        \psset{nodesep=1pt}
        \ncline{-}{A}{B} 
        \ncline{-}{B}{C}
        \ncline{-}{C}{D}
        \ncline{-}{D}{A}
      \end{pspicture}
      \end{center}
    \end{minipage}
    \begin{minipage}{.45\textwidth}
      \begin{center}
      \begin{pspicture}(2,1.5)
        \cnodeput(0,0){A}{1} 
        \cnodeput(1,0){B}{2} 
        \cnodeput(1,1){C}{3}
        \cnodeput(0,1){D}{4} 
        \psset{nodesep=1pt}
        \ncline{-}{A}{B} 
        \ncline{-}{B}{C}
        \ncline{-}{D}{A}
      \end{pspicture}
      \end{center}
    \end{minipage}
    \end{center}
    \caption{A cycle of $4$ agents (left) and the resulting graph (right)
      after the deletion of a link from agent $3$ to agent $4$.}
    \label{fig:four_cycle}
    \end{figure}
    
    If agent $3$ removes a link to agent $4$, then her net gain is $c -
    \delta - \delta^3$. For the range of costs of $\delta - \delta^2 < c
    < \delta - \delta^3 < \delta$, she will never do it. If agent $3$
    adds a link to agent $1$, Fig. (\ref{fig:four_cycle_short_cut}), the
    net gain is $\delta - \delta^2 < 0$. Thus, for $n=4$ and $\delta -
    \delta^2 < c < \delta - \delta^3$, then there are at least two
    pairwise stable networks: the star and the cycle.

    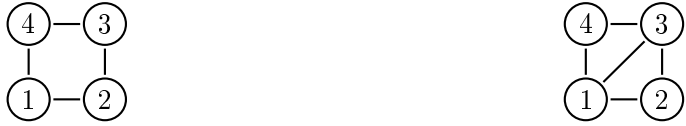
\begin{figure}
      \begin{center}
      \begin{minipage}{.45\textwidth}
        \begin{center}
      \begin{pspicture}(2,1.5)
        \cnodeput(0,0){A}{1} 
        \cnodeput(1,0){B}{2} 
        \cnodeput(1,1){C}{3}
        \cnodeput(0,1){D}{4} 
        \psset{nodesep=1pt}
        \ncline{-}{A}{B} 
        \ncline{-}{B}{C}
        \ncline{-}{C}{D}
        \ncline{-}{D}{A}
      \end{pspicture}
       \end{center}
    \end{minipage}
    \begin{minipage}{.45\textwidth}
      \begin{center}
      \begin{pspicture}(2,1.5)
        \cnodeput(0,0){A}{1} 
        \cnodeput(1,0){B}{2} 
        \cnodeput(1,1){C}{3}
        \cnodeput(0,1){D}{4} 
        \psset{nodesep=1pt}
        \ncline{-}{A}{B} 
        \ncline{-}{B}{C}
        \ncline{-}{C}{D}
        \ncline{-}{D}{A}
        \ncline{-}{C}{A}
      \end{pspicture}
       \end{center}
    \end{minipage}
    \end{center}
    \caption{A cycle of $4$ agents (left) and the resulting graph (right)
      after the creation of a link from agent $3$ to agent $1$.}
    \label{fig:four_cycle_short_cut}
    \end{figure}

  \item For $\delta < c$ the star is not a pairwise stable network
    because the agent in the center of the star would gain $c - \delta$
    from deleting a link. Moreover, it can be shown
    \citep{jackson96:_strat_model_social_econom_networ} that any
    connected agent has at least $2$ links.
  
\end{enumerate}
  
\end{proof}

One can see, from the two propositions described above, that a pairwise
stable network is not necessarily efficient. For high cost, i.e. $c >
\delta$ there are non-empty pairwise stable networks but they are not
efficient.

We now come to the evolution of the network as described in
\citet{jackson02:_evolution_social_economic_networks}. The network
changes when agents create or delete a link. At every time step an agent
is chosen at random and tries to establish a new link or delete an
already existing one. If a link is added, then the two agents involved
must both agree to its addition, with at least one of them strictly
benefiting (in terms of a higher utility) of the new link. Similarly a
deletion of a link can only be in a mutual agreement. This adding and
deleting of links creates a sequence of networks. A sequence of networks
created by agents myopically adding and deleting links is called an
\textit{improving path}\footnote{Each network in the sequence of network
  updates differs in one link from the previous one. An improving path is
  a finite set of networks $G_1,...,G_k$ in which one agents is better
  off by deleting a link ($G_{k+1}$ has one link less than $G_k$) or two
  agents are better off by adding a link ($G_{k+1}$ has one link more than
  $G_k$).}  \citep{jackson02:_evolution_social_economic_networks}.

There is a small probability, $\epsilon$, that a mistake occurs
(trembling hand) and the link is deleted if present or added if absent.
$\epsilon$ goes to zero in the long run, $\lim_{t \to \infty} \epsilon(t)
= 0$.  By introducing this decreasing error $\epsilon$ in the agent's
decisions, the evolution of the network becomes a Markov
process\footnote{A Markov process is a random process whose future states
  are determined by its present state and not on the past states, i.e. it
  is conditionally independent on the past states given the present
  state.} with a unique limiting stationary distribution of networks
visited \citep{jackson02:_evolution_social_economic_networks}. The
following definition is important to describe the stochastic evolution of
the network.
\begin{defn} A network is \textit{evolutionary stable} if it is in the
  limiting stationary distribution of networks of the above mentioned
  Markov process.
\end{defn}
We have already investigated the structure and stability of the star,
Fig. (\ref{fig:star}), and the cycle, Fig. (\ref{fig:four_cycle}). In
\citet{jackson02:_evolution_social_economic_networks} it is shown that
for the case of $4$ agents, the evolutionary stable networks indeed are
the stars and cycles. So the network of agents evolves into a quite
simple equilibrium configuration.

\subsection{The Connections Model and Small-World Networks}
\label{sec:small-world}

\citet{carayol05:_strat_model_compl_networ_format,
  carayol03:_self_organizing_innovation_networks} propose a model of
innovation networks in which networks emerge that show the properties of
a ``small world''\footnote{A small-world network combines high clustering
  (high probability that your acquaintances are also acquaintances to
  each other) with a short characteristic path length (small average
  distance between two
  nodes)\citep{strogatz98:_collective_dynamics_small_world_networks}.}.
This model is an extension of the above described connections model, sec.
(\ref{sec:connections-model}), and it uses the same notion of pairwise
stability and efficiency\footnote{A network is pairwise stable if and
  only if (i) removing any link does not increase the utility of any
  agent, and (ii) adding a link between any two agents, either does not
  increase the utility of any of the two agents, or if it does increase
  one of the two agents' utility then it decreases the other agent's
  utility. It is efficient if it maximizes total utility.}.

We now give a sketch of the model. Agents are localized on a cycle and
benefit from knowledge flows from their direct and indirect neighbors.
Knowledge transfer decays along paths longer than one link. This means
that less knowledge is received, the longer the path between the not
directly connected agents is. The transfer rate is controlled by an
exogenous parameter, $\delta$. Each agent has a probability to innovate,
that is dependent on her amount of knowledge. The knowledge level of an
agent is dependent on two factors. (i) the in-house innovative
capabilities of the agent and (ii) the knowledge flows coming directly
from the neighbors or indirectly (with a certain attenuation factor) from
those agents that are connected to the neighbors.

Agent $i$ supports costs, $c_i(t)$, for direct connections which are
linearly increasing with geographic distance, that is the distance on the
cycle on which they recede. Agent $i$'s utility $u_i$ at a time $t$ is
given by the following expression.
\begin{equation}
  \label{eq:carayol_utility}
  u_i(G(t)) = \sum_{j \in N_i} \delta^{d_{ij}} - c \sum_{j \in N_i} d'_{ij},
\end{equation}
where $d_{ij}$ is the geodesic distance between agent $i$ and agent $j$.
$\delta \in (0,1)$ is a knowledge decay parameter and $\delta^{d_{ij}}$
gives the payoffs resulting from the direct or indirect connection
between agent $i$ and agent $j$. $d'_{ij}$ describes the geographic
distance between agent $i$ and agent $j$, that is the distance on the
cycle. This is the main difference in the assumptions compared to the
connections model discussed in sec. (\ref{sec:connections-model}).

Agents are able to modify their connections. This is were the network
becomes dynamic. 
Pairs of agents are randomly selected. If the selected two agents are
directly connected they can jointly decide to maintain a link or
unilaterally decide to sever the link. If they are not connected, they
can jointly decide to form a link. The decision is guided by the
selfishness of the agents, which means that they only accept links from
which they get a higher utility.

The stochastic process of adding links to the network can be seen as a
Markov process where each state is the graph structure at a certain time
step. The evolution of the system is a discrete time stochastic process
with the state space of all possible graphs. A small random perturbation
where the agents make mistakes in taking the optimal decision to form a
link or not is introduced. Agents are making errors with a probability
$\epsilon(t)$. This error term decreases with time, $\lim_{t \to \infty}
\epsilon(t) = 0$. 

The introduction of $\epsilon$ enables us to find long-run stationary
distributions that are independent of initial conditions (the ergodicity
of the system) \citep{jackson02:_evolution_social_economic_networks}.
Simulations are used in order to find these stationary distributions.
Agents are forming and severing links until the network reaches a
pairwise stable configuration where the agents have no incentive to
create or delete links any more.  The set of stochastically stable
networks selected in the long run is affected by the rate of knowledge
transfer, $\delta$. The authors find critical values of this parameter
for which stable ``small world'' networks are dynamically selected. This
is the main difference in the resulting equilibrium network structure to
the connections model, in which simpler network configurations are
obtained.

\subsection{Introducing Heterogeneous Knowledge}
\label{sec:innov-sett-evol}

\citet{ricottilli06:_const_freed_action,
  ricottilli05:_firms_network_formation} studies the evolution of a
network of agents that improve their technological capabilities through
interaction while knowledge is heterogeneously distributed among agents.
In addition to the sharing of knowledge, each agent is assumed to have an
``in-house'' innovative capability. Considerable effort is necessary for
this ``in-house'' research and as research is not always successful, it
is assumed to change stochastically.

An agent $i$'s innovation capability, $V_i$, is given by
\begin{equation}
V_i(t)=
\sum_{j=1}^{N}
a_{ij}b_{ij}(t)V_j(t)
+
C_i(t)
\label{eq:inno_cap}
\end{equation}
with an economy consisting of $n$ agents. $a_{ij}=$const. is the
broadcasting capacity of agent $j$ to agent $i$ and $a_{ii}=0$ since no
agent can broadcast information to herself. The matrix $A$ with elements
$a_{ij}$ indicates the total technological information broadcasting
capability of this economy. The proximity matrix elements $b_{ij}(t)$ are
either $0$ or $1$ according to whether agent $j$ is identified from agent
$i$ as an information supplier. This is the neighborhood of agent $i$.
$C_i(t) \in (0,1)$ is the in-house capability of agent $i$. This is a
stochastic variable.

Each agent $i$ assesses the value of knowledge of its neighbors (where
$b_{ij} \neq 0$), which are the addends of the first term in
(\ref{eq:inno_cap}).  From this function the least contributing
one, denoted by $\gamma_i(t)$, is selected.
\begin{equation}
\gamma_i(t)=
\mathrm{min}_{1 \le j \le N}
\{
a_{ij} b_{ij}(t-1)
V_j(t-1)
\}
\end{equation}
In a random replacement procedure (search routine) an agent selects
either its neighbors and second neighbors (local, weak bounded
rationality) or the entire economy excluding its first and second
neighbors (global, strong bounded rationality). By doing so, agent $i$
assigns a new member $j$ to the set of information suppliers, setting
$b_{ij}$ from $0$ to $1$. This selection is only accepted if
\begin{equation}
V_i(t)>V_i(t-1)
\end{equation}
The population of agents is classified according to the size of the set
of other agents by which they are observed.  Global paradigm setters are
agents that are observed by almost all agents in the economy. Local
paradigm setters are observed by almost all agents belonging to the same
component.

Simulations of the evolution of the network show that stable patterns
emerge. When the knowledge-heterogeneity of the economy is not very high,
global paradigm setters emerge. For high levels of heterogeneity the
economy becomes partitioned in two separate halves. In each homogeneous
one, local paradigm setters emerge.
\citet{ricottilli06:_const_freed_action,
  ricottilli05:_firms_network_formation} shows that the highest
technological capabilities are achieved neither with a local search
routine in which only the second neighbors are included nor in global
search routines that span the whole economy. Rather a combination of both
improves the system's innovative efficiency the most.

\subsection{Emerging Cyclic Network Topologies}
\label{sec:net_evol_with_cycles}

When studying multi-sector trading economies and input-output systems
\citet{rosenblatt57:_linear_model_graph_minkow_leont_matric} already
identified the importance of circular flows and ``feed-back'' input
dependencies between industries (realized by subgraphs called ``cyclic
nets''). A sufficient condition for a strongly connected network (in
which there exits a path from every agent to every other agent and that
has an irreducible adjacency matrix) is the existence of a cycle.
Subsequent works \citep{maxfield94:_gener_equil_theor_direc_graph,
  baldry05:_irred} have further incorporated the conditions of strong
connectedness and cycles for the existence for a competitive economy.
More recently, \citet{kima07:_networ} study a generalized model of
\citet{goyal00:_Noncooperative_model_network_formation} and find that the
equilibrium networks consist of cycles (so called ``sub-wheel
partitions'').

In the following sections we will focus on some recent network models of
knowledge transaction and innovation (the creation of new knowledge) in
which cyclic interactions of agents emerge. In section
\ref{sec:evolving_networks} we will study a new model of evolving
innovation networks. Similarly to the above mentioned authors we find
that the existence of equilibrium networks with a positive knowledge
production depends critically on the existence of cycles in the network.

\subsubsection{Production Recipes and Artifacts}
 \label{sec:prod-recip-artif}

In the following sections we will focus on network models of knowledge
transaction and innovation, the creation of new knowledge, in which
\emph{cyclic interactions} of agents emerge. We start by reviewing a
model by
\citet{lane05:_theory_based_dynamical_model_innovation_processes} in
which agents try to produce and sell artifacts.  These artifacts can be
manufactured according to a production recipe.  Such a recipe can either
be found independently or through the sharing of knowledge with other
agents, which in turn can lead to an innovation, that is the discovery of
a new recipe.

Let us denote with $r_{ik}$ the $k^{th}$ recipe of agent $i$. There is
an external environment which consists of external agents (customers) and
artifacts which are not produced in the model. At each time period $t$
one agent $i$ is randomly chosen. Then the following steps are taken:
\begin{enumerate}
\item The agent tries to get the input required for each recipe $r_{ik}$.
  If it is in the agent's own stock then she can produce immediately. If
  it is not, she buys it from another agent and if it cannot be bought she
  moves to another recipe.

\item 
  The agent chooses a goal, i.e. the product she wants to produce (one
  that gives high sales).  Therefore she has to find the right recipe for
  the goal. She produces the product if a successful recipe is found.
  This can be achieved in two ways. The agent either can try to innovate
  by herself or she can try to innovate together with another agent.

\item The wealth of agent $i$ at time $t+1$, $w_i(t)$, is calculated
  according to
  \begin{equation}
    w_i(t+1)=
    w_i(t)
    +
    \sum_{k=1}^{N_k} n_{ik}(t)
    -
    w_i(t)
    \sum_{l=1}^{N_l} p_{il} c_{il}
    -
    \lambda w_i(t)
  \end{equation}
where $N_k$ is the number of products sold, $n_{ik}(t)$ is the number of
units sold of product $k$ belonging to agent $i$, $p_{il}$ is the number
products produced with recipe $r_{il}$ and $c_{il}$ is the production
cost.

The last term $-\lambda w_i(t)$ guarantees that the wealth of an agent
that has not sold any products and does not have any active recipes
vanishes.

\item The recipes that could not be successfully used to produce products
  are canceled.

\item The set of acquaintances of an agent is enlarged. This is possible
  when two or more agents, that have goals which are close in artifact
  space (i.e. they require similar inputs) cooperate to produce that
  artifact.

\item With a certain probability dead agents are substituted.
\end{enumerate}
The basic dynamics, absent innovation, is one of production and sales,
where the supply of raw materials is external as well as final product
demand. There are two main differences to most agent-based innovation
models. First, here the agents try to develop new recipes in order to
produce products with high sales, as opposed to many agent-based models
where the generation of novelty is driven by some stochastic process.
Second, in simulations
\citet{lane05:_theory_based_dynamical_model_innovation_processes} shows
that the network of customers and suppliers often forms closed,
self-sustaining cycles.

\subsubsection{An Autocatalytic Model with Hypercycles}
\label{sec:an-autoc-model}

\citet{Padgett03:_economic_production_chemistry,
  padgett96:_emerg_simpl_ecolog_skill} introduces an autocatalytic model,
based on a hypercycle\footnote{A hypercycle is a system which connects
  self-replicative units through a cycle linkage
  \citep{eigenschuster79:_hyper}.} model. Here agents are represented as
skills and these skills are combined in order to produce.  Skills, like
chemical reactions, are rules that transform products into other
products.

In the following we will give a short overview of the model\footnote{This
  agent based model is publicly available on the website
  http://repast.sourceforge.net/examples/index.html under the application
  module hypercycle.}. There are two main aspects in the dynamics
interaction of the agents: The process of production and the process of
learning.
The process of production includes three entities: skills, products and
agents.  Skills transform products into other products. The skills are
features of the agents. On a spatial grid the agents are arrayed with
periodic boundaries. Each agent has eight possible neighbors. At each
asynchronous iteration a random skill is chosen. An agent with that skill
randomly chooses an input product. If this product fits to the skill then
the product is transformed.  The transformed product is passed randomly
to the neighbors of the agent.  If the trading partner has the necessary
skill it transforms the product further and passes it on. If the agent
doesn't have the compatible skill, the product is ejected into the output
environment and a new input product is selected.

One can look at the production process from a wider perspective. An input
product comes from the environment, then passes through production chains
of skills until it is passed back as output to the environment. These
chains self-organize because of a feedback mechanism of the agents. This
mechanism is learning through the trade of products.

The process of learning is modelled as learning by doing. If a skill
transforms a product and then passes it on to another transforming skill,
then the skill is reproduced (learned). Whenever one skill is reproduced
anywhere in the system then another one is deleted at random to keep the
overall number of skills constant.  The agents are able to learn new
skills by doing and they can forget skills they didn't use for a certain
period of time. This procedure of learning introduces a feedback
mechanism. When an agent loses all its skills, then it is assumed to
never recover.

In \citet{Padgett03:_economic_production_chemistry,
  padgett96:_emerg_simpl_ecolog_skill} the emergence of self-reinforcing
hypercycle production chains is shown. In these hypercycles agents
reproduce each other through continuous learning. Such cycles generate a
positive growth effect on the reproduction of skills. Thus, even in a
competitive environment the sharing of knowledge is crucial to the
long-run performance of the system.

\section{A New Model of Evolving Innovation Networks}
\label{sec:evolving_networks}

\subsection{Outline of the Modeling Framework}
\label{outline3}

In this section we study the evolution of networks of agents exchanging
knowledge\footnote{See also the chapter of Robin Cowan and Nicolas Jonard
  in this book as well as \citet{cowan04:_networ,
    cowan04:_knowl_dynam_networ_indus}.} in a novel framework.  The
network can evolve over time either, by an external selection mechanism
that replaces the worst performing agent with a new one or, by a local
mechanism, in which agents take decisions on forming or removing a link.
In the latter case, we investigate different modalities of interaction
between agents, namely bilateral interactions, representing R\&D
collaborations \citep{hagedoorn00:_resear,
  hagedoorn06:_and_patter_in_inter_firm_r} or informal knowledge trading
\citep{von87:_cooper_rival}, versus unilateral interactions (similar to
\citet{goyal00:_Noncooperative_model_network_formation} agents decide
unilaterally whom to connect to), representing a generalization of
informal knowledge trading. We further study the impact of varying costs
for maintaining links and the impact of augmenting or diminishing effects
on the value of knowledge with the number of users associated with
different types of knowledge. Our model exhibits equilibrium networks and
we compare their structure and performance.  Similar to the models
discussed in the last section we will show that cyclic patterns in the
interactions between agents play an important role for the stability
(permanence) and performance of the system.

We study different assumptions on the behavior of agents. In most simple
case, denoted by \textit{Extremal Dynamics}, agents form links at random
and, through an external market selection mechanism, the worst performing
agent (this is where the denotation extremal stems from) is replaced with
a new one. In this setting agents are completely passive and they are
exposed to a least-fit selection mechanism.

In a more realistic setting, called \textit{Utility Driven Dynamics},
agents choose with whom to interact, but their behavior is still
boundedly rational and does not consider strategic interaction. The way
in which agents create or delete links to other agents is a trial and
error process for finding the right partner. Here we study two different
modes of interaction. In the first interaction mode, agents are creating
bilateral links. Bilateral links represent formal R\&D collaborations
among agents \citep{hagedoorn00:_resear}, or informal knowledge trading
\citep{von87:_cooper_rival}.  In the second interaction mode, agents are
transferring knowledge unilaterally, which means that one agent may
transfer her knowledge to another but the reverse is not mandatory. In
this setting, the transfer of knowledge may be reciprocated, but
knowledge can also be returned from a third party.  In the latter case,
we speak of indirect reciprocity. If knowledge is transferred
unilaterally, the innovation network can be represented as a directed
graph comprising unilateral links, while if all interactions are
bilateral, the innovation network can be represented as an undirected
graph.

In the setting of unilateral links we also investigate the impact of
additional benefits from network externalities. These benefits consider
specific structural properties of networks which have an augmenting
effect on the value of knowledge. We study two different types of network
properties which increase the value of knowledge. We call these types
\textit{Positive Network Externalities}. The first Positive Network
Externality considers the factor that, the more the centrality of an
agent rises with the creation of a link, the higher is the benefit from
that link. A high centrality indicates that an agent is connected to
other agents through short paths. This means that, when knowledge travels
along short distances between agents, it has a higher value than
knowledge that has to be passed on between many agents. This effect can
be captured by introducing an attenuation of knowledge with the distance
it has to travel (by getting passed on from one agent to the next) until
reaching an agent. The second Positive Network Externality captures
an opposite effect when knowledge is passed on from one agent to the
next. Here the value of knowledge increases with the number of
transmitters (who are also user) of that knowledge. More precisely, we
assume that feedback loops create an increase in the value of knowledge
of the agents that are part of the loop. The more agents absorb and pass
on knowledge the higher is the value of that knowledge.  This means that
a link that is part of a long feedback loop increases the value of the
knowledge passed on from one agent to the next\footnote{We study closed
  loops, because we assume that knowledge issued from one agent has to
  return to that agent in order for her to take advantage of this added
  value of knowledge (created by the multiplicity of other users).}.

We can summarize the different settings that are studied in this section
as follows. We investigate the performance and evolution under the two
aforementioned assumptions on the behavior of agents, namely Extremal
Dynamics and Utility Driven Dynamics. In the latter setting, we further
study the effect of different modes of interaction, i.e.  bilateral and
unilateral knowledge transactions among agents. When studying unilateral
interaction among agents, we introduce different augmenting processes on
the value of knowledge depending on the structure of the network, called
Positive Network Externality. We study the impact of an
attenuation of the value of knowledge by the distance from the giver to
the receiver as well as the contrary effect of an increase of the value
of knowledge with the number of users of that knowledge depending on the
type of knowledge under investigation.  Finally, we discuss the networks
obtained under these different settings with respect to their topologies
and performance.

\subsubsection{Bilateral versus Unilateral Knowledge Exchange}
\label{sec:bilateral_unilateral}

We interpret bilateral interactions as R\&D collaborations on a formal or
informal basis \citep{hagedoorn00:_resear}. Both parties involved share
their knowledge in a reciprocal way, that means one agent is giving
knowledge to another if and only if the other agent is doing this as well
and both agents benefit from this transaction.

We then compare bilateral interaction with the case of agents sharing
knowledge in a unidirectional way with other agents. They then maintain
only those interactions that are in some form reciprocated (and this way
lead to an increase in their knowledge levels after a certain time) but
not necessarily from the agent they initially gave their knowledge to
(indirect reciprocity). The latter is referred to unilateral knowledge
exchange which can be seen as a generalization of informal knowledge
trading.

In the case of informal knowledge trading agents exchange knowledge if
both strictly benefit. Instead, in the case of generalized informal
knowledge trading, one agent transfers knowledge to another one without
immediately getting something back. After a certain time (time horizon
$T$) an agent evaluates its investment by assessing its total net
increase in knowledge. By introducing unilateral knowledge exchange we
relax two requirements: (i) we do not require that the investment in
sharing ones knowledge has to be reciprocated instantaneously and in a
mutually concerted way. And (ii) the reciprocation does not necessarily
have to come from the same agent. With this generalization we introduce
that (i) agents have only limited information on the value of knowledge
of others and on the network of interactions. (ii) Agents proceed in a
trial and error fashion to find the right partners for exchanging their
knowledge. In this setting reciprocity emerges either directly or
indirectly.

If the total knowledge level of an agent at the time horizon $T$ is
higher than it was when the agent started to share her knowledge with
another agent, this interaction is evaluated beneficial, otherwise it is
not.  Only if the interaction is evaluated beneficial, the agent
continues sharing its knowledge with the other agent, otherwise it stops
the interaction. This procedure requires only limited information on the
other agents, since the agent cares for its own total increase in
knowledge and does not need to evaluate the individual knowledge levels
of others. We describe this link formation mechanism in more detail in
section (\ref{sec:local_link_formation}).

\subsection{Unilateral Knowledge Exchange and Reciprocity}
\label{sec:unilateral_reciprocity}

If the interaction of agents are unilateral then agents invest into
innovation by sharing knowledge with other agents. An investment is an
advance payment with the expectation to earn future profits. When one
agent transfers knowledge to another one without immediately getting
something back, this can be regarded as an investment. There are usually
two ways in which an investment can be expected to bring in reasonable
returns.

One way is the creation of contracts. As a precondition for contracts
technologies must be protectable (IPR).  Otherwise agents can not trade
them (once the technology is offered, i.e.  made public, everybody can
simply copy it and there is no more need to pay for it). Contracts must
be binding and complete \citep{dickhaut01:_inves_game}. The contract has
to be binding or agents may not meet their agreement after the payment
has been made.  It has to be complete, or uncertain agreements may lead
agents to interpret it in a way most favorable to their position and this
can cause agents to retreat from the contract.

The requirements for contracts can be difficult to realize. Another way
is to expect reciprocative behavior to the investment. The beneficiary
can either directly or indirectly reciprocate the benefit.  Direct
reciprocity means to respond in kind to the investor, and indirect
reciprocity to reward someone else than the original investor.

One of the possible explanations for reciprocal behavior \citep{nowak2005, nowak1998,
  fehr2003:nature_of_altruism, fehr99:_theor_fairn_compet_cooper,
  bolton00:_erc}, \citep[see e.g.][for a survey]{dieckmann04:_power_recip}
is to assume the existence of reputation. Agents believe that if they
invest into another agent they will increase their reputation and then
realize a reasonable return coming back to them directly or indirectly
(``strategic reputation building'').

In reality, only partial information about reputation is available and
experimental works show that, even in the absence of reputation, there
is a non-negligible amount of reciprocal cooperative behavior among
humans \citep{bolton05:_cooper_stran_limit_infor_reput}. As
\citet{dickhaut01:_inves_game} put it, ``...investment occurs even though
agents cannot create binding contracts nor create reputation.''  Thus,
agents invest into each other by transferring their knowledge even if
they cannot immediately evaluate the benefit from this investment.

We assume that agents are not a priori reciprocating if they receive
knowledge from others. But they perceive, that interactions that are
reciprocated in some way are beneficial (increasing their own knowledge)
and these are the interactions that they maintain in the long run.

The problems associated with bilateral exchange of knowledge (direct
reciprocity) and experimental evidence suggest that unilateral knowledge
exchange, in which indirect reciprocity can emerge, is a relevant mode of
interaction between agents. Moreover, the fact that interactions between
anonymous partners become increasingly frequent in global markets and
tend to replace the traditional long-lasting mutual business
relationships poses a challenge to economic theory and is one of the
reason for the growing interest about indirect reciprocity in the
economic literature.

\subsubsection{Indirect Reciprocity, Directed Graphs and Cycles}
\label{sec:indir-recipr-direct}

An R\&D network can be described as a graph in which agents are
represented by nodes, and their interactions by directed links. Indeed,
as mentioned above, if agent $i$ transfers knowledge to agent $j$ (e.g.
by providing a new technology), the reverse process, i.e. that agent $j$
in turn transfers knowledge to $i$, is in principle not mandatory. This
means that the links representing the transfer of knowledge are directed.
The underlying graph can be represented by an adjacency matrix,
$\mathbf{A}$ with elements $a_{ij} \in [0, 1]$, which is not symmetric,
$a_{ij}\neq a_{ji}$. In other words, directed means that we distinguish
the pairs $(i,j)$ and $(j,i)$ representing the links from $i$ to $j$ and
from $j$ to $i$, respectively. On the other hand, if the adjacency matrix
is symmetric, it means that any two agents are connected both by a link
from $i$ to $j$ and by a link from $j$ to $i$.  We say, in this case,
that they are connected by a \textit{bidirectional} link.  Notice that
the symmetry also implies that the two links have identical weights.

Reciprocity requires the presence of cycles.  In particular, direct
reciprocity corresponds to a cycle of order $k=2$, while indirect
reciprocity corresponds to a cycle of order $k \geq 3$ (see Fig.
\ref{fig:reciprocity_and_cycles}). Therefore, the emergence and
permanence of direct/indirect reciprocity is deeply connected to the
existence of cycles and in the graph of interactions.
\begin{figure}[htpb]
\begin{minipage}{.45\textwidth}
\begin{center}
  \begin{pspicture}(2,3) 
    \cnodeput(0,1){A}{1} 
    \cnodeput(2,1){B}{2}
    \psset{nodesep=1pt} 
    \ncarc[arcangle=-50]{->}{A}{B}
    \ncarc[arcangle=-50]{->}{B}{A}
  \end{pspicture}
  \caption{A cycle of length $2$ represents an interaction between agents
    that is direct reciprocal.}
\end{center}
\end{minipage}
\hfill
\begin{minipage}{.45\textwidth}
\begin{center}
  \begin{pspicture}(2,3) 
    \cnodeput(0,1){A}{1} 
    \cnodeput(2,1){B}{2}
    \cnodeput(1,2.732){C}{3} 
    \psset{nodesep=1pt}
    \ncarc[arcangle=-50]{->}{A}{B} 
    \ncarc[arcangle=-50]{->}{B}{C}
    \ncarc[arcangle=-50]{->}{C}{A}
  \end{pspicture}
  \caption{A cycle of length $3$ (or longer) represents an interaction
    between agents that is indirect reciprocal.}
\end{center}
\end{minipage}
\label{fig:reciprocity_and_cycles}
\end{figure}

\subsection{Formal Modeling Framework}
\label{sec:model} 

In this section, we formalize the general framework for the investigation
of evolving networks of selfish agents engaged in knowledge production
via the sharing of knowledge. In such a framework it is possible to
investigate how the the emergence and permanence of different structures
in the network is affected by (1) the form of the \textbf{growth
  function} of the value of knowledge, (2) the length of \textbf{time
  horizon} after which interactions are evaluated and (3) the
\textbf{link formation/deletion rules}.  At a first glance, this problem
includes a multitude of dimensions, as the space of utility functions and
link formation/deletion rules is infinite.  However, some natural
constraints limit considerably the number of possibilities and make a
systematic study possible. In the following, we present the general
framework. We then focus on a subset of the space of utility functions
and link formation rules. For these, we present briefly some analytical
results, but since the value of knowledge of an agent is assumed to be a
nonlinear function of the neighboring agents, we illustrate them in terms
of computer simulations. We finally summarize the results and discuss
them in relation to the context of innovation.

We consider a set of agents, $N=\{1,...,n\}$, represented as nodes of a
network $G$, with an associated variable $x_i$ representing the value of
knowledge of agent $i$. The value of knowledge is measured in the units
of profits an agent can make in a knowledge-intensive market. It has been
shown that the growth of such knowledge-intensive industries is highly
dependent on the number and intensity of strategic alliances in R\&D
networks \citep{w.06:_oxfor_handb_innov}. In our model we bring the value
of knowledge of an agent, denoted by $x_i(t)$, at time $t$ in relation
with the values of knowledge of the other agents $x_j(t)$ at time $t$ in
the economy, that are connected to the current agent $i$. A link from $i$
to $j$, $e_{ij}$, takes into account that agent $i$ transfers knowledge
to agent $j$. The idea is, that through interaction, agents transfer
knowledge to each other which in turn increases their values of
knowledge.

We focus here only on the network effects on the value of knowledge of an
agent. We therefore neglect the efforts of agents made to innovate on
their own, without the interaction with others\footnote{The ``in-house''
  R\&D capabilities of an agent could be introduced by an additional
  (stochastic) term $S_i(x_i)$. Similar to
  \citet{ricottilli06:_const_freed_action} in section
  (\ref{sec:prod-recip-artif}) $S_i(x_i)$ captures the innovation
  activities of agent $i$ without the interaction with other agents. We
  assume that the ``in-house'' capabilities of agents are negligible
  compared to network effects. Thus we concentrate only on network
  effects on the increase or decrease in the value of knowledge.}. In
particular we assume that the growth of the value of knowledge of agent
$i$ depends only on the value of knowledge of the agents, $j$, with
outgoing links pointing to him (those who transfer knowledge to her), $j
\in V(G)$ such that $e_{ji} \in E(G)$.

In a recent study on the dynamics of R\&D collaboration networks in the
US IT industry \citet{hanaki07:_dynam_r_d_collab_it_indus} have shown
that firms form R\&D collaborations in order to maximize their net
knowledge (information) flow. \citet{cassiman02:_r_d_cooper_spill}
suggested that this knowledge flow can be decomposed in incoming and
outgoing spillovers capturing the positive and negative effects of R\&D
collaborations.

We try to incorporate these positive and negative effects into a
differential equation that describes the change (increase or decrease) in
the value of knowledge of an agent through R\&D collaborations with other
agents. We assume that the knowledge growth function can be decomposed in
a decay term a benefit term and a cost term depending on the interactions
of an agent. The equation for knowledge growth reads
\begin{equation}
\label{eq:general_form_productivity_growth}
  \frac{dx_i}{dt} = 
  - D_i(x_i) 
  + B_i(\mathbf{A}, \mathbf{x})
  - C_i(\mathbf{A}, \mathbf{x}) 
\end{equation}
where
\begin{center}
\begin{tabular}{rl}
  $\dot{x}_i$ & growth of the value of knowledge of agent $i$\\
  $\mathbf{A}$ & adjacency matrix (representing the network)\\
  $\mathbf{x}$ & vector of agents' values of knowledge\\
  $D_i(x_i)$ & knowledge decay (obsolescence of knowledge)\\
  $B_i(\mathbf{A}, \mathbf{x})$ & interaction benefits of agent $i$\\
  $C_i(\mathbf{A}, \mathbf{x})$ & interaction costs of agent $i$
\end{tabular}
\end{center}
$\mathbf{B} \ge 0$ and $\mathbf{C} \ge 0$ are benefit and cost terms,
respectively, while $\mathbf{D} \ge 0$ is a decay term which includes the
fact that a technology loses its value over time (obsolescence). In our
setting, only through R\&D collaborations with other agents, an agent can
overcome the obsolescence of knowledge. This ensures that agents who do
not interact with others have necessarily vanishing value of knowledge in
our model (since $B_i=C_i=0 \Leftrightarrow a_{ij}=0 \text{ } \forall j$
and thus $\dot{x}_i<0$). In other words we investigate an R\&D intensive
economy in which an agent's performance is critically depending on its
R\&D collaborations.

Interaction is described by the adjacency matrix $\mathbf{A}$ that
contains the elements $a_{ij}$ in terms of $0$ and $1$. This dynamics can
be interpreted as a \emph{catalytic network} of R\&D interactions
(passing a technology to another agent, R\&D collaborations), where the
different agents are represented by nodes, and their interaction by
links between these nodes, cf. Fig. (\ref{example_dirnet}). More
precisely,
\begin{equation}
a_{ij}=\left\{
\begin{array}{l}
1 \mbox{  if agents $i$ transfers knowledge to agent $j$} \\
0 \mbox{  otherwise}
\end{array}\right.
\label{adj}
\end{equation}
We noted already that the network of interactions is modeled on a
directed graph, which means that the adjacency matrix is not generally
symmetric: $a_{ij} \not= a_{ji}$.

The benefit term, $B_i(\mathbf{x},\mathbf{A})$, accounts for the fact
that an agent's value of knowledge increases by receiving knowledge from
other agents. The cost term, $C_i(\mathbf{x},\mathbf{A})$, accounts for
the fact that transferring knowledge to other agents is costly. Such a
cost can vary in magnitude depending on the technological domain, but, in
general, to make someone else proficient in whatever new technology
requires a non-null effort.

In the following we will further specify the growth of the value of in
 (\ref{eq:general_form_productivity_growth}).  We will make simple
assumptions on benefits, $B_i(\mathbf{x},\mathbf{A})$, and costs,
$C_i(\mathbf{x},\mathbf{A})$, which allow us to derive some analytical
results and thus gain some insight on the behavior of the system.

\subsubsection{Pairwise Decomposition}
\label{sec:pairw-decomp}

Networks are sets of pairwise relationships. In systems of interacting
units in physics, a superposition principle holds, such that the force
perceived by a unit is due to the sum of pairwise interactions with other
units. Similarly, one could think of decomposing both benefits and costs
of each agent $i$ in a sum of terms related to the agents $j$ interacting
with $i$. However, this would imply to ignore network
externalities\footnote{In our model we define a network externality as a
  function of the network that affects the utility of an agent.} (it is
very important to note this fact). We will see in the following that
externality does play an important role. So far the literature of Complex
Networks have considered only the pairwise interaction term, while the
literature on economic networks has focused on some simple externalities
such as the network size, or the distance from other agents, see section
(\ref{sec:net_evol_with_cycles}).

Our approach is to assume that benefit and cost are each decomposable in
two terms: one term related to the direct interaction, further
decomposable in pairwise terms, and another term related to externality
(corresponding to positive and negative externality):
\begin{eqnarray}
  B_i(\mathbf{A}, \mathbf{x}) =  \sum_j  b_{ji}(x_j,a_{ji}) +
  b^e_{ji}(x_j,\mathbf{A})\\
  C_i(\mathbf{A}, \mathbf{x}) =  \sum_j  c_{ij}(x_i,a_{ij}) +
  c^e_{ij}(x_i,\mathbf{A}, \mathbf{x})
  \label{eq:utility terms decompos}
\end{eqnarray}
where $b$ stands for benefit, $c$ for cost, $e$ for externality. The
effect of network externalities will be explained in section
(\ref{sec:externalities}). Benefit, $b_{ji}(x_j,a_{ji})$, and cost,
$c_{ij}(x_i,a_{ij})$, terms are monotonically increasing with
the value of knowledge, $x_i$ . They have the the following properties
\begin{eqnarray}
  \label{eq:benefit_cost}
  b_{ji}(x_j,a_{ji}) = \begin{cases} 0 & \text{if } a_{ji}=0 \vee x_j=0
    \\ >0 & \text{if } a_{ji}=1 \wedge x_j>0 \end{cases} \\
  c_{ij}(x_i,a_{ij}) = \begin{cases} 0 & \text{if } a_{ij}=0 \vee
    x_i=0 \\ >0 & \text{if } a_{ij}=1 \wedge x_i>0 \end{cases}
\end{eqnarray}
We assume that benefits are linear functions of the value of knowledge of
agent $i$ which shares its knowledge with agent $j$. We the linear
assumption $b_{ji}(x_j,a_{ji}) = a_{ji} x_j$.

In the most simple case costs for transferring knowledge can be
neglected, $c_{ij}(x_i,a_{ij})=0$. This means that knowledge is fully
codified \citep{foray04:_econom_knowl} and it can be transferred to
another agent without any losses. Further, null costs imply that
knowledge is non-rivalrous, meaning that the value of knowledge is not
reduced by the use of that knowledge by another agent. When costs are
neglected, the growth in the value of knowledge of agent $i$ is given by
the following equation (the case of \textbf{Null Interaction Costs},
further analyzed in section (\ref{sec:null_cost_analysis})).
\begin{equation}
  \frac{dx_i}{dt}= -d x_i + b \sum_{j=1}^{n} a_{ji} x_j 
  \label{eq:growth_null_costs}
\end{equation}
In more realistic setting, costs cannot be neglected. In order to come up
with a reasonable expression for these costs we make some further
assumptions.  We assume that the higher the value of knowledge of an
agent is, the more complex it is.  Moreover the more complex knowledge
is, the more difficult is it to transfer it \citep{sorenson06:_compl,
  rivkin00}.  The coordination and processing capabilities of agents are
constrained (``managerial breakdown''). Thus, the more complex knowledge
gets the higher are the costs for transferring it. The cost,
$c_{ij}(x_i,a_{ij})$, for transferring knowledge from agent $i$ to agent
$j$ is an increasing function of the value of the knowledge that is to be
transferred, $x_i$. We assume that costs increase by more than a
proportional change in the value of knowledge that is being transferred.
\begin{equation}
  c_{ij}(\alpha x_i) > \alpha c_{ij}(x_i)
\end{equation}
This characteristic is closely related to decreasing returns to scale and
convex cost functions\footnote{In the standard economic theory of the
  agent the extent to which a given input can increase output is usually
  assumed to be a decreasing function of the input. The output increases
  at a decreasing rate when the input in production increases
  \citep{hausman03:_inexac_separ_scien_econom}.}. The most simple setting
for such a function is a quadratic term of the form $c_{ij}(x_i, a_{ij})
= c a_{ij} x_i^2$. The growth in the value of knowledge of agent $i$ is
then governed by the following equation (the case of \textbf{Increasing
  Interaction Costs}, further analyzed in section
(\ref{sec:increasing_cost})).
\begin{equation}
  \frac{dx_i}{dt}= -d x_i + b \sum_{j=1}^{n} a_{ji} x_j - c \sum_{j=1}^{n} a_{ij} x_i^2
  \label{eq:growth_quadratic_costs}
\end{equation}
This is an ordinary differential equation with a linear decay, a linear
benefit and quadratic costs.

(\ref{eq:growth_quadratic_costs}) can be interpreted as an extension
of a logistic equation. In a complete graph every agent shares her
knowledge with every other agent. Starting with the same initial values
this symmetry implies, that all knowledge values are identical, i.e.
$x_i=x$. (\ref{eq:growth_quadratic_costs}) then becomes
\begin{eqnarray}
  \begin{array}{ll}
    \frac{dx}{dt} & = - d x + b (n-1) x -  c (n-1) x^2 \\
    & \xrightarrow{\frac{d}{b} \ll n} b(n-1) x \left( 1 - \frac{c}{b} x \right)\\
  \end{array}
  \label{eq:logistic_function}
\end{eqnarray}
(\ref{eq:logistic_function}) is similar to the logistic function
$\dot{x}=\alpha x(1-\frac{x}{\beta})$ with parameters $\alpha=b(n-1)$ and
$\beta=b/c$.

In the following section we relate the topology (cyclic topologies in
particular) of the network with the long-run values of knowledge of the
agents.

\subsection{Non-Permanence of Directed Acyclic Graphs}
\label{sec:dag}

The study of the relation between the performance of an economy and the
underlying network of interactions has already a long tradition, see e.g.
\citet{rosenblatt57:_linear_model_graph_minkow_leont_matric} (``cyclic
nets'').  More recently \citet{maxfield94:_gener_equil_theor_direc_graph}
has shown that the existence of a competitive equilibrium is related to
the strong connectedness of the network of relations between users and
producers in a market economy. Strong connectedness means that there
exists a closed walk or a cycle in the network. On the other hand, if
there does not exist such a cycle, then the network is not strongly
connected. In a similar way in our model strong connectedness is
critically influencing the performance of the agents. In the main result
of this section (\ref{prop:non-permanence}) we show that in our model all
values of knowledge vanish if the underlying network of interactions does
not contain a cycle.

For the general equation (\ref{eq:general_form_productivity_growth}) we
can identify the topology of the network in which agents cannot be
permanent. \citet{hofbauer98:_evolut_games_popul_dynam} give the
following definition of permanence:
\begin{defn}
  A dynamical system is said to be permanent if there exists a
  $\delta > 0$ such that $x_i(0)>0$ for $i=1,...,n$ implies $\lim_{t \to
    \infty}$ $\inf$ $x_i(t) > \delta$.
\end{defn}
First, we have to introduce the definition of graphs which do no contain
any closed walks or cycles.
\begin{definition} A directed acyclic graph is a directed graph
  with no directed cycles.
\end{definition} 
More general, if a graph is a directed acyclic graph then it does not
contain a closed walk. 

For several proofs in this section we need the following lemma (denoted
by the \textit{comparison principle} \citep{khalil95:_nonlin_system}).
\begin{lemma}
  If we consider two time-dependent variables, $x(t)$ and $y(t)$ with
  different growth functions $g(x)$ and $f(x)$ (continuous,
  differentiable) 
 \begin{eqnarray}
   \dot{x}=f(x) \\
   \dot{y}=g(x) \\
   x(0) = y(0)
 \end{eqnarray}
 and $g(x) \ge f(x)$ then it follows that $y(t) \ge x(t)$. Similarly if
 $g(x) \le f(x)$ then $y(t) \le x(t)$.
\label{lem:dominating}
\end{lemma}
\begin{proof}
  Using Cauchy's mean value theorem for the two continuous,
  differentiable functions, $x(t)$ and $y(t)$, we have
  \begin{equation}
    \frac{x'(\tau)}{y'(\tau)} = \frac{x(t)-x_0}{y(t)-y_0} \ge 1
  \end{equation}
  with $\tau \in (0,t)$. The inequality holds since $x'(\tau)=f(x(\tau))
  \ge y'(\tau)=g(y(\tau))$ $\forall \tau \in (0,t)$. It follows that
  \begin{eqnarray}
    x(t) - x_0 \ge y(t) - y_0 \\
    x_0 = y_0 
  \end{eqnarray}
  and thus $x(t) \ge y(t)$. $\Box$
\end{proof}
If a network is a directed acyclic graph then it does not contain a
closed walk. For a directed acyclic graph we can make the following
observation
\begin{prop} In every directed acyclic graph, there is at least one node
  $v$ with no incoming links, i.e. a source.
    \label{prop:source} 
\end{prop} 
\begin{proof}
  \citep{chris01:_algeb_graph_theor} We give a proof by contradiction. We
  assume that every node has an incoming link. We start with some node
  $u$ and find an incoming link $(x,u)$ - by assumption every node has at
  least one incoming link. We go to the destination of the link, $x$.
  Again we can find an incoming link $(y,x)$. We then proceed to node
  $y$. There is an incoming link $(z,y)$. We consider node $z$. After at
  most $n+1$ steps, we will visit some node in the graph twice. This is a
  contradiction to the assumption that the graph is acyclic. $\Box$
\end{proof}
We can partition the nodes in the network into specific sets which take
into account from which other nodes there exists an incoming path to
these nodes. We will show that this is important to obtain a result on
the permanence of the values of knowledge of the agents.
\begin{defn}
  We denote the set of sources of a directed acyclic graph $G$ by $S_0$.
  We say that $S_0$ is the $0$-th order sources of $G$. The nodes that
  have only incoming links from $S_0$ are denoted by $S_1$, the $1$-st
  order sources of $G$. We consider the graph $G \backslash S_0$. The
  nodes that have only incoming links from $S_1$ in $G \backslash S_0$
  (obtained by removing the nodes in $S_0$ and their incident links from
  $G$) are denoted by $S_2$. Accordingly, the nodes having only incoming
  links from $S_{k-1}$ in the graph $G \backslash (S_{k-2} \cup ... \cup
  S_0)$ are denoted by $S_k$, the $k$-th order sources of $G$, where $k
  \le n$.
\label{def:higher_order_sources}
\end{defn}
We can have at most $n$ such sets in the graph $G$ with $n$ nodes. In
this case $G$ is a directed path $P_k$. Moreover we have that
\begin{prop}
  The nodes in a directed acyclic graph $G$ can be partitioned in the
  sets $S_0,S_1,...,S_k$, $k \le n$ defined in
  (\ref{def:higher_order_sources}).
\label{prop:source_partition}
\end{prop}
\begin{proof}
  From proposition (\ref{prop:source}) we know that the directed acyclic
  graph $G$ has at least one source node. All the sources form the set
  $S_0$. If we remove the nodes in $S_0$ (as well as their incident
  links) from $G$ then we obtain again a directed acyclic graph $G_1 := G
  \backslash S_0$ (since the removal of links cannot create cycles).
  Therefore proposition (\ref{prop:source}) also holds for $G_1$. We
  consider the source nodes in $G_1$. These nodes have not been sources
  in $G$ and they have become sources by removing the incident links of
  the sources in $G$. Thus, the source nodes in $G_1$ have only incoming
  links from nodes in $S_0$. Further on, the sources in $G_1$ form the
  set $S_1$. We can now remove the nodes $S_1$ from $G_1$ and obtain the
  graph $G_2$ with new sources $S_2$. We can consider the $k-th$ removal
  of source nodes. We make the induction hypothesis that the sources of
  $G_{k-1}$ form the set $S_{k-1}$. Removing the sources from $G_{k-1}$
  gives a directed acyclic graph $G_k$ which contains the sources $S_k$.
  One can continue this procedure until all nodes have been put into sets
  $S_0,S_1,...,S_k$ with at most $k=n$ sets. $\Box$
\end{proof}
There exists a relationship between the set (defined in
(\ref{def:higher_order_sources})) a node belongs to and the nodes from
which there exists an incoming path to that node.
\begin{corollary}
  Consider a node $i \in S_j$. Then there does not exist a path from
  nodes $k \in S_m$, $m \ge j$, to node $i$. Conversely node $i$ has only
  incoming path from nodes in the sets $S_0,...,S_{j-1}$.
\label{prop:incoming_path}
\end{corollary}
\begin{proof}
  Assume for contradiction that there exists such a path from a node $k
  \in S_m$, $m \ge j$ to a node $i \in S_j$. By the construction of the
  sets $S_j$ (\ref{def:higher_order_sources}) node $i$ must be a source
  with no incoming links after the removal of the sets $S_0,...,S_{j-1}$
  from $G$. But this is a contradiction to the assumption that node $j$
  has an incoming link from a node $k \in S_m$, $m \ge j$. $\Box$
\end{proof}
From the above definition and observations we can derive an upper bound
on the values of knowledge of the nodes in a directed acyclic graph.
\begin{prop}
  Consider (\ref{eq:general_form_productivity_growth}) with a linear
  decay $D_i(x_i) = d x_i$, a linear benefit
  $B_i(\mathbf{A}(G),\mathbf{x}) = b \sum_{j \in N_i^-} x_j$ and a
  non-negative cost $C_i(\mathbf{A}(G),\mathbf{x}) \ge 0$ where $d\ge0$,
  $b\ge0$. Then for every node $i$ in $G$ there exists a $k \le n$ such
  that
  \begin{equation}
    x_i(t) \le (a_{k} t^{k} + a_{k-1} t^{k-1} + ... + a_0) e^{-dt}
  \end{equation}
  \label{prop:sources_bounded}
\end{prop}
\begin{proof}
  From proposition (\ref{prop:source_partition}) we know the the directed
  acyclic graph $G$ has a partition of nodes into sources $S_0,...,S_k$,
  $k \le n$.
  Consider a node $x_0 \in S_0$.  With
  (\ref{eq:general_form_productivity_growth}) the time evolution of her
  value of knowledge is given by
  \begin{equation}
    \dot{x}_0 = - d x_0 - C_0 \le -d x_0 
  \end{equation}
  Here we use the fact that $C_0 \ge 0$. The function solving the
  equation $\dot{x}=-dx$ is an upper bound for $x_0(t)$ (with identical
  initial conditions), see (\ref{lem:dominating}).
  From proposition (\ref{prop:incoming_path}) we know that there are
  first-order sources $S_1$ in $G$ that have only incoming links from
  nodes in $S_0$. The evolution of the value of knowledge for a node $x_1
  \in S_1$ is given by
  \begin{equation}
    \dot{x}_1 = - d x_1 + \sum_{j \in S_0} x_j  - C_1 
    \label{eq:first_sources}
  \end{equation}
  The second term on the right hand side of the above equation contains
  the sum of all values of knowledge of all nodes in $S_0$. We know that
  they are bounded from above by $x(t) \le x(0) e^{-dt}$. Thus,
  (\ref{eq:first_sources}) has an upper bound
  \begin{equation}
    \dot{x}_1 \le - d x_1 + a_1 e^{-dt}
  \end{equation}
  with an appropriate constant $a_1$. The solution of the equation
  $\dot{x} = - d x + a_1 e^{-dt}$ is given by $x(t) = (a_1 + a_0 t)
  e^{-dt}$. It follows that
  \begin{equation}
    x_1 \le (a_1 + a_0 t) e^{-dt}
  \end{equation}
  In the following we make a strong induction. We have the induction
  hypothesis that for the $(k-1)$-th order sources there exists an upper
  bound
  \begin{equation}
    \dot{x}_{k-1} \le (a_{k-1} t^{k-1} + a_{k-2} t^{k-2} + ... + a_0)
    e^{-dt}
    \label{eq:k-1_bound}
  \end{equation}
  and this holds also for all nodes in the sets of sources with order
  less than $k-1$. We consider the nodes in $S_k$ with $l \in S_k$. We
  have that 
  \begin{equation}
    \dot{x}_l(t) = -d x_l + b \sum_{j \in N_l^-} x_j - C_l 
    \label{eq:k_bound}
  \end{equation}
  where the in-neighborhood $N_l^-$ contains only nodes in the sets
  $S_0,...,S_{k-1}$. For these nodes an upper bound is given by
  (\ref{eq:k-1_bound}) and thus we get an upper bound for
  (\ref{eq:k_bound})
  \begin{equation}
    \dot{x}_l(t) \le -d x_l + (a_{k-1} t^{k-1} + a_{k-2} t^{k-2} + ... + a_0)
    e^{-dt}
  \end{equation}
  We can now use the following lemma
  \begin{lemma}
    For an ordinary differential equation of the form
    \begin{equation}
      \dot{y} + d y = (a_k t^k + a_{k-1} t^{k-1} + ... + a_2 t + a_1)
      e^{-dt}
      \label{eq:general-ode}
    \end{equation}
    there exists a solution of the form
    \begin{equation}
      y(t) = \left( \frac{a_k}{k+1} t^{k+1}+...+a_0 \right) e^{-dt}
      \label{eq:sol}
    \end{equation}
    with the limit $\lim_{t \to \infty} y(t) = 0$
    \label{lem:ode}
  \end{lemma}
  Solving for the upper bound from above gives the desired result.
  \begin{equation}
    x_l(t) \le (a_{k} t^{k} + a_{k-1} t^{k-1} + ... + a_0) e^{-dt}
  \end{equation}
  $\Box$
\end{proof}
With the last proposition (\ref{prop:sources_bounded}) it is
straightforward to obtain the following corollary, which is the main
result of this section.
\begin{corollary}
  Consider (\ref{eq:general_form_productivity_growth}) with a linear
  decay $D_i(x_i) = d x_i$, a linear benefit
  $B_i(\mathbf{A}(G),\mathbf{x}) = b \sum_{j \in N_i^-} x_j$ and a
  non-negative cost $C_i(\mathbf{A}(G),\mathbf{x}) \ge 0$ where $d\ge0$,
  $b\ge0$. If the network $G$ is a directed acyclic graph then the values
  of knowledge vanish. This means that $G$ is not permanent.
\label{prop:non-permanence}
\end{corollary}
\begin{proof}
  From proposition (\ref{prop:sources_bounded}) we know that each node
  $k$ in the graph $G$ has a value of knowledge which is bounded by
  $x_k(t) \le (a_{k} t^{k} + a_{k-1} t^{k-1} + ... + a_0) e^{-dt}$ for
  some finite $k \le n$. Since any finite polynomial grows less than an
  exponential function we have that $\lim_{t \to \infty} x_k(t)=0$.  This
  holds for all nodes in $G$. This completes the proof that for all
  $i=1,...,n$ in a directed acyclic graph $G$ we have that $\lim_{t \to
    \infty} x_i(t)=0$ and therefore $G$ is not permanent. $\Box$
\end{proof}
Thus, if agents are permanent, the graph contains a closed walk (or a
cycle). If agents get their links attached at random, only those survive,
who are part of a cycle. If agents can chose, whom to transfer their
knowledge to, then they have to form cycles, in order to survive. Others
\citep{goyal00:_Noncooperative_model_network_formation, kima07:_networ}
have found similar results in which the equilibrium network consists of
cycles.

There exist a convenient way to identify if a network contains a cycle
without actually looking at the permanence of the network which would
require to compute the long-run values of knowledge (usually by numerical
integration). Instead, from the eigenvalues of the adjacency matrix,
$\mathbf{A}(G)$, of a graph, $G$, one can determine if $G$ contains a
cycle. The Perron-Frobenius eigenvalue of a graph $G$, denoted by
$\lambda_{\text{PF}}(G)$, is the largest real eigenvalue of
$\mathbf{A}(G)$. The following properties hold
\citep{chris01:_algeb_graph_theor}
\begin{prop}
  If a graph $G$
  \begin{enumerate}
  \item has no closed walk, then $\lambda_{\text{PF}}(G) = 0$,
  \item has a closed walk, then $\lambda_{\text{PF}}(G) > 1$. 
  \end{enumerate}
\end{prop}
Thus, if the graph contains permanent agents, then
$\lambda_{\text{PF}}(G) > 1$.  \citet{stadler96permanence,
  hofbauer98:_evolut_games_popul_dynam} have found similar conditions
under which populations are permanent in a network of
replicators\footnote{The replicator equation (in continuous form) is
  given by: $\dot{x}_i = x_i \left( f_i(\mathbf{x}) - \phi(\mathbf{x})
  \right)$ where $\phi(\mathbf{x}) = \sum_i x_i f_i(\mathbf{x})$ and
  $f_i(\mathbf{x})$ is the fitness of species $i$.}.

Finally, we can compute the probability of a network to contain a cycle
if links were attached at random.
\begin{prop}
  The probability of a random graph $G(n,p)$ with $n$ nodes containing a
  cycle is given by \citep{jain02:_graph_theor_evolut_autoc_networ}
  \begin{equation}
    P=\left(1-(1-p)^{n-1}\right)^n
    \label{eq:prob_acs_rand_graph}
  \end{equation}
   which is $0$ if $p=0$ and $1$ if $p=1$.
  \label{prop:prob_acs}
\end{prop}
\begin{proof}
  We can compute the probability of having a closed walk in a random
  graph $G(n,p)$. Each link is created with probability $p$. Thus we have
  a Bernoulli process for the adjacency matrix elements $a_{ij}$ (which
  indicate if an link exists or not).
  \begin{equation}
    a_{ij}=\begin{cases} 1 & \text{with probability } p \\ 0 & \text{with
        probability } 1-p \end{cases}
  \end{equation}
  For every node we have $n-1$ events to create an link and we are asking
  for the probability of having at least one of them being created (every
  node should have at least one incoming link). This is a binomial
  cumulative function of the form \citep{durrett04:_probab,
    casella01:_statis_infer}
  \begin{equation}
    P=\sum_{k=1}^{n-1} {n \choose k} p^k (1-p)^{n-k}
  \end{equation}
  which is equivalent to eq. (\ref{eq:prob_acs_rand_graph}), if we use
  the Binomial theorem
  \begin{equation}
    (x+y)^n = \sum_{i=1}^n {n \choose i} x^i y ^{n-i}
  \end{equation}
  $\Box$
\end{proof}
A similar result to (\ref{prop:non-permanence}) has been found by
\citet{kima07:_networ}. The authors study a generalized version of the
network formation model introduced by
\citet{goyal00:_Noncooperative_model_network_formation}\footnote{For a
  further study of
  \citet{goyal00:_Noncooperative_model_network_formation} applied to
  information networks see \citet{haller05:_nash_networ_heter_links,
    haller07:_non_nash}.}. The equilibrium networks in their model are so
called ``minimal'' graphs which are graphs that maximize the number of
agents that are connected while maintaining only as few links as
possible. It is intuitively clear, that the most sparse connected graph
is a cycle. Thus, the authors find stable equilibrium networks that
consist of cycles. However, in section \ref{sec:source_squared} we will
show that the network evolution can reduce the set of possible cycles in
the equilibrium network such that only the smallest cycles survive.

Thus, cycles play an important role in the evolution of the network and
the ability of agents to have non-vanishing knowledge levels. Before we
define the evolution of the network in section
(\ref{sec:evolution-links}) we study the dynamics of the values of
knowledge for a static network in the next section
(\ref{sec:static_network_analysis}). There we will further specify the
cost functions under investigation: null costs and nonlinear costs for
maintaining links.

\subsection{Static Network Analysis}
\label{sec:static_network_analysis}

In the following we analyze the growth functions for the value of
knowledge and study two cases separately. In the first, costs are set to
zero while in the second costs are a quadratic function of the values of
knowledge of the agents.

\subsubsection{Null Interaction Costs}
\label{sec:null_cost_analysis}

The most simple case of our general framework is the one of linear
benefit and null costs\footnote{This model has been studied by
  \citet{jain98:_emerg_growt_compl_networ_adapt_system,
    krishna03:_form_destr_acs_evolv_netw_mod} to explain the origin of
  life from the perspective of interacting agents. The model of Jain and
  Krishna intends to describe the catalytic processes in a network of
  molecular species (which we will denote in the following by agents).
  However, it was very soon suggested to be applicable to an economic
  innovation context of interacting agents. In the next sections we will
  present a more general framework encompassing some of the limitation of
  the present one. In their model the $\mathbf{x}$ were interpreted as
  concentrations of chemical species.  The $a_{ij}$ are the kinetic
  coefficients that describe the replication of agents $i$ resulting from
  \emph{binary} interactions with other agents $j$.}.
\begin{eqnarray}
  \frac{dx_i}{dt}= -d x_i + \sum_{i=1}^{n} a_{ji} x_j
\label{eq:growth_null_cost}
\end{eqnarray}
In vector notation (\ref{eq:growth_null_cost}) reads:
\begin{equation}
\mathbf{\dot{x}} = ( \mathbf{A}^T - d \mathbf{I} ) \mathbf{x} 
\label{eq:growth_null_cost_matrix}
\end{equation}
where $A^T$ is the transposed of the adjacency matrix and $\mathbf{I}$ is
the identity matrix.  The solution of the set of equations
(\ref{eq:growth_null_cost_matrix}) depends on the properties of the
matrix ${A}$ and has the general form (matrix exponential):
\begin{equation}
\mathbf{x}(t) = e^{-dt} e^{\mathbf{A}^Tt} \mathbf{x}(0)
\end{equation}
representing an exponential increase in time of the vector of knowledge
values. The relative values of knowledge (shares) are given by
\begin{equation}
  \label{eq:knowledge_shares}
 y_i=\frac{x_i}{\sum_j x_j}\;; \quad \sum_j y_j = 1
\end{equation}
Rewriting (\ref{eq:growth_null_cost}) by means of 
(\ref{eq:knowledge_shares}) gives us the dynamics of the shares:
\begin{equation}
\dot{y_i} = \sum_{j}^{n} a_{ji} y_j - y_i \sum_{k,j}^{n} a_{jk} y_j
\label{eq:knowledge_shares_dynamics}
\end{equation}
(\ref{eq:knowledge_shares_dynamics}) has the property of preserving the
normalization of $\mathbf{y}$. Note that the decay term does not appear
in this equation for the relative values. It can be shown
\citep{horn90:_matrix_analy, boyd:_linear_dynam_system,
  krishna03:_form_destr_acs_evolv_netw_mod} that the eigenvector to the
largest real eigenvalue of $\mathbf{A}^T$ ($\mathbf{A}$ respectively) is
the stable fixed point of
(\ref{eq:knowledge_shares_dynamics})\footnote{If the largest real
  eigenvalue has multiplicity more than one then the stable fixed point
  can be written as a linear combination of the associated eigenvector
  and generalized eigenvectors
  \citep{braun93:_differ_equat_their_applic}.}. If we consider an
eigenvector $\mathbf{y}^{(\lambda)}$ associated with the largest real
eigenvalue $\lambda$ of matrix $\mathbf{A}^T$ (identical to the largest
real eigenvalue of $\mathbf{A}$) we have
\begin{equation}
  \sum_{j=1}^n a_{ji} y_j^{(\lambda)} = \lambda y_i^{(\lambda)} 
\end{equation}
Inserting $\mathbf{y}^{(\lambda)}$ into
(\ref{eq:knowledge_shares_dynamics}) yields
\begin{eqnarray}
  \dot{y}_i^{(\lambda)} = \sum_{j}^{n} a_{ji} y_j^{(\lambda)} - y_i^{(\lambda)}
  \sum_{k,j=1}^{n} a_{jk} y_j^{(\lambda)} \\
  = \lambda y_i^{(\lambda)} - y_i^{(\lambda)} \underbrace{\sum_{k,j=1}^{n}
    a_{jk} y_j^{(\lambda)}}_{\lambda \sum_{k}^{n} y_k^{(\lambda)} = \lambda }
  \\
  = \lambda y_i^{(\lambda)} - \lambda y_i^{(\lambda)} = 0
\end{eqnarray}
Thus, $y_i^{(\lambda)}$ is a stationary solution of 
(\ref{eq:knowledge_shares_dynamics}). For the proof of stability see e.g.
\citet{krishna03:_form_destr_acs_evolv_netw_mod}.

\subsubsection{Increasing Interaction Costs}
\label{sec:increasing_cost}

In the following we study the evolution of the values of knowledge under
a given network structure and we try to compute the fixed points where
ever possible.  We first show that the values of knowledge are
non-negative and bounded.  For graphs with $2$ nodes, for regular graphs
(including the complete graph), cycles and stars with an arbitrary number
of nodes, we can compute the equililibrium points analytically. For
generic graphs with $n \ge 3$ nodes we have to rely on numerical
integrations.

The nonlinear (quadratic) dynamical system is given by
\begin{equation}
  \label{eq:dyn_sys}
  \dot{x_i} = -d x_i + b \sum_{j=1}^n a_{ji} x_j - c \sum_{j=1}^n a_{ij} x_i^2
\end{equation}
with initial conditions, $x_{i}(0) > 0$. $a_{ij}$ are the elements of the
adjacency matrix, $\mathbf{A}$, of a graph $G$. This can be written as
\begin{equation}
  \dot{x_i} = -d x_i + b \sum_{j=1}^n a_{ji} x_j - c d_i^+ x_i^2
\end{equation}
where $d_i^+ = \sum_{j=1}^n a_{ij}$ is the outdegree of node $i$. In the
case of increasing costs we know that the values of knowledge are
bounded. We have that
\begin{prop}
  For the dynamical system (\ref{eq:dyn_sys}) the values of knowledge are
  non-negative and finite, i.e. $0 \le x_i < \infty$, $i=1,...,n$.
\label{prop:bounded}
\end{prop}
\begin{proof}
  For the lower bound $x_i \ge 0$ we observe that 
  \begin{equation}
    \dot{x}_i \ge -d x_i -c(n-1)x_i^2
  \end{equation}
  The lower bound is the solution of the equation
  $\dot{x}=-dx-c(n-1)x^2$. The solution of this equation can be found by
  solving the corresponding equation for the transformed variable
  $z=\frac{1}{x}$. We get $x(t)=\frac{de^{da}}{e^{dt}-c(n-1)e^{da}}$ with
  an appropriate constant $a=\frac{1}{d} \ln \frac{x(0)}{d+(n-1)c}$.
  Starting from non-negative initial values $x(0) \ge 0$ this lower bound
  is non-negative as well and approaches null for large $t$, i.e.
  $\lim_{t \to \infty} x(t)=0$. We conclude that $x_i(t) \ge 0$.

  In order to compute an upper bound, $x_i \le$ const.$<\infty$ we first
  make the following observation. The nodes of a graph, $G=(V,E)$, can
  be partitioned in nodes without outgoing links, $V_f \subseteq V$
  (``free-riders''), and nodes with at least one outgoing link, $V_s
  \subseteq V$ (sources).

  Since the ``free-riders'' in $V_f$ have no outgoing links, the benefit
  terms of the sources in $V_s$ are independent of the values of
  knowledge of the free-riders. Accordingly a source node $i \in V_s$ has
  the following knowledge dynamics.
  \begin{equation}
    \dot{x}_i = - dx_i + b \sum_{j \in V_s \backslash i} a_{ji} x_j - c x_i^2 d^+_i
  \end{equation}
  where $d_i^+$ is the out-degree of node $i$. We can give an upper bound
  of
  \begin{equation}
    \dot{x}_i \le -d x_i + b \sum_{j \in V_s} x_j - c x_i^2
  \end{equation}
  This upper bound has a (finite) fixed point and so does $x_i(t)$. The
  fixed point is given by
  \begin{equation}
    d x_i + c x_i^2 = b \sum_{j \in V_s} x_j
  \end{equation}
  This is a symmetric equation and therefore all $x_i$ are identical,
  $x_i=x$. For contradiction assume that there would be $x_i \ne x_j$.
  Then we have that
  \begin{equation}
    \underbrace{d x_i + c x_i^2}_{b \sum_{k=1}^n x_k} \ne \underbrace{d x_j + c x_j^2}_{b \sum_{k=1}^n x_k}
  \end{equation}
  But the left and right side of the equation are identical and so two
  different $x_i,x_j$ cannot exist. 

  When all solutions are identical we get $x_i=x=\frac{bn-d}{c}$ $\forall
  i$. Thus, we have shown that there exists an upper bound with a finite
  fixed point for the source nodes, that is $x_i(t)\le \infty$, $i \in
  V_s$.

  We now consider the nodes with no outgoing links (``free-riders''). A
  node $i \in V_f$ follows the dynamics
  \begin{equation}
    \dot{x}_i = - dx_i + b \sum_{j \in V_s} a_{ji}  x_j
  \end{equation}
  We have shown already that the source nodes are bounded by some
  constant, $\sum_{j \in V_s} x_j \le$ const.. Thus we have that
  \begin{equation}
    \dot{x}_i \le - d x_i + \text{const.}
  \end{equation}
  We have an upper bound of the $x_i$, $i \in V_f$, given by
  \begin{equation}
    x_i(t) \le x_0 e^{-dt} + \frac{\text{const.}}{d}
  \end{equation}
  with $\lim_{t \to \infty} x_i(t) =\frac{\text{const.}}{d}$. We have
  shown that for all nodes (sources $V_s$ as well as ``free-riders''
  $V_f$) $0 \le x_i < \infty$, $i \in V(G)$. $\Box$
\end{proof}
For special types of graphs we can deduce further results on the values
of knowledge of the agents. First, we can compute the fixed points (given
by $\dot{x}_i=0$) for regular graphs.
\begin{prop}
  For any $k-$regular graph $G$ the fixed point of the values of
  knowledge is given by $x^*=\frac{kb-d}{kc}$. In particular the complete
  graph $K_n$ has the highest total value of knowledge among all regular
  graphs with $x^*=\frac{(n-1)b-d}{(n-1)c}$.
\label{prop:regular}
\end{prop}
\begin{proof}
  The dynamics of the values of knowledge of the nodes in a regular graph
  with degree $d_i^+=d_i^-=k$ is given by
  \begin{equation}
    \dot{x}_i = -d x_i+b\sum_{j \in N_i} x_j -ckx_i^2
  \end{equation}
  Starting with homogeneous initial conditions we make the Ansatz $x_i=x$
  $i=1,...,n$. We get the positive stable fixed points
  $x^*=\frac{kb-d}{kc}$.  $\Box$
\end{proof}
Second, we can compute the fixed points for cycles.
\begin{prop}
  For any cycle $C_n$ the fixed point of the values of knowledge is given
  by $x^*=\frac{b-d}{c}$.
  \label{prop:cycle}
\end{prop}
\begin{proof}
  The dynamics of the values of knowledge of a cycle $C_k$ of length $k$
  is given by
  \begin{equation}
    \dot{x}_i = -d x_i + b x_{i-1} -ckx_i^2
  \end{equation}
  Starting with homogeneous conditions we make the Ansatz $x_i=x$
  $i=1,...,n$. We get the positive stable fixed points
  $x^*=\frac{b-d}{c}$.  $\Box$
\end{proof}
Third, the fixed points for a star can be computed (the proof can be
found in the appendix).
\begin{prop}
  For a star $K_{n,n-1}$ there exists a fixed point which increases with
  the number of nodes. For $d=0$ the star has a fixed point of
  $x^*=\frac{b}{c}$.
  \label{prop:star}
\end{prop}
\begin{proof}
  The dynamics of the values of knowledge of a star $K_{1,n-1}$ is given by
  \begin{eqnarray}
    \dot{x}_1 = +b \sum_{i=2}^n x_i - c (n-1) x_i^2\\
    (\dot{x}_{i})_{i > 1} = -dx_i + b x_1 - c x_i^2
  \end{eqnarray}
  where we assume that all links are bidirectional. Starting with
  homogeneous initial conditions we make the Ansatz $x_i=x_2$
  $i=2,...,n$. Then $x_2$ is determined by the root of the polynomial
  \begin{equation}
    x_2^3+\frac{2d}{c}x_2^2+\frac{d(cb+(n-1)cd)}{(n-1)c^3}x_2+\frac{b(d^2-(n-1)b^2)}{(n-1)c^3}=0
  \end{equation}
  And $x_1=\frac{d}{b}x_2+\frac{c}{b}x_2^2$. For $d=0$ we obtain
  $x_1^*=x_2^*=\frac{b}{c}$.  $\Box$
\end{proof}
The fixed point increases with the benefit $b$ and decreases with the
decay $d$ and the cost $c$. 

We observe that, for vanishing decay, $d=0$, the fixed point of the
system is identical for the regular graph, the cycle and the star and
given by $\frac{b}{c}$. As expected this fixed point is increasing with
the benefit and decreasing with the cost. In a regular graph the fixed
point is increasing with the degree $k$ and the asymptotic value (for
large $k$) is $\frac{b}{c}$. Thus, in a regular graph the fixed point
ranges for increasing $k$ from $\frac{b-d}{c}$ to $\frac{b}{c}$.
Similarly, for the star the fixed point also increases with the number of
nodes (i.e. the degree of the central node) but we can not provide an
analytical expression here. On the other hand, the fixed point of the
cycle is independent of the length of the cycle. This means that there is
no incentive for nodes to be part of larger cycles. And, as we will see
in the next section, this limits the growth of the network.

\begin{example}
  We numerically integrate (\ref{eq:dyn_sys}) for $n=2$ nodes.  We set
  $d=0.5$, $c=0.5$ and $b=1$. Fixed points are denoted $x_i^*$ for
  $i=1,2$. $x_i^*=0$ is a fixed point for all graphs.
\end{example}
\begin{minipage}{.5\textwidth}
  \scalebox{0.6}{ \begin{pspicture}(0,-1)(3,2)
      \psset{nodesep=0.5pt,linewidth=1.5pt,arrowsize=5pt 2}
      \rput(0,1){\Large (1)} \cnodeput(0,0){1}{\Large 1}
      \cnodeput(3,0){2}{\Large 2}
      \ncarc[linewidth=1pt,arcangle=10]{->}{1}{2}
      \ncarc[linewidth=1pt,arcangle=10]{->}{2}{1}
    \end{pspicture}}
  \begin{displaymath}
    A_1=\left(
      \begin{array}{cc}
        0 & 1 \\
        1 & 0 \\
      \end{array}
    \right)
  \end{displaymath}
\end{minipage}
\begin{minipage}{.4\linewidth}
  \psfrag{x1}[c][][3][0]{$x_1$} 
  \psfrag{x2}[c][][3][0]{$x_2$} 
  \psfrag{xi}[c][][4][0]{$x_i$} 
  \psfrag{time}[c][][4][0]{$t$} 
  \centerline{\scalebox{0.3}{\includegraphics[angle=0]
      {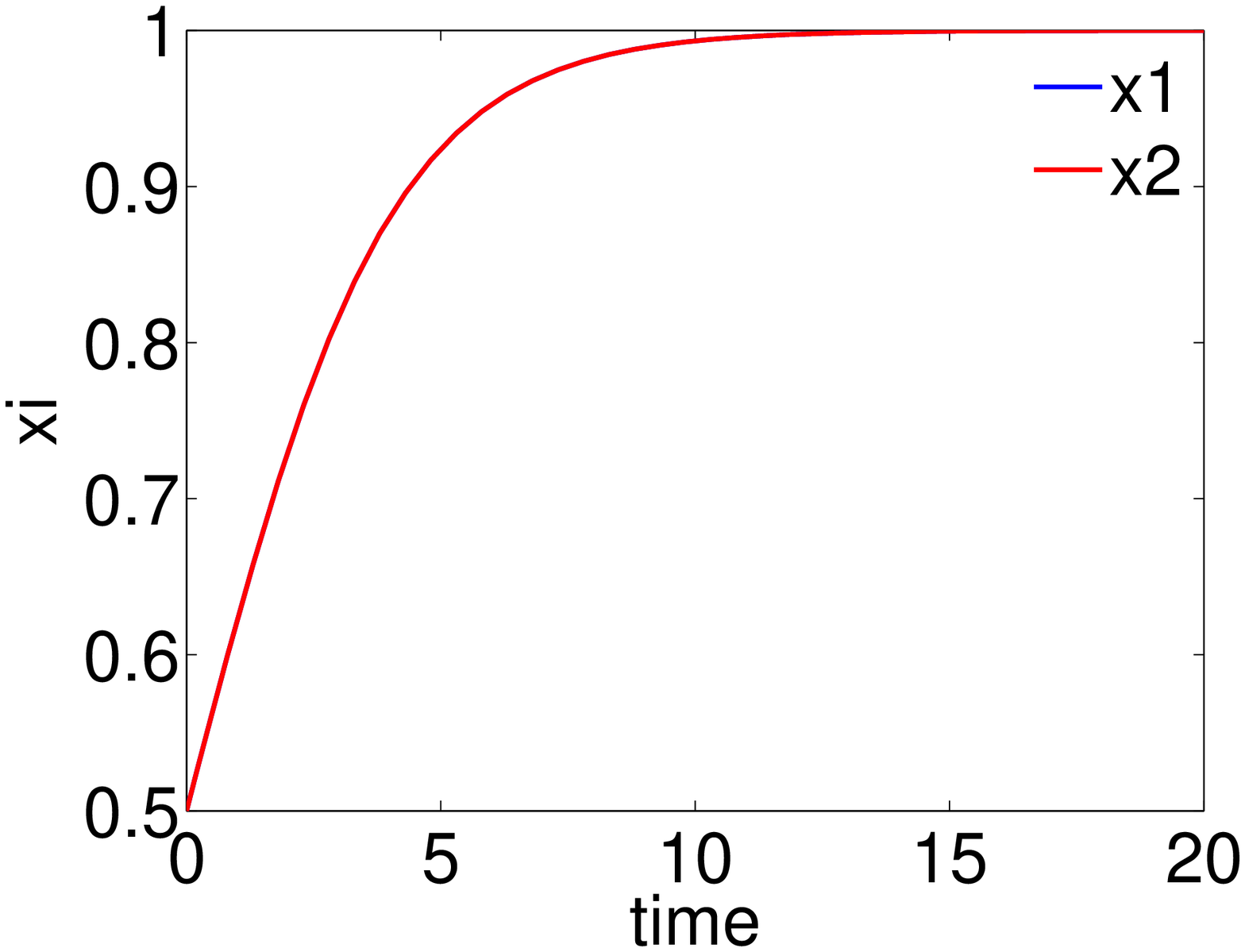}}}
\end{minipage}

The fixed points are given by
\begin{equation}
  \label{eq:fixed_point_2G1}
  x_i^*=\frac{b-d}{c} \text{,   } i=1,2,3
\end{equation}

\begin{minipage}{.5\textwidth}
\scalebox{0.6}{\begin{pspicture}(0,-1)(3,2) 
  \psset{nodesep=0.5pt,linewidth=1.5pt,arrowsize=5pt 2} 
  \rput(0,1){\Large (2)}
  \cnodeput(0,0){1}{\Large 1}
  \cnodeput(3,0){2}{\Large 2} 
  \ncarc[linewidth=1pt,arcangle=10]{->}{1}{2}
\end{pspicture}}
\begin{displaymath}
A_2=\left(
\begin{array}{cc}
0 & 1 \\
0 & 0 \\
\end{array}
\right)
\end{displaymath}
\end{minipage}
\begin{minipage}{.4\linewidth}
  \psfrag{x1}[c][][3][0]{$x_1$} 
  \psfrag{x2}[c][][3][0]{$x_2$} 
  \psfrag{xi}[c][][4][0]{$x_i$} 
  \psfrag{time}[c][][4][0]{$t$}
    \centerline{\scalebox{0.3}{\includegraphics[angle=0]
        {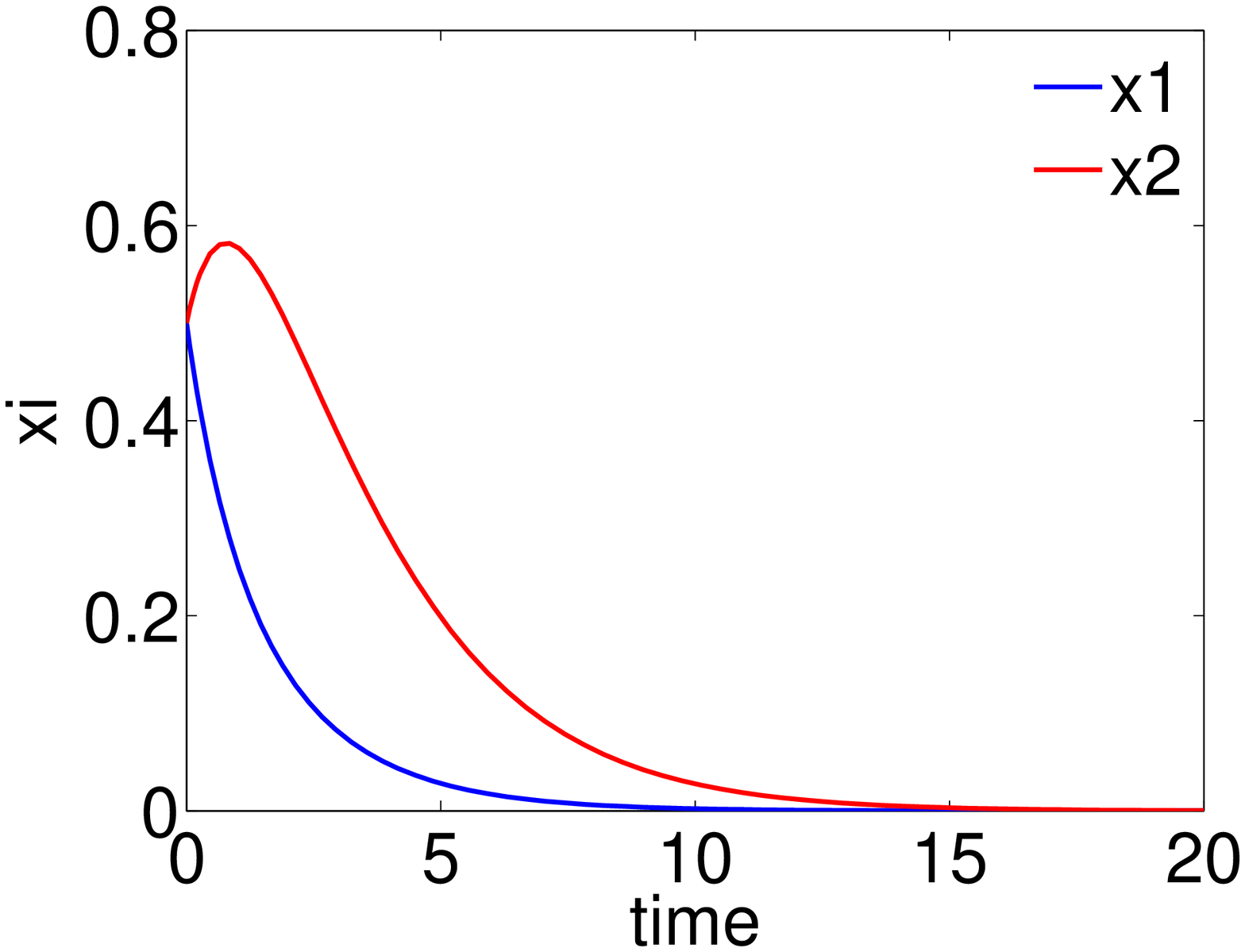}}}
\end{minipage}

The fixed points are given by
\begin{equation}
  \label{eq:fixed_point_2G2}
  x_i^*=0 \text{,   } i=1,2,3
\end{equation}

\begin{minipage}{.5\textwidth}
\scalebox{0.6}{\begin{pspicture}(0,-1)(3,2) 
  \psset{nodesep=0.5pt,linewidth=1.5pt,arrowsize=5pt 2} 
  \rput(0,1){\Large (3)}
  \cnodeput(0,0){1}{\Large 1}
  \cnodeput(3,0){2}{\Large 2} 
\end{pspicture}}
\begin{displaymath}
A_3=\left(
\begin{array}{cc}
0 & 0 \\
0 & 0 \\
\end{array}
\right)
\end{displaymath}
\end{minipage}
\begin{minipage}{.4\linewidth}
  \psfrag{x1}[c][][3][0]{$x_1$} 
  \psfrag{x2}[c][][3][0]{$x_2$} 
  \psfrag{xi}[c][][4][0]{$x_i$} 
  \psfrag{time}[c][][4][0]{$t$}
    \centerline{\scalebox{0.3}{\includegraphics[angle=0]
        {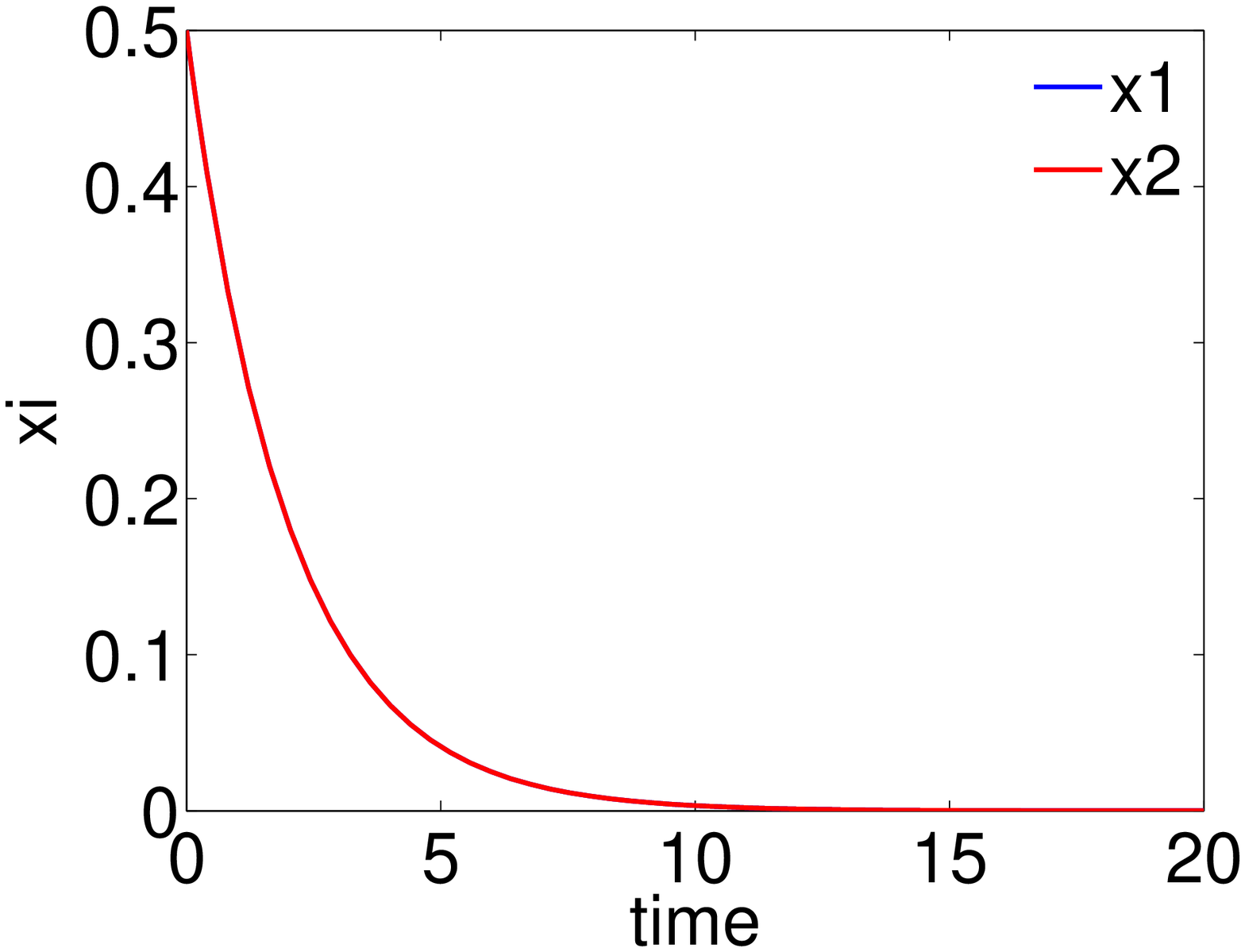}}}
\end{minipage}

The fixed points are given by
\begin{equation}
  \label{eq:fixed_point_2G3}
  x_i^*=0 \text{,   } i=1,2,3
\end{equation}
In general the fixed points of (\ref{eq:dyn_sys}) can only be computed
numerically. As an example, we compute the fixed points for all graphs
with $n=3$ nodes for a specific choice of parameters. The results can be
found in the appendix in section (\ref{sec:stationary_solutions_G3}). In
our model we numerically integrate (\ref{eq:dyn_sys}) for a large
time $T$ (and we find that in our simulations the system always reaches a
stable fixed point).

\section{Dynamics of Network Evolution}
\label{sec:evolution-links}

\subsection{Network Evolution as an Iterative Process}

After providing the static equilibrium analysis, in this section we turn
now to the \emph{dynamics} of the network evolution by investigating
different assumptions for the \emph{creation} and \emph{deletion} of
links in the network. In particular, we compare two different scenarios,
namely the so-called extremal dynamics, where agents do not decide
themselves about the link creation and deletion, and the utility driven
dynamics, where agents make this decision themselves based on different
rules discussed below.

We first define the utility of the agents in our model for a given
network $G$.

\begin{defn}
  Consider a (static) network $G$. The utility of agent $i$ is given by
  \begin{equation}
    u_i=
    \begin{cases}
      y_i(T), & \text{for Null Interaction Costs} \\
      x_i(T), & \text{for Increasing Interaction Costs}
    \end{cases}
  \end{equation}
  where the value of knowledge $x_i(t)$ is given by (\ref{eq:dyn_sys})
  and $\mathbf{A}(G)$, the relative value of knowledge $y_i(t)$ by
  (\ref{eq:knowledge_shares_dynamics}) and $\mathbf{A}(G)$. $T$ is called
  the time horizon.
  \label{defn:utility}
\end{defn}

We assume that the accumulation of knowledge is faster than the frequency
of the agents creating or deleting links\footnote{This means that the
  value of knowledge on the market (which is not explicitly modeled here)
  reaches a stationary state determined by the R\&D collaborations of
  each agent (and her neighbors). Only after this adaptation of the
  evaluation of the stocks of knowledge is finished, i.e. it has reached
  a stationary state, agents asynchronously change their links.}. With
this assumption, we can introduce a \emph{time-scale separation} between
the accumulation of knowledge and the evolution of the network.

\begin{figure}[htpb]
\begin{center}
  \begin{pspicture}(0,1)(6,4)
   \rput(1,3.5){\rnode{A}{\psframebox{
         \begin{tabular}{c}
           \textbf{initialization}
         \end{tabular}
   }}}
   \rput(1,2){\rnode{B}{\psframebox{
	\begin{tabular}{c}
		$x_i$ reach \\ 
		\textbf{quasi-equilibrium}
	\end{tabular}
   }}}
   \rput(5,2){\rnode{C}{\psframebox{
	\begin{tabular}{c}
		\textbf{perturbation}\\
		of $ a_{ij}$
	\end{tabular}
   }}}
   \ncline[linewidth=0.5pt,linestyle=dotted]{->}{A}{B}
   \ncarc[linewidth=1pt,arcangle=40]{->}{B}{C}
   \ncarc[linewidth=1pt,arcangle=40]{->}{C}{B}
  \end{pspicture}
\end{center}
\caption{Schematic representation of the network evolution as an
  iterative process. \label{fig:schema}}
\end{figure}
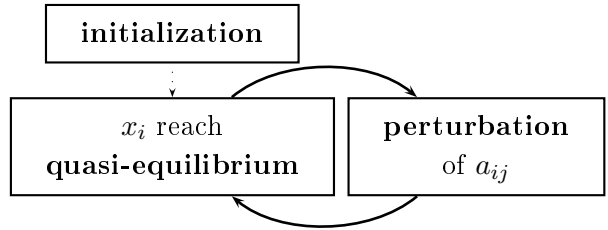
The evolution of the system is then defined by an alternating sequence of
knowledge accumulation, where we keep the network fixed for a given time
$T$, $\mathbf{A}(G)=$const., and changes in the links (asynchronous
updating of the nodes) (see Fig. \ref{fig:schema}). When the knowledge
accumulation has reached time $T$, the network structure is changed. A
change in the network takes place by either link addition between two
agents $i$ and $j$, $a_{ij}=0 \rightarrow a_{ij}=1$, or by link removal,
$a_{ij}=1 \rightarrow a_{ij}=0$. When the network has changed, the new
utility, determined by (\ref{defn:utility}), can be computed for time
$2T$. This iterative procedure of knowledge accumulation and link changes
continues for $3T,4T,...$ and so on until the network reaches an
equilibrium. One can schematically represent this iteration by the
following algorithm:
\begin{itemize}
\item[\textbf{1}] initialization: Random graph $G(n,p)$.
 
\item[\textbf{2}] \textbf{quasi-equilibrium}: fast knowledge
  growth/decline
 
  With $\mathbf{A}$ fixed, agents evolve according to
  (\ref{eq:general_form_productivity_growth}) for a given (large) time
  $T$.
  
\item[\textbf{3}] \textbf{perturbation}: slow network evolution
 
  After time $T$, the network evolves according to two alternative
  selection processes:
  \begin{enumerate}
  \item \textit{Extremal Dynamics\footnote{See
        \citet{bak93:_punct_equil_critic_simpl_model_evolut}}.}

    The agent with the minimum utility is chosen (if there are more than
    one agent with the same minimum value, then one of them is chosen at
    random). The utility of that agent is set to its initial value and
    all its outgoing and ingoing links are replaced with new random links
    drawn with probability $p$ from and to all other agents in the
    system.
       
  \item \textit{Utility Driven Dynamics}

    An agent is randomly chosen to create or delete one link
    (unidirectional or bidirectional link formation mechanisms, see
    section \ref{sec:local_link_formation}). More specifically:
 
    \begin{itemize}
      
    \item[(i)] Either a pair or a single agent is randomly chosen to
      create or remove a link.
      
    \item[(ii)] The effect of this link decision (creation or deletion)
      is evaluated at time $T$. The evaluation can have the following
      consequences on the link decision.
        
      \begin{itemize}
      \item If the utility has increased, then sustain the link
        decision.
      \item If the utility has decreased, then undo the link
        decision.
      \end{itemize}

    \end{itemize}

  \end{enumerate}

\item[\textbf{4}] Stop the evolution, if the network is stable (stability
  is defined in section (\ref{sec:local_link_formation}), otherwise go to
  \textbf{2}
\end{itemize}

\subsection{Extremal Dynamics versus Utility Driven Dynamics}
\label{sec:global-vs-local}

Extremal dynamics intends to mimic natural selection (the extinction of
the weakest) and the introduction of novelty, which is a global selection
mechanism. In contrast, utility driven dynamics is a local selection
mechanism that mimics the process by which selfish agents improve their
utility through a trial and error process.

The decision upon to add or to remove a link implies a certain level of
information processing capabilities (IPC) of the agents. IPC is usually
bounded in a complex environment consisting of many other agents and a
complex structure of interactions between these agents. In our approach
we assume that the agents have no information on the knowledge values of
the other agents and only limited information on their links (alliances).
They only know with whom they interact directly (their neighborhood). In
table (\ref{tab:ipc}) we give a short overview of levels of increasing
IPC.

\begin{center}
  \begin{table}[htbp]
    \begin{center}
      \begin{tabular}{|l||l|}
        \hline
        \textbf{0} & Least fit addition/removal of links, e.g
        \citet{jain98:_emerg_growt_compl_networ_adapt_system} \\
        \hline
        \textbf{1} & Reactive (passive) acceptation/refusal of link changes.
        \\
        \hline
        \textbf{2} & Deliberate decision upon to add/remove a link based on
        a utility function \\
        & depending on the network, e.g.
        \citet{bornholdt02:_evolutionary_games}, without considering \\
        &  the possible decision of others.\\
        \hline
        \textbf{3} & Strategic interaction, e.g.
        \citet{goyal00:_Noncooperative_model_network_formation}, considering
        the possible \\
        & actions of others. \\
        \hline
      \end{tabular}
    \end{center}

    \label{tab:ipc}
    \caption{Increasing levels of
      information processing capabilities (IPC) of agents.}
  \end{table}
\end{center}

  Extremal Dynamics refers to a situation in which agents are exposed to
  link changes they cannot influence and thus to level $0$ in table
  (\ref{tab:ipc}). Utility Driven Dynamics instead requires a higher
  level of IPC than a mere acceptation or refusal of link changes. But it
  requires less IPC than an approach assuming strategic interactions of
  agents. This follows from the fact that in our model, agents do not
  estimate how other agents could react on their decisions to change
  their links. This situation refers to level $2$ in table
  (\ref{tab:ipc}).
  In this chapter we compare two different settings, level $0$ and level
  $2$. In the following paragraphs we describe them in more detail.

\begin{itemize}

\item[\textbf{0}] \textbf{Extremal Dynamics:} At time $T$ the agent with
  the smallest utility is removed from the system and replaced with a new
  one (market entry). The new agent is randomly connected to the already
  existing agents and a small initial value of knowledge is assigned to
  it. This process is a least fit replacement (extinction of the weakest)
  and the new agent introduces a kind of novelty in the system
  (innovation).

\item[\textbf{2}] \textbf{Utility Driven Dynamics:} The main difference
  between local link formation (Utility Driven Dynamics) compared to
  global link formation (Extremal Dynamics) is that agents are now
  individually taking decisions upon their interactions and they do that
  on the basis of a utility function (their values of knowledge at time
  $T$). Agents are bounded rational since they explore their possible
  interaction partners in a trial and error process.

  At every period, that is after time $T$, an agent is selected at random
  to create and delete links (asynchronous update). We distinguish two
  possible link formation mechanisms which we study separately, namely
  unilateral and bilateral link formation. In the former, unilateral link
  formation (i), the agent optimally deletes an old link and randomly
  creates a new link.  Optimal means that either for creation or deletion
  of links the action is taken only if it increases the value of
  knowledge of the agent at time $T$ in the range of all possible
  actions. In the latter, bilateral formation (ii), the selected agent
  optimally deletes a bilateral connection that she currently has or she
  randomly creates a new bilateral connection. Here optimal (in the range
  of all possible actions) means, that links are deleted if the initiator
  of the deletion, i.e. the selected agent, can increase its value of
  knowledge at time $T$ with the deletion of the link, while for the
  bilateral creation both agents involved have to strictly benefit from
  the creation of the mutual connection\footnote{This behavior is
    individually optimal and thus may also be called rational.}. In the
  following two sections we give a description of mechanisms (i) and
  (ii).

\end{itemize}

To compare the two levels, for Utility Driven Dynamics the evolution of
the network follows from local, utility driven, actions, as opposed to
Extremal Dynamics, where the evolution follows from a global stochastic
process (least fit selection plus random link formation). To be more
specific about the latter, the rules for the network evolution, i.e. the
creation and deletion of links under Extremal Dynamics, are the
following:

\begin{center}
  \begin{tabular}[c]{lp{10cm}l}
    \textit{Step 1} & After a given time $T$ the \textit{least fit agent}, i.e., the one with the
    smallest $u_i=y_i(T)$, is determined. This agent is removed from the network
    along with all its incoming and outgoing links. \\
    \textit{Step 2} &  A new agent is added to the network with some small initial
    value of knowledge $y_0$. The new agent will take the place of the old one (it
    gets the same label), and randomly links itself to the other nodes in
    the network with the same probability $p$. Each of the other nodes can
    in turn link itself to the newcomer node with a probability $p$. \\
  \end{tabular}
\end{center}

These rules for the network evolution are intended to capture two key
features: \textit{natural selection}, in this case, the extinction of
the weakest; and the \textit{introduction of novelty}. Both of these
can be seen as lying at the heart of natural evolution. The particular
form of selection used in this model has been inspired by what
\textit{Bak and Sneppen} have called ``extremal dynamics''
\citep{bak93:_punct_equil_critic_simpl_model_evolut}.

\subsection{Rules for Link Creation and Deletion Using Utility Driven
  Dynamics}
\label{sec:local_link_formation}

In this section we introduce the process of the formation and deletion of
links by agents that maximize a local utility function (depending on the
agent and its neighbors). After time $T$, long enough such that the
system reaches a quasi-equilibrium in the values of knowledge, an agent
is randomly chosen to create or delete a link, either unidirectional or
bidirectional.

\subsubsection{Unilateral Link Formation}
\label{par:unilateral_link_formation}

If agents unilaterally delete or create links it is possible that the
interactions they form create a feed back loop, i.e. a closed cycle of
knowledge sharing agents, that involves more than $2$ agents. This
introduces the concept of indirect reciprocity (see section
\ref{sec:unilateral_reciprocity}). Unilateral formation of links (we
then have a directed network) is necessary for indirect reciprocity to
emerge, since, if all interactions were bilateral they would be direct
reciprocal by definition. We now describe the procedure of unilateral
link creation and deletion.

 \emph{1. Random Unilateral Creation}

 An agent creates a link to another one to which it is not already
 connected at random and evaluates the creation of the link by comparing
 the change in their values of knowledge before and after the creation.
 Only if the change is positive, the link is maintained, otherwise the
 agent does not create the link. In this way agents explore possible
 partners for sharing their knowledge in a trial and error procedure.

 \begin{center}
   \begin{tabular}[c]{lp{10cm}l}
     \textit{Step 1} & An agent $i$ is selected at random. \\
     \textit{Step 2} & Another agent $j$ is selected at random which is
     not already an out-neighbor of $i$.\\
     \textit{Step 3} & Agent $i$ creates an outgoing link to agent $j$.\\
     \textit{Step 4} & The new utility (for the old network plus
     the new link $e_{ij}$) of 
     agent $i$ is computed and compared with the utility before
     the creation.\\ 
     \textit{Step 5} & Only if agent $i$'s utility strictly
     increases compared to her old utility, then the link is created.\\
   \end{tabular}
 \end{center}

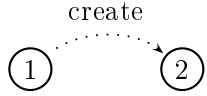
\begin{figure}
  \centering
    \begin{pspicture}(0,0.5)(2,1.5)
      \cnodeput(0,1){A}{1} 
      \cnodeput(2,1){B}{2} 
      \psset{nodesep=1pt}
      \ncarc[linestyle=dotted,arcangle=40]{->}{A}{B} 
      \naput{create}
  \end{pspicture}
\caption{Random unilateral creation.\label{fig:ruc}}
\end{figure}

\emph{2. Optimal Unilateral Deletion}

An agent deletes one outgoing link if this increases her utility.

\begin{center}
  \begin{tabular}[c]{lp{10cm}l}
    \textit{Step 1} & Agent $i$ is selected at random s.t. it has at least one
    outgoing link.\\
    \textit{Step 2} &  Agent $i$ deletes separately each
    of its outgoing links to its neighbors $v_j \in N_i^{+}$ and
    records the change in her utility, $\Delta u_i$. Before a new link is
    deleted, the old one is recreated.\\
    \textit{Step 3} & Agent $i$ computes the maximum change $\Delta u_i$ and if it
    is positive, deletes the referring link. This means that only one
    link is finally deleted. The deletion only takes place if the
    current agent strictly increases her utility.\\
  \end{tabular}
\end{center}

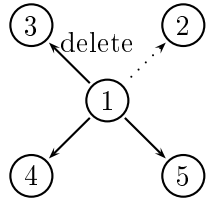
\begin{figure}
  \centering
   \begin{pspicture}(2,3)
     \cnodeput(1,1.5){A}{1} 
     \cnodeput(2,2.5){B}{2} 
     \cnodeput(0,2.5){C}{3}
     \cnodeput(0,0.5){D}{4} 
     \cnodeput(2,0.5){E}{5} 
     \psset{nodesep=1pt}
     \ncline[linestyle=dotted]{->}{A}{B} \naput{delete}
     \ncline{->}{A}{C} 
     \ncline{->}{A}{D}
     \ncline{->}{A}{E}
  \end{pspicture}
\caption{Optimal unilateral deletion.}
\end{figure}

To characterize the equilibrium networks under this link formation and
deletion mechanism we introduce the following characterization of
stability\footnote{Compare this to the definition of bilateral stability
  (\ref{defn:bilaterally_stable}).}.
\begin{defn}
  A network is \textbf{unilaterally stable} if and only if (i) no agent
  can create a link to (strictly) increase her utility and (ii) no agent
  can remove a link to (strictly) increase her utility.
\label{defn:unilaterally_stable}
\end{defn}

\subsubsection{Bilateral Link Formation}
\label{par:bilateral_link_formation}

If agents form links bilaterally then all interactions are direct
reciprocal by definition. We describe the process of bilateral link
creation and deletion in the following paragraphs.

 \emph{1. Random Bilateral Creation}

  In this link creation process a pair of agents is selected at random and
  given the possibility to form a bilateral connection.

  \begin{center}
    \begin{tabular}[c]{lp{10cm}l}
      \textit{Step 1} & Two agents are uniformly selected at random such
      that they are not connected already. \\
      \textit{Step 2} & Both agents create an outgoing link to each other and therewith
      create a $2$-cycle.\\
      \textit{Step 3} & The new utilities (for the old network plus the new
      $2$-cycle) of both agents are computed and compared with the
      utilities before the creation.\\
      \textit{Step 4} & Only if both agents strictly benefit in terms of their
      utilities compared to their old utilities, then the
      bilateral connection is created.\\
    \end{tabular}
  \end{center}

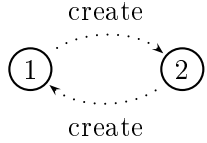
\begin{figure}[htbp]
  \centering
    \begin{pspicture}(2,1.5)
      \cnodeput(0,1){A}{1} 
      \cnodeput(2,1){B}{2} 
      \psset{nodesep=1pt}
      \ncarc[linestyle=dotted,arcangle=40]{->}{A}{B} 
      \naput{create}
      \ncarc[linestyle=dotted,arcangle=40]{->}{B}{A} 
      \naput{create}
  \end{pspicture}
\caption{Random bilateral Creation.}
\end{figure}

\emph{2. Optimal Bilateral Deletion}

  An agent deletes one of its outgoing links to another agent from which the
  agent also has an incoming link if this deletion increases her utility.

  \begin{center}
    \begin{tabular}[c]{lp{10cm}l}
      \textit{Step 1} & Agent $i$ is selected at random such that it has at least one
      mutual link to another agent.\\
      \textit{Step 2} & From all bilaterally connected neighbors agent $i$ deletes
      separately each of its outgoing links to its neighbors (and so does
      each neighbor $j$ to agent $i$). For each, the change in the utility,
      $\Delta u_i$ is recorded. Before a new links are deleted, the old
      ones are recreated.\\
      \textit{Step 3} & Agent $i$ computes the maximum change $\Delta u_i$ and, if
      it is positive, the referring bilateral connection is deleted. The deletion
      only takes place if agent $i$ strictly increases her utility.\\
    \end{tabular}
  \end{center}

  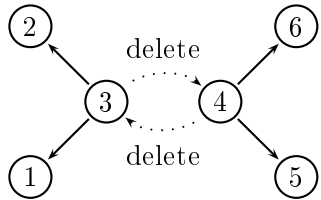
\begin{figure}
 \begin{center}
  \begin{pspicture}(4,3) 
    \cnodeput(0,0.5){A}{1} 
    \cnodeput(0,2.5){B}{2}
    \cnodeput(1,1.5){C}{3} 
    \cnodeput(2.5,1.5){D}{4}
    \cnodeput(3.5,0.5){E}{5} 
    \cnodeput(3.5,2.5){F}{6}
    \psset{nodesep=1pt}
    \ncline{->}{C}{A}
    \ncline{->}{C}{B} 
    \ncline{->}{D}{E} 
    \ncline{->}{D}{F}
    \ncarc[linestyle=dotted,arcangle=40]{->}{C}{D} \naput{delete}
    \ncarc[linestyle=dotted,arcangle=40]{->}{D}{C} \naput{delete}
  \end{pspicture}
 \end{center}
 \caption{Optimal bilateral deletion.}
\end{figure}

In order to characterize the equilibrium outcomes of our simulations we
will introduce a characterization of network stability. This definition
has been introduced by
\citet{jackson03:_survey_models_network_formation_stability_efficiency}

\begin{defn}
  A network $G$ is \textit{pairwise stable} if (i) removing any link does
  not increase the utility of any agent and (ii) adding a link between
  any two agents, either doesn't increase the utility of any of the two
  agents, or if it does increase one of the two agents' utility then it
  decreases the other agent's utility.
\label{defn:bilaterally_stable}
\end{defn}

\subsection{The Role of the Time Horizon for Unilateral Link Formation}
\label{sec:time_horizon}

So far we have assumed that the time horizon $T$ (after which agents
evaluate their decisions to create or delete links) is long enough such
that the values of knowledge reach a stationary state and the utilities
of the agents are given by the fixed points of the values of knowledge.
In this section we discuss the effect of a time horizon that is smaller
than the time to convergence to the stationary state of the values of
knowledge. For related works that incorporate a finite time horizon in
the evaluation of the actions of agents see e.g.
\citet{huberman94:_belief} or \citet{lane97:_fores_compl_strat}.

If we consider Utility Driven Dynamics, we will show that permanent
networks with positive values of knowledge emerge if agents wait long
enough in evaluating their decisions. This is a necessary condition.
Otherwise networks are not able to emerge or, if a network with positive
knowledge values is existing already, it gets destroyed over time
(network breakdown). This effect is important in the case of null as
well as increasing costs.

\begin{center}
  \begin{figure}[htpb]
    \begin{minipage}{0.5\linewidth}
      \psfrag{x1}[c][][3][0]{$x_1$} 
      \psfrag{x2}[c][][3][0]{$x_2$} 
      \psfrag{x3}[c][][3][0]{$x_3$}
      \psfrag{x4}[c][][3][0]{$x_4$}
      \psfrag{x5}[c][][3][0]{$x_5$}
      \psfrag{xi}[c][][4][0]{$x_i$} 
      \psfrag{time}[c][][4][0]{$t$}
      \psfrag{1/n}[c][][3][0]{$1/n$}
      \scalebox{0.3}{\includegraphics[angle=0]
        {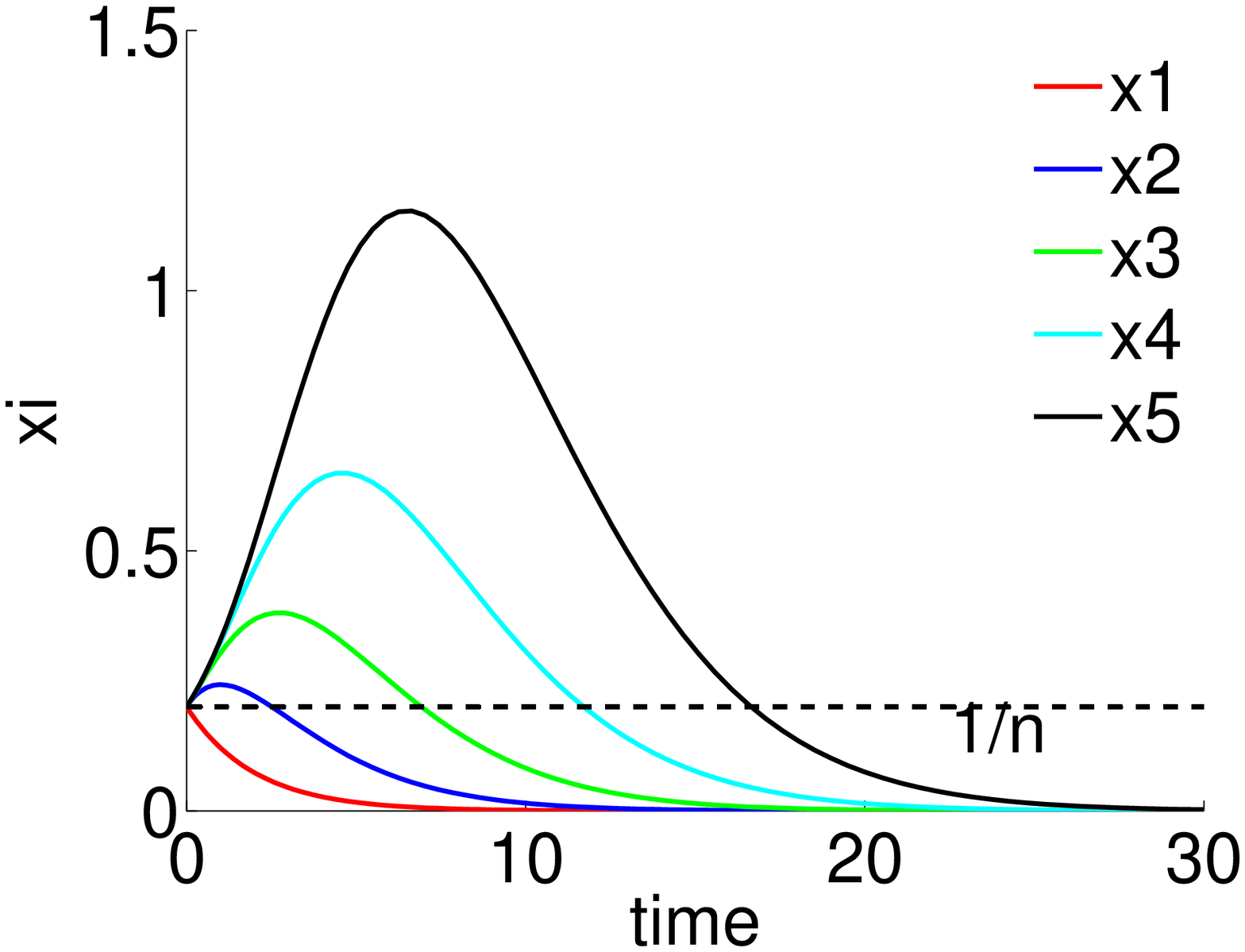}}
    \end{minipage}
    \begin{minipage}{0.5\linewidth}
      \psfrag{x1}[c][][3][0]{$x_1$} 
      \psfrag{x2}[c][][3][0]{$x_2$} 
      \psfrag{x3}[c][][3][0]{$x_3$}
      \psfrag{x4}[c][][3][0]{$x_4$}
      \psfrag{x5}[c][][3][0]{$x_5$}
      \psfrag{xi}[c][][4][0]{$x_i$} 
      \psfrag{time}[c][][4][0]{$t$}
      \psfrag{(b-d)/c}[c][][3][0]{$(b-d)/c$}
      \scalebox{0.3}{\includegraphics[angle=0]
        {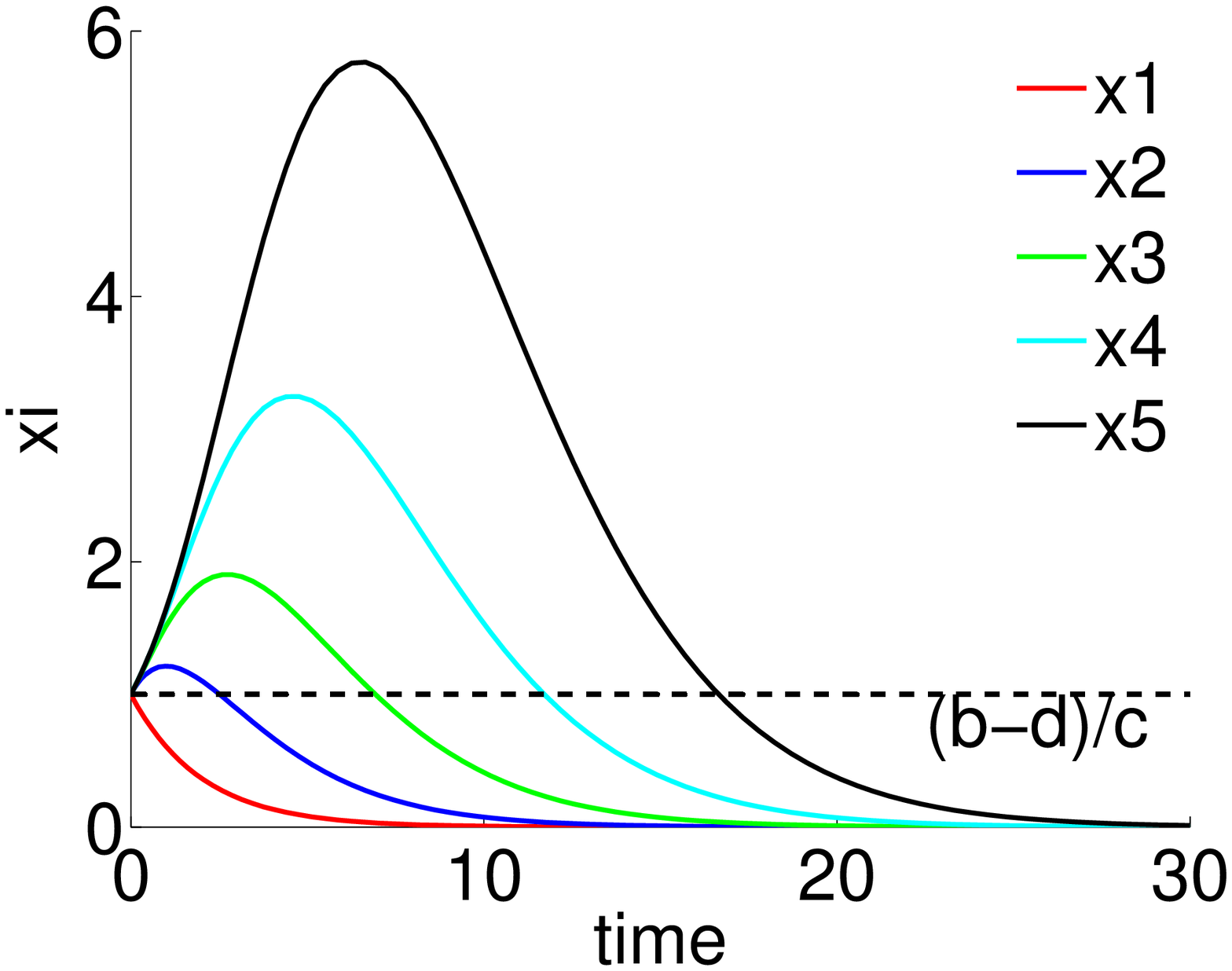}}
    \end{minipage}
    \caption{Numerical integration of the value of knowledge for
      $d=0.5$, $b=1.0$, null cost $c=0.0$ (left) and cost $c=0.5$
      (right): evolution of knowledge values for a linear chain of $5$
      nodes (obtained from the $C_{5}$ by removing an link). The agent
      that removes the link (black upper curve) initially experiences an
      increase in the value of knowledge. After an initial increase she
      experiences a decline and at a certain time her value of knowledge
      reaches her initial value ($\frac{1}{n}$ in the case of null cost
      and $\frac{b-d}{c}$ in the case of increasing cost) and then it
      further decreases.  After a time long enough her value of knowledge
      vanishes completely.}
    \label{fig:linear_chain}
\end{figure}
\end{center}

To illustrate this point, we consider a $5$-cycle of agents and the
deletion of one link in this cycle which creates a linear chain of $5$
nodes, Fig. (\ref{fig:c5_chain}). The evolution of value of knowledge
for null costs and for costs $c=0.5$ can be seen in Fig.
(\ref{fig:linear_chain}).

\begin{center}
  \begin{figure}[htpb]
    \begin{center}
        \begin{minipage}{.3\textwidth}
          \begin{center}
            \scalebox{0.75}{\begin{pspicture}(0,0)(2,2)
              \cnodeput(1.309,2.451){1}{\Large1} 
              \cnodeput(0.11,2.088){2}{\Large 2}
              \cnodeput(0.11,0.912){3}{\Large 3}
              \cnodeput(1.309,0.549){4}{\Large 4}
              \cnodeput(2.0,1.5){5}{\Large 5}
              \ncarc[linewidth=1.pt]{->}{2}{1}
              \ncarc[linewidth=1.pt]{->}{3}{2}
              \ncarc[linewidth=1.pt]{->}{4}{3}
              \ncarc[linewidth=1.pt]{->}{5}{4}
              \ncarc[linewidth=1.pt]{->}{1}{5}
            \end{pspicture}}
          \end{center}
        \end{minipage}
        {\Large \textbf{$\rightarrow$}}
        \begin{minipage}{.3\textwidth}
          \begin{center}
            \scalebox{0.75}{\begin{pspicture}(0,0)(2,2) 
              \cnodeput(1.309,2.451){1}{\Large1} 
              \cnodeput(0.11,2.088){2}{\Large 2}
              \cnodeput(0.11,0.912){3}{\Large 3}
              \cnodeput(1.309,0.549){4}{\Large 4}
              \cnodeput(2.0,1.5){5}{\Large 5}
              \ncarc[linewidth=1.pt]{->}{2}{1}
              \ncarc[linewidth=1.pt]{->}{3}{2}
              \ncarc[linewidth=1.pt]{->}{4}{3}
              \ncarc[linewidth=1.pt]{->}{1}{5}
            \end{pspicture}}
          \end{center}
        \end{minipage}
      \caption{A $5$-cycle and a linear chain of $5$ nodes (obtained from
        the cycle be removing one link).}
      \label{fig:c5_chain}
    \end{center}  
    \end{figure}
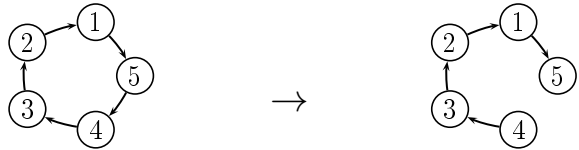
  \end{center}

More formally we can give the following proposition.
\begin{prop}
  Consider the dynamical system (\ref{eq:growth_quadratic_costs}). For a
  directed path $P_k$ of length $k$ the value of knowledge of node $k$ is
  larger than $\epsilon$ for $t \le \tau(\epsilon)$, i.e.  $x_k(t \le
  \tau(\epsilon)) \ge \epsilon$ while $\lim_{t \to \infty} x_k(t)=0$.
\label{prop:lower_bound_tau}
\end{prop}
\begin{proof}
  Consider a directed path $P_k$ of length $k$.

  \begin{center}
    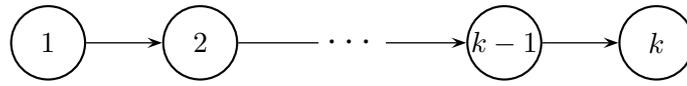
\begin{figure}[htpb]
      \begin{center}
        \begin{pspicture}(0,0)(8,2) 
          \cnode(0,1){0.5cm}{A}
          \rput(0,1){$1$} 
          \cnode(2,1){0.5cm}{B} 
          \rput(2,1){$2$}
          \cnode(6,1){0.5cm}{C} 
          \rput(6,1){$k-1$} 
          \cnode(8,1){0.5cm}{D}
          \rput(8,1){$k$} 
          \ncline[linewidth=0.5pt,arrowsize=3pt 2]{->}{A}{B}
          \ncline[linewidth=0.5pt,arrowsize=3pt 2]{->}{B}{C} 
          \ncput*{\Large $\cdots$}
          \ncline[linewidth=0.5pt,arrowsize=3pt 2]{->}{C}{D}
        \end{pspicture}
      \end{center}
      \caption{A directed path $P_k$ of length $k$.}
    \end{figure}
  \end{center}

  For node $1$ (the source has no incoming links) in
  (\ref{eq:growth_quadratic_costs}) we get
  \begin{equation}
    \dot{x}_1(t) = - d x_1 - c x_1^2
  \end{equation}
  By introducing the variable $z=\frac{1}{x_1}$ and solving for $z$ one can
  find the solution for $x_1$
  \begin{equation}
    x_1(t) = \frac{d e^{da}}{e^{dt}-ce^{da}}
  \end{equation}
  with a constant $a=\frac{1}{d} \ln \frac{x_1(0)}{d+c}$ and the limit
  $\lim_{t \to \infty} x_1(t) = 0$. Accordingly for the $k$-th node we
  have that
  \begin{equation}
    \dot{x}_k = - d x_k + b x_{k-1} - c x_k^2
  \end{equation}
  Since $x_k \ge 0$, from proposition (\ref{prop:bounded}), the following
  inequality holds
  \begin{equation}
    \dot{x}_k \ge - d x_k - c x_k^2
  \end{equation}
  and
  \begin{equation}
    x_k(t) \ge \frac{d e^{da'}}{e^{dt}-ce^{da'}}
  \end{equation}
  with a proper constant $a'=\frac{1}{d} \ln \frac{x_k(0)}{d+c}$.
  Equating with $\epsilon(t)$ at $t=\tau$ we get
  \begin{equation}
    \epsilon = \frac{d }{e^{dt-da'}-c}
  \end{equation}
  which yields
  \begin{equation}
    \tau(\epsilon) = \frac{\ln \left( \frac{d}{\epsilon} + c \right)
      + a'd}{d}  
  \end{equation}
  Thus we have found an $\epsilon(\tau)$ such that for $t \le
  \tau(\epsilon)$ $x_k(t) \ge \epsilon$. The limit $\lim_{t \to \infty}
  x_k(t)=0$ follows directly from the fact that the directed path $P_k$
  is a directed acyclic graph and we can apply proposition
  (\ref{prop:non-permanence}). $\Box$
\end{proof}

With proposition (\ref{prop:lower_bound_tau}) one can readily infer the
following. If an agent in a cycle $C_k$ of length $k$ removes a link
unilaterally then a path $P_k$ is created. If the time horizon after
which the agent evaluates this link removal is smaller then
$\tau(\epsilon))$ the agent's value of knowledge satisfies $x_k(t \le
\tau(\epsilon)) \ge \epsilon$ (this gives the utility of the agent, see
(\ref{defn:utility})).  From proposition (\ref{prop:cycle}) we know that
the value of knowledge of the agent in the cycle is given by
$x_k(0)=\frac{b-d}{c}$.  Choosing $\tau(\epsilon)$ such that $\epsilon >
\frac{b-d}{c}$ gives $x_k(t \le \tau(\epsilon)) \ge x_k(0)$ and the agent
experiences an increase in her utility by removing the link.  The agent
removes the link in order to increase her utility. This destroys the
cycle. The time horizon of the agent in this case is too short in order
to anticipate the vanishing long-run values of knowledge of all the
agents in the resulting path, $\lim_{t \to \infty} x_(t)=0$.

From this observation we conclude that if the time horizon is too short,
then all cycles would get destroyed and no network would ever be able to
emerge nor sustain, since only cyclic networks can be permanent. The
free-riding behavior of agents leads to the breakdown of the economy.

\subsection{Simple Equilibrium Networks for Unilateral Link Formation}
\label{sec:simple_equilibrium_networks}

In this section we identify the most simple equilibrium networks for
Unilateral Link Formation. There exists a multitude of other equilibrium
networks which usually cannot be computed analytically and which depend
on the parameter values for decay, benefit and cost. 

The most simple equilibrium network is the empty network.
\begin{prop}
  The empty graph is unilaterally stable.
\end{prop}
\begin{proof}
  In an empty graph all nodes have vanishing values of knowledge.
  Creating a link does not create a cycle (which would be the case
  however if links were formed bilaterally) and thus the empty graph plus
  a link is a directed acyclic graph with vanishing values of knowledge,
  see proposition (\ref{prop:non-permanence}). The creation of a link
  does not increase the utility of an agent. Thus, the agents
  do not form any links. $\Box$
\end{proof}
Moreover, if all agents form disconnected cycles then we have an
equilibrium network.
\begin{prop}
  The set of disconnected cycles $\{ C^1,...,C^k \}$, and possibly
  isolated nodes is unilaterally stable.
\label{prop:cycles_stable}
\end{prop}
\begin{proof}
  We give a proof for two-cycles. The proof can easily be extended to
  cycles of any length. Consider the two cycles $C_2^1$ and $C_2^2$ in Fig.
  (\ref{fig:two_C_2}).

\begin{center}
    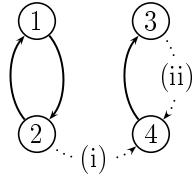
\begin{figure}[htpb]
      \begin{center}
        \scalebox{0.75}{\begin{pspicture}(0,-1)(3,3) 
          \cnodeput(0,2){1}{\Large 1}
          \cnodeput(0,0){2}{\Large 2}
          \cnodeput(2,2){3}{\Large 3}
          \cnodeput(2,0){4}{\Large 4}
          \ncarc[linewidth=1pt,arcangle=40]{->}{1}{2}
          \ncarc[linewidth=1pt,arcangle=40]{->}{2}{1}
          \ncarc[linewidth=1pt,arcangle=40,linestyle=dotted]{->}{3}{4}
          \ncput*{\Large (ii)}
          \ncarc[linewidth=1pt,arcangle=40]{->}{4}{3}
          \ncarc[linewidth=1pt,arcangle=-40,linestyle=dotted]{->}{2}{4}
          \ncput*{\Large (i)}
        \end{pspicture}}
       \end{center}
       \caption{Two cycles $C_2^1$ and $C_2^2$ and the cases of link
         creation (i) and deletion (ii).}
       \label{fig:two_C_2}
    \end{figure}
  \end{center}

  From proposition (\ref{prop:cycle}) we know that the fixed points are
  given by $x_i=\frac{b-d}{c}$, $i=1,...,4$. In order to show that we
  have a unilaterally stable equilibrium we (i) first show that no link is
  created and in the following (ii) that no link is deleted.

  \begin{itemize}
  \item[(i)] If a link is created (w.l.o.g.) from node $2$ to $4$ we
    get from (\ref{eq:growth_quadratic_costs})
    \begin{equation}
      \begin{array}{l}
        \dot{x}_1 = -d x_1 + b x_2 - c x_1^2 \stackrel{!}= 0 \\
        \dot{x}_2 = -d x_1 + b x_1 - c x_2^2 \stackrel{!}= 0 \\
      \end{array}
    \end{equation}
    From the first order conditions for the fixed points we get for node
    $1$
    \begin{equation}
      x_1=\frac{bx_2-c}{d}
    \end{equation}
    And inserting this into the fixed point of node $2$ gives
    \begin{equation}
      x_2 = \frac{ b^2 - d^2 + \sqrt{ b^4 - 8bc^3d - 2b^2d^2 + d^4 } }{
        4cd } 
    \end{equation}
    If the last inequality is fulfilled, then the creation of the link
    would decrease the utility of agent $2$. The inequality
    holds if $c^3 \ge \frac{(d-b)^3}{b}$ which is certainly true for
    $b>d$ and $c>0$. Thus, no link is created between the cycles.
  \item[(ii)] If an link is deleted in a $C_2$ then the we get vanishing
    steady state values of knowledge. Since $\frac{b-d}{c} \ge 0$ this
  would reduce the utility of the agent. Therefore the link is
  not removed.
\end{itemize}

If there are $k \le \lfloor \frac{n}{2} \rfloor$ $2$-cycles in $G$ the
the above argument holds for any pair of cycles. Similarly no isolated
node can create a link in order to increase her utility nor can a node in
a cycle create a link to an isolated node. Neither link creation nor
removal increases the utility of the initiating agent and so the set of
two-cycles is unilaterally stable.  $\Box$

\end{proof}

We further conjecture that a set of disconnected autocatalytic sets,
where an autocatalytic set is defined as a set of nodes each having an
incoming link from a node of that set \citep{jain01}, stays disconnected
under unilateral link formation. Thus, the size (in terms of nodes) is
stable.

With (\ref{prop:cycles_stable}) we know that a cycle is unilaterally
stable. In section (\ref{sec:time_horizon}) however we have shown that
this result is critically depending on the time horizon $T$ after which
the action of an agent is evaluated (and it is true for cycles of any
length only if $T \to \infty$).

For parameters values $d=0.5$, $b=0.5$ and $c=0.1$ also the
complete graph with three nodes $K_3$ and the path $P_3$ is unilateral
stable. We observe this in simulations in Fig.
(\ref{fig:source_squared_unilateral}). However, by computing the fixed
points numerically for $d=0.5$, $b=0.5$ and $c=0.1$ in section
(\ref{sec:stationary_solutions_G3}) one can see that $K_3$ is no longer
unilaterally stable (because removing a link increases the utility of an
agent).

In the next section we investigate if the dynamic processes of link
formation and deletion lead to the simple equilibrium structures
suggested above (and indeed we show that they are not obtained).

\section{Simulation Studies Using Different Growth Functions}
\label{sec:simulation_studies}

In the remainder of this chapter, we study simulations with different
growth functions (for the value of knowledge) and different link
formation mechanisms. We assume that the time horizon $T$ is long enough
such that the values of knowledge reach their stationary state. The
dynamics of the value of knowledge is given by
(\ref{eq:growth_null_costs}) with null costs or by
(\ref{eq:growth_quadratic_costs}) with increasing costs. The different
link formation mechanisms are described in section
(\ref{sec:global-vs-local}). We compare the equilibrium networks obtained
from different costs and link formation rules in terms of their structure
and performance. Finally, we study the effect of different positive
network externalities on the equilibrium networks.

Table (\ref{tab:simulation_studies}) gives an overview of the simulations
that we study in the following. We set $d=0.5$, $b=0.5$ and $c=0.1$. The
complete set of parameter values used throughout this section can be
found in table (\ref{tab:parameters}) in the appendix.

\begin{center}
  \begin{table*}
    \centering
    \begin{tabular}[c]{|l|l|c|
}
      \hline
      \textsc{Knowledge dynamics} & \textsc{Network dynamics} &
                                \textsc{Section} 
\\
      \hline
      \hline
      Null costs, $c_{ij}=0$: & least fit replacement 
& (\ref{sec:null_cost_simulation}) 
\\
      &    & 
\\
      $\frac{dx_i}{dt}= -d x_i + b \sum_{j=1}^{n} a_{ji} x_j$ &   & 
\\
      & & 
\\ 
      \hline
      Quadratic cost, $c_{ij} \propto x_i^2$: & least fit replacement 
& (\ref{sec:source_squared})
\\
      & unilateral link formation & 
\\
      $\frac{dx_i}{dt}= -d x_i + b \sum_{j=1}^{n} a_{ji} x_j - c
      \sum_{j=1}^{n} a_{ij} x_i^2$ & bilateral link formation & 
\\
      &  & 
\\
      \hline
      Quadratic cost, $c_{ij} \propto x_i^2$, and & unilateral link
      formation 
& (\ref{sec:externalities}) 
\\
      externality, $w_{ji}$:  & & 
\\
      & & 
\\
      $ \frac{dx_i}{dt}= -d x_i + \sum_{j=1}^{n} (b a_{ji} + b_e
      w_{ji}) x_j - c \sum_{j=1}^{n} a_{ij} x_i^2 $ & & 
\\
      & & 
\\
      \hline
    \end{tabular}
    \caption{Overview of the simulation studies in this section with different knowledge
      growth functions and different link formation mechanisms.}
    \label{tab:simulation_studies}
  \end{table*}
\end{center}

\subsection{Null Interaction Costs}
\label{sec:null_cost_simulation}

In the following we briefly discuss the evolution of the network with
least fit link formation and null link costs. This model has been studied
in detail by \citet{seufert-fs-07, jain01:_cooperation_interdependence_structure}. Later
\citet{saurabh07:_growt_knowl_compl_evolv_networ} have applied it to an
innovation model where new ideas are created and destroyed in a network
of ideas. 

In this model agents do not have to pay costs for maintaining
interactions. Accordingly, the dynamics on the values of knowledge is
given by (\ref{eq:growth_null_costs})
\begin{displaymath}
  \frac{dx_i}{dt}= -d x_i + b \sum_{j=1}^{n} a_{ji} x_j
\end{displaymath}
and the dynamics in the shares of the values of knowledge
$y_i=x_i / \sum_{j=1}^n x_j$ is given by
(\ref{eq:knowledge_shares_dynamics}).
\begin{displaymath}
\dot{y_i} = \sum_{j}^{n} a_{ji} y_j - y_i \sum_{k,j}^{n} a_{jk} y_j
\end{displaymath}
The utility is given by $u_i=y_i(T)$. We have described in section
(\ref{sec:null_cost_analysis}) that the fixed point (stationary solution)
of the relative values of knowledge in
(\ref{eq:knowledge_shares_dynamics}) exists and is given by the
eigenvector to the largest real eigenvalue of the adjacency matrix.  We
assume that the time horizon $T$ (after which links get created or
deleted) is large enough such that the system has reached this stationary
state before links are changed.

\subsubsection{Extremal Dynamics: least fit replacement}

After time $T$ the worst performing agent (in terms of her share of value
of knowledge $y_i(T)$) is replaced with a new one. We have described this
global link formation mechanism in section (\ref{sec:global-vs-local}).
\citet{seufert-fs-07, jain01:_cooperation_interdependence_structure,
  jain98:_autoc_sets_growt_compl_evolut} have extensively studied the
behavior of the dynamics on $\mathbf{y}$ and the network $G$ represented
by $\mathbf{A}(G)$.  They show that strongly connected sets of nodes with
free-riders (that are receiving knowledge from the strong component but
are not contributing knowledge back to the strong component)
attached\footnote{\citet{jain01} denote this set of nodes the
  \textit{autocatalytic set (ACS)}: it is a subgraph of nodes in which
  every node has at least one incoming link from that subgraph.} appear
and get destroyed in a process of repeatedly removing the worst
performing node (with minimum $y_i$) and replacing it with a new one.

\begin{figure}[htpb]
  \centering
  \begin{minipage}{.45\linewidth}
    \psfrag{<u>}[][][4][0]{$\langle u \rangle$}           
    \psfrag{T}[][][4][0]{$\times T$}  
    \centerline{\scalebox{0.3}{\includegraphics[angle=0]
        {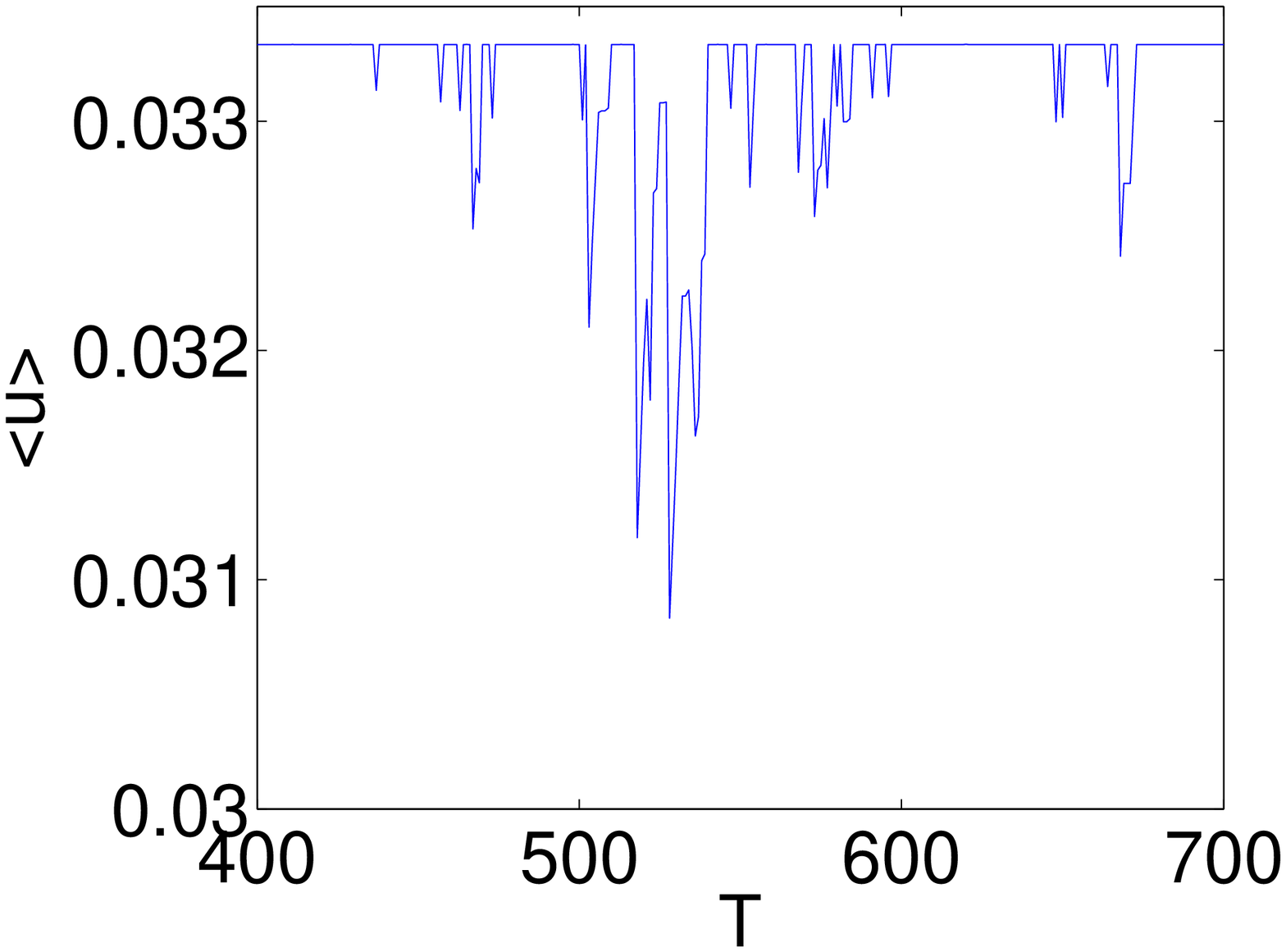}}} \bf{a}
  \end{minipage}
  \hfill
  \begin{minipage}{.45\linewidth}
    \psfrag{<d>}[][][4][0]{$\langle d \rangle$}             
    \psfrag{T}[][][4][0]{$\times T$}
    \centerline{\scalebox{0.3}{\includegraphics[angle=0]
        {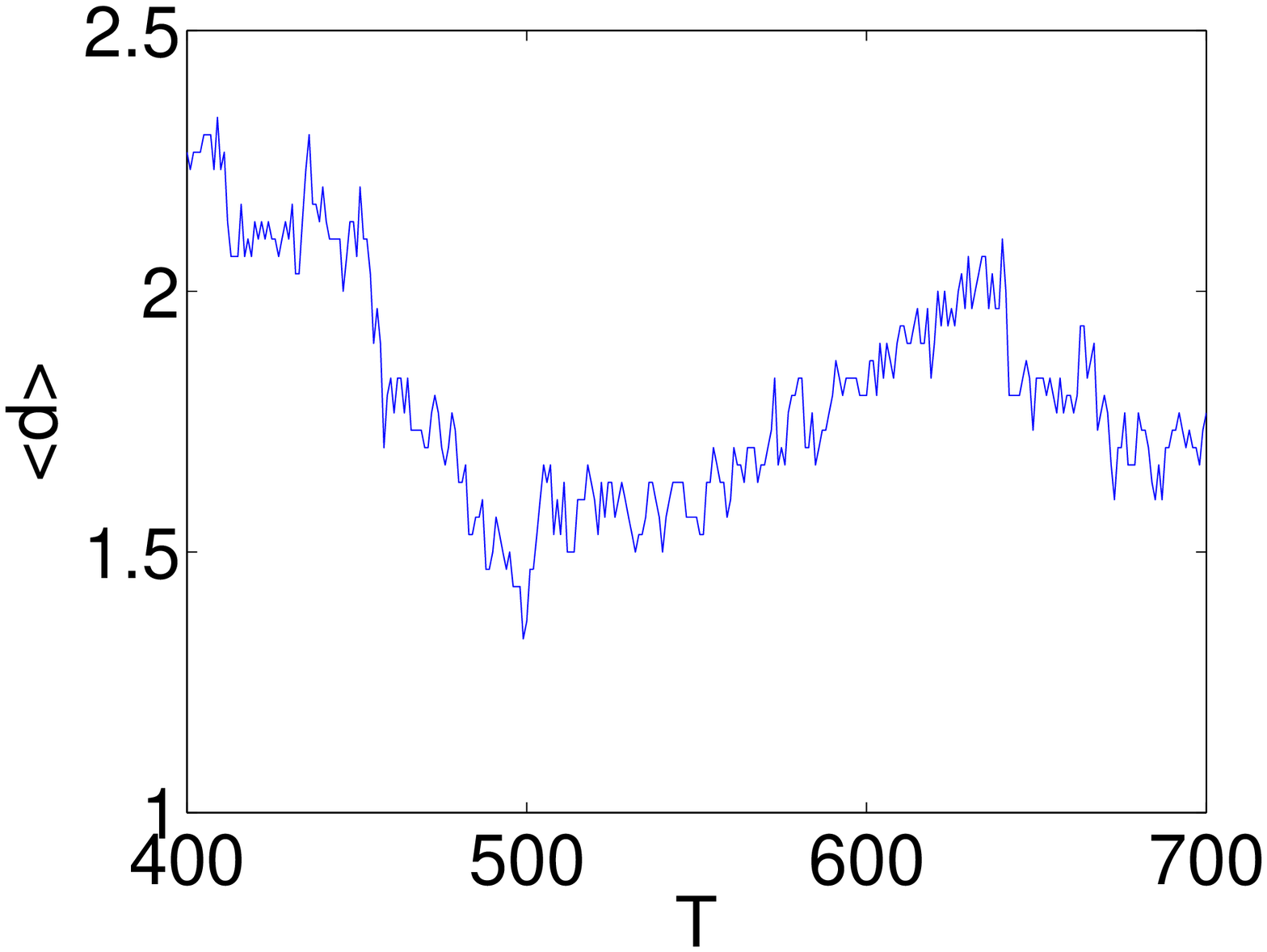}}} \bf{b}
  \end{minipage}
  \begin{minipage}{.45\linewidth}
    \centerline{\scalebox{0.3}{\includegraphics[angle=0]
        {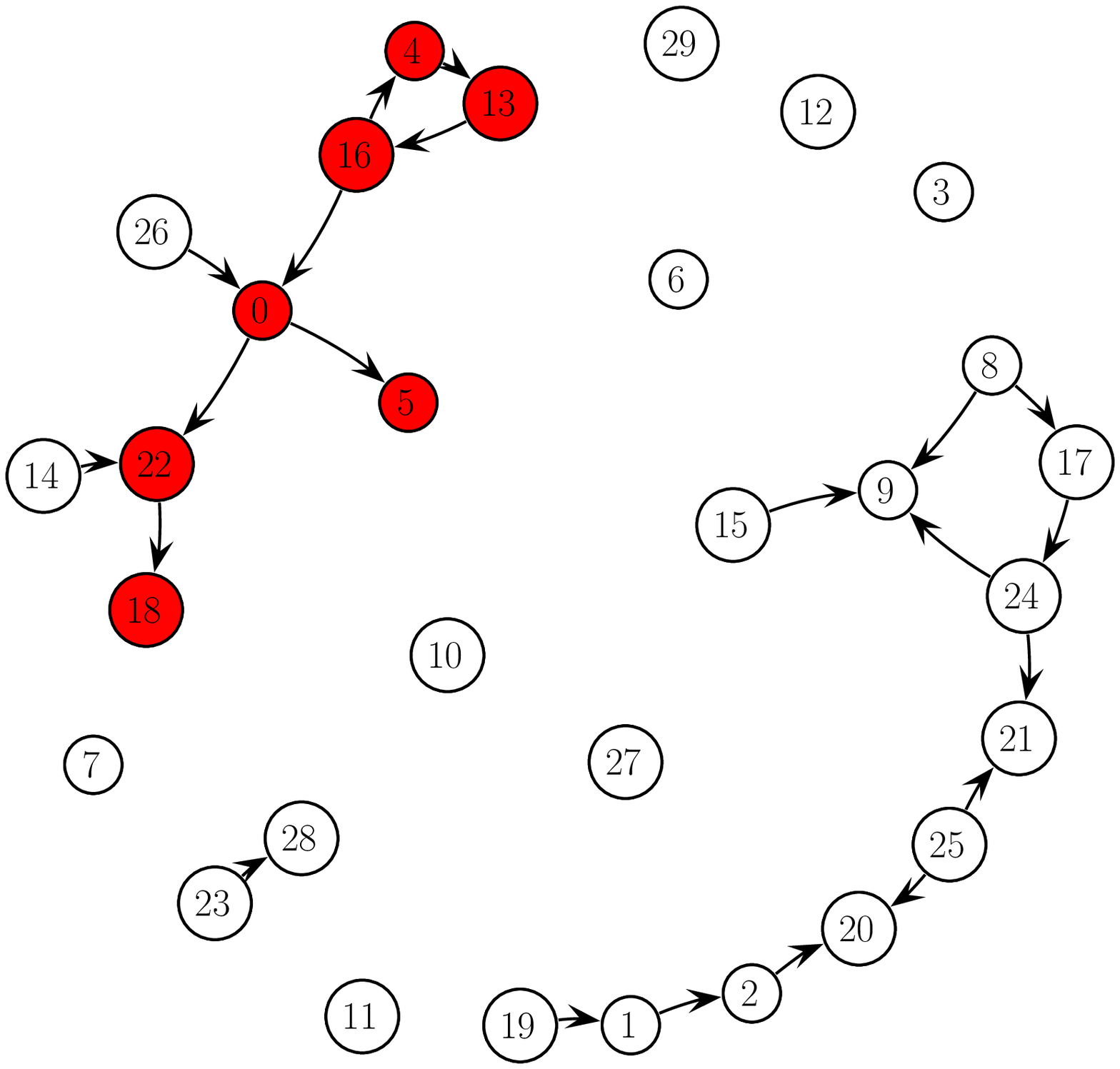}}} \bf{c}
  \end{minipage}
  \hfill
  \begin{minipage}{.45\linewidth}
    \centerline{\scalebox{0.3}{\includegraphics[angle=0]
        {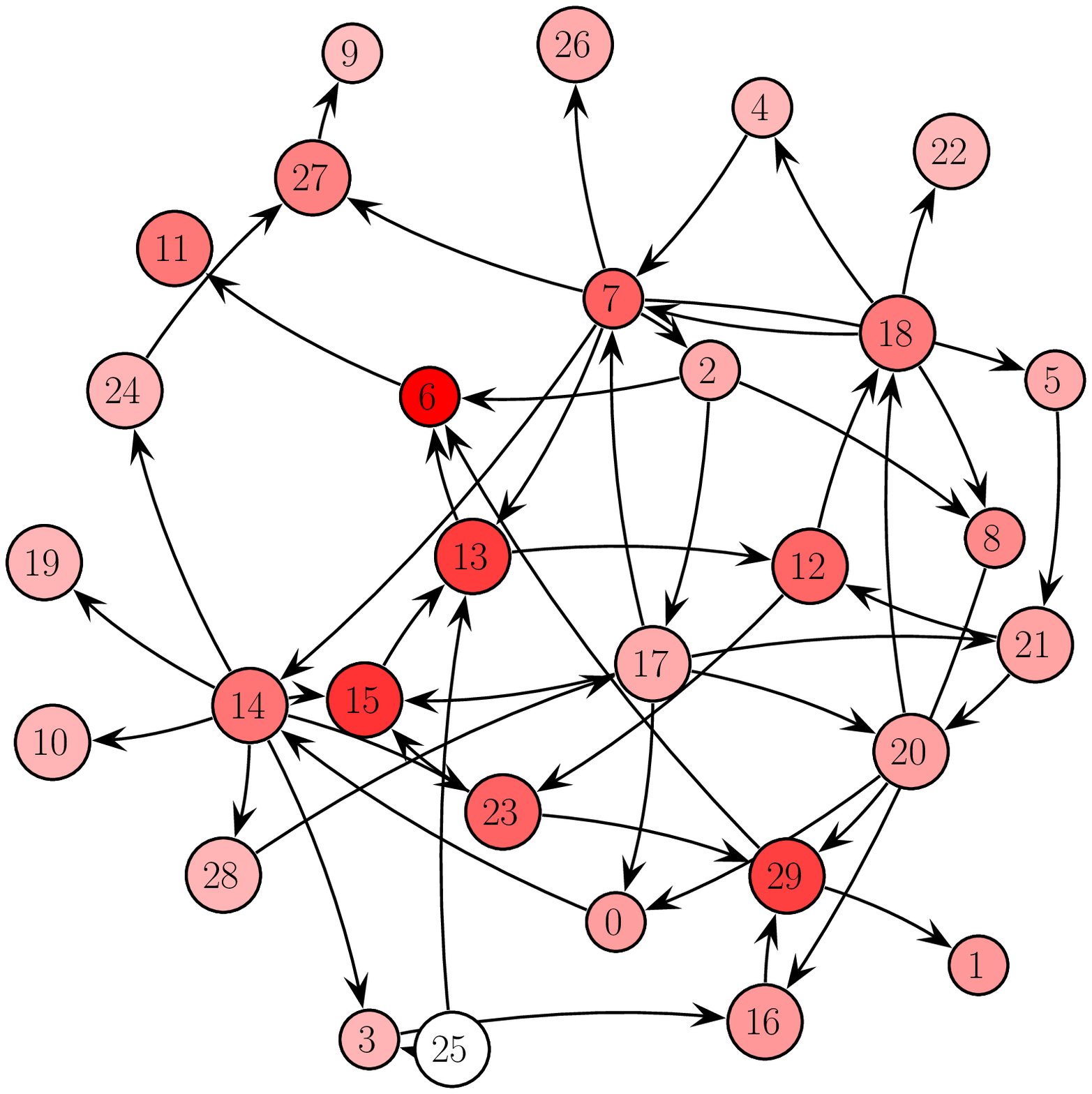}}} \bf{d}
  \end{minipage}
  \caption{Least Fit replacement: \textbf{(a)} Average
    utility. \textbf{(b)} Average degree. \textbf{(c)} Initial
    random graph. \textbf{(d)} Graph after $5000$ iterations.}
\label{fig:jain_krishna_leastFit}
\end{figure}

In computer simulations we can reproduce the results of
\citet{seufert-fs-07, jain01:_cooperation_interdependence_structure}.  We observe
crashes and recoveries in the average utility and degrees of
the agents over time as can be seen in Fig.
(\ref{fig:jain_krishna_leastFit}). Thus, no stable equilibrium network
can be realized with this type of network dynamics.

In the model of \citet{jain01:_cooperation_interdependence_structure}
links are costless. In the next section (\ref{sec:source_squared}) we
assume that links have a cost attached that is an increasing function of
the value of knowledge that is being transfered (section
(\ref{sec:source_squared})).

Moreover, The least fit network dynamics treats agents as completely
passive units that are exposed to an external selection mechanism. In a
more realistic approach one should take into account that agents are
deliberately deciding upon with whom to engage in an R\&D collaboration
or to share their knowledge with. These decisions are taken on the basis
of increasing a utility function, that is their value of
knowledge\footnote{A model in which the eigenvector associated with the
  largest real eigenvalue is used as a utility function is studied in
  \citet{ballester06:_who_networ}.}. We introduce local link formation
rules in section (\ref{sec:local_link_formation}). Moreover, as a further
extension we study the effect of positive network externalities in
section (\ref{sec:externalities}).

\subsection{Increasing Interaction Costs} 
\label{sec:source_squared}

In this section we study the effect of increasing costs for maintaining
interactions with other agents on the resulting equilibrium networks. The
evolution of the value of knowledge is given by \ref{eq:dyn_sys} and the
utility of the agents by (\ref{defn:utility}). The cost of a link depends
quadratically on the value of knowledge of the agent that initiates the
interaction. We study three different link formation mechanisms. The
first is a least fit replacement. We will compare the results of the
simulation with the preceding section where links were costless. In the
following two sections link formation mechanisms are studied in which
agents decide locally upon to create or delete links either unilaterally
or bilaterally based on their utility (\ref{defn:utility}). We assume
that the time horizon is long enough such that the utility of the agents
is given by the fixed points of the value of knowledge. We will show that
least fit replacement of agents leads to a total network breakdown
eventually from which the system cannot recover. Moreover, we show that
bilateral link formation leads to a complete graph while with unilateral
link formation this is not the case. For unilateral link formation only a
small number of agents have non-vanishing knowledge values in the
resulting equilibrium network and these cluster together in bilateral
connections.  Depending on the link formation mechanism and the parameter
values (for decay, benefit and cost) the equilibrium networks can vary
considerably.

The evolution of the value of knowledge of agent $i$ 
(\ref{eq:dyn_sys})

\begin{displaymath}
    \frac{dx_i}{dt}= - d x_i + b \sum_{i=1}^{n} a_{ji} x_j - c
    \sum_{i=1}^{n} a_{ij} x_i^2
  \label{eq:source_squared}
\end{displaymath}
 
and her utility is given by $ u_i=x_i(T)$.

\subsubsection{Extremal Dynamics: least fit replacement}

Similarly to the preceding section links are formed and removed by a
least-fit selection mechanism (introduced in section
(\ref{sec:global-vs-local})). The agent with the smallest utility
(\ref{defn:utility}) is replaced with a new agent. But in this section
costs for maintaining links are an increasing function of the knowledge
value of the transmitting agent.

\begin{figure}[htpb]
  \centering
  \begin{minipage}{.45\linewidth}
    \psfrag{<u>}[][][4][0]{$\langle u \rangle$}
    \psfrag{T}[][][4][0]{$\times T$}
    \centerline{\scalebox{0.3}{\includegraphics[angle=0]    
        {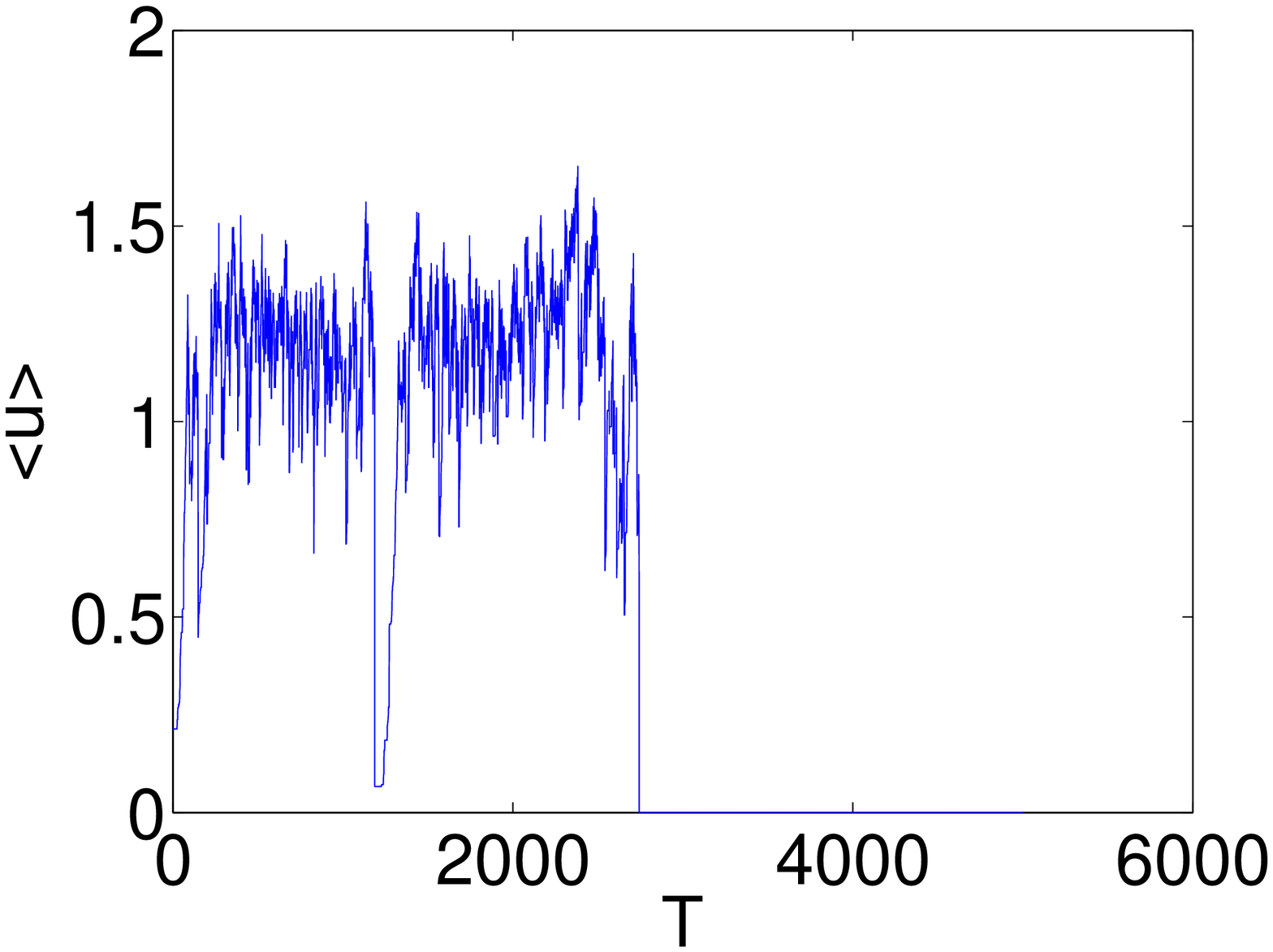}}} \bf{a}
  \end{minipage}
  \hfill
  \begin{minipage}{.45\linewidth}
    \psfrag{<d>}[][][4][0]{$\langle d \rangle$}
    \psfrag{T}[][][4][0]{$\times T$}
    \centerline{\scalebox{0.3}{\includegraphics[angle=0]    
        {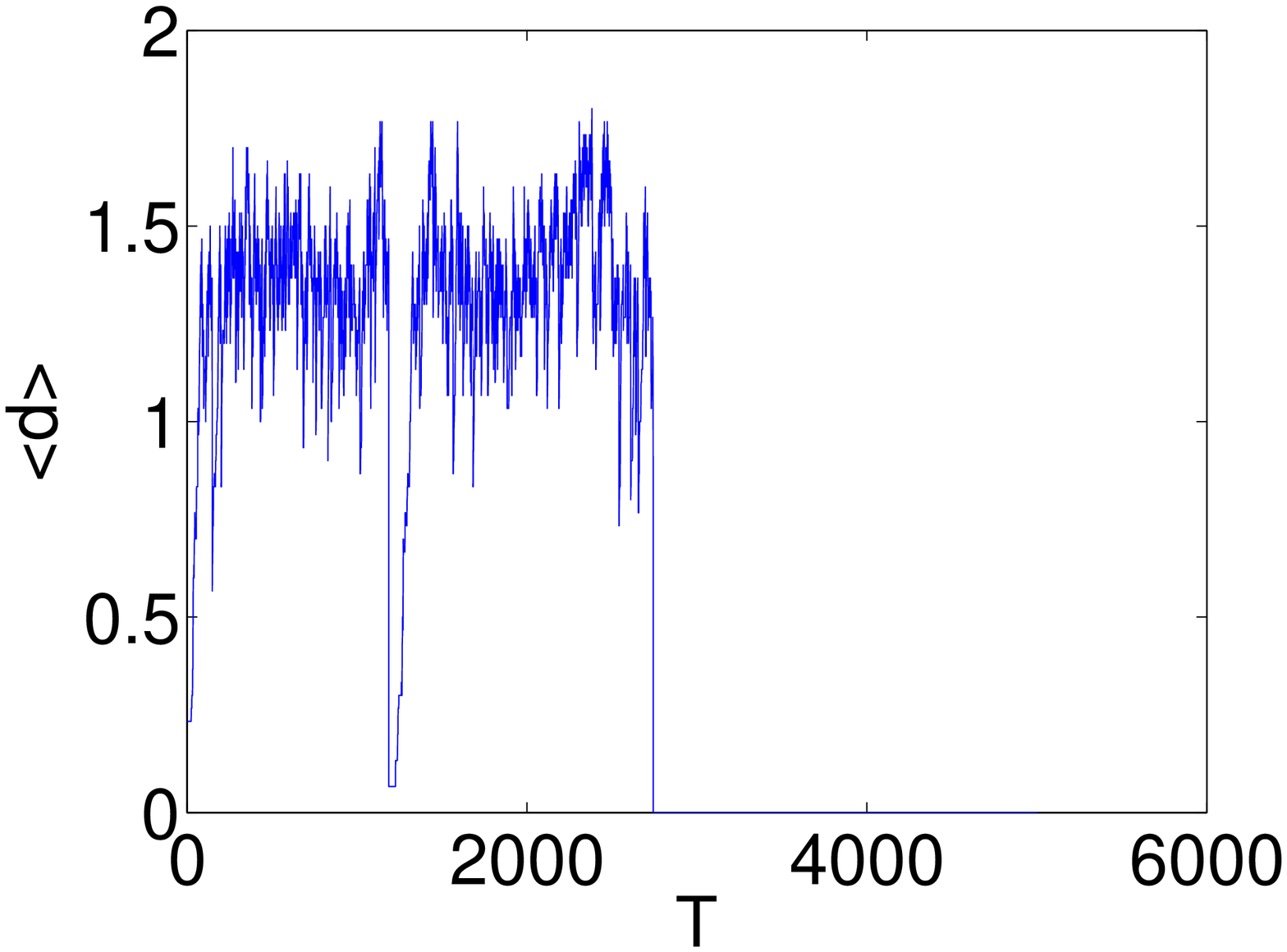}}} \bf{b}
  \end{minipage}
  \begin{minipage}{.45\linewidth}
    \centerline{\scalebox{0.3}{\includegraphics[angle=0]
        {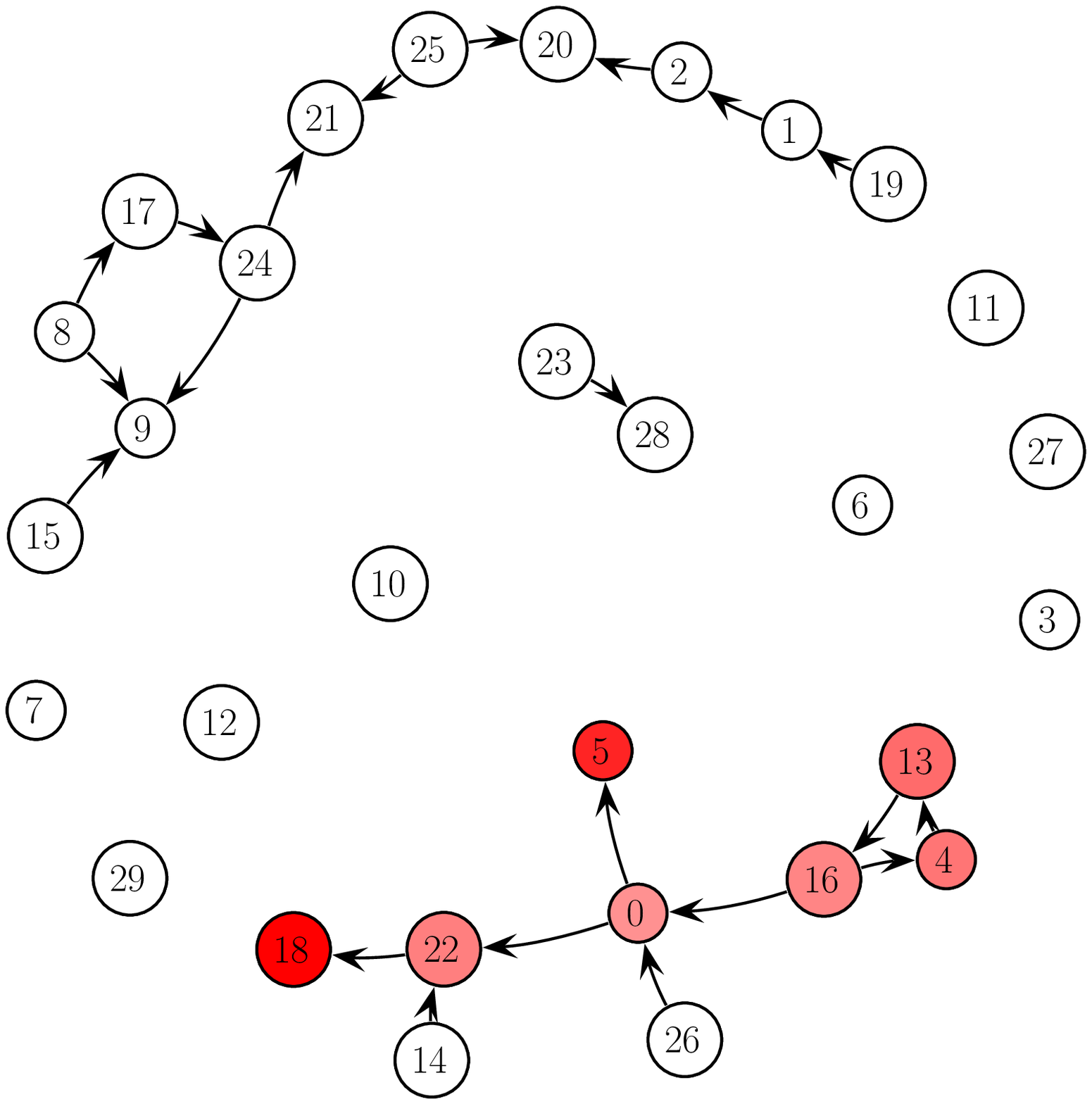}}} \bf{c}
  \end{minipage}
  \hfill
  \begin{minipage}{.45\linewidth}
    \centerline{\scalebox{0.3}{\includegraphics[angle=0]
        {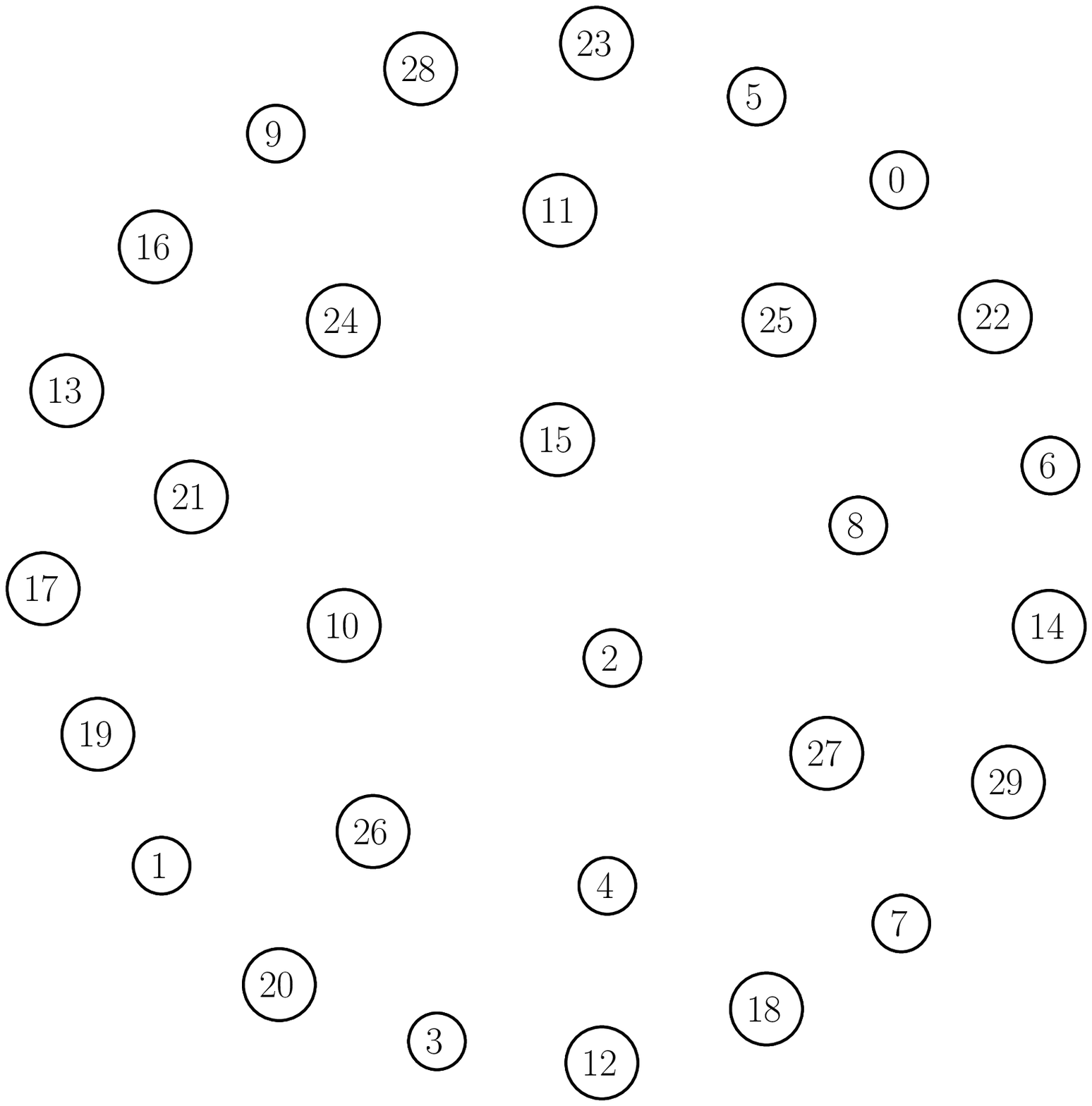}}} \bf{d}
  \end{minipage}
  \caption{Extremal Dynamics: \textbf{(a)} Average utility. \textbf{(b)}
    Average degree. \textbf{(c)} Initial random graph. \textbf{(d)} Graph
    after $5000$ iterations (in the equilibrium). The network experiences
    a total breakdown eventually.}
\label{fig:source_squared_leastFit}
\end{figure}

In this setting, it is possible that the system breaks down completely. A
simulation run exhibiting such a crash can be seen in Fig.
(\ref{fig:source_squared_leastFit}). If the network is sparse enough the
link removal mechanism can destroy the cycles in the network and thus
creates a directed acyclic graph. As soon as the network evolution hits a
directed acyclic graph all value of knowledge vanish (and accordingly the
utilities of the agents) and the network entirely breaks down.

We do not experience a breakdown of the network in the case of null costs
in the last section since there we were considering relative values of
knowledge only. The normalization of the relative values, $\sum_{i=1}^n
y_i=1$ prevents all the shares to become $0$ at the same time, $y_i=0$
$\forall i$. Thus we do not get a total breakdown of the network in which
all values of knowledge vanish. Instead, there the system can always
recover from a crash of the network.

\subsubsection{Utility Driven Dynamics: bilateral link formation}

\begin{figure}[htpb]
  \centering
  \begin{minipage}{.45\linewidth}
    \psfrag{<u>}[][][4][0]{$\langle u \rangle$}
    \psfrag{T}[][][4][0]{$\times T$}
    \centerline{\scalebox{0.3}{\includegraphics[angle=0]
        {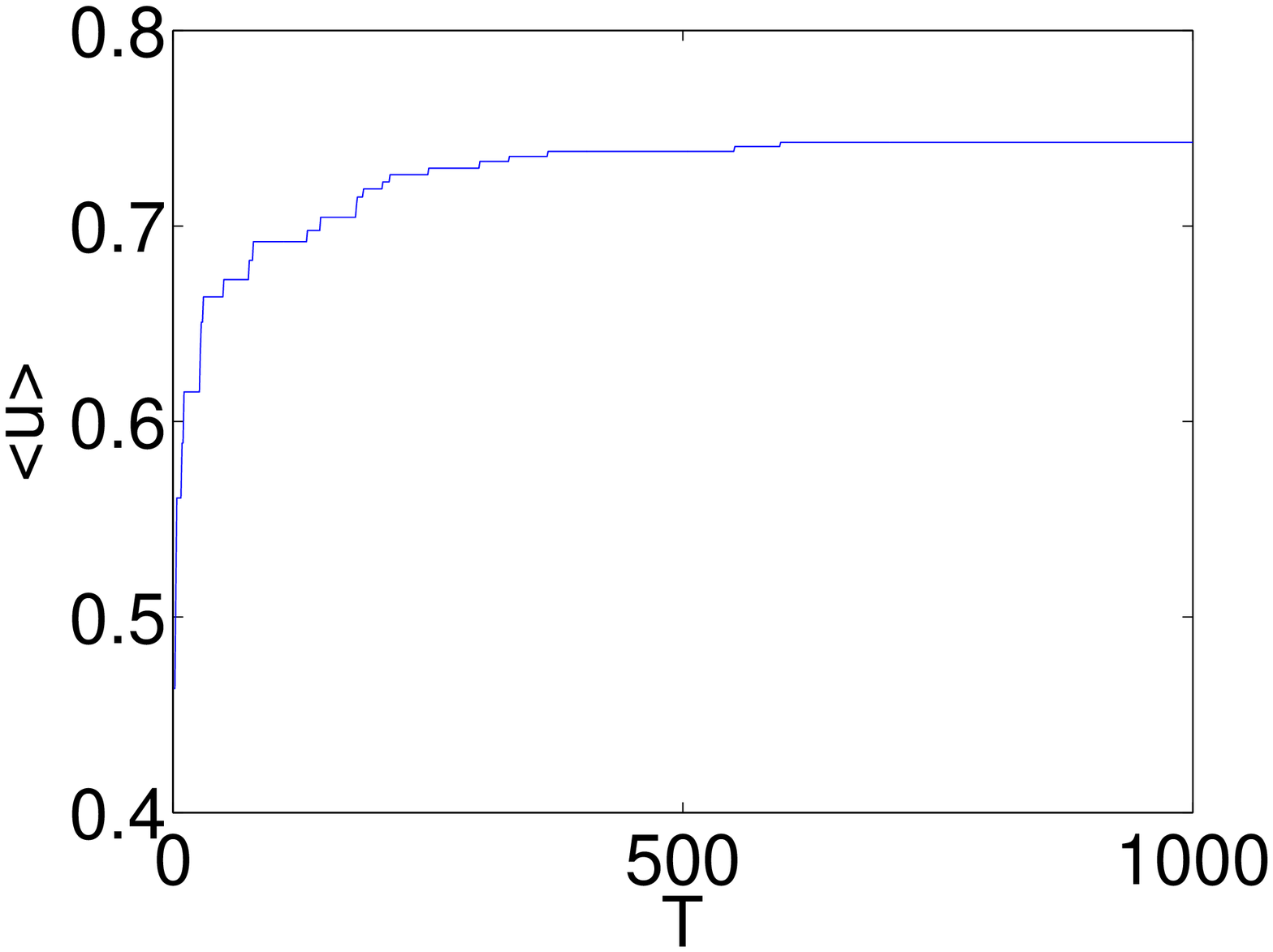}}} \bf{a}
  \end{minipage}
  \hfill
  \begin{minipage}{.45\linewidth}
    \psfrag{<d>}[][][4][0]{$\langle d \rangle$}
    \psfrag{T}[][][4][0]{$\times T$}
    \centerline{\scalebox{0.3}{\includegraphics[angle=0]
        {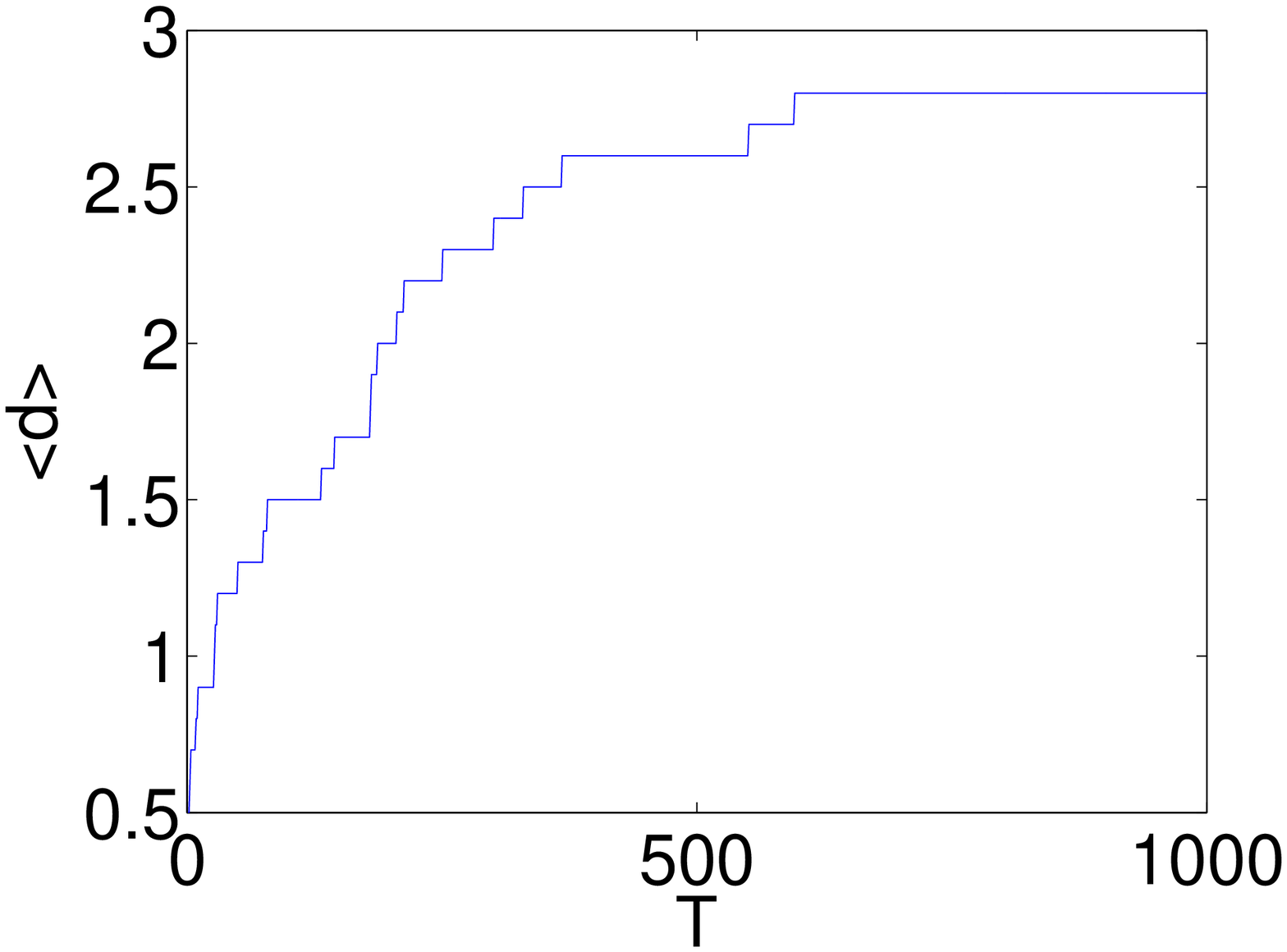}}} \bf{b}
  \end{minipage}
  \begin{minipage}{.45\linewidth}
    \centerline{\scalebox{0.5}{\includegraphics[angle=0]
        {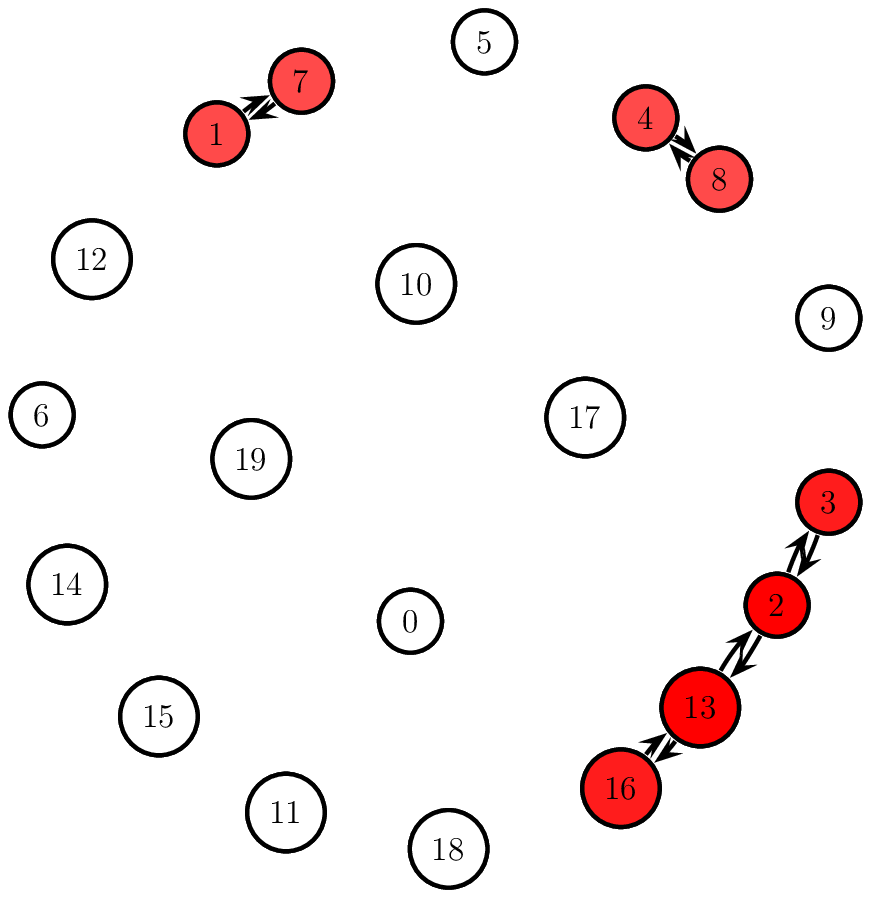}}} \bf{c}
  \end{minipage}
  \hfill
  \begin{minipage}{.45\linewidth}
    \centerline{\scalebox{0.5}{\includegraphics[angle=0]
        {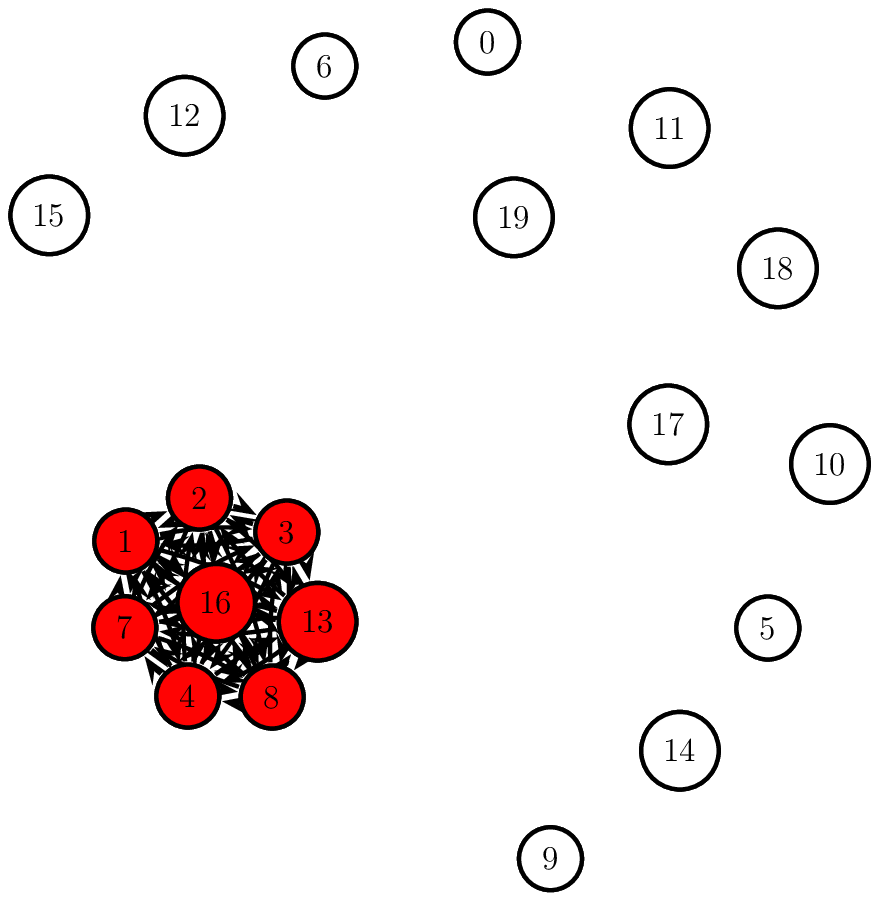}}} \bf{d}
  \end{minipage}
  \caption{Bilateral link formation: \textbf{(a)} Average value of
    knowledge. \textbf{(b)} Average degree.  \textbf{(c)} Initial random
    graph (for reasons of visualization we have chosen a rather sparse
    random graph). \textbf{(d)} Graph after $1000$ iterations (in the
    equilibrium).}
\label{fig:source_squared_bilateral}
\end{figure}

In this section agents are creating or deleting links bilaterally. All
interactions are therefore direct reciprocal. In simulations we observe
the following effect. Bilateral creation and deletion results in a
complete subgraph (The average degree is $1/n \sum d_i = 1/20 \cdot
8\cdot7=2.8$, see Fig.  \ref{fig:source_squared_bilateral}) of the agents
that were part of a permanent set in the initial graph\footnote{the
  creation of the initial random graph with a given link creation
  probability has been chosen rather small such that only a few nodes are
  permanent.} (the stability criterion which defines an equilibrium
network is given in (\ref{defn:bilaterally_stable})).

\subsubsection{Utility Driven Dynamics: unilateral link formation}

The mechanisms of unilateral creation and deletion of links has been
introduced in section (\ref{sec:local_link_formation}). In our
simulations we observe the following observe the following effect. When
we allow for unilateral link formation, large cycles get reduced to a
small set of $2$-cycles. In the equilibrium network (the stability
criterion which defines an equilibrium network is given in
(\ref{defn:unilaterally_stable})) most of the agents are isolated nodes
and thus have vanishing values of knowledge. Only a few of them are
organized in $2$-cycles and small subgraphs consisting of multiple
$2$-cycles. As we will show, the reason for this is that as soon as there
exists a shortcut (a smaller cycle) in a larger cycle agents try to
free-ride and, after the other agents have realized that and sopped
sharing their knowledge with them, they get isolated and experience
vanishing values of knowledge. One can interpret this result as follows:
Even though agents could in principal form indirect reciprocal
interactions the resulting equilibrium network consists only of direct
reciprocal interactions ($2$-cycles and clusters of $2$-cycles).

We can give an example of the process of the reduction of cycles in a
graph $G$ with $3$ nodes for parameter values $d=0.5$, $b=1$ and $c \in
(0,1)$. By numerically comparing utilities (the fixed points of the value
of knowledge) before and after a link is created or deleted we show that
there exists a sequence of link deletions and creations which transform a
$3$-cycle into a $2$-cycle while every link change is associated with an
increase in the utility (the fixed point in the value of knowledge) of
the initiating agent
(\citet{jackson03:_survey_models_network_formation_stability_efficiency}
calls this sequence of graphs an ``improving path'').

\begin{figure}
  \begin{minipage}{.25\textwidth}
    \scalebox{0.6}{\begin{pspicture}(0,-1)(3,4)
      \psset{nodesep=0.5pt,linewidth=1.5pt,arrowsize=5pt 2}
      \cnodeput(0,0){2}{\Large 2}
      \cnodeput(1.5,3){1}{\Large 1} \cnodeput(3,0){3}{\Large 3}
      \ncarc[linewidth=1pt,arcangle=0]{->}{2}{1}
      \ncarc[linewidth=1pt,arcangle=0]{->}{1}{3}
      \ncarc[linewidth=1pt,arcangle=0]{->}{3}{2}
    \end{pspicture}}
  \end{minipage} {\Large\textbf{$\rightarrow$}}
  \begin{minipage}{.25\textwidth}
    \scalebox{0.6}{\begin{pspicture}(0,-1)(3,4)
      \psset{nodesep=0.5pt,linewidth=1.5pt,arrowsize=5pt 2}
      \cnodeput(0,0){2}{\Large 2}
      \cnodeput(1.5,3){1}{\Large 1} \cnodeput(3,0){3}{\Large 3}
      \ncarc[linewidth=1pt,arcangle=10]{->}{3}{1}
      \ncarc[linewidth=1pt,arcangle=10]{->}{1}{3}
      \ncarc[linewidth=1pt,arcangle=0]{->}{3}{2}
      \ncarc[linewidth=1pt,arcangle=0]{->}{2}{1}
    \end{pspicture}}
  \end{minipage}
  \begin{minipage}{.45\linewidth}
    \centerline{\scalebox{0.7}{\includegraphics[angle=0]
        {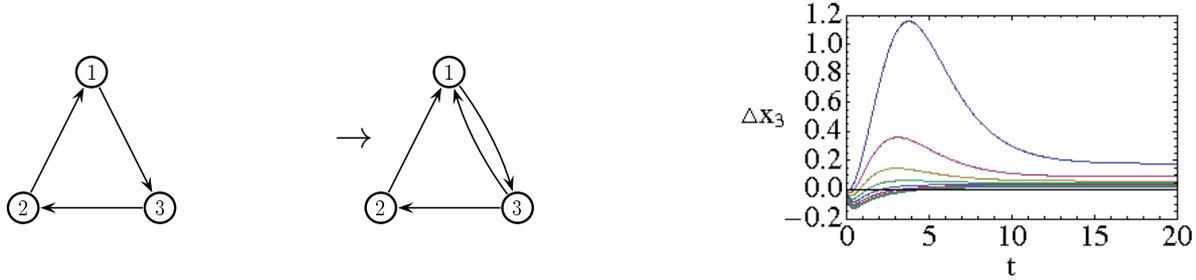}}}
  \end{minipage}
  \caption{Agent $3$ forms a link to agent $1$, $e_{31}$, and thus a
    $2$-cycle is created inside a $3$-cycle.}
\label{fig:9_10}
\end{figure}

In Fig. (\ref{fig:9_10}) (left) agent $3$ creates a link to agent $1$
because in this range of parameters this increases her value of
knowledge. This can be seen in Fig. (\ref{fig:9_10}) (right), where the
increase $\Delta x_3$ different costs $c\in(0,1)$ are plotted and $\Delta
u_3 = \lim_{t \to infty} \Delta x_3>0$.

\begin{figure}
  \begin{minipage}{.25\textwidth}
    \scalebox{0.6}{\begin{pspicture}(0,-1)(3,4)
      \psset{nodesep=0.5pt,linewidth=1.5pt,arrowsize=5pt 2}
      \cnodeput(0,0){2}{\Large 2}
      \cnodeput(1.5,3){1}{\Large 1} \cnodeput(3,0){3}{\Large 3}
      \ncarc[linewidth=1pt,arcangle=10]{->}{3}{1}
      \ncarc[linewidth=1pt,arcangle=10]{->}{1}{3}
      \ncarc[linewidth=1pt,arcangle=0]{->}{3}{2}
      \ncarc[linewidth=1pt,arcangle=0]{->}{2}{1}
    \end{pspicture}}
  \end{minipage} {\Large\textbf{$\rightarrow$}}
  \begin{minipage}{.25\textwidth}
    \scalebox{0.6}{\begin{pspicture}(0,-1)(3,4)
      \psset{nodesep=0.5pt,linewidth=1.5pt,arrowsize=5pt 2}
      \cnodeput(0,0){2}{\Large 2}
      \cnodeput(1.5,3){1}{\Large 1} \cnodeput(3,0){3}{\Large 3}
      \ncarc[linewidth=1pt,arcangle=10]{->}{1}{3}
      \ncarc[linewidth=1pt,arcangle=10]{->}{3}{1}
      \ncarc[linewidth=1pt,arcangle=0]{->}{3}{2}
    \end{pspicture}}
  \end{minipage}
  \begin{minipage}{.45\linewidth}
    \centerline{\scalebox{0.7}{\includegraphics[angle=0]
        {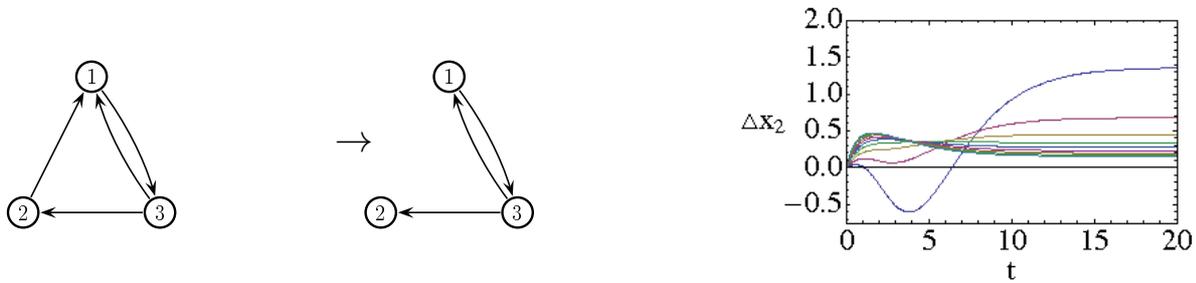}}}
  \end{minipage}
  \caption{Deletion of the link $e_{21}$. Agent $2$ is not sharing any
    knowledge with others but only receiving knowledge from agent $3$.
    Thus, agent $2$ is freeriding.}
\label{fig:10_3}
\end{figure}

In Fig. (\ref{fig:10_3}) (left) agent $2$ removes her link to agent $1$
and thus she stops contributing knowledge but instead is only receiving
knowledge from agent $3$. We say that agent $2$ is free-riding. This
increases her utility, since $\Delta u_2 = \lim_{t \to infty} \Delta
x_2>0$, as can be seen in Fig. (\ref{fig:10_3}) (right) for different
costs $c$.

\begin{figure}
  \begin{minipage}{.25\textwidth}
    \scalebox{0.6}{\begin{pspicture}(0,-1)(3,4)
      \psset{nodesep=0.5pt,linewidth=1.5pt,arrowsize=5pt 2}
      \cnodeput(0,0){2}{\Large 2}
      \cnodeput(1.5,3){1}{\Large 1} \cnodeput(3,0){3}{\Large 3}
      \ncarc[linewidth=1pt,arcangle=10]{->}{1}{3}
      \ncarc[linewidth=1pt,arcangle=10]{->}{3}{1}
      \ncarc[linewidth=1pt,arcangle=0]{->}{3}{2}
    \end{pspicture}}
  \end{minipage} {\Large\textbf{$\rightarrow$}}
  \begin{minipage}{.25\textwidth}
    \scalebox{0.6}{\begin{pspicture}(0,-1)(3,4)
      \psset{nodesep=0.5pt,linewidth=1.5pt,arrowsize=5pt 2}
      \cnodeput(0,0){2}{\Large 2}
      \cnodeput(1.5,3){1}{\Large 1} \cnodeput(3,0){3}{\Large 3}
      \ncarc[linewidth=1pt,arcangle=10]{->}{1}{3}
      \ncarc[linewidth=1pt,arcangle=10]{->}{3}{1}
    \end{pspicture}}
  \end{minipage}
  \begin{minipage}{.45\linewidth}
    \centerline{\scalebox{0.7}{\includegraphics[angle=0]
        {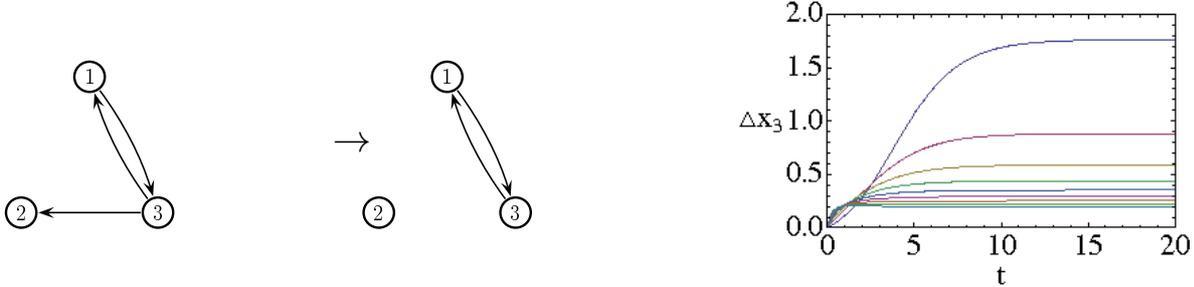}}}
  \end{minipage}
  \caption{Deletion of the link $e_{32}$ by agent $3$. Agent $3$ realizes
    that she is better off by not sharing her knowledge with agent $2$.
    Agent $2$, that was free-riding before now gets isolated and
    experiences a vanishing value of knowledge in the long run.}
\label{fig:3_16}
\end{figure}

Finally, in Fig. (\ref{fig:3_16}) (left) agent $3$ removes her link to
agent $2$ because she is better off, as illustrated in Fig.
(\ref{fig:3_16}) (right), when she stops contributing knowledge to an
agent that is nothing contributing in return. This is actually true for
agent $2$.  Agent $2$ therefore gets isolated and experiences a vanishing
value of knowledge in the long-run, $\lim_{t \to \infty} x_2 =0$. Her
utility is null.

\begin{figure}[htpb]
  \centering
  \begin{minipage}{.45\linewidth}
    \psfrag{<u>}[][][4][0]{$\langle u \rangle$}
    \psfrag{T}[][][4][0]{$\times T$}
    \centerline{\scalebox{0.3}{\includegraphics[angle=0]
        {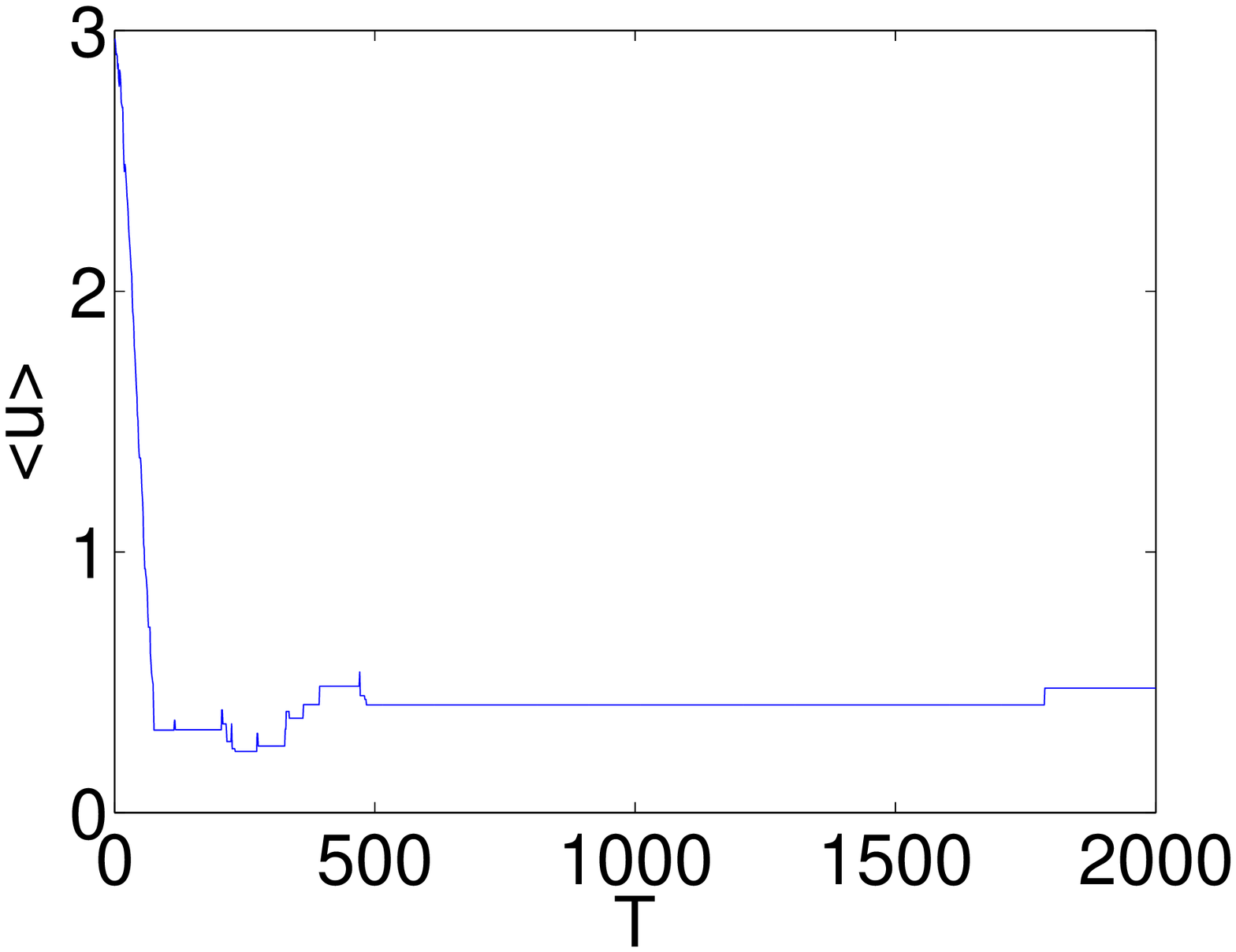}}} \bf{a}
  \end{minipage}
  \hfill
  \begin{minipage}{.45\linewidth}
    \psfrag{<d>}[][][4][0]{$\langle d \rangle$}
    \psfrag{T}[][][4][0]{$\times T$}
    \centerline{\scalebox{0.3}{\includegraphics[angle=0]
        {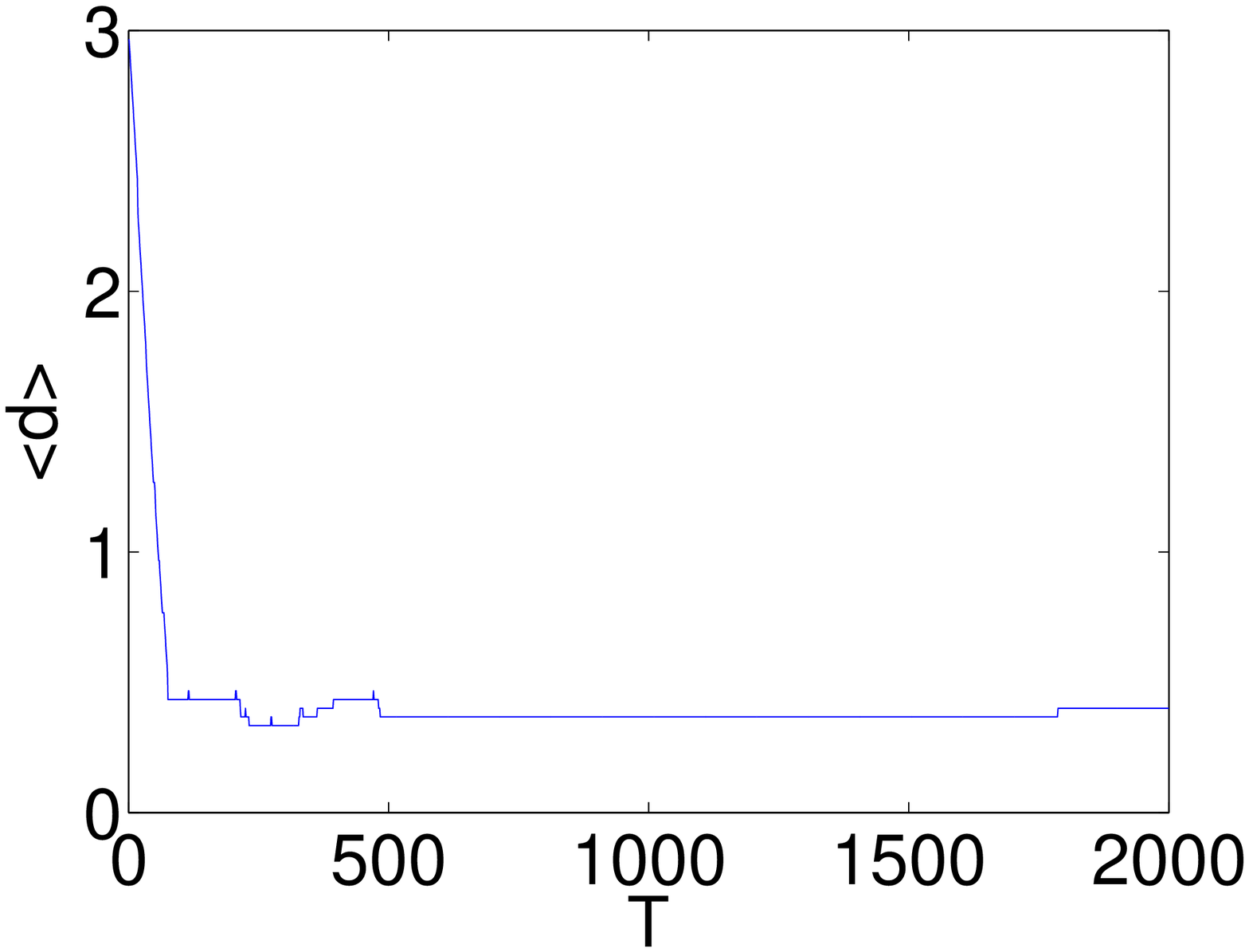}}} \bf{b}
  \end{minipage}
  \begin{minipage}{.45\linewidth}
    \centerline{\scalebox{0.3}{\includegraphics[angle=0]
        {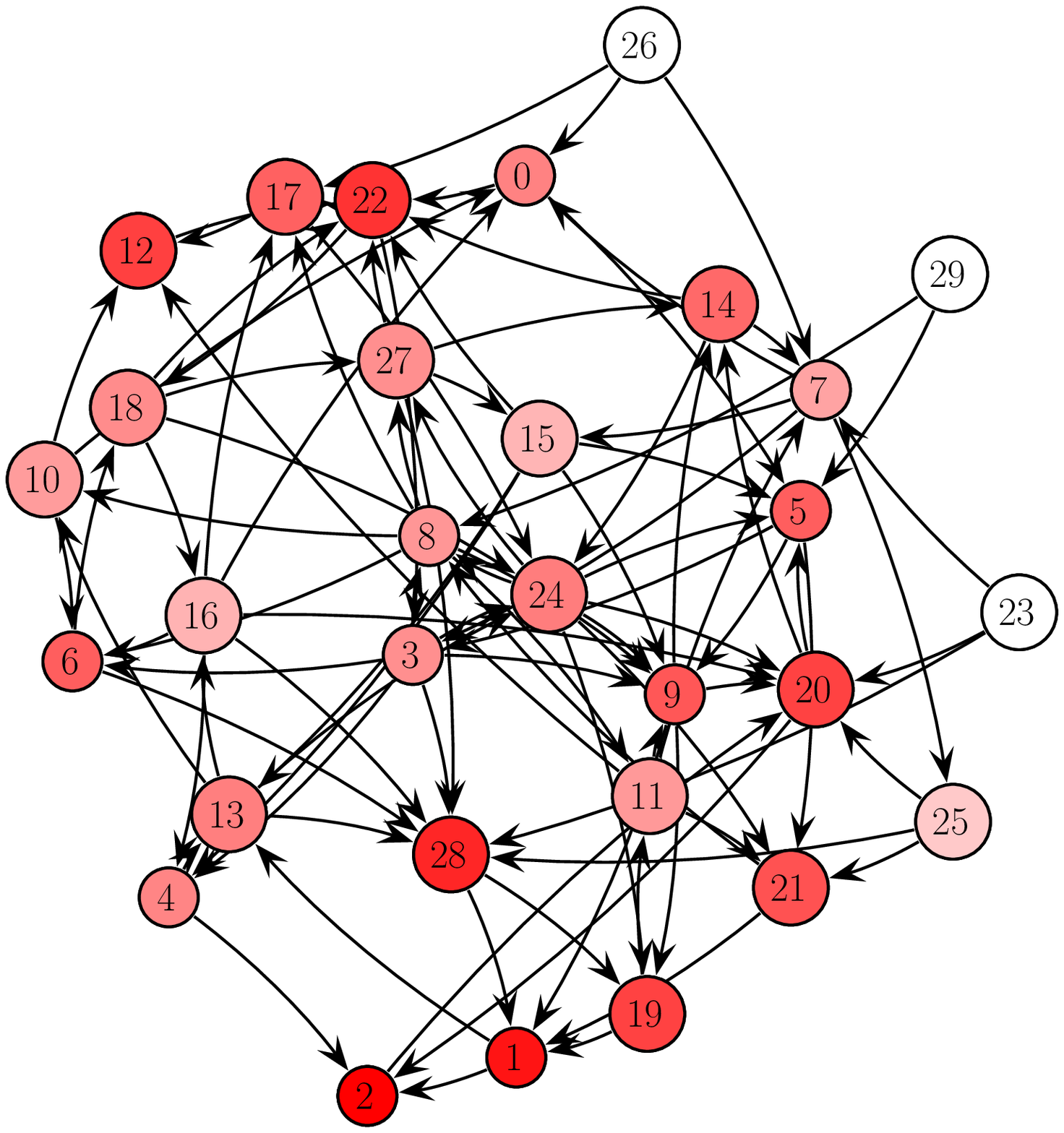}}} \bf{c}
  \end{minipage}
  \hfill
  \begin{minipage}{.45\linewidth}
    \centerline{\scalebox{0.3}{\includegraphics[angle=0]
        {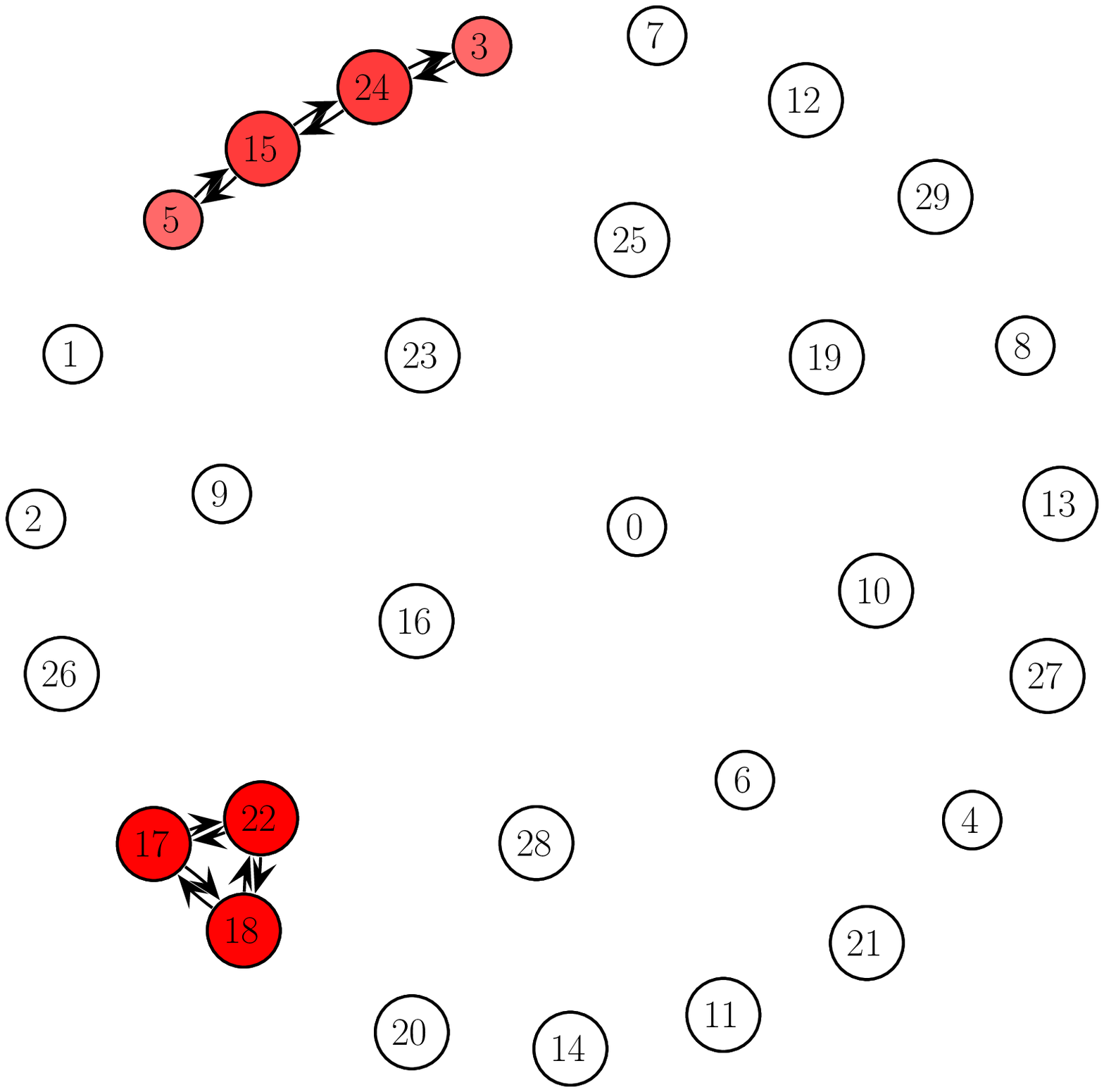}}} \bf{d}
  \end{minipage}
  \caption{Unilateral link formation: \textbf{(a)} Average utility.
    \textbf{(b)} Average degree. \textbf{(c)} Initial random graph.
    \textbf{(d)} Graph after $2000$ iterations (in the equilibrium). For
    the parameter values $d=0.5$, $b=0.5$, $c=0.1$, used in this
    simulation, the complete graph $K_3$ is an equilibrium. Note from
    section (\ref{sec:stationary_solutions_G3}) one can see that for
    parameter values $d=0.5$, $b=1$, $c=0.5$ this is no longer the case
    and $K_3$ would be reduced to a $2$-cycle $C_2$.}
\label{fig:source_squared_unilateral}
\end{figure}

We end up in a setting where out of a cooperation of many (the sharing of
knowledge) only a small set of cooperators remains and all the remaining
agents vanish, i.e. have vanishing values of knowledge and utility. We
can see this in a simulation starting from an initial random graph with
$30$ agents and the resulting equilibrium network in Fig.
(\ref{fig:source_squared_unilateral}) (bottom right).

Since the performance of the system in terms of the total value of
knowledge is very low, we investigate in the next section the conditions
under which the performance can be increased (with more agents being
permanent in the equilibrium). We find that the existence of a positive
network externality (explained in the next section) can enhance the
performance of the system.

\subsection{Introducing Positive Network  Externalities}
\label{sec:externalities}

In this section we study the growth of the value of knowledge which
includes an additional benefit term contributing to an increase in the
value of knowledge. This additional benefit depends on the network
structure itself. In the economic literature
\citep{tirole88:_theor_indus_orgniz,mas-colell95:_microec}, ``positive
network externalities arise when a good is more valuable to a user the
more users adopt the same good or compatible goods''. In our model we
define a network externality simply as a function of the network
structure that affects the utility of an agent. Including the externality
in the benefit can yield more complex structures with non vanishing
knowledge values as equilibrium networks. The growth of the value of
knowledge of agent $i$ is given by the following equation:
\begin{equation}
  \frac{dx_i}{dt}= - d x_i + b \sum_{i=1}^{n} a_{ji} x_j +
  \underbrace{b_e \sum_{i=1}^{n} w_{ji}
    x_j}_{\text{positive network externality}}  - c \sum_{i=1}^{n}
  a_{ij} x_i^2
  \label{eq:centrality}
\end{equation}
and the utility is again given by $u_i=\lim_{t \to \infty} x_i(t)$. Link
changes are based on the increase in utility. The network benefit
incorporates the fact that the value of knowledge can change with the
number of users of that knowledge, (\ref{eq:centrality}).  But the number
of users can either enhance or diminish the value of knowledge that is
being transferred between agents, depending on the type of knowledge
under investigation.  On one hand, the value can decrease with the number
of agents that pass on that knowledge.  Knowledge is attenuated with the
distance from the creator to the receiver. We study this type of
knowledge with a link weight defined in (\ref{eq:centrality}) and denoted
by $w_{ji}^{\text{c}}$. In the next section we study the opposite effect:
the value of knowledge increases with the number of users. This holds for
example for general purpose technologies that get more valuable the more
they are applied and used in different contexts (and users). The link
weights used for this type of knowledge in (\ref{eq:centrality}) are
denoted by $w_{ji}^{\text{ccn}}, w_{ji}^{\text{cce}}$, where the first
measures the number of agents using that knowledge and the second the
number of interactions.

We introduce different link weights, denoted by $w_{ji}^{\text{c}},
w_{ji}^{\text{ccn}}, w_{ji}^{\text{cce}}$. Moreover, agents are creating
and deleting links unilaterally (Utility Driven Dynamics). We then study
the effect of different weights on the equilibrium networks obtained.

\subsubsection{Centrality}
\label{sec:centrality}

The growth function of the value of agent $i$ is given by
\begin{equation}
  \frac{dx_i}{dt}= -d x_i + \sum_{j=1}^{n} (b a_{ji} + b_e
  w_{ji}^{\text{c}}) x_j - c \sum_{j=1}^{n} a_{ij} x_i^2 
\end{equation}
The utility is given by $u_i=\lim_{t \to \infty} x_i(t)$ (large $T$).
Link changes are accepted on the basis of an increase in utility. The
\textit{centrality} measure computes the sum of the inverse lengths of
all the shortest paths containing the link for which the centrality is
computed. If two agents are not connected then the length of the path is
assumed to be infinity and thus its weight is zero.  Instead, if two
agents are directly connected via an link, then the weight is one. The
weight values links that brings agents closer to each other higher. This
is a similar approach to the Connections Model introduced in section
\ref{sec:connections-model} with a utility given by
(\ref{eq:utility_connections_model}). The centrality link weight,
$w_{ij}^{\text{c}}$, is then computed as follows.
\begin{equation}
  w_{ij}^{\text{c}} = \sum_{v \in V} \frac{1}{(d_{jv}+1)},  w_{ij}^{\text{c}} \in [0,1]
\end{equation}
$d_{jv}$ is the shortest path between node $j$ and node $v$. If there
exists no path between two nodes, then the distance between them is
infinity\footnote{We use a standard \textit{depth-first-search} algorithm
  to compute the shortest paths. More details on this algorithm and
  further discussion is given in \citet{steger01:_diskr_struk,
    steger02:_diskr_struk, ahuja93:_networ, cormen01:_introd}}.

\begin{figure}[htpb]
  \centering
  \begin{minipage}{.45\linewidth}
    \psfrag{<u>}[][][4][0]{$\langle u \rangle$}
    \psfrag{T}[][][4][0]{$\times T$}
    \centerline{\scalebox{0.3}{\includegraphics[angle=0]
        {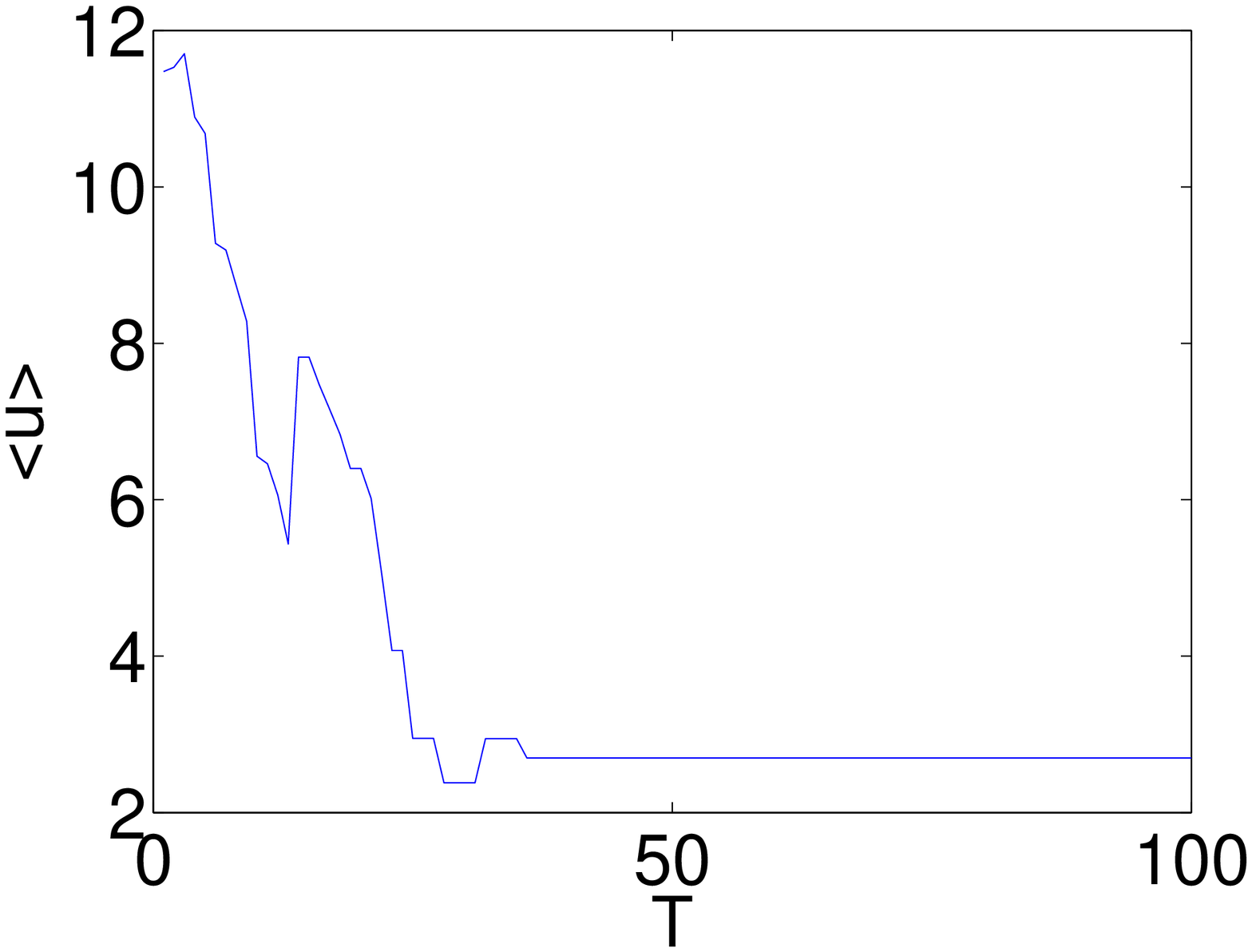}}}
    \bf{a}
  \end{minipage}
  \hfill
  \begin{minipage}{.45\linewidth}
    \psfrag{<d>}[][][4][0]{$\langle d \rangle$}
    \psfrag{T}[][][4][0]{$\times T$}
    \centerline{\scalebox{0.3}{\includegraphics[angle=0]
        {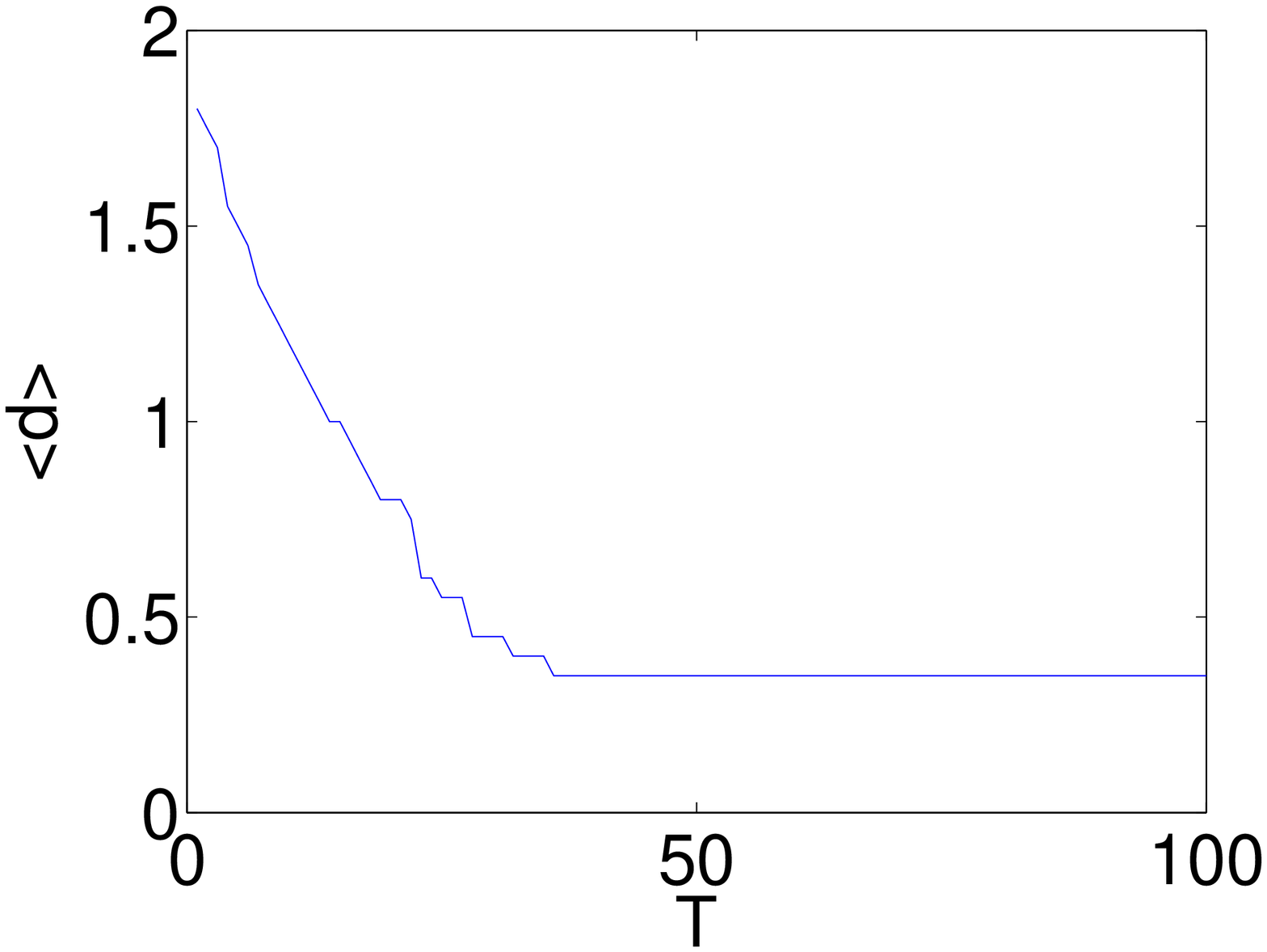}}}
    \bf{b}
  \end{minipage}
  \begin{minipage}{.45\linewidth}
    \centerline{\scalebox{0.5}{\includegraphics[angle=0]
        {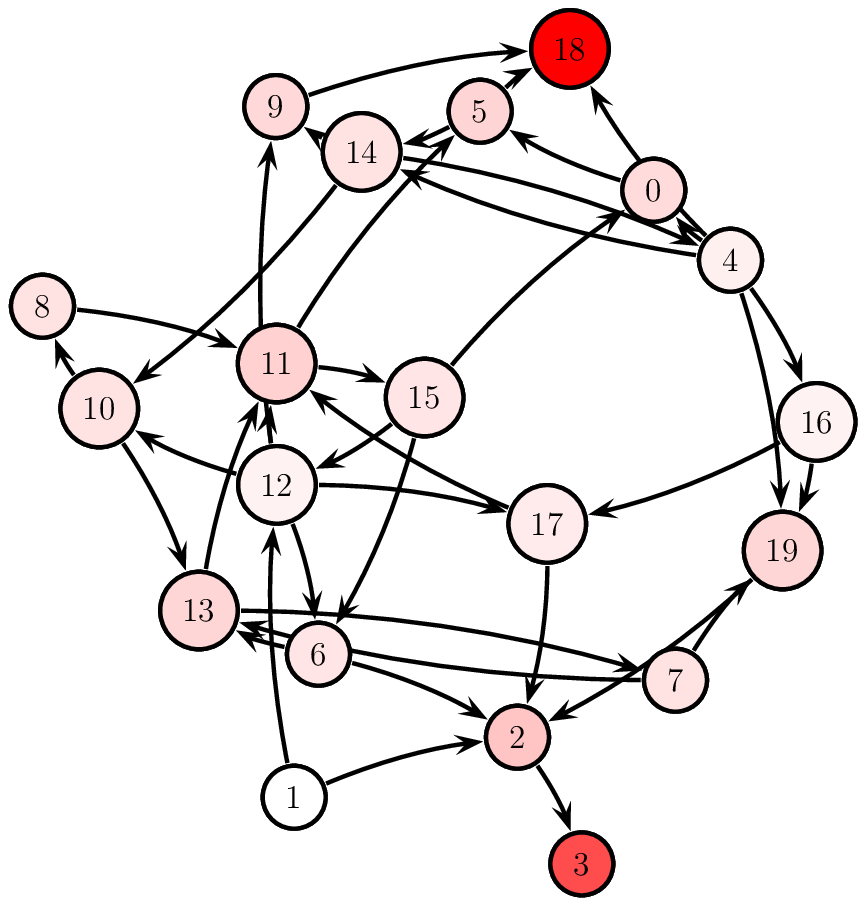}}}
    \bf{c}
  \end{minipage}
  \hfill
  \begin{minipage}{.45\linewidth}
    \centerline{\scalebox{0.5}{\includegraphics[angle=0]
        {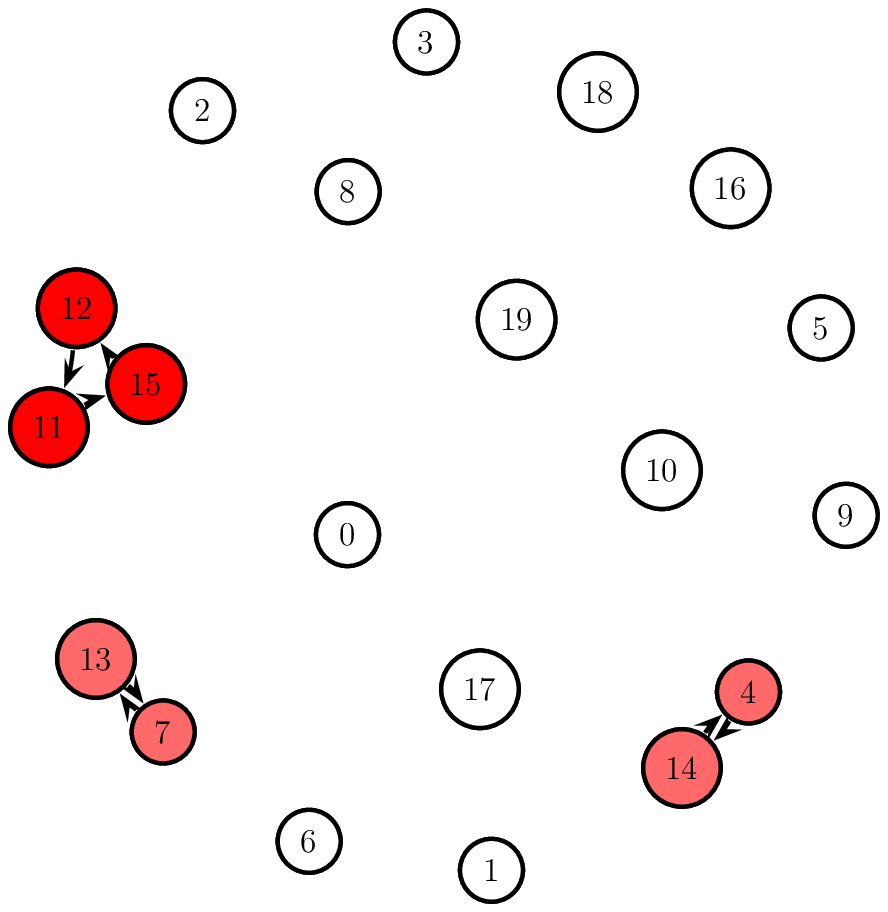}}}
    \bf{d}
  \end{minipage}
  \caption{Centrality (\ref{sec:centrality}): \textbf{(a)} Average
    utility. \textbf{(b)} Average degree.  \textbf{(c)} Initial random
    graph. \textbf{(d)} Graph after $500$ iterations (in the
    equilibrium).}
\label{fig:centrality}
\end{figure}

In simulations, Fig. (\ref{fig:centrality}), we observe, that in the
equilibrium network only a few agents have non-vanishing utilities (the
asymptotic knowledge values) and most of them are isolated nodes with
zero utilities. This result does not differ too much from the studies in
section (\ref{sec:source_squared}) where no externality is considered.
Apparently, if more agents should have non-vanishing utilities induced by
an additional benefit depending on the network structure, this cannot be
realized with the centrality link weight.

\subsubsection{Circuit-Centrality}
\label{sec:circuit_centrality}

The \textit{circuit-centrality} measure puts a weight on the links that
depends on the number of distinct nodes that are contained in all the
circuits going through the link under consideration. The motivation is
that, if many agents are involved in the transfer of knowledge and this
knowledge then comes back to the agent (thus creating a feedback on the
technology issued by the agent), it gets an added value (e.g. for general
purpose technologies (GPT) \citep{bresnahan1995:general_purpose_tech,
  stoneman95:_technological_diffusion,
  cohen95:_empirical_innov_activity})
The more agents use a technology the more it is improved and so the more
agents are involved in such a feedback loop the higher is the value of
the technology. We can either count the number of different agents
involved in this feedback loop or the number of interactions (links).
Either possibility is explored in the next sections. This is an
alternative way to study the emergence of indirect reciprocity where
others \citet{nowak2005} have studied it by introducing a (global)
reputation mechanism.

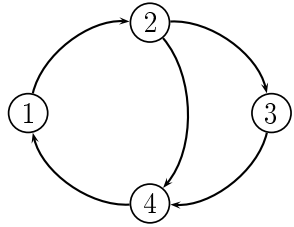
\begin{figure}
    \centering
    \scalebox{0.8}[0.8]{
      \begin{pspicture}(0,1)(5,5) \cnodeput(1,3) {1}{\Large 1}
        \cnodeput(3,4.5){2}{\Large 2} \cnodeput(3,1.5){3}{\Large 4}
        \cnodeput(5,3) {4}{\Large 3}
        \ncarc[linewidth=1pt,arcangle=40]{->}{1}{2}
        \ncarc[linewidth=1pt,arcangle=40]{->}{2}{3}
        \ncarc[linewidth=1pt,arcangle=40]{->}{3}{1}
        \ncarc[linewidth=1pt,arcangle=40]{->}{2}{4}
        \ncarc[linewidth=1pt,arcangle=40]{->}{4}{3}
      \end{pspicture}}
    \label{fig:circuit_example}
    \caption{The link $e_{12}$ is contained in two circuits with nodes
      ${1,2,4}$ and ${1,2,3,4}$. The number of distinct nodes in these
      circuits is $4$ and the number of distinct links is $5$.
      Accordingly $w_{ij}^{\text{ccn}}=\frac{4}{4}=1$ and
      $w_{ij}^{\text{cce}}=\frac{5}{12}=0.42$.}
  \end{figure}

We then define the weight of an link $w_{ij}$ as (i) the number $m_n$ of
distinct \textit{nodes} that are in the circuits from node $i$ to $j$,
  \begin{equation}
    w_{ij}^{\text{ccn}}=\frac{m_n}{n}, \text{   } w_{ij}^{\text{ccn}} \in [0,1]
\label{eq:cc-nodes}
  \end{equation}
  and (ii) the number  $m_e$ of distinct \textit{links} that are in the circuits
  from node $i$ to $j$,
  \begin{equation}
    w_{ij}^{\text{cce}}=\frac{m_e}{n(n-1)}, \text{   }
    w_{ij}^{\text{cce}} \in [0,1] 
\label{eq:cc-edges}
  \end{equation}
  An example of the different link weights can be seen in Fig.
  (\ref{fig:circuit_example}).

  In order to compute all circuits in a directed graph $G$ one needs to
  compute the trails in $G$. The closed trails then are the circuits in
  $G$. We use an algorithm to compute all trails in $G$ from a given
  source node $s$. An explanation of the algorithm is given in appendix
  (\ref{sec:algorithm}).

  By introducing the circuit-centrality externality, we will show that
  more agents are permanent in the equilibrium network. The performance
  of the system is increased compared to the equilibrium networks that
  emerge with unilateral link formation without this externality.

\begin{figure}[htpb]
  \centering
  \begin{minipage}{.45\linewidth}
    \psfrag{<u>}[][][4][0]{$\langle u \rangle$}
    \psfrag{T}[][][4][0]{$\times T$}
    \centerline{\scalebox{0.3}{\includegraphics[angle=0]
        {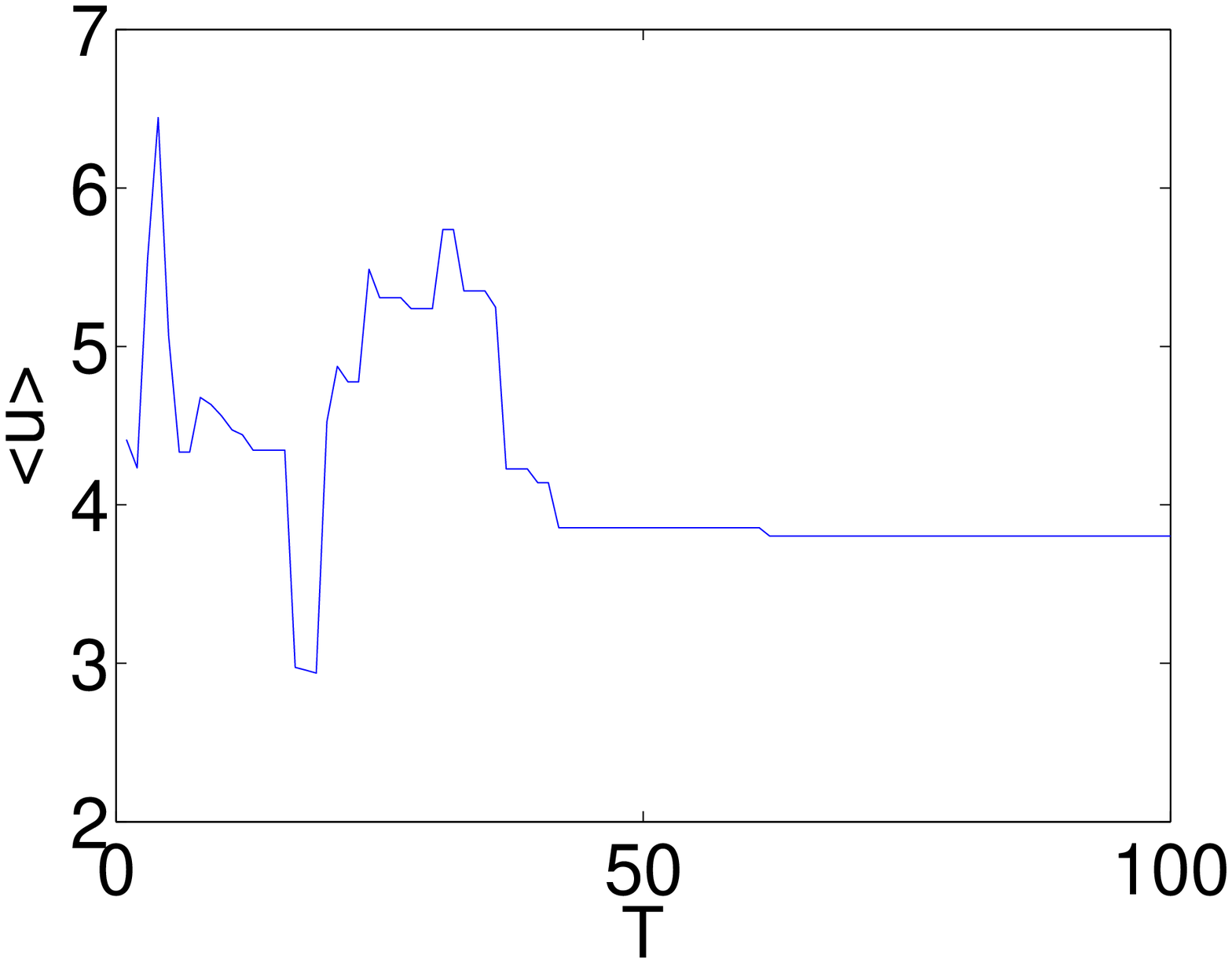}}}
    \bf{a}
  \end{minipage}
  \hfill
  \begin{minipage}{.45\linewidth}
    \psfrag{<d>}[][][4][0]{$\langle d \rangle$}
    \psfrag{T}[][][4][0]{$\times T$}
    \centerline{\scalebox{0.3}{\includegraphics[angle=0]
        {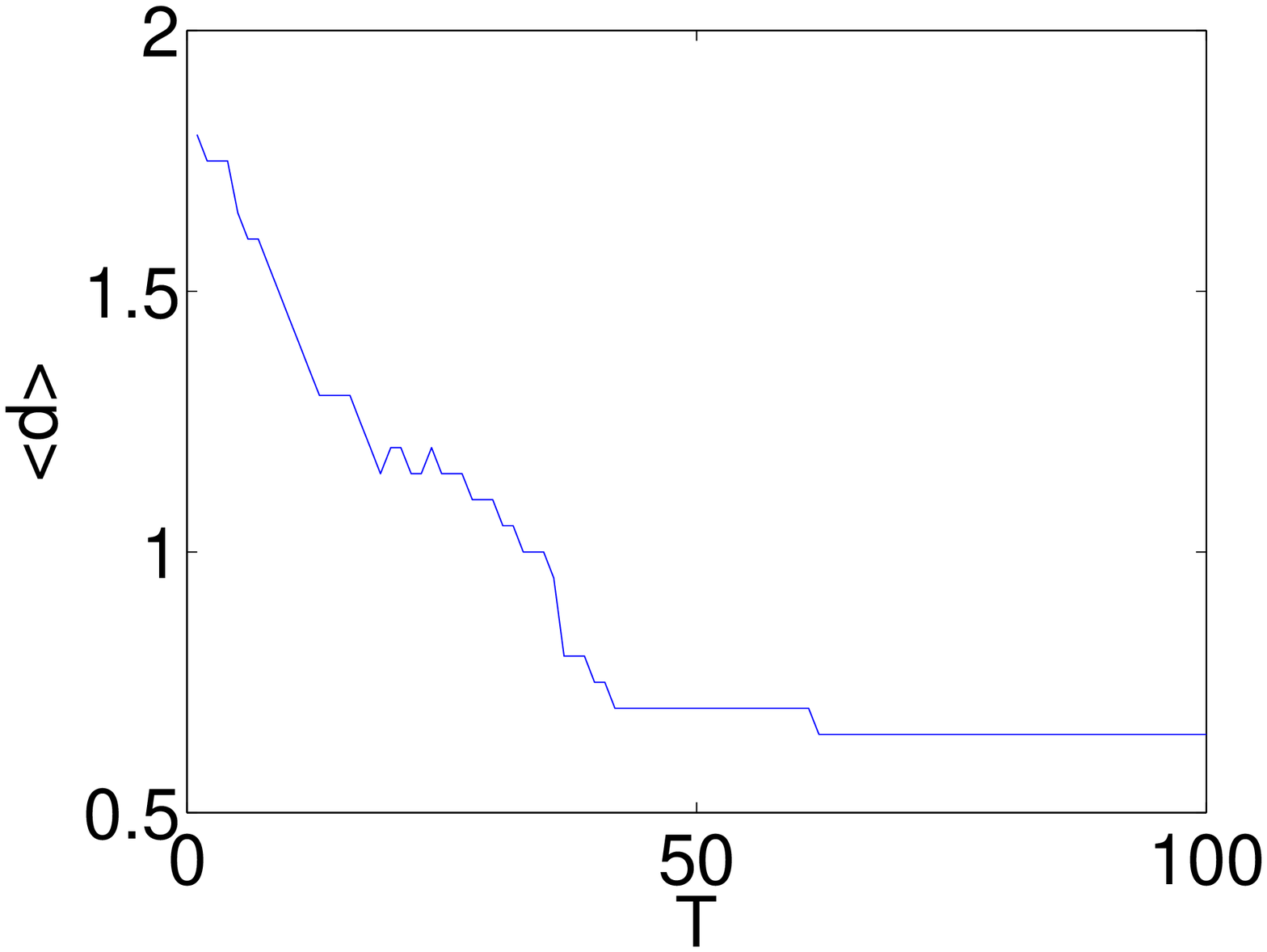}}}
    \bf{b}
  \end{minipage}
  \begin{minipage}{.45\linewidth}
    \centerline{\scalebox{0.5}{\includegraphics[angle=0]
        {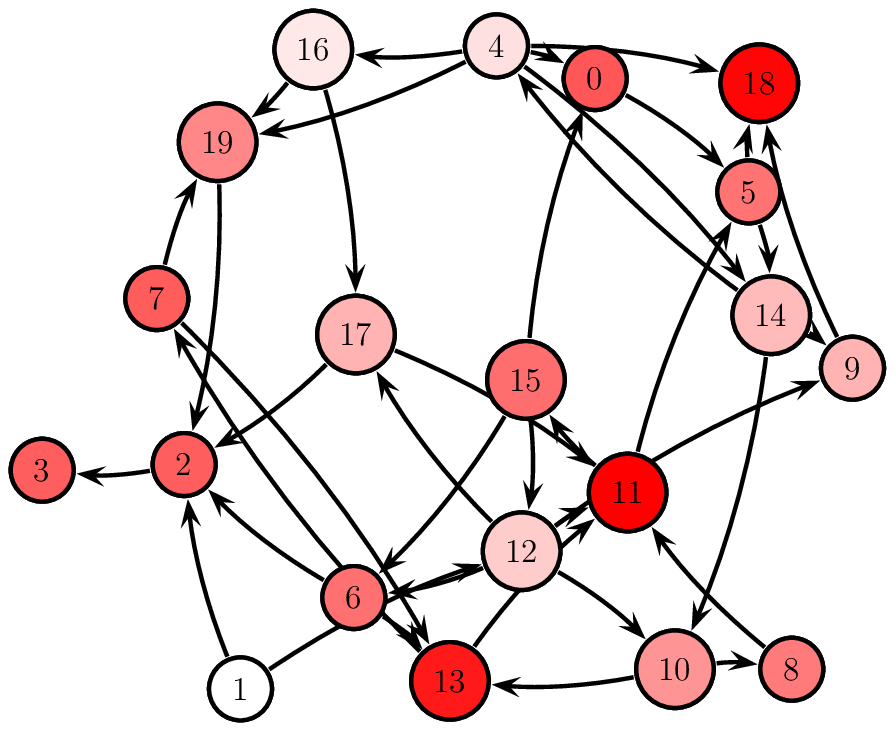}}}
    \bf{c}
  \end{minipage}
  \hfill
  \begin{minipage}{.45\linewidth}
    \centerline{\scalebox{0.5}{\includegraphics[angle=0]
        {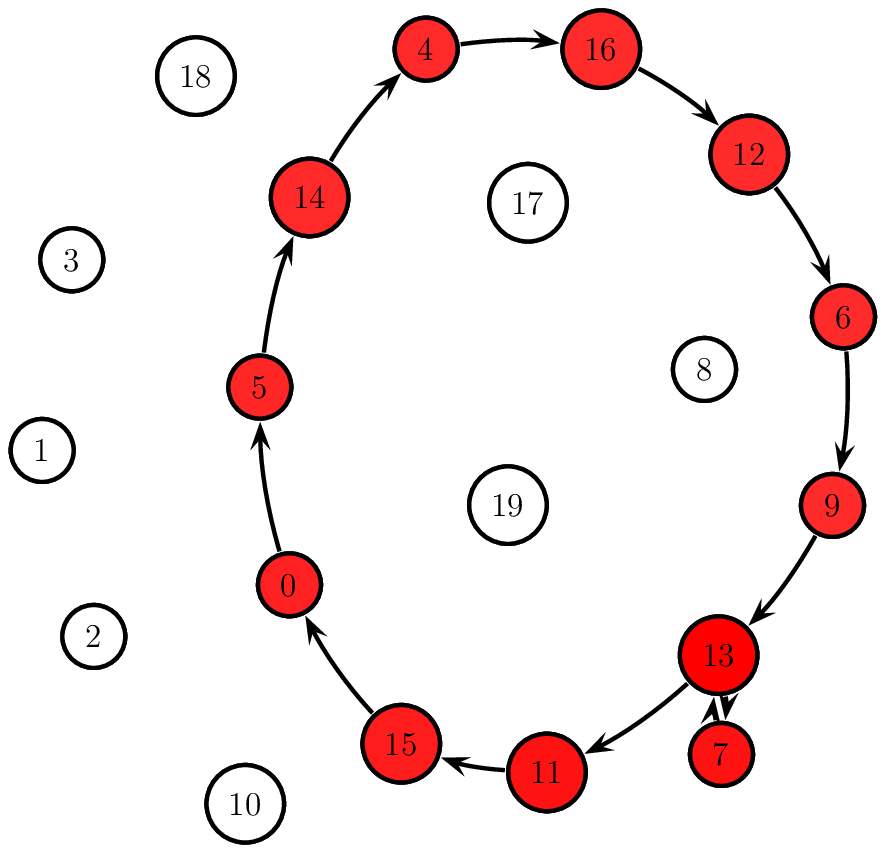}}}
    \bf{d}
  \end{minipage}
  \caption{Circuit number of nodes externality: \textbf{(a)} Average
    utility. \textbf{(b)} Average degree.  \textbf{(c)} Initial random
    graph. \textbf{(d)} Graph after $500$ iterations (in the
    equilibrium).}
\label{fig:circuitNNodes}
\end{figure}

  Using circuit-centrality measures the number of nodes
  (\ref{eq:cc-nodes}) and the growth of the value of knowledge is given
  by
\begin{equation}
  \frac{dx_i}{dt}= -d x_i + \sum_{j=1}^{n} (b a_{ji} + b_e
  w_{ji}^{\text{ccn}}) x_j - c \sum_{j=1}^{n} a_{ij} x_i^2
\end{equation}
The utility is given by $u_i=\lim_{t \to \infty} x_i(t)$ (large $T$).
Link changes are accepted on the basis of an increase in utility.
Different to the centrality externality (\ref{sec:centrality}) we observe
larger cycles as the equilibrium networks. This can be seen in the
simulation run in Fig.  (\ref{fig:circuitNNodes})
This positive externality allows for more agents to be permanent in the
equilibrium network than without an externality or with the
centrality-externality. We thus obtain a higher performance of the
system.

\begin{figure}[htpb]
  \centering
  \begin{minipage}{.45\linewidth}
    \psfrag{<u>}[][][4][0]{$\langle u \rangle$}
    \psfrag{T}[][][4][0]{$\times T$}
    \centerline{\scalebox{0.3}{\includegraphics[angle=0]
        {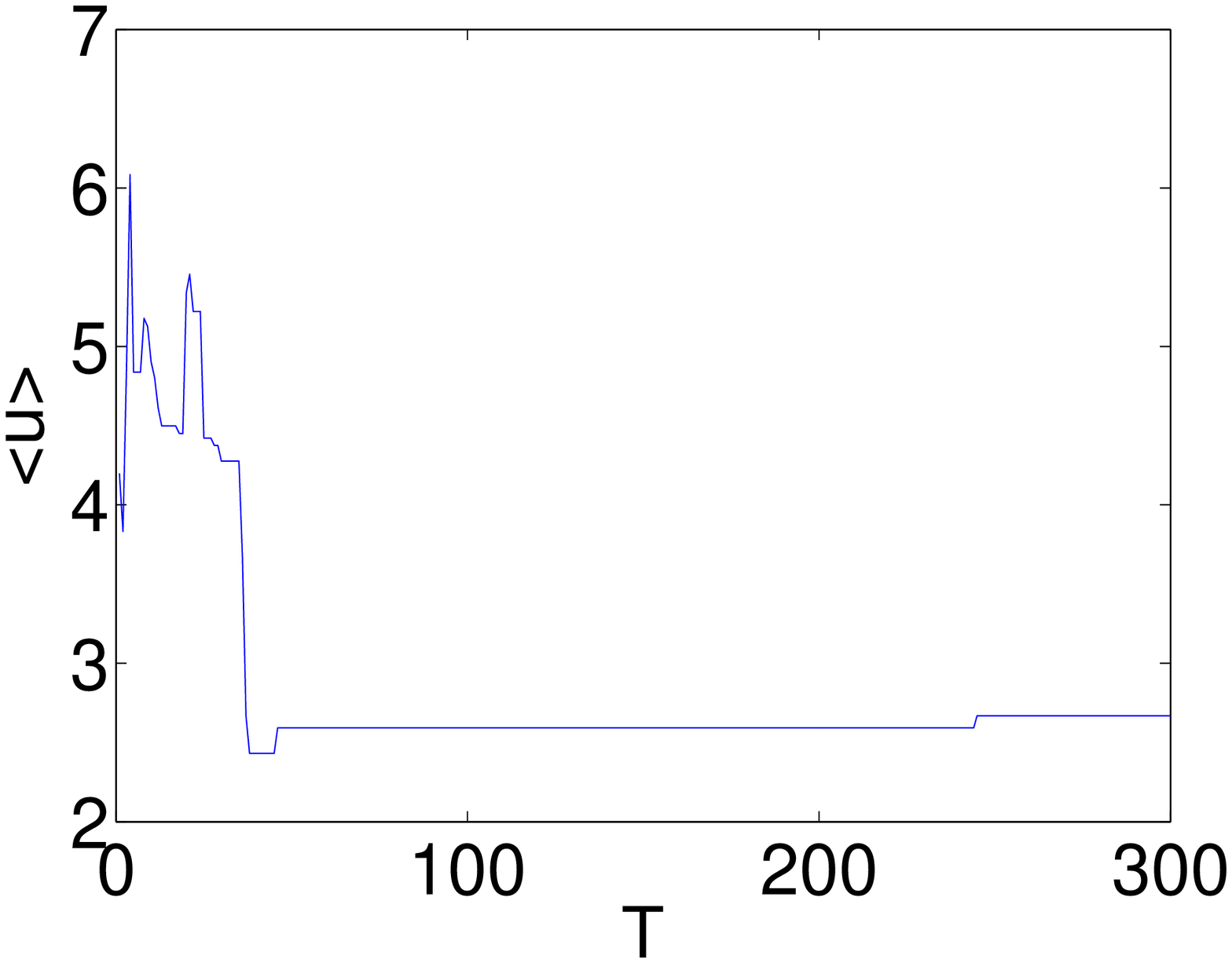}}}
    \bf{a}
  \end{minipage}
  \hfill
  \begin{minipage}{.45\linewidth}
    \psfrag{<d>}[][][4][0]{$\langle d \rangle$}
    \psfrag{T}[][][4][0]{$\times T$}
    \centerline{\scalebox{0.3}{\includegraphics[angle=0]
        {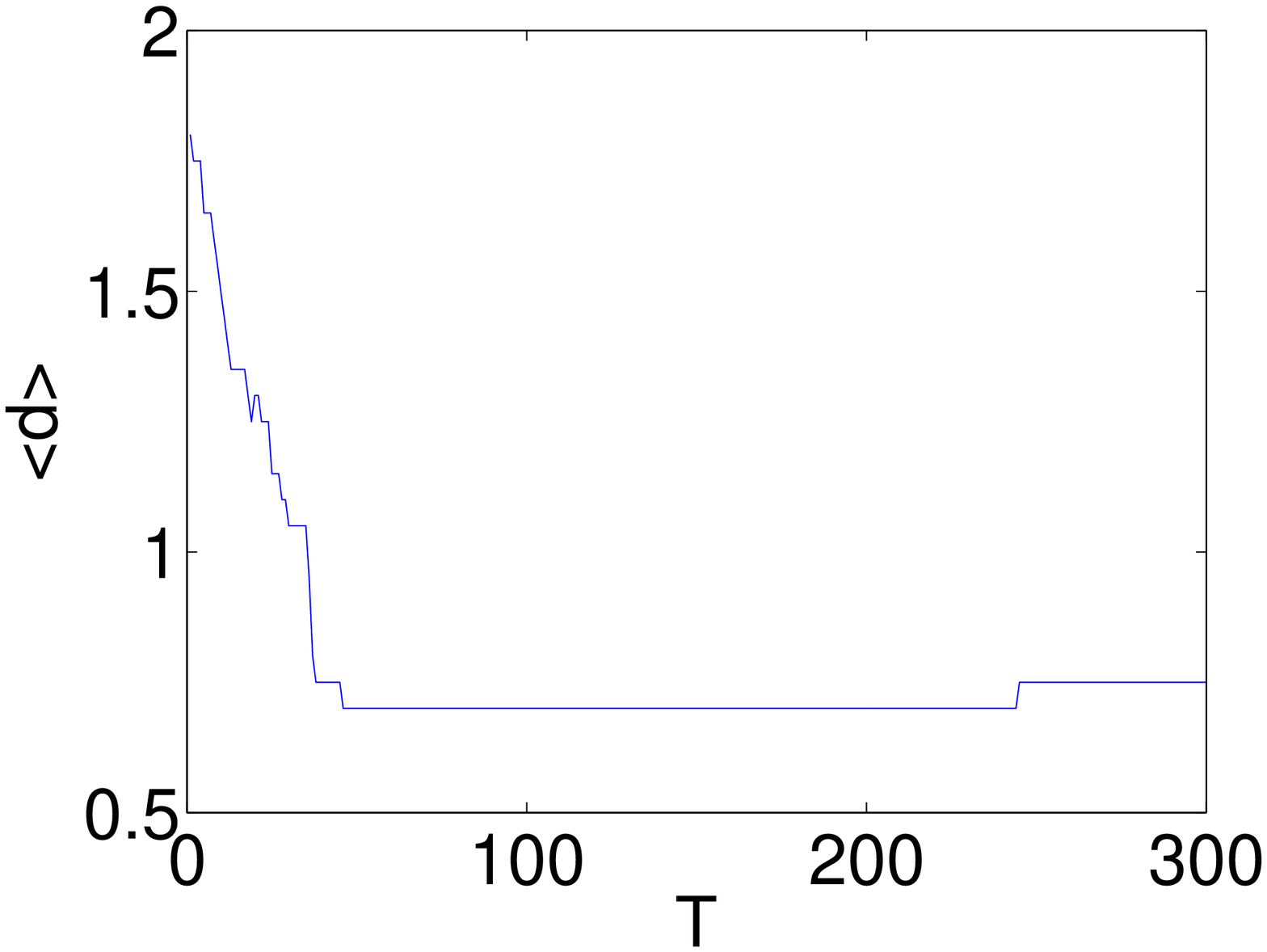}}}
    \bf{b}
  \end{minipage}
  \begin{minipage}{.45\linewidth}
    \centerline{\scalebox{0.5}{\includegraphics[angle=0]
        {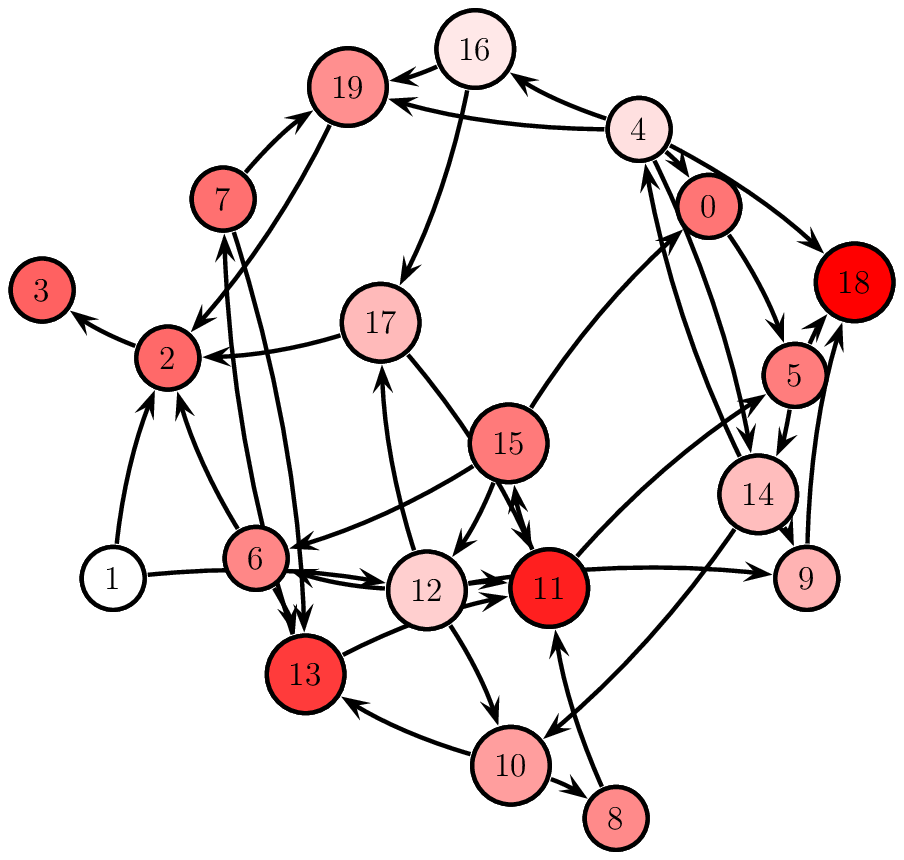}}}
    \bf{c}
  \end{minipage}
  \hfill
  \begin{minipage}{.45\linewidth}
    \centerline{\scalebox{0.5}{\includegraphics[angle=0]
        {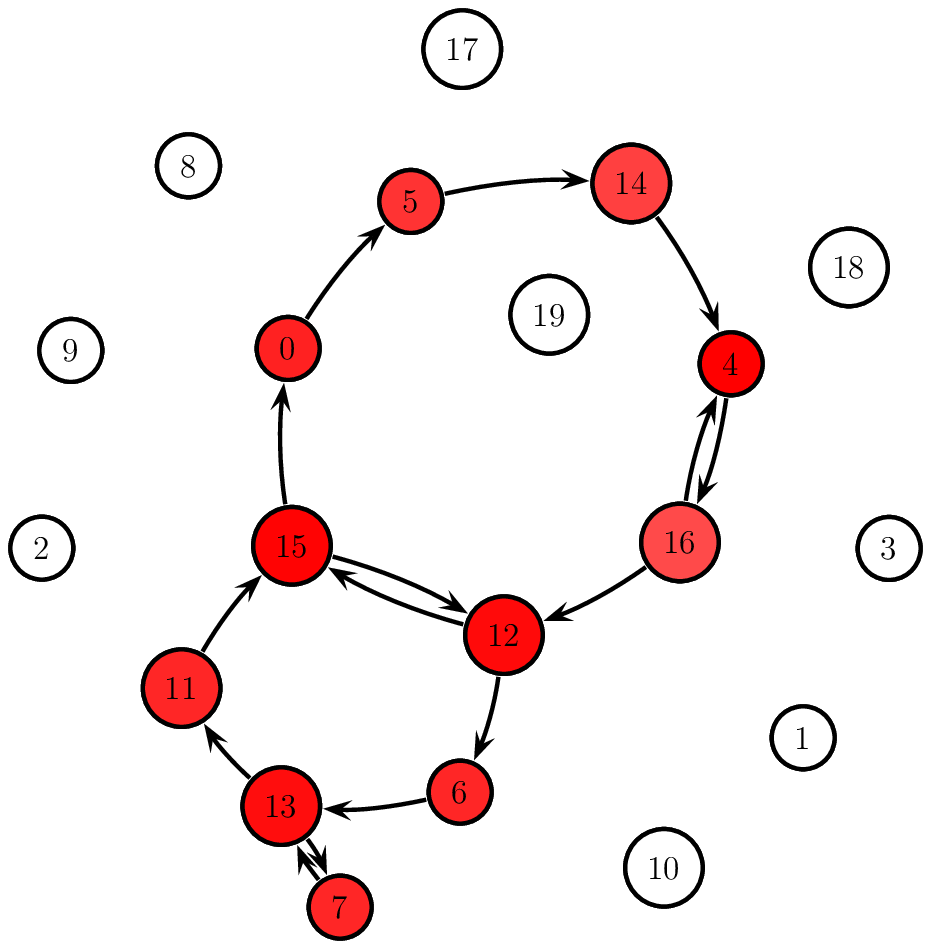}}}
    \bf{d}
  \end{minipage}
  \caption{Circuit number of links externality: \textbf{(a)} Average
    utility. \textbf{(b)} Average degree.  \textbf{(c)} Initial random
    graph. \textbf{(d)} Graph after $500$ iterations (in the
    equilibrium).}
\label{fig:circuitNEdges}
\end{figure}

Using the circuit-centrality measure with the number of links
(\ref{eq:cc-edges}) the growth of the value of knowledge is given by:
\begin{equation}
  \frac{dx_i}{dt}= -d x_i + \sum_{j=1}^{n} (b a_{ji} + b_e
  w_{ji}^{\text{cce}}) x_j - c \sum_{j=1}^{n} a_{ij} x_i^2
\end{equation}
The utility is given by $u_i=\lim_{t \to \infty} x_i(t)$ (large $T$).
Link changes are accepted on the basis of an increase in utility. The
circuit centrality with the number of links values the number of
interactions instead the number of agents, that take part in the transfer
of knowledge. We still observe (fig. \ref{fig:circuitNEdges}) the
emergence of circuits as equilibrium networks and a similar level of
performance (in terms of the total value of knowledge of the system). But
now there are more circuits (more links) in the subgraph containing the
set of permanent agents. The equilibrium network has a higher level of
redundancy (since its has more circuits and links) and is therefore more
robust against the destruction of a single circuit (induced by a node or
link failure).

\section{Discussion and Conclusion}
\label{sec:discussion}

\subsection{Results From the Novel Modeling Approach}
\label{sec:summary}

In the following we summarize the results found by studying our model of
innovation dynamics, as described in sections
(\ref{sec:evolving_networks}), (\ref{sec:evolution-links})
(\ref{sec:simulation_studies}).  Let us start by looking at the dynamics
of the value of knowledge in a static network, in section
(\ref{sec:static_network_analysis}). If we assume that growth occurs only through
interaction among agents (thus neglecting ``in-house'' R\&D
capabilities), then the network sustains itself only through cycles (more
precisely through closed walks or strongly connected components). Agents
survive and grow only if they are part of a cycle (strongly connected
component) or if they are connected to such a cycle through an incoming
path\footnote{\citet{jain01} have denoted this set of nodes the
  autocatalytic set (ACS).}. We have shown that an innovation network
which is acyclic will have vanishing knowledge values for all agents in
the network. However, if agents form cycles they have permanent knowledge
values.

Considering the evolution of the network we have studied two different
settings, Extremal Dynamics and Utility Driven Dynamics. If the network
evolves according to a least fit selection mechanism (Extremal Dynamics)
then we observe crashes and recoveries of the knowledge values of the
agents and the network itself. Thus, an extremal market selection
mechanism which replaces the worst performing agent with a new agent
cannot generate equilibrium networks nor does it sustain a high
performance in the value of knowledge of the individual agents or the
economy as a whole. Notice also that Extremal Dynamics means that agents
are completely passive and have no control on whom they interact with.

In a more realistic setting (Utility Driven Dynamics), agents decide with
whom they interact and they do so in order to increase their utility. In
the context of innovation this corresponds to their value of knowledge.
The information processing capabilities of agents may be limited,
especially if there is a large number of agents in the economy. Thus, we
allow agents to decide themselves to create or delete links
on a trial and error basis. Those interactions that prove to
be beneficial are maintained while un-beneficial ones are severed.  We
find that, under these conditions, the evolution of the network depends
on the cost, $c_{ij}$, of an interaction between the agents, the type of
link formation (unilateral versus bilateral) and the time horizon $T$
after which interactions are evaluated.

In the case of null cost, agents always form new links and thus the
complete graph is eventually realized.

We have shown in section (\ref{sec:dag}) that the knowledge values of
agents vanish if the underlying network does not contain a cycle (similar
to the results obtained by
\citet{rosenblatt57:_linear_model_graph_minkow_leont_matric,
  maxfield94:_gener_equil_theor_direc_graph}. The equilibrium networks
are contain cycles similar to
\citet{goyal00:_Noncooperative_model_network_formation, kima07:_networ}).
However, the evolution of the network driven by the selfish linking
process of agents can lead to the destruction of theses vital cycles. For
a short time horizon $T$, and unilateral link formation, cycles get
destroyed because agents free-ride and delete their outgoing links as it
is beneficial in the short run to save the costs of supporting other
agents.  As a result, the whole innovation network is destroyed. On the
other hand, if the time horizon is long enough, agents do not delete the
cycles they are part of.

However, even when the time horizon is long, large cycles get destroyed
in favor of smaller ones when agents unilaterally form or delete links.
The network, starting from an initial state of high density, evolves into
an absorbing state in which most of the nodes are isolated and few pairs
of nodes are connected by bilateral links. These pairs are trivial cycles
of length $k=2$\footnote{This is different to the results obtained in
  \citet{kima07:_networ} since there the benefit term in the utility of
  the agents depends on the size of the connected component but not on
  its structure.}.

Recall, that pairwise connections are direct reciprocal interactions.
This means that, even though agents are unilaterally forming links and
therefore indirect reciprocal interactions would be possible in principle
(this is equivalent to interaction taking place on a cycle of length $k
\geq 3$) no relation of indirect reciprocity is able to emerge nor to
survive. From the point of view of the global performance of the
innovation network, this is a very unsatisfactory situation. 

In section (\ref{sec:externalities}) we have studied situations in which
even unilateral knowledge exchange can have a higher performance in terms
of the number of permanent agents and their total value of knowledge. We
introduce an externality in the knowledge growth function which increases
the value of knowledge of the agent depending on their position in the
network. We study a type of technology where the value of knowledge
decreases with the number of agents transferring the knowledge. Here
unilateral knowledge exchange still leads to equilibrium networks with a
low performance and only a few permanent agents.

However, for a type of knowledge where its value increases with the
number of agents that transfer and use it, more agents can be permanent
in the equilibrium network and the system performance is increased.
Moreover, if the number of interactions instead of the number of users
determines the added value of the knowledge that is being transferred,
then the equilibrium has not only a higher performance than in the
setting, where knowledge is attenuated with the number of users or where
no externalities are considered, but it is also more robust against node
or connection failures. We observe that, in our framework, indirect
reciprocity emerges if it is associated with a positive externality,
taking into account the structure of the network.

If agents form or delete links bidirectionally, that means, every
exchange of knowledge is direct reciprocal, the network evolves into a
complete graph. This equilibrium network has a high performance and all
agents are permanent. In our study we find that unilateral knowledge
exchange is always inferior to bilateral knowledge exchange. But the
above discussion has shown that, when bilateral knowledge exchange is not
possible and agents are sharing their knowledge unilaterally, innovation
networks are still able to emerge.

The different cases studied in this section have shown that the
equilibrium innovation network that is realized in the evolution of the
system depends critically on the assumptions made on the behavior of
agents (Extremal Dynamics versus Utility Driven Dynamics), on their time
horizon for evaluating their decisions and on the cost associated with
the sharing of knowledge. These results are summarized in table
(\ref{tab:summary_equilibrium_networks})

\begin{center}
  \begin{table}
    \centering
      \begin{tabular}{|l||c|c|c|c|c|}
        \hline
        \multirow{4}{*}{\backslashbox{Cost Func.}{Netw. Evol.}} & \multirow{4}{*}{Extremal Dynamics} &
        \multicolumn{3}{c}{Utility Driven Dynamics} \\
        \cline{3-5}
        & & \multirow{3}{*}{Bilateral} & \multicolumn{2}{c}{Unilateral}\\
        \cline{4-5}
        & & & without & with \\
        & & & externality & externality\\
        \hline
        \hline
        \multirow{2}{*}{$c_{ij} = c x_i^2 $} & network &
        \multirow{2}{*}{$K_{n}$} & set of & set of \\
        & breakdown & & $C_2$ & $C_{k \ge 2}$ \\
        \hline
      \end{tabular}
      \caption{Overview of the equilibrium networks that are realized under
        different assumptions on the network evolution and a quadratic cost function.}
      \label{tab:summary_equilibrium_networks}
  \end{table}
\end{center}

\subsection{General Conclusions}

In this chapter we studied a variety of different models for innovation
networks. We started by discussing the importance of networks in
economics and emphasized that these networks are intrinsically dynamic
and composed of heterogeneous units.  The notion of a \textit{complex
  network} was used in the beginning to briefly explain how statistical
physics can be involved to study them. We tried to classify different
approaches to modeling economic networks, in particular we considered the
connection between the state variables associated with the nodes of a
network, e.g.  the productivity level of a firm, and the dynamics of the
network itself, i.e. the interactions between firms.

Before developing our own modeling framework, we discussed some basic
models of economic networks with agents engaged in knowledge production.
These models show that the economy can evolve into equilibrium networks
which are not necessarily efficient.  Moreover, the equilibrium networks
that emerge in these models are rather simple. We briefly introduced some
models in which more complicated network structures emerge, which may be
closer to real-world innovation networks. We then discussed models in
which cycles, i.e.  closed feedback loops, play an important role in the
network formation and the performance of the system (similar to
\citet{rosenblatt57:_linear_model_graph_minkow_leont_matric}).

The major part of the chapter was devoted to the development of our own
modeling framework which is based on catalytic knowledge interactions. In
this setting there are permanent agents (with non-vanishing knowledge
values) only if the underlying network contains a cycle.  We investigated
the evolution and performance of the system under different selection
mechanisms, i.e., a least fit selection mechanism, denoted by Extremal
Dynamics, versus Utility Driven Dynamics in which agents decide upon
their interaction partners in a trial and error procedure. We observe
that a least fit mechanism cannot generate stable networks nor sustain
high performance in knowledge production.  Moreover such a mechanism
assumes that agents are completely passive entities. In the case of
Utility Driven Dynamics, agents choose their actions in order to increase
their utility but their information processing capabilities are limited.
If agents are evaluating their interactions after a time long enough, we
obtain equilibrium networks with non-vanishing (permanent) knowledge
production.

In our framework, we investigated different assumptions about the
behavior of agents, that is, we either assume that agents share knowledge
bilaterally or unilaterally. If all interactions are bilateral, the
equilibrium network is a complete graph and it has the highest
performance. However, if direct reciprocal interactions cannot be
enforced (which means that links are not necessarily bilateral), we still
observe the emergence of networks of knowledge sharing agents. But in the
equilibrium network only bilateral interactions remain. Moreover, only a
few agents are permanent and the system has a low performance compared to
the case of purely bilateral interactions.  However, for unilateral
interactions, the number of permanent agents can be significantly
increased, for a type of technology where the number of users increases
its value.

Our studies show that the range of innovation networks that can emerge in
this general framework is affected by various parameters. Amongst these
are information processing capabilities of agents, their time horizon,
their interactional behavior, the cost associated with the sharing of
knowledge and the type of technology which agents produce and transfer.

The variety of possible networks is quite large and the network model
appropriate for a given application should be determined based on the
specificities of the problem under investigation.

\section*{Appendix}
\begin{appendix}

\section{Stationary Solutions for $G(n = 3)$}
\label{sec:stationary_solutions_G3}

\begin{example}
  We compute the fixed points for all graphs (auto-morphisms) with $n=3$
  nodes and initial values $\mathbf{x}(0) =
  (\frac{1}{n},\frac{1}{n},\frac{1}{n})^T =
  (\frac{1}{3},\frac{1}{3},\frac{1}{3})^T$.  For the numerical
  integration we set $d=0.5$, $c=0.5$ and $b=1$.  The fixed points
  (stationary solutions are denoted $x_i^*$ for $i=1,2,3$).  Where
  possible, we give the analytical solutions for the positive fixed
  points. $x_i^*=0$ is a fixed point for all graphs.
\end{example}

\begin{minipage}{.2\textwidth}
\scalebox{0.6}{\begin{pspicture}(0,-1)(3,4) 
  \psset{nodesep=0.5pt,linewidth=1.5pt,arrowsize=5pt 2} 
  \rput(0,3){\Large (1)}
  \cnodeput(0,0){2}{\Large 2}
  \cnodeput(1.5,3){1}{\Large 1} 
  \cnodeput(3,0){3}{\Large 3}
  \ncarc[linewidth=1pt,arcangle=0]{->}{1}{2}
  \ncarc[linewidth=1pt,arcangle=0]{->}{1}{3}
\end{pspicture}}
\end{minipage}
\begin{minipage}{.2\textwidth}
\begin{displaymath}
A_1=\left(
\begin{array}{ccc}
0 & 1 & 1 \\
0 & 0 & 0 \\
0 & 0 & 0 \\
\end{array}
\right)
\end{displaymath}
\end{minipage}
\begin{minipage}{.5\linewidth}
  \psfrag{x1}[c][][3][0]{$x_1$} 
  \psfrag{x2}[c][][3][0]{$x_2$} 
  \psfrag{x3}[c][][3][0]{$x_3$} 
  \psfrag{xi}[c][][4][0]{$x_i$} 
  \psfrag{time}[c][][4][0]{$t$}
   \centerline{\scalebox{0.3}{\includegraphics[angle=0]
        {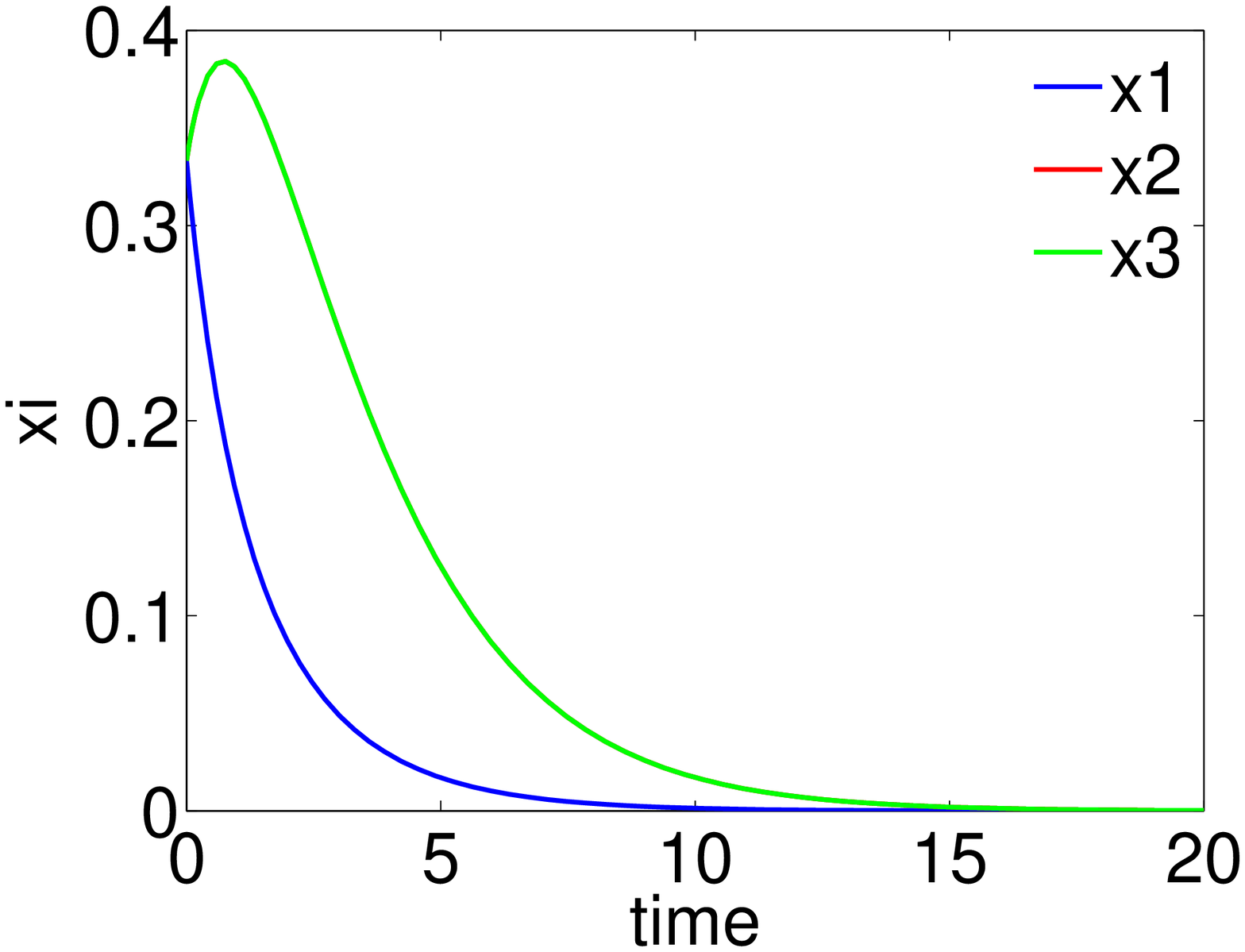}}}
\end{minipage}

The fixed points are given by
\begin{equation}
  \label{eq:fixed_point_3Graph1}
  x_i^*=0
\end{equation}

\begin{minipage}{.2\textwidth}
\scalebox{0.6}{\begin{pspicture}(0,-1)(3,4) 
  \psset{nodesep=0.5pt,linewidth=1.5pt,arrowsize=5pt 2} 
  \rput(0,3){\Large (2)}
  \cnodeput(0,0){1}{\Large 1}
  \cnodeput(1.5,3){2}{\Large 2} 
  \cnodeput(3,0){3}{\Large 3}
  \ncarc[linewidth=1pt,arcangle=0]{->}{1}{2}
  \ncarc[linewidth=1pt,arcangle=0]{->}{2}{3}
\end{pspicture}}
\end{minipage}
\begin{minipage}{.2\textwidth}
\begin{displaymath}
A_2=\left(
\begin{array}{ccc}
0 & 1 & 0 \\
0 & 0 & 1 \\
0 & 0 & 0 \\
\end{array}
\right)
\end{displaymath}
\end{minipage}
\begin{minipage}{.5\linewidth}
  \psfrag{x1}[c][][3][0]{$x_1$} 
  \psfrag{x2}[c][][3][0]{$x_2$} 
  \psfrag{x3}[c][][3][0]{$x_3$} 
  \psfrag{xi}[c][][4][0]{$x_i$} 
  \psfrag{time}[c][][4][0]{$t$}
    \centerline{\scalebox{0.3}{\includegraphics[angle=0]
        {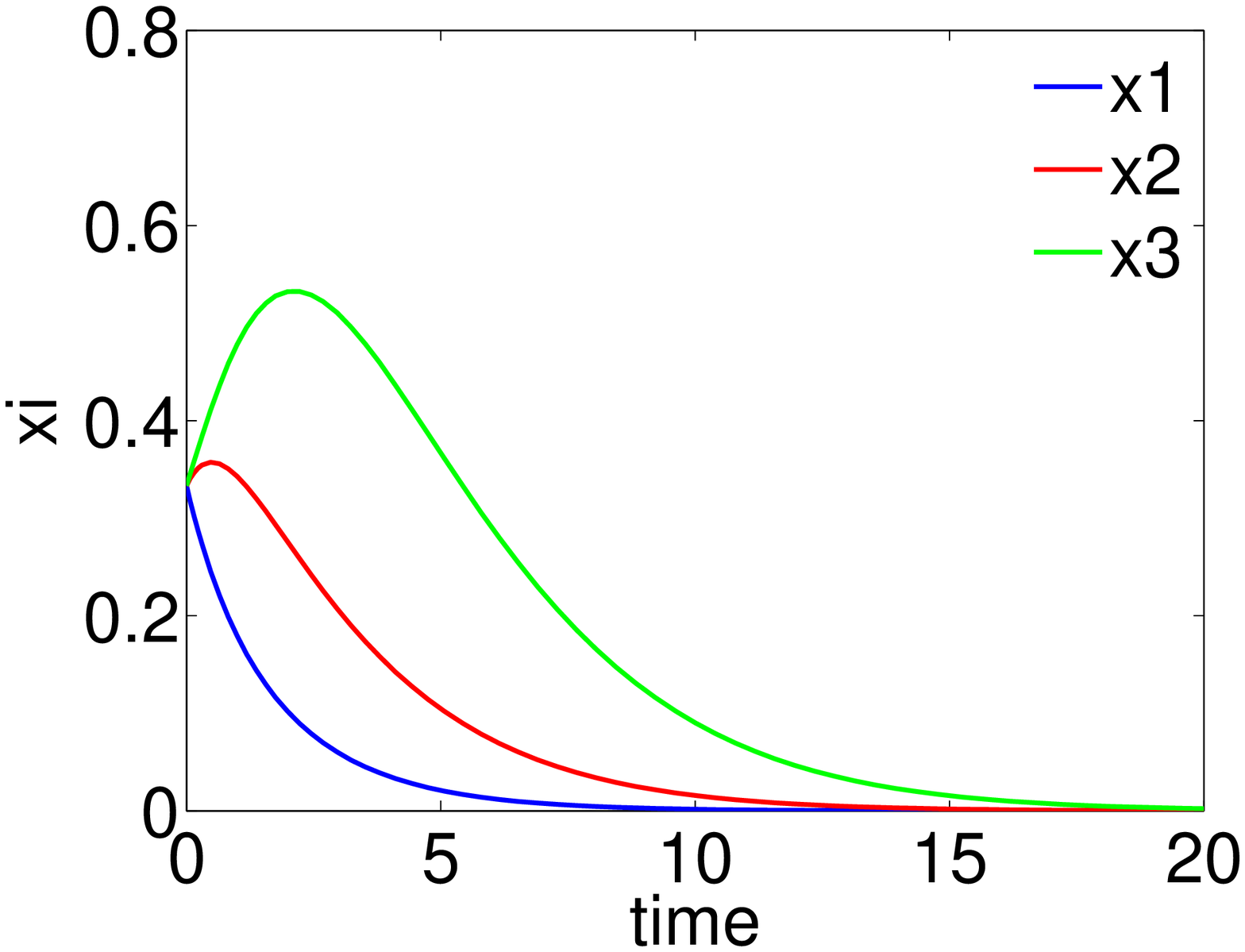}}}
\end{minipage}

The fixed points are given by
\begin{equation}
  \label{eq:fixed_point_3Graph2}
  x_i^*=0
\end{equation}

\begin{minipage}{.2\textwidth}
\scalebox{0.6}{\begin{pspicture}(0,-1)(3,4) 
  \psset{nodesep=0.5pt,linewidth=1.5pt,arrowsize=5pt 2} 
  \rput(0,3){\Large (3)}
  \cnodeput(0,0){1}{\Large 1}
  \cnodeput(1.5,3){2}{\Large 2} 
  \cnodeput(3,0){3}{\Large 3}
  \ncarc[linewidth=1pt,arcangle=10]{->}{1}{2}
  \ncarc[linewidth=1pt,arcangle=10]{->}{2}{1}
  \ncarc[linewidth=1pt,arcangle=0]{->}{2}{3}
\end{pspicture}}
\end{minipage}
\begin{minipage}{.2\textwidth}
\begin{displaymath}
A_3=\left(
\begin{array}{ccc}
0 & 1 & 0 \\
1 & 0 & 1 \\
0 & 0 & 0 \\
\end{array}
\right)
\end{displaymath}
\end{minipage}
\begin{minipage}{.5\linewidth}
  \psfrag{x1}[c][][3][0]{$x_1$} 
  \psfrag{x2}[c][][3][0]{$x_2$} 
  \psfrag{x3}[c][][3][0]{$x_3$} 
  \psfrag{xi}[c][][4][0]{$x_i$} 
  \psfrag{time}[c][][4][0]{$t$}
    \centerline{\scalebox{0.3}{\includegraphics[angle=0]
        {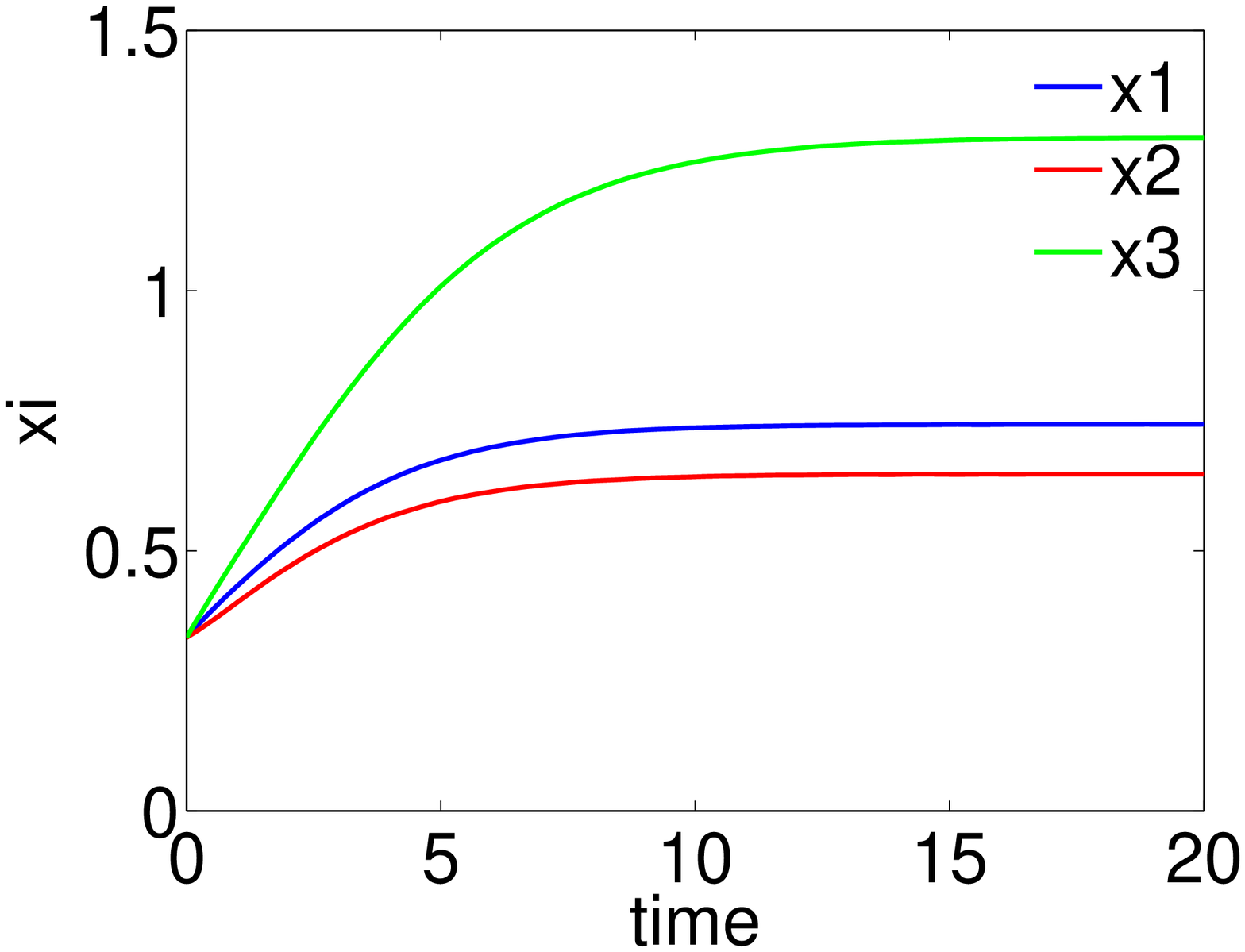}}}
\end{minipage}

\begin{minipage}{.2\textwidth}
\scalebox{0.6}{\begin{pspicture}(0,-1)(3,4) 
  \psset{nodesep=0.5pt,linewidth=1.5pt,arrowsize=5pt 2} 
  \rput(0,3){\Large (4)}
  \cnodeput(0,0){1}{\Large 1}
  \cnodeput(1.5,3){3}{\Large 3} 
  \cnodeput(3,0){2}{\Large 2}
  \ncarc[linewidth=1pt,arcangle=0]{->}{1}{3}
  \ncarc[linewidth=1pt,arcangle=0]{->}{2}{3}
\end{pspicture}}
\end{minipage}
\begin{minipage}{.2\textwidth}
\begin{displaymath}
A_4=\left(
\begin{array}{ccc}
0 & 0 & 1 \\
0 & 0 & 1 \\
0 & 0 & 0 \\
\end{array}
\right)
\end{displaymath}
\end{minipage}
\begin{minipage}{.5\linewidth}
  \psfrag{x1}[c][][3][0]{$x_1$} 
  \psfrag{x2}[c][][3][0]{$x_2$} 
  \psfrag{x3}[c][][3][0]{$x_3$} 
  \psfrag{xi}[c][][4][0]{$x_i$} 
  \psfrag{time}[c][][4][0]{$t$}
    \centerline{\scalebox{0.3}{\includegraphics[angle=0]
        {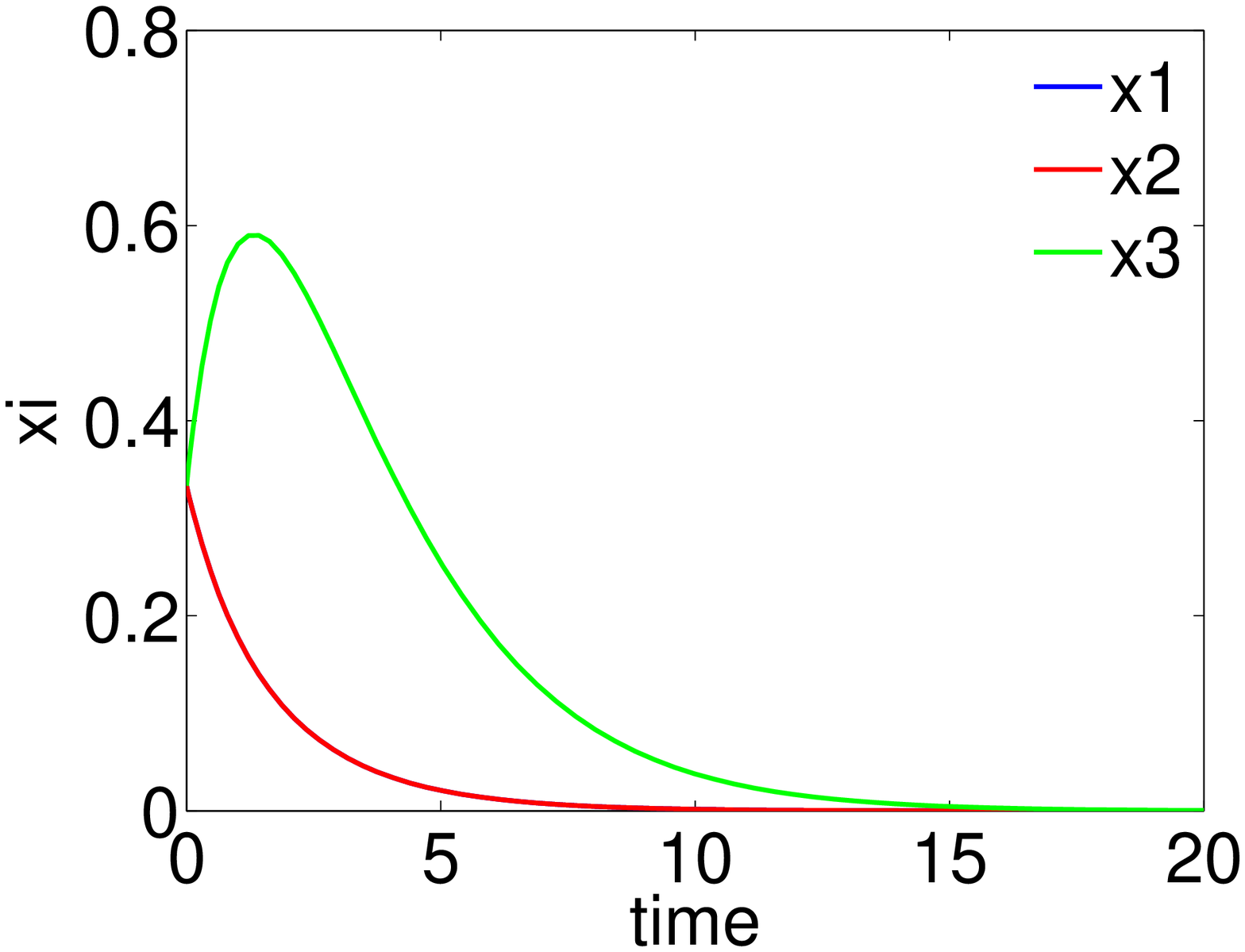}}}
\end{minipage}

The fixed points are given by
\begin{equation}
  \label{eq:fixed_point_3Graph4}
  x_i^*=0
\end{equation}

\begin{minipage}{.2\textwidth}
\scalebox{0.6}{\begin{pspicture}(0,-1)(3,4) 
  \psset{nodesep=0.5pt,linewidth=1.5pt,arrowsize=5pt 2} 
  \rput(0,3){\Large (5)}
  \cnodeput(0,0){2}{\Large 2}
  \cnodeput(1.5,3){1}{\Large 1} 
  \cnodeput(3,0){3}{\Large 3}
  \ncarc[linewidth=1pt,arcangle=0]{->}{2}{1}
  \ncarc[linewidth=1pt,arcangle=0]{->}{1}{3}
  \ncarc[linewidth=1pt,arcangle=0]{->}{2}{3}
\end{pspicture}}
\end{minipage}
\begin{minipage}{.2\textwidth}
\begin{displaymath}
A_5=\left(
\begin{array}{ccc}
0 & 0 & 1 \\
1 & 0 & 1 \\
0 & 0 & 0 \\
\end{array}
\right)
\end{displaymath}
\end{minipage}
\begin{minipage}{.5\linewidth}
  \psfrag{x1}[c][][3][0]{$x_1$} 
  \psfrag{x2}[c][][3][0]{$x_2$} 
  \psfrag{x3}[c][][3][0]{$x_3$} 
  \psfrag{xi}[c][][4][0]{$x_i$} 
  \psfrag{time}[c][][4][0]{$t$}
    \centerline{\scalebox{0.3}{\includegraphics[angle=0]
        {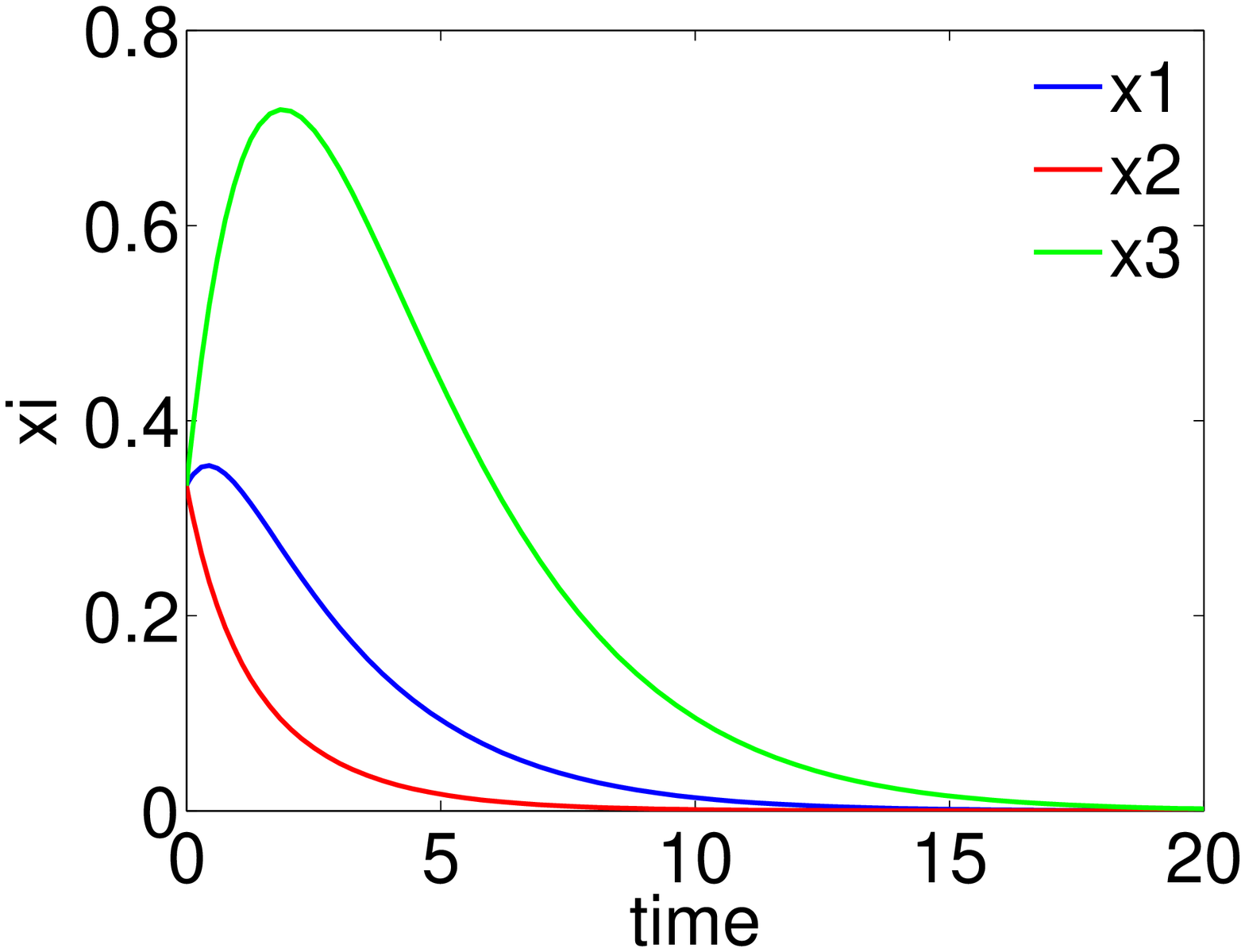}}}
\end{minipage}

The fixed points are given by
\begin{equation}
  \label{eq:fixed_point_3Graph5}
  x_i^*=0
\end{equation}

\begin{minipage}{.2\textwidth}
\scalebox{0.6}{\begin{pspicture}(0,-1)(3,4) 
  \psset{nodesep=0.5pt,linewidth=1.5pt,arrowsize=5pt 2} 
  \rput(0,3){\Large (6)}
  \cnodeput(0,0){2}{\Large 2}
  \cnodeput(1.5,3){1}{\Large 1} 
  \cnodeput(3,0){3}{\Large 3}
  \ncarc[linewidth=1pt,arcangle=10]{->}{1}{2}
  \ncarc[linewidth=1pt,arcangle=10]{->}{2}{1}
  \ncarc[linewidth=1pt,arcangle=0]{->}{1}{3}
  \ncarc[linewidth=1pt,arcangle=0]{->}{2}{3}
\end{pspicture}}
\end{minipage}
\begin{minipage}{.2\textwidth}
\begin{displaymath}
A_6=\left(
\begin{array}{ccc}
0 & 1 & 1 \\
1 & 0 & 1 \\
0 & 0 & 0 \\
\end{array}
\right)
\end{displaymath}
\end{minipage}
\begin{minipage}{.5\linewidth}
  \psfrag{x1}[c][][3][0]{$x_1$} 
  \psfrag{x2}[c][][3][0]{$x_2$} 
  \psfrag{x3}[c][][3][0]{$x_3$} 
  \psfrag{xi}[c][][4][0]{$x_i$} 
  \psfrag{time}[c][][4][0]{$t$}
    \centerline{\scalebox{0.3}{\includegraphics[angle=0]
        {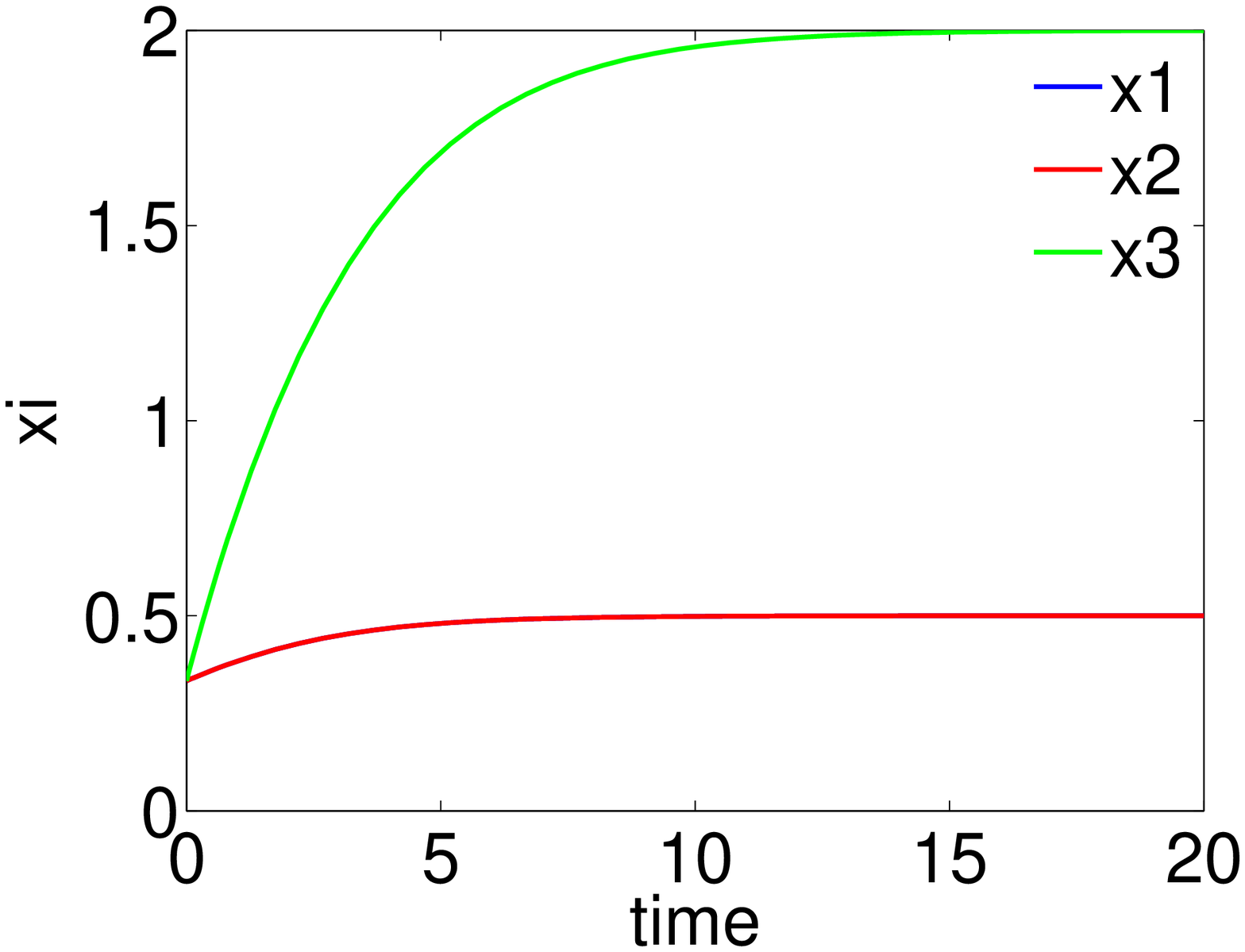}}}
\end{minipage}

The fixed points are given by
\begin{eqnarray}
  \label{eq:fixed_point_3Graph6}
  x_1^*=x_2^*=\frac{b-d}{2c} \\
  x_3^*=\frac{b(b-d)}{cd}
\end{eqnarray}

\begin{minipage}{.2\textwidth}
\scalebox{0.6}{\begin{pspicture}(0,-1)(3,4) 
  \psset{nodesep=0.5pt,linewidth=1.5pt,arrowsize=5pt 2} 
  \rput(0,3){\Large (7)}
  \cnodeput(0,0){1}{\Large 1}
  \cnodeput(1.5,3){2}{\Large 2} 
  \cnodeput(3,0){3}{\Large 3}
  \ncarc[linewidth=1pt,arcangle=10]{->}{1}{2}
  \ncarc[linewidth=1pt,arcangle=10]{->}{2}{1}
  \ncarc[linewidth=1pt,arcangle=0]{->}{3}{2}
\end{pspicture}}
\end{minipage}
\begin{minipage}{.2\textwidth}
\begin{displaymath}
A_7=\left(
\begin{array}{ccc}
0 & 1 & 0 \\
1 & 0 & 0 \\
0 & 1 & 0 \\
\end{array}
\right)
\end{displaymath}
\end{minipage}
\begin{minipage}{.5\linewidth}
  \psfrag{x1}[c][][3][0]{$x_1$} 
  \psfrag{x2}[c][][3][0]{$x_2$} 
  \psfrag{x3}[c][][3][0]{$x_3$} 
  \psfrag{xi}[c][][4][0]{$x_i$} 
  \psfrag{time}[c][][4][0]{$t$}
    \centerline{\scalebox{0.3}{\includegraphics[angle=0]
        {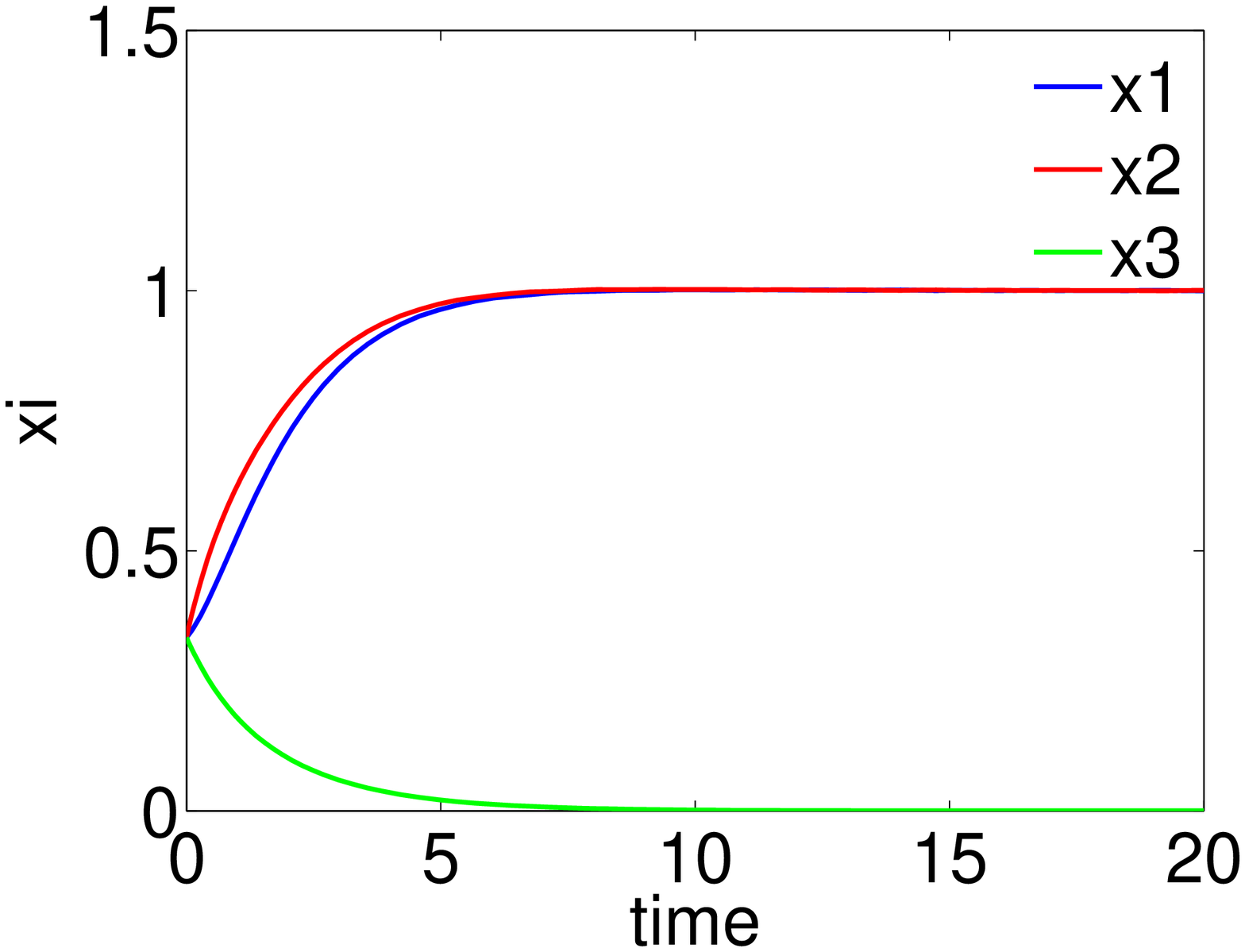}}}
\end{minipage}

The fixed points are given by
\begin{eqnarray}
  \label{eq:fixed_point_3Graph7}
  x_1^*=x_2^*=\frac{b-d}{c} \\
  x_3^*=0
\end{eqnarray}

\begin{minipage}{.2\textwidth}
\scalebox{0.6}{\begin{pspicture}(0,-1)(3,4) 
  \psset{nodesep=0.5pt,linewidth=1.5pt,arrowsize=5pt 2} 
  \rput(0,3){\Large (8)}
  \cnodeput(0,0){1}{\Large 1}
  \cnodeput(1.5,3){3}{\Large 3} 
  \cnodeput(3,0){2}{\Large 2}
  \ncarc[linewidth=1pt,arcangle=10]{->}{1}{3}
  \ncarc[linewidth=1pt,arcangle=10]{->}{3}{1}
  \ncarc[linewidth=1pt,arcangle=10]{->}{3}{2}
  \ncarc[linewidth=1pt,arcangle=10]{->}{2}{3}
\end{pspicture}}
\end{minipage}
\begin{minipage}{.2\textwidth}
\begin{displaymath}
A_8=\left(
\begin{array}{ccc}
0 & 0 & 1 \\
0 & 0 & 1 \\
1 & 1 & 0 \\
\end{array}
\right)
\end{displaymath}
\end{minipage}
\begin{minipage}{.5\linewidth}
  \psfrag{x1}[c][][3][0]{$x_1$} 
  \psfrag{x2}[c][][3][0]{$x_2$} 
  \psfrag{x3}[c][][3][0]{$x_3$} 
  \psfrag{xi}[c][][4][0]{$x_i$} 
  \psfrag{time}[c][][4][0]{$t$}
    \centerline{\scalebox{0.3}{\includegraphics[angle=0]
        {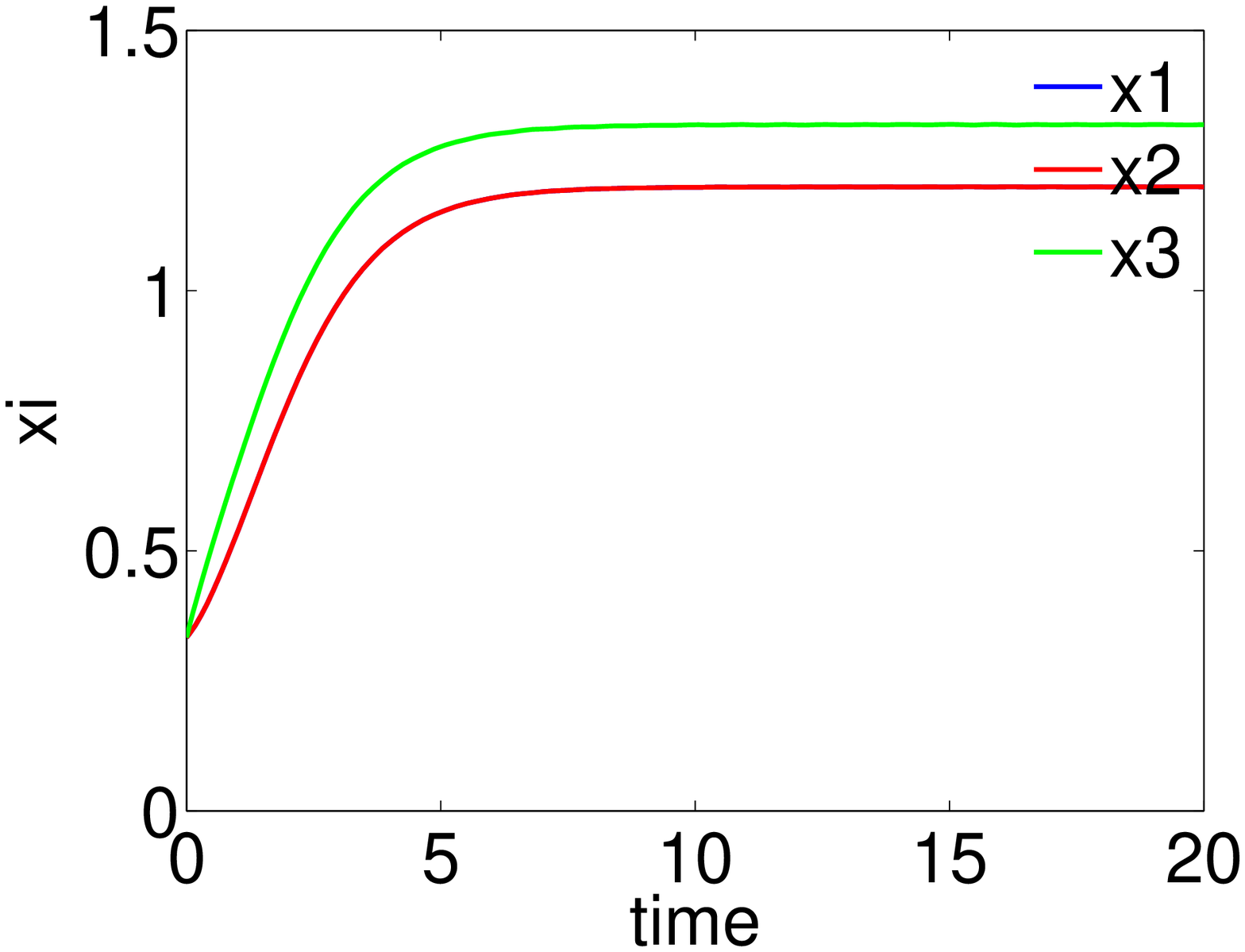}}}
\end{minipage}

$x_1^*=x_2^*$ are the roots of the polynomial
\begin{displaymath}
  x^3 + \frac{2d}{c}x^2+\frac{d(b+2d)}{2c^2}x+\frac{b(d^2-2b^2)}{2c^3}=0
\end{displaymath}
and $x_3^*=\frac{d}{b}x+\frac{c}{b}x^2$.

\begin{minipage}{.2\textwidth}
\scalebox{0.6}{\begin{pspicture}(0,-1)(3,4) 
  \psset{nodesep=0.5pt,linewidth=1.5pt,arrowsize=5pt 2} 
  \rput(0,3){\Large (9)}
  \cnodeput(0,0){2}{\Large 2}
  \cnodeput(1.5,3){1}{\Large 1} 
  \cnodeput(3,0){3}{\Large 3}
  \ncarc[linewidth=1pt,arcangle=0]{->}{1}{2}
  \ncarc[linewidth=1pt,arcangle=0]{->}{2}{3}
  \ncarc[linewidth=1pt,arcangle=0]{->}{3}{1}
\end{pspicture}}
\end{minipage}
\begin{minipage}{.2\textwidth}
\begin{displaymath}
A_9=\left(
\begin{array}{ccc}
0 & 1 & 0 \\
0 & 0 & 1 \\
1 & 0 & 0 \\
\end{array}
\right)
\end{displaymath}
\end{minipage}
\begin{minipage}{.5\linewidth}
  \psfrag{x1}[c][][3][0]{$x_1$} 
  \psfrag{x2}[c][][3][0]{$x_2$} 
  \psfrag{x3}[c][][3][0]{$x_3$} 
  \psfrag{xi}[c][][4][0]{$x_i$} 
  \psfrag{time}[c][][4][0]{$t$}
    \centerline{\scalebox{0.3}{\includegraphics[angle=0]
        {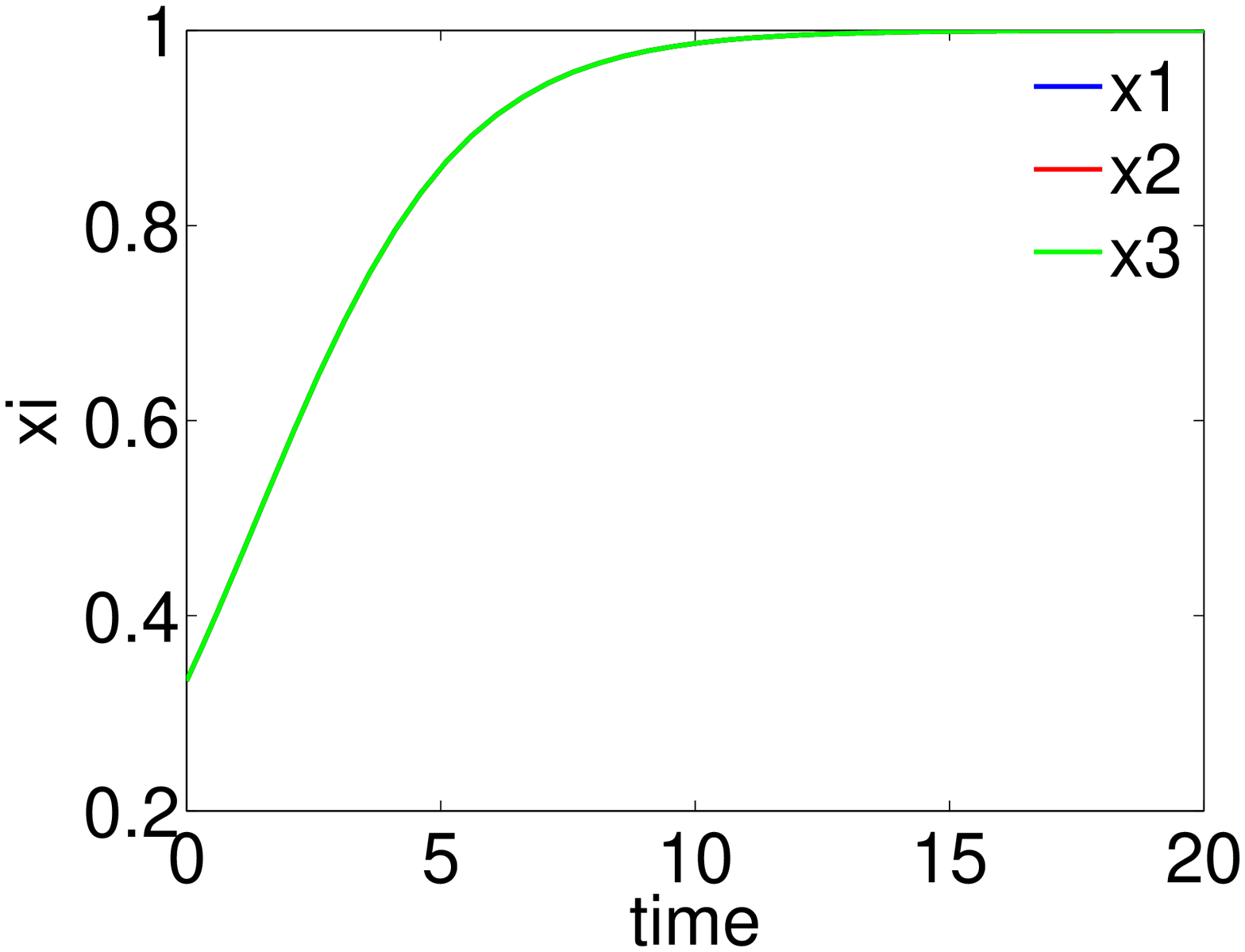}}}
\end{minipage}

The fixed points are given by
\begin{equation}
  \label{eq:fixed_point_3Graph9}
  x_1^*=x_2^*=x_3^*=\frac{b-d}{c} 
\end{equation}

\begin{minipage}{.2\textwidth}
\scalebox{0.6}{\begin{pspicture}(0,-1)(3,4) 
  \psset{nodesep=0.5pt,linewidth=1.5pt,arrowsize=5pt 2} 
  \rput(0,3){\Large (10)}
  \cnodeput(0,0){2}{\Large 2}
  \cnodeput(1.5,3){1}{\Large 1} 
  \cnodeput(3,0){3}{\Large 3}
  \ncarc[linewidth=1pt,arcangle=0]{->}{2}{1}
  \ncarc[linewidth=1pt,arcangle=0]{->}{1}{3}
  \ncarc[linewidth=1pt,arcangle=10]{->}{2}{3}
  \ncarc[linewidth=1pt,arcangle=10]{->}{3}{2}
\end{pspicture}}
\end{minipage}
\begin{minipage}{.2\textwidth}
\begin{displaymath}
A_{10}=\left(
\begin{array}{ccc}
0 & 0 & 1 \\
1 & 0 & 1 \\
0 & 1 & 0 \\
\end{array}
\right)
\end{displaymath}
\end{minipage}
\begin{minipage}{.5\linewidth}
  \psfrag{x1}[c][][3][0]{$x_1$} 
  \psfrag{x2}[c][][3][0]{$x_2$} 
  \psfrag{x3}[c][][3][0]{$x_3$} 
  \psfrag{xi}[c][][4][0]{$x_i$} 
  \psfrag{time}[c][][4][0]{$t$}
    \centerline{\scalebox{0.3}{\includegraphics[angle=0]
        {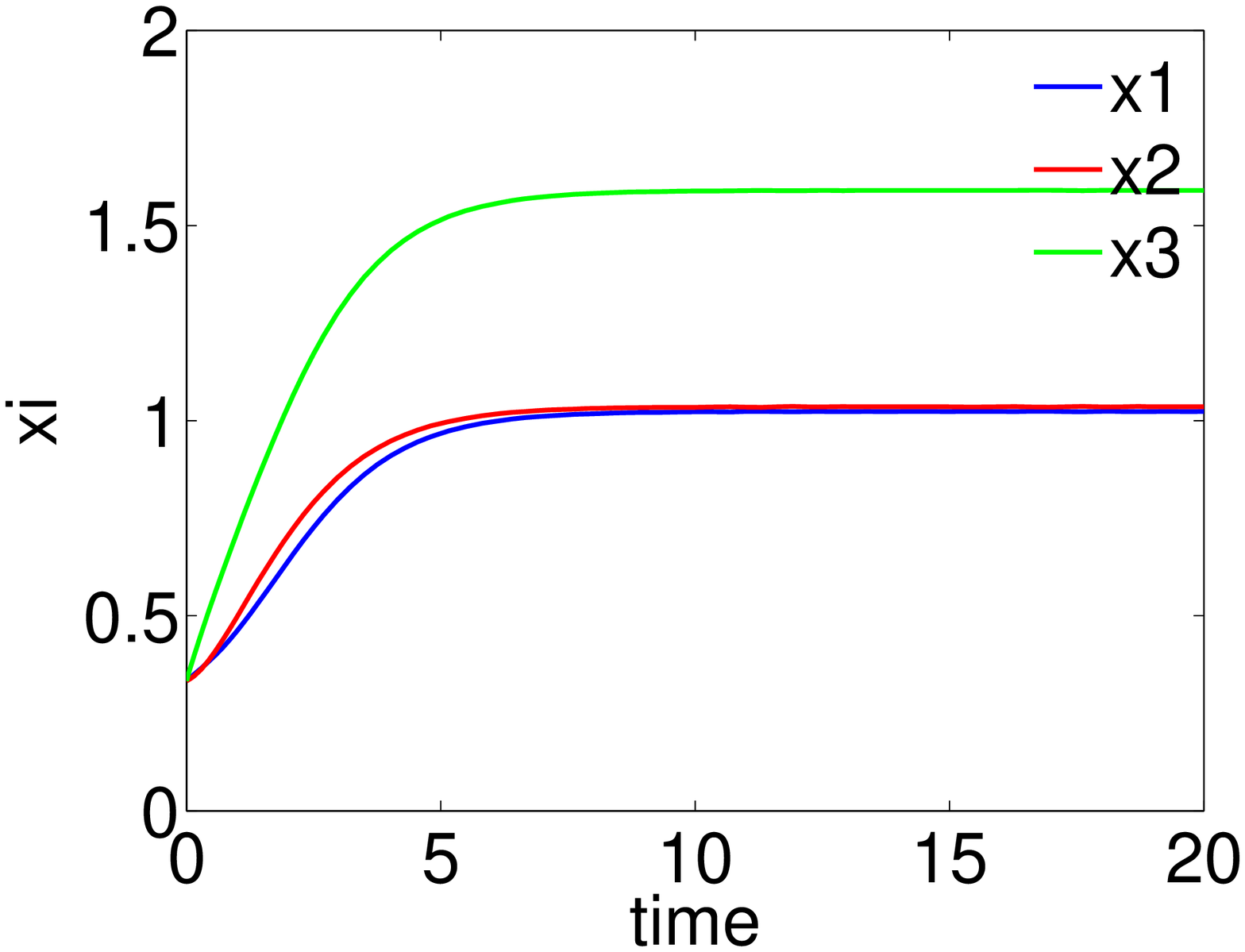}}}
\end{minipage}

\begin{minipage}{.2\textwidth}
\scalebox{0.6}{\begin{pspicture}(0,-1)(3,4) 
  \psset{nodesep=0.5pt,linewidth=1.5pt,arrowsize=5pt 2} 
  \rput(0,3){\Large (11)}
  \cnodeput(0,0){2}{\Large 2}
  \cnodeput(1.5,3){1}{\Large 1} 
  \cnodeput(3,0){3}{\Large 3}
  \ncarc[linewidth=1pt,arcangle=0]{->}{1}{2}
  \ncarc[linewidth=1pt,arcangle=0]{->}{1}{3}
  \ncarc[linewidth=1pt,arcangle=10]{->}{2}{3}
  \ncarc[linewidth=1pt,arcangle=10]{->}{3}{2}
\end{pspicture}}
\end{minipage}
\begin{minipage}{.2\textwidth}
\begin{displaymath}
A_{11}=\left(
\begin{array}{ccc}
0 & 1 & 1 \\
0 & 0 & 1 \\
0 & 1 & 0 \\
\end{array}
\right)
\end{displaymath}
\end{minipage}
\begin{minipage}{.5\linewidth}
  \psfrag{x1}[c][][3][0]{$x_1$} 
  \psfrag{x2}[c][][3][0]{$x_2$} 
  \psfrag{x3}[c][][3][0]{$x_3$} 
  \psfrag{xi}[c][][4][0]{$x_i$} 
  \psfrag{time}[c][][4][0]{$t$}
    \centerline{\scalebox{0.3}{\includegraphics[angle=0]
        {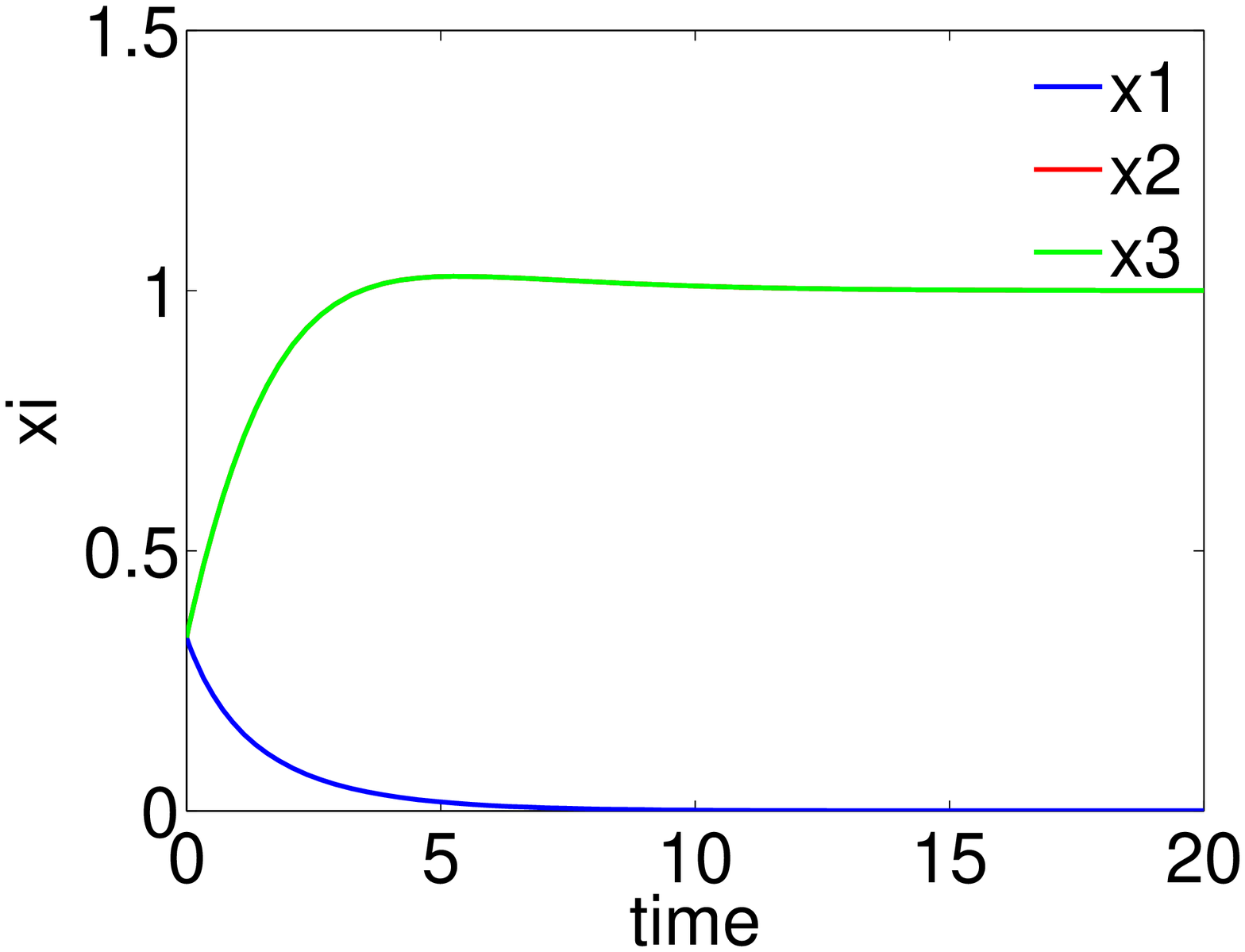}}}
\end{minipage}

The fixed points are given by
\begin{eqnarray}
  \label{eq:fixed_point_3Graph11}
  x_2^*=x_3^*=\frac{b-d}{c} \\
  x_1^*=0
\end{eqnarray}

\begin{minipage}{.2\textwidth}
\scalebox{0.6}{\begin{pspicture}(0,-1)(3,4) 
  \psset{nodesep=0.5pt,linewidth=1.5pt,arrowsize=5pt 2} 
  \rput(0,3){\Large (12)}
  \cnodeput(0,0){2}{\Large 2}
  \cnodeput(1.5,3){1}{\Large 1} 
  \cnodeput(3,0){3}{\Large 3}
  \ncarc[linewidth=1pt,arcangle=0]{->}{1}{3}
  \ncarc[linewidth=1pt,arcangle=10]{->}{2}{3}
  \ncarc[linewidth=1pt,arcangle=10]{->}{3}{2}
  \ncarc[linewidth=1pt,arcangle=10]{->}{1}{2}
  \ncarc[linewidth=1pt,arcangle=10]{->}{2}{1}
\end{pspicture}}
\end{minipage}
\begin{minipage}{.2\textwidth}
\begin{displaymath}
A_{12}=\left(
\begin{array}{ccc}
0 & 1 & 1 \\
1 & 0 & 1 \\
0 & 1 & 0 \\
\end{array}
\right)
\end{displaymath}
\end{minipage}
\begin{minipage}{.5\linewidth}
  \psfrag{x1}[c][][3][0]{$x_1$} 
  \psfrag{x2}[c][][3][0]{$x_2$} 
  \psfrag{x3}[c][][3][0]{$x_3$} 
  \psfrag{xi}[c][][4][0]{$x_i$} 
  \psfrag{time}[c][][4][0]{$t$}
    \centerline{\scalebox{0.3}{\includegraphics[angle=0]
        {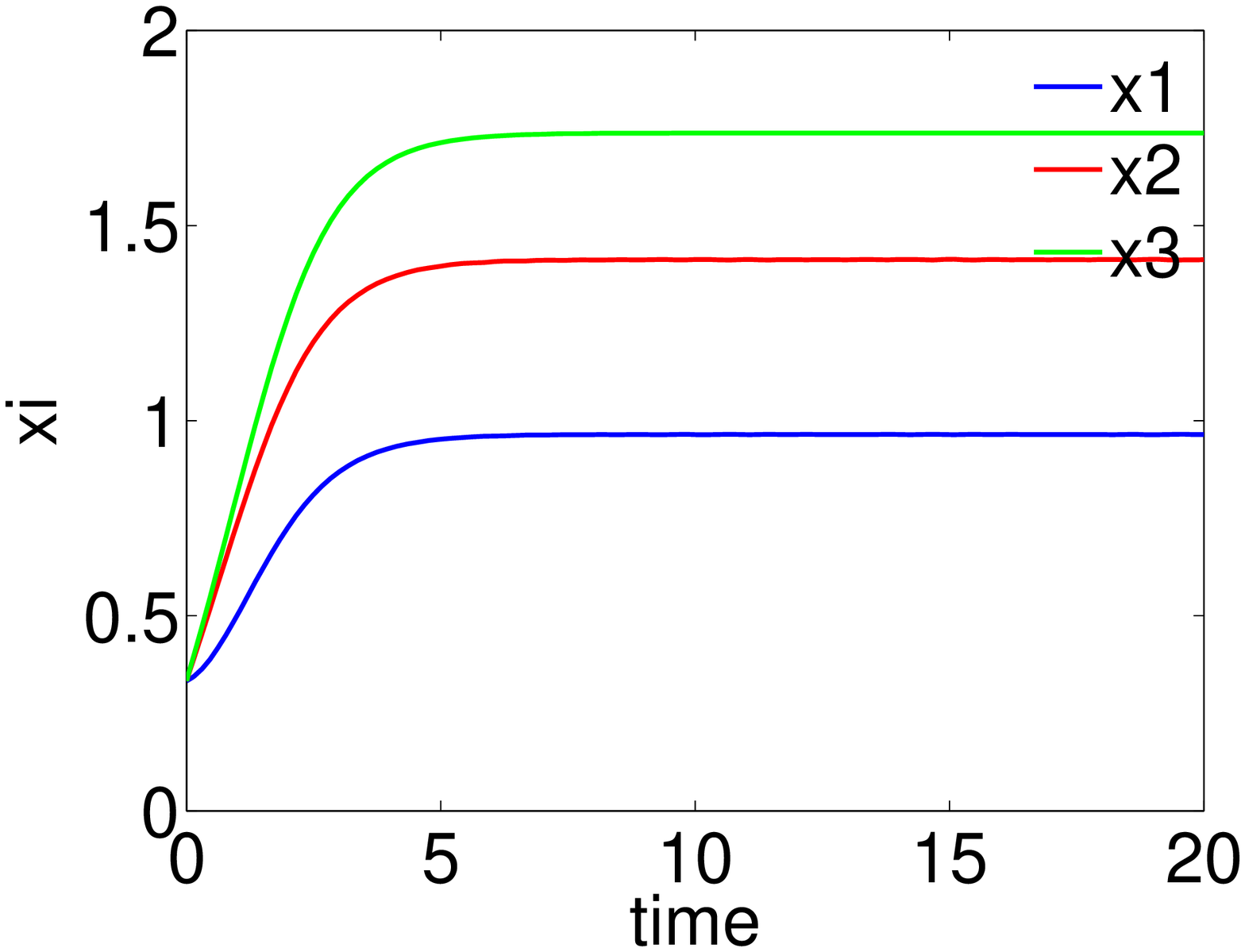}}}
\end{minipage}

\begin{minipage}{.2\textwidth}
\scalebox{0.6}{\begin{pspicture}(0,-1)(3,4) 
  \psset{nodesep=0.5pt,linewidth=1.5pt,arrowsize=5pt 2} 
  \rput(0,3){\Large (13)}
  \cnodeput(0,0){2}{\Large 2}
  \cnodeput(1.5,3){1}{\Large 1} 
  \cnodeput(3,0){3}{\Large 3}
  \ncarc[linewidth=1pt,arcangle=10]{->}{1}{3}
  \ncarc[linewidth=1pt,arcangle=10]{->}{3}{1}
  \ncarc[linewidth=1pt,arcangle=10]{->}{2}{3}
  \ncarc[linewidth=1pt,arcangle=10]{->}{3}{2}
  \ncarc[linewidth=1pt,arcangle=10]{->}{1}{2}
  \ncarc[linewidth=1pt,arcangle=10]{->}{2}{1}
\end{pspicture}}
\end{minipage}
\begin{minipage}{.2\textwidth}
\begin{displaymath}
A_{13}=\left(
\begin{array}{ccc}
0 & 1 & 1 \\
1 & 0 & 1 \\
1 & 1 & 0 \\
\end{array}
\right)
\end{displaymath}
\end{minipage}
\begin{minipage}{.5\linewidth}
  \psfrag{x1}[c][][3][0]{$x_1$} 
  \psfrag{x2}[c][][3][0]{$x_2$} 
  \psfrag{x3}[c][][3][0]{$x_3$} 
  \psfrag{xi}[c][][4][0]{$x_i$} 
  \psfrag{time}[c][][4][0]{$t$}
    \centerline{\scalebox{0.3}{\includegraphics[angle=0]
        {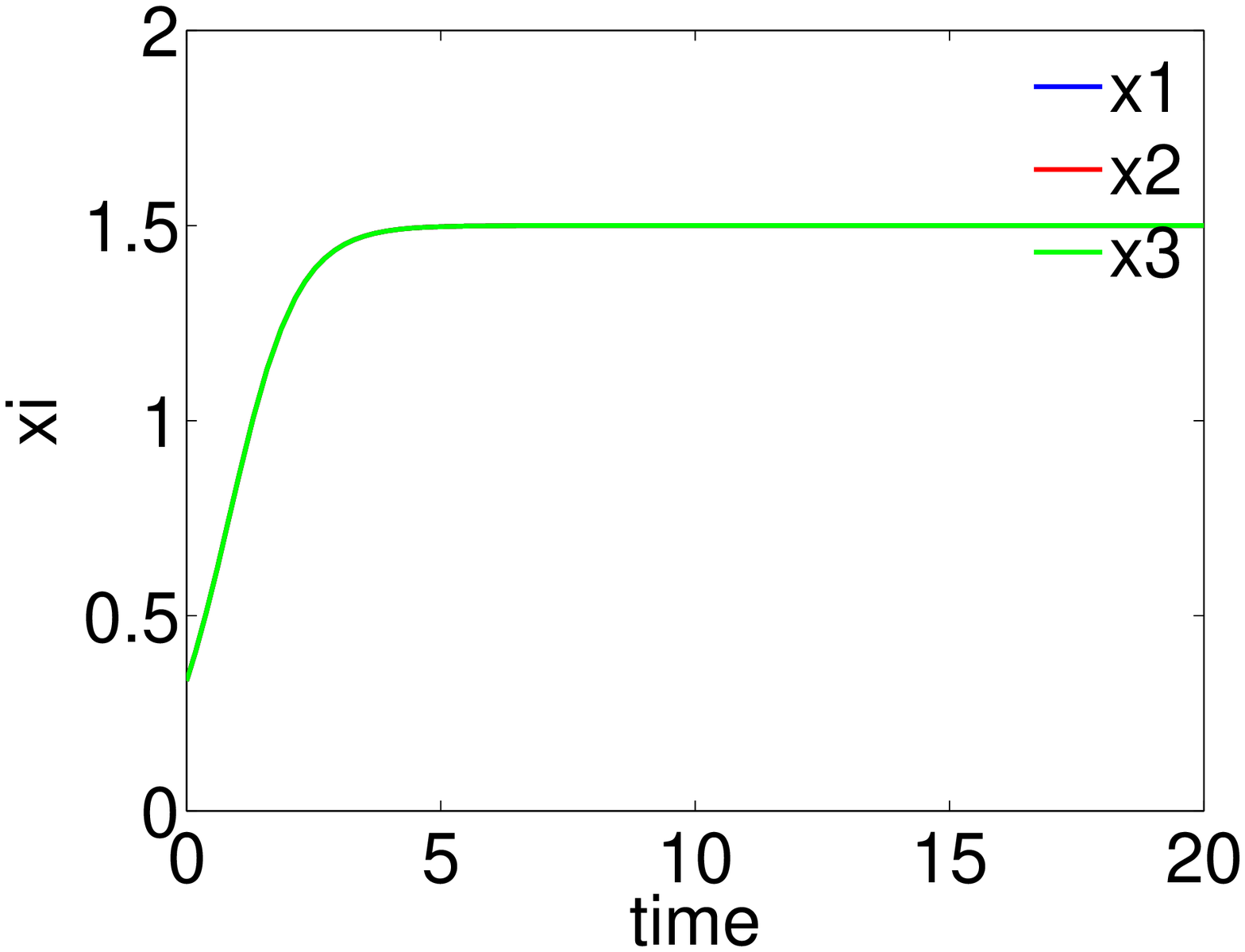}}}
\end{minipage}

The fixed points are given by
\begin{equation}
  \label{eq:fixed_point_3Graph13}
   x_1^*=x_2^*=x_3^*=\frac{2b-d}{2c}
\end{equation}

\begin{minipage}{.2\textwidth}
\scalebox{0.6}{\begin{pspicture}(0,-1)(3,4) 
  \psset{nodesep=0.5pt,linewidth=1.5pt,arrowsize=5pt 2} 
  \rput(0,3){\Large (14)}
  \cnodeput(0,0){3}{\Large 3}
  \cnodeput(1.5,3){2}{\Large 2} 
  \cnodeput(3,0){1}{\Large 1}
  \ncarc[linewidth=1pt,arcangle=0]{->}{3}{2}
\end{pspicture}}
\end{minipage}
\begin{minipage}{.2\textwidth}
\begin{displaymath}
A_{14}=\left(
\begin{array}{ccc}
0 & 0 & 0 \\
0 & 0 & 0 \\
0 & 1 & 0 \\
\end{array}
\right)
\end{displaymath}
\end{minipage}
\begin{minipage}{.5\linewidth}
  \psfrag{x1}[c][][3][0]{$x_1$} 
  \psfrag{x2}[c][][3][0]{$x_2$} 
  \psfrag{x3}[c][][3][0]{$x_3$} 
  \psfrag{xi}[c][][4][0]{$x_i$} 
  \psfrag{time}[c][][4][0]{$t$}
    \centerline{\scalebox{0.3}{\includegraphics[angle=0]
        {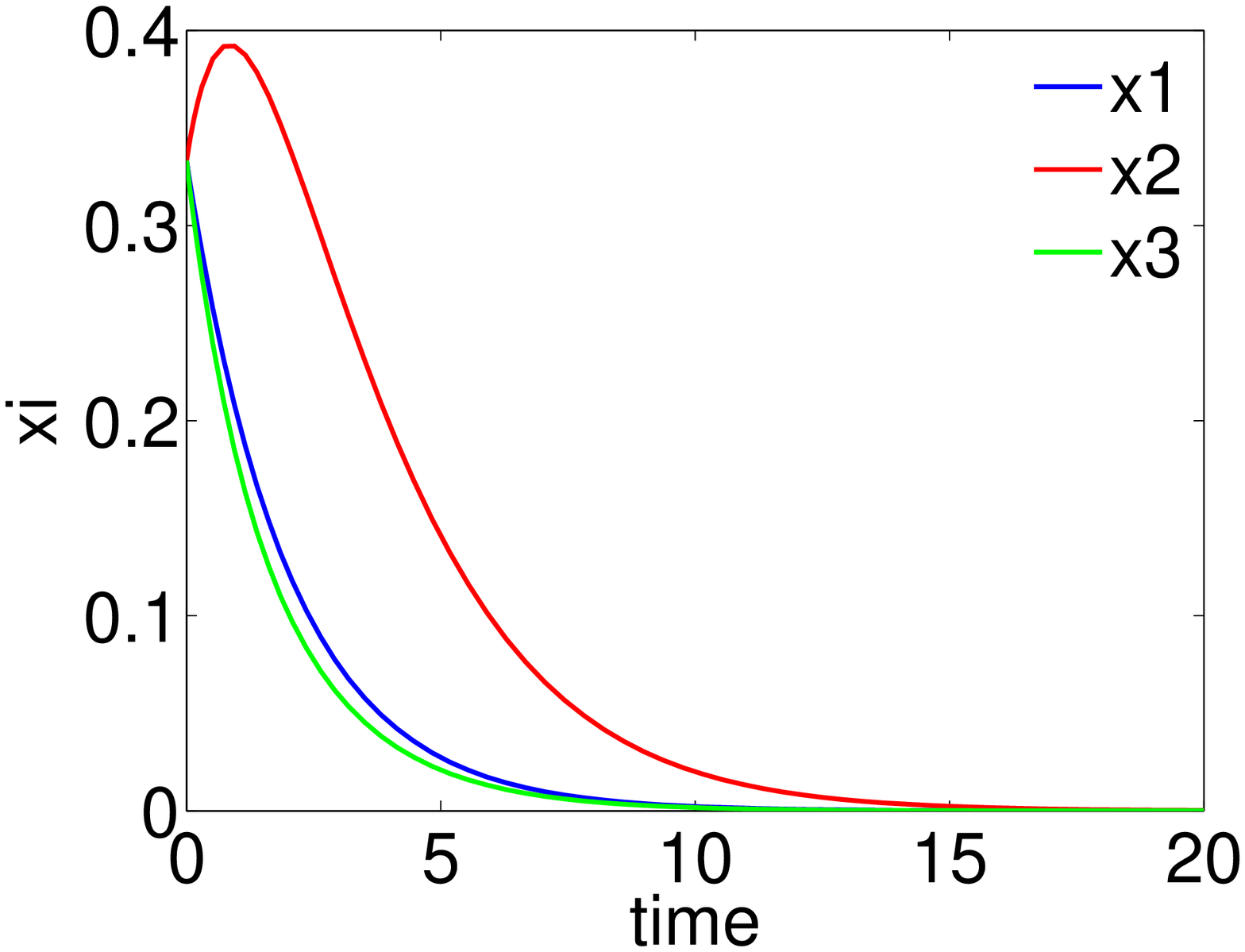}}}
\end{minipage}

The fixed points are given by
\begin{equation}
  \label{eq:fixed_point_3Graph14}
  x_i^*=0
\end{equation}

\begin{minipage}{.2\textwidth}
\scalebox{0.6}{\begin{pspicture}(0,-1)(3,4) 
  \psset{nodesep=0.5pt,linewidth=1.5pt,arrowsize=5pt 2} 
  \rput(0,3){\Large (15)}
  \cnodeput(0,0){3}{\Large 3}
  \cnodeput(1.5,3){2}{\Large 2} 
  \cnodeput(3,0){1}{\Large 1}
\end{pspicture}}
\end{minipage}
\begin{minipage}{.2\textwidth}
\begin{displaymath}
A_{15}=\left(
\begin{array}{ccc}
0 & 0 & 0 \\
0 & 0 & 0 \\
0 & 0 & 0 \\
\end{array}
\right)
\end{displaymath}
\end{minipage}
\begin{minipage}{.5\linewidth}
  \psfrag{x1}[c][][3][0]{$x_1$} 
  \psfrag{x2}[c][][3][0]{$x_2$} 
  \psfrag{x3}[c][][3][0]{$x_3$} 
  \psfrag{xi}[c][][4][0]{$x_i$} 
  \psfrag{time}[c][][4][0]{$t$}
    \centerline{\scalebox{0.3}{\includegraphics[angle=0]
        {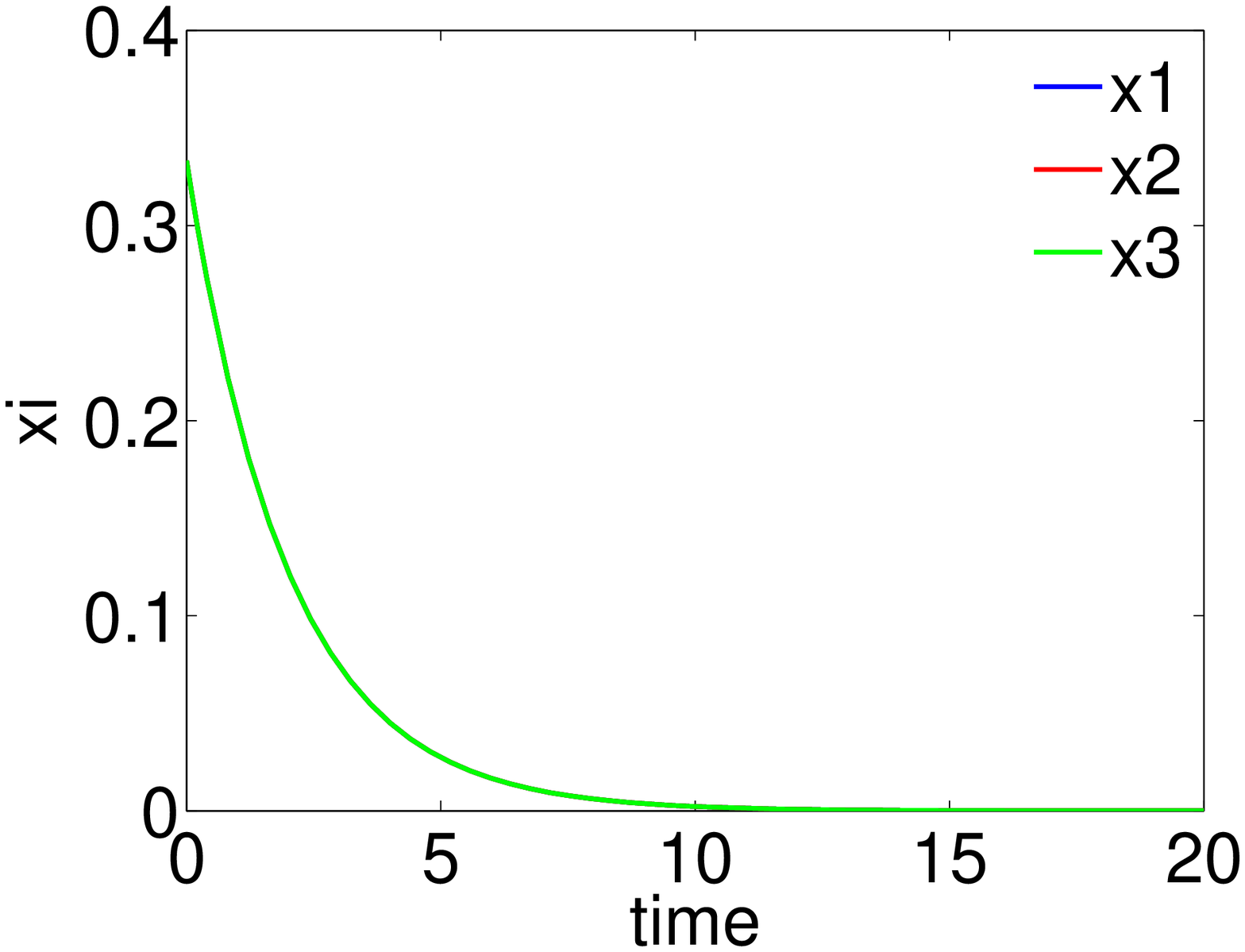}}}
\end{minipage}

The fixed points are given by
\begin{equation}
  \label{eq:fixed_point_3Graph15}
  x_i^*=0
\end{equation}

\begin{minipage}{.2\textwidth}
\scalebox{0.6}{\begin{pspicture}(0,-1)(3,4) 
  \psset{nodesep=0.5pt,linewidth=1.5pt,arrowsize=5pt 2} 
  \rput(0,3){\Large (16)}
  \cnodeput(0,0){2}{\Large 2}
  \cnodeput(1.5,3){1}{\Large 1} 
  \cnodeput(3,0){3}{\Large 3}
  \ncarc[linewidth=1pt,arcangle=10]{->}{1}{2}
  \ncarc[linewidth=1pt,arcangle=10]{->}{2}{1}
\end{pspicture}}
\end{minipage}
\begin{minipage}{.2\textwidth}
\begin{displaymath}
A_{16}=\left(
\begin{array}{ccc}
0 & 1 & 0 \\
1 & 0 & 0 \\
0 & 0 & 0 \\
\end{array}
\right)
\end{displaymath}
\end{minipage}
\begin{minipage}{.5\linewidth}
  \psfrag{x1}[c][][3][0]{$x_1$} 
  \psfrag{x2}[c][][3][0]{$x_2$} 
  \psfrag{x3}[c][][3][0]{$x_3$} 
  \psfrag{xi}[c][][4][0]{$x_i$} 
  \psfrag{time}[c][][4][0]{$t$}
    \centerline{\scalebox{0.3}{\includegraphics[angle=0]
        {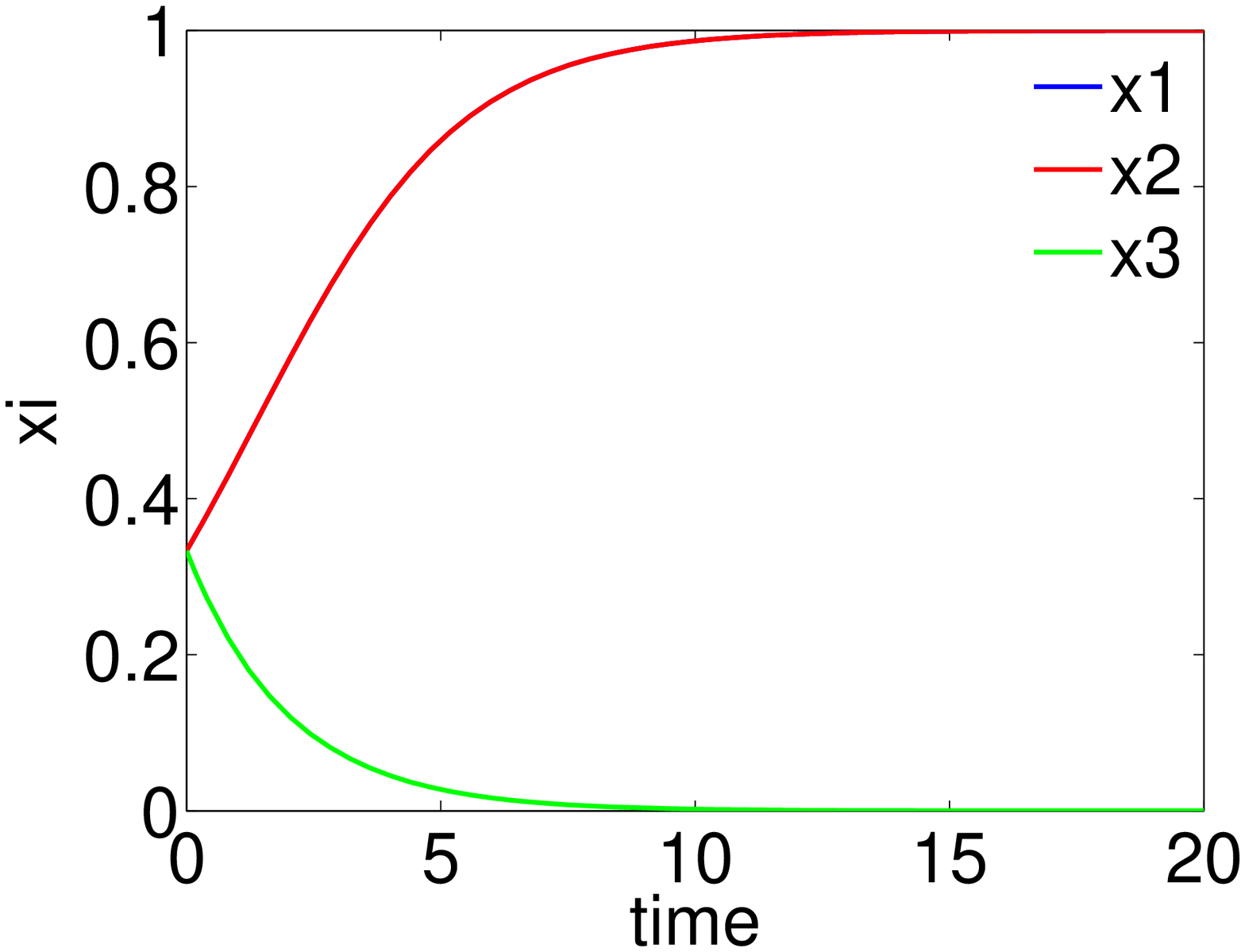}}}
\end{minipage}

The fixed points are given by
\begin{eqnarray}
  \label{eq:fixed_point_3Graph16}
  x_3^*=0\\
  x_1^*=x_2^*=\frac{b-d}{c}
\end{eqnarray}

\section{All-Trails-Single-Source Algorithm}
\label{sec:algorithm}

In this section we introduce an algorithm ALL-TRAILS-SINGLE-SOURCE that
computes all trails from a given node $s \in V$ to all other nodes in a
directed graph $G=(V,E)$.  From these trails we can extract the trails
which end in node $s$ and thus form circuits. 
In the following we will give a short description of the algorithm.

\begin{algorithm}
\caption{ALL-TRAILS-SINGLE-SOURCE}
\begin{algorithmic}
\STATE $S \leftarrow new Stack()$;
\STATE $v \leftarrow s$;
\STATE $W \leftarrow \{\};$  \COMMENT{initialization of empty
  list of trails}
\LOOP                             
 \IF{$\exists u \in N^+(v) \backslash \{s\}$ s.t. the link $e_{vu}$ cannot
   be appended to $W$ to create a new trail}
  \STATE $S.push(v)$;
  \STATE $W[u].addEdge((v,u))$;
  \STATE $v \leftarrow u$;
 \ELSIF{$S.isEmpty()$ == \FALSE}
  \STATE $v \leftarrow s.pop()$;
 \ELSE
  \STATE $break$;
 \ENDIF
\ENDLOOP
\RETURN $W$;
\label{alg:trails}
\end{algorithmic}
\end{algorithm}

Similar to a depth-first-search algorithm, links are explored out of the
most recent discovered node $v$ that still has unexplored links leaving
it. This procedure of exploring links can be represented by a search tree
$T$. The tree $T$ explored by the algorithm contains all trails starting
at the source $s$ to every node in $G$.

At every node $i$ a list $W[i]$ of trails leading from the source $s$ to
$i$ is assigned. When the next link $e_{ij}$ from $i$ to $j$ is
processed, to all trails in $W[i]$ the link $e_{ij}$ is appended (if this
is possible, meaning that no link repetition is allowed), denoted by
$W[i]+e_{ij}$. At node $j$ these trails are added, that is $W'[j]=W[j]
\cup \{W[i]+e_{ij}\}$. This procedure is continued until the algorithm
terminates. The algorithm terminates, if the are no further links
available for exploration. The progress of the algorithm on a directed
graph with $4$ nodes and $2$ circuits containing the nodes $(2,3,4,1)$
and $(2,4,1)$, respectively, is shown in Fig.  (\ref{fig:alg_progress}).

\begin{center}
  \begin{figure}
    \centering
    \begin{minipage}{.45\textwidth}
      \scalebox{0.6}[0.6]{
        \begin{pspicture}(7,7) 
          \psset{nodesep=0.5pt, linewidth=1.5pt, arrowsize=10pt 2} 
          \rput(1,6){\textit{\Large step $1$}}
          \cnodeput(3,6){1}{\Large 1}
          \nput{0}{1}{$W[1]=\{\}$} 
          \cnodeput[fillstyle=solid,fillcolor=red](0,3){2}{\Large 2}
          \nput{180}{2}{$W[2]=\{\}$} 
          \cnodeput(3,0){3}{\Large 3}
          \nput{0}{3}{$W[3]=\{\}$} 
          \cnodeput(6,3){4}{\Large 4}
          \nput{45}{4}{$W[4]=\{\}$}
          \ncarc[linewidth=1pt,arcangle=0]{->}{1}{2}
          \ncarc[linewidth=1pt,arcangle=0]{->}{2}{3}
          \ncarc[linewidth=1pt,arcangle=0]{->}{3}{4}
          \ncarc[linewidth=1pt,arcangle=0]{->}{4}{1}
          \ncarc[linewidth=1pt,arcangle=0]{->}{1}{2}
          \ncarc[linewidth=1pt,arcangle=0]{->}{2}{4}
        \end{pspicture}}
    \end{minipage}
    \begin{minipage}{.45\textwidth}
      \scalebox{0.6}[0.6]{
        \begin{pspicture}(7,7) 
          \psset{nodesep=0.5pt, linewidth=1.5pt, arrowsize=10pt 2} 
          \rput(1,6){\textit{\Large step $2$}}
          \cnodeput(3,6){1}{\Large 1}
          \nput{0}{1}{$W[1]=\{\}$} 
          \cnodeput(0,3){2}{\Large 2}
          \nput{225}{2}{$W[2]=\{\}$} 
          \cnodeput[fillstyle=solid,fillcolor=red](3,0){3}{\Large 3}
          \nput{0}{3}{$W[3]=\{(e_{23})\}$} 
          \cnodeput(6,3){4}{\Large 4}
          \nput{0}{4}{$W[4]=\{\}$}
          \ncarc[linewidth=1pt,arcangle=0]{->}{1}{2}
          \ncarc[linewidth=1pt,arcangle=0]{->}{2}{3}
          \ncarc[linewidth=1pt,arcangle=0]{->}{3}{4}
          \ncarc[linewidth=1pt,arcangle=0]{->}{4}{1}
          \ncarc[linewidth=1pt,arcangle=0]{->}{1}{2}
          \ncarc[linewidth=1pt,arcangle=0]{->}{2}{4}
        \end{pspicture}}
    \end{minipage}
    \begin{minipage}{.45\textwidth}
      \scalebox{0.6}[0.6]{
        \begin{pspicture}(7,7) 
          \psset{nodesep=0.5pt, linewidth=1.5pt, arrowsize=10pt 2} 
          \rput(1,6){\textit{\Large step $3$}}
          \cnodeput(3,6){1}{\Large 1}
          \nput{0}{1}{$W[1]=\{\}$} 
          \cnodeput(0,3){2}{\Large 2}
          \nput{180}{2}{$W[2]=\{\}$} 
          \cnodeput(3,0){3}{\Large 3}
          \nput{0}{3}{$W[3]=\{(e_{23})\}$} 
          \cnodeput[fillstyle=solid,fillcolor=red](6,3){4}{\Large 4}
          \nput{45}{4}{$W[4]=\{(e_{23},e_{34})\}$}
          \ncarc[linewidth=1pt,arcangle=0]{->}{1}{2}
          \ncarc[linewidth=1pt,arcangle=0]{->}{2}{3}
          \ncarc[linewidth=1pt,arcangle=0]{->}{3}{4}
          \ncarc[linewidth=1pt,arcangle=0]{->}{4}{1}
          \ncarc[linewidth=1pt,arcangle=0]{->}{1}{2}
          \ncarc[linewidth=1pt,arcangle=0]{->}{2}{4}
        \end{pspicture}}
    \end{minipage}
    \begin{minipage}{.45\textwidth}
      \scalebox{0.6}[0.6]{
        \begin{pspicture}(7,7) 
          \psset{nodesep=0.5pt, linewidth=1.5pt, arrowsize=10pt 2} 
          \rput(1,6){\textit{\Large step $4$}}
          \cnodeput[fillstyle=solid,fillcolor=red](3,6){1}{\Large 1}
          \nput{0}{1}{$W[1]=\{(e_{23},e_{34},e_{41})\}$} 
          \cnodeput(0,3){2}{\Large 2}
          \nput{200}{2}{$W[2]=\{\}$} 
          \cnodeput(3,0){3}{\Large 3}
          \nput{0}{3}{$W[3]=\{(e_{23})\}$} 
          \cnodeput(6,3){4}{\Large 4}
          \nput{0}{4}{$W[4]=\{(e_{23},e_{34})\}$}
          \ncarc[linewidth=1pt,arcangle=0]{->}{1}{2}
          \ncarc[linewidth=1pt,arcangle=0]{->}{2}{3}
          \ncarc[linewidth=1pt,arcangle=0]{->}{3}{4}
          \ncarc[linewidth=1pt,arcangle=0]{->}{4}{1}
          \ncarc[linewidth=1pt,arcangle=0]{->}{1}{2}
          \ncarc[linewidth=1pt,arcangle=0]{->}{2}{4}
        \end{pspicture}}
    \end{minipage}
    \begin{minipage}{.45\textwidth}
      \scalebox{0.6}[0.6]{
        \begin{pspicture}(0,-1)(7,7) 
          \psset{nodesep=0.5pt, linewidth=1.5pt, arrowsize=10pt 2} 
          \rput(1,6){\textit{\Large step $5$}}
          \cnodeput(3,6){1}{\Large 1}
          \nput{0}{1}{$W[1]=\{(e_{23},e_{34},e_{41})\}$} 
          \cnodeput[fillstyle=solid,fillcolor=red](0,3){2}{\Large 2}
          \nput{200}{2}{\begin{tabular}{c}
            $W[2]=$\\
            $\{(e_{23},e_{34},e_{41},e_{12})\}$\\
          \end{tabular}} 
          \cnodeput(3,0){3}{\Large 3}
          \nput{0}{3}{$W[3]=\{(e_{23})\}$} 
          \cnodeput(6,3){4}{\Large 4}
          \nput{45}{4}{$W[4]=\{(e_{2,3},e_{3,4})\}$}
          \ncarc[linewidth=1pt,arcangle=0]{->}{1}{2}
          \ncarc[linewidth=1pt,arcangle=0]{->}{2}{3}
          \ncarc[linewidth=1pt,arcangle=0]{->}{3}{4}
          \ncarc[linewidth=1pt,arcangle=0]{->}{4}{1}
          \ncarc[linewidth=1pt,arcangle=0]{->}{1}{2}
          \ncarc[linewidth=1pt,arcangle=0]{->}{2}{4}
        \end{pspicture}}
    \end{minipage}
    \begin{minipage}{.45\textwidth}
      \scalebox{0.6}[0.6]{
        \begin{pspicture}(0,-1)(7,7) 
          \psset{nodesep=0.5pt, linewidth=1.5pt, arrowsize=10pt 2} 
          \rput(1,6){\textit{\Large step $6$}}
          \cnodeput(3,6){1}{\Large 1}
          \nput{0}{1}{$W[1]=\{(e_{23},e_{34},e_{41})\}$} 
          \cnodeput(0,3){2}{\Large 2}
          \nput{230}{2}{\begin{tabular}{c}
            $W[2]=$\\
            $\{(e_{23},e_{34},e_{41},e_{12})\}$\\
          \end{tabular}} 
          \cnodeput(3,0){3}{\Large 3}
          \nput{0}{3}{$W[3]=\{(e_{23})\}$} 
          \cnodeput[fillstyle=solid,fillcolor=red](6,3){4}{\Large 4}
          \nput{0}{4}{\begin{tabular}{c}
              $W[4]=\{(e_{24}),(e_{23},e_{34})$,\\
              $(e_{23},e_{34},e_{41},e_{12},e_{24})\}$ \\
            \end{tabular}}
          \ncarc[linewidth=1pt,arcangle=0]{->}{1}{2}
          \ncarc[linewidth=1pt,arcangle=0]{->}{2}{3}
          \ncarc[linewidth=1pt,arcangle=0]{->}{3}{4}
          \ncarc[linewidth=1pt,arcangle=0]{->}{4}{1}
          \ncarc[linewidth=1pt,arcangle=0]{->}{1}{2}
          \ncarc[linewidth=1pt,arcangle=0]{->}{2}{4}
        \end{pspicture}}
    \end{minipage}
    \caption{The progress of the algorithm ALL-TRAILS-SINGLE-SOURCE on a
      directed graph with node $2$ as the source node. The node indicated
      in red is visited in the succeeding steps. After step $6$ no
      further trails are added to the list of trails.}
    \label{fig:alg_progress}
  \end{figure}
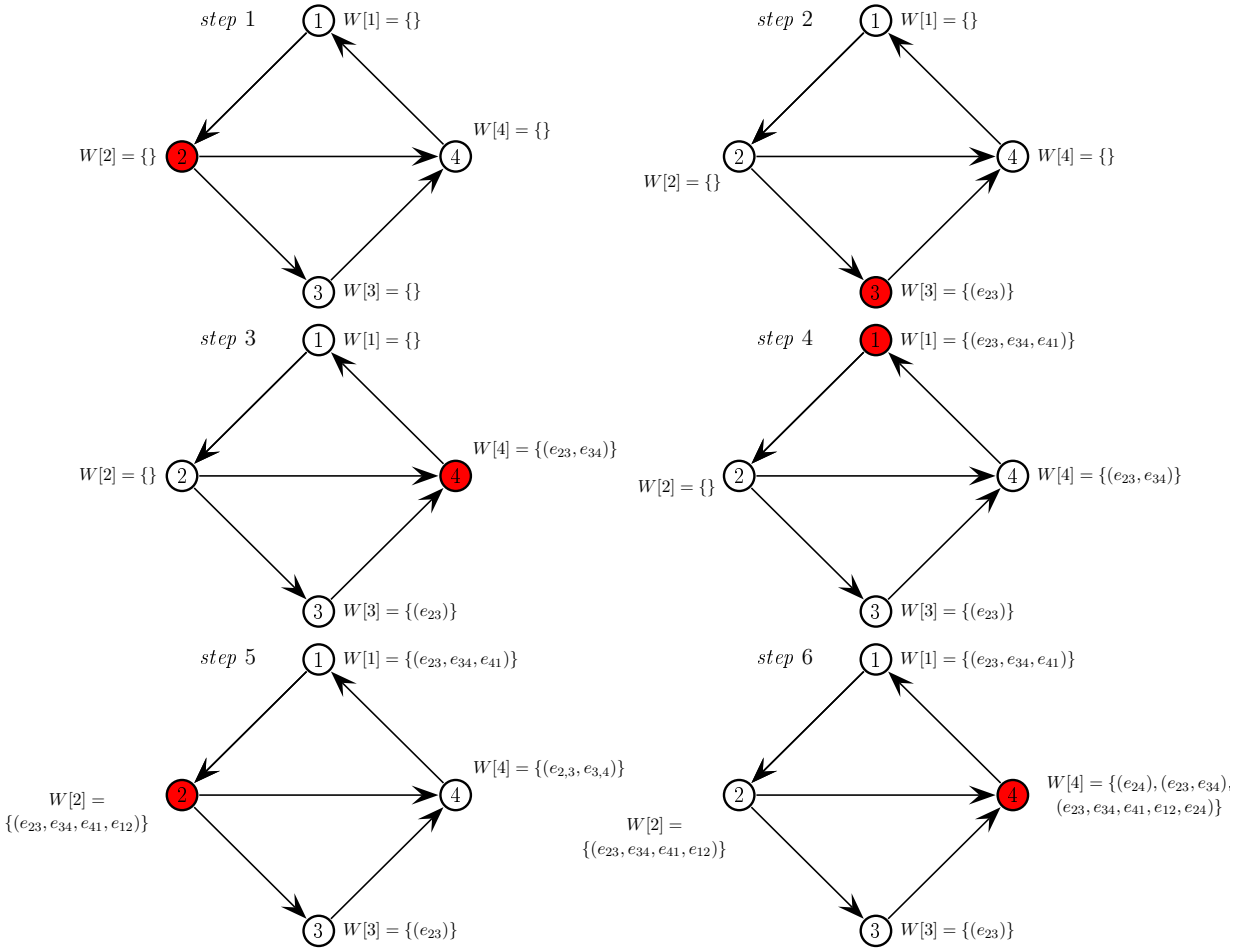
\end{center}

\section{Simulation Parameters}
\label{sec:parameters}

The parameters shown in table (\ref{tab:parameters}) have been used for
the simulation runs presented in section (\ref{sec:simulation_studies}).

\begin{table*}
  \centering
  \begin{tabular}{|l|c|c|}
    \hline
    \textsc{Description} & \textsc{Variable} & \textsc{Value} \\
    \hline
    \hline
    initial link creation probability & $p$ & $0.1$ \\
    \hline
    initial value of knowledge & $\mathbf{x}(0)$ & $1.0$ \\
    \hline
    number of agents (without externality) & $n$ & $30$ \\
    \hline
    number of agents (with externality) & $n$ & $20$ \\
    \hline
    max. numerical integration time (time horizon) & $T$ & $100$ \\
    \hline
    numerical integration time step & $\Delta t$ & $0.05$ \\
    \hline
    max. number of network updates & $N$ & $[100,5000]$ \\
    \hline
    benefit & $b$ & $0.5$ \\
    \hline
    decay & $d$ & $0.5$ \\
    \hline
    cost & $c$ & $0.1$ \\
    \hline
  \end{tabular}
  \caption{Simulation parameters.}
  \label{tab:parameters}
\end{table*}

\end{appendix}

\section*{Acknowledgements}
\label{sec:acknowledgements}

We are much obliged to Koen Frenken who helped a lot to improve the model
presented in this chapter. Moreover we would like to thank Giorgio
Fagiolo for the clarifying discussions. In writing the chapter we have
always been conscious of our debt to colleagues who have helped us in
bringing this work to the present state.  Among these are in particular
Kerstin Press and Mauro Napoletano who have pointed us to some
inconsistencies and sections which needed a more extended explanation of
the material treated there in the early versions of this chapter. The
discussion with Dimo Brockhoff on the algorithms were of great help. We
would like to thank the discussant and the critical audience at the 1st
International Conference on Economic Sciences with Heterogeneous
Interacting Agents in Bologna, 2006.  Finally, we would like to thank
Peter Howitt, Robert Axtell, Herbert Gintis and Leigh Tesfatsion for
taking so much time in discussing the model at the 7th Trento Summer
School on Agent-Based Computational Economics in Trento, 2006 and Axel
Leijonhufvud for making these discussions possible.

\bibliography{BookChapter}
\end{document}